\documentstyle[preprint,aps,eqsecnum]{revtex}

\def \ovr{\over}
\def \en{\end{eqnarray}}
\def \bg{\begin{eqnarray}}
\def \enm{\end{mathletters}}
\def \bgm{\begin{mathletters}}

\def \Orc2{{1\ovr c^2}}

\def \Oc2{O(1/c^2)}

\begin{document}
\title{The Rest-Frame Darwin Potential from the Lienard-Wiechert Solution in the
Radiation Gauge}
\author{Horace Crater}
\address{The University of Tennessee Space Institute\\
Tullahoma, Tennessee\\
37388 USA\\
e-mail HCrater@utsi.edu}
\author{and}
\author{Luca Lusanna}
\address{Sezione INFN di Firenze\\
Largo E.Fermi 2,\\
50124 Firenze, Italy\\
e-mail: LUSANNA@FI.INFN.IT}
\maketitle

\begin{abstract}
In the semiclassical approximation in which the electric charges of scalar
particles are described by Grassmann variables ($Q_i^2=0,\ Q_iQ_j\ne 0$), it
is possible to re-express the Lienard-Wiechert potentials and electric
fields in the radiation gauge as phase space functions, because the
difference among retarded, advanced, and symmetric Green functions is of
order $Q_i^2$. By working in the rest-frame instant form of dynamics, the
elimination of the electromagnetic degrees of freedom by means of suitable
second classs contraints leads to the identification of the Lienard-Wiechert
reduced phase space containing only $N $ charged particles with mutual
action-at-a-distance vector and scalar potentials. A Darboux canonical basis
of the reduced phase space is found. This allows one to re-express the
potentials for arbitrary $N$ as a unique effective scalar potential
containing the Coulomb potential and the complete Darwin one, whose $1/c^2$
component agrees for with the known expression. The effective potential
gives the classical analogue of all static and non-static effects of the
one-photon exchange Feynman diagram of scalar electrodynamics.

\vskip 1truecm \noindent \today

\vskip 1truecm
\end{abstract}


\vfill\eject

\section{\protect\bigskip Introduction}

\bigskip

Recently the rest-frame Wigner-covariant instant form of dynamics has been
developed in Ref.\cite{lu1} for isolated systems in Minkowski spacetime $%
M^{4}$ starting from the case of $N$ scalar charged particles plus the
electromagnetic field. The charges of the particles are described by
bilinears in Grassmann variables following the scheme (called
pseudo-classical mechanics) which uses a semiclassical approximation to
quantum operators with a finite discrete spectrum like spin\cite{lu2,bar},
that otherwise would have no strict classical limit; Grassmann variables
give fermionic oscillators after quantization. The extension of this scheme
to the electric charge is based on the experimental fact that all measurable
charges are multiples of $\pm e$, the electron and positron charges.
Therefore, even if it is not clear whether the electric charge has to be
considered as a quantum operator in the standard sense (except in the case
of the existence of magnetic charges; in this case there is the Dirac
quantization rule for the product of the electric and magnetic charges), one
can consider it as a two-level system [which becomes a six-level system ($%
\pm e$, $\pm {\frac{1}{3}} e$, $\pm {\frac{2}{3}} e$) at the quark-lepton
level] described by an operator with quantum $e$ instead of $\hbar$. Then
one can define a semiclassical approximation with Grassmann variables like
in the case of spin. As shown in Ref.\cite{lu1}, this semiclassical
approximation automatically implies the regularization of the Coulomb
self-energies (the $i\not= j$ rule). Therefore, this semiclassical
approximation may be considered as an alternative to the extended electron
models, which were introduced for regularization aims.

The idea leading to the rest-frame instant form is to consider an arbitrary
3+1 splitting of Minkowski spacetime by means of a foliation with spacelike
hypersurfaces $\Sigma (\tau )$ diffeomorphic to $R^{3}$. The parameter $\tau
$ labelling the leaves is used as a Lorentz scalar mathematical time
parameter. For each $\tau $ the leaf $\Sigma (\tau )$ is defined through the
embedding $R^{3}\mapsto \Sigma (\tau )\subset M^{4}$, $(\tau ,\vec{\sigma}%
)\mapsto z^{\mu }(\tau ,\vec{\sigma})$, where $\vec{\sigma}$ are curvilinear
coordinates on $R^{3}$. Then one considers the Lagrangian describing the
coupling of the given isolated system to an external gravitational field and
replaces the 4-metric with the induced metric on $\Sigma (\tau )$, which is
a functional of $z^{\mu }(\tau ,\vec{\sigma})$. In this way one gets the
Lagrangian for the description of the isolated system on arbitrary spacelike
hypersurfaces (i.e. in arbitrary accelerated reference frames in Minkowski
spacetime) with the embedding functions $z^{\mu }(\tau ,\vec{\sigma})$ as
extra configuration variables describing the hypersurface. However, there
are four first class constraints at each point implying the independence of
the description from the chosen 3+1 splitting. Thus the $z^{\mu }(\tau ,\vec{%
\sigma})$'s are gauge variables. Therefore, one can restrict the description
of the isolated system to spacelike hyperplanes $\Sigma _{H}(\tau )$, $%
z^{\mu }(\tau ,\vec{\sigma})=x_{s}^{\mu }(\tau )+b_{\check{r}}^{\mu }(\tau
)\sigma ^{\check{r}}$ (inertial reference frames in Minkowski spacetime; $%
x_{s}^{\mu }(\tau )$ is an arbitrary origin).

Then, if one selects all the configurations of the isolated system with
total timelike 4-momentum (they are dense in the space of all
configurations), one finds that each timelike configuration identifies a
privileged family of hyperplanes: those orthogonal to its total 4-momentum
[Wigner hyperplanes $\Sigma _{W}(\tau )$]. At this stage one has obtained
the analogue of the nonrelativistic center-of-mass separation and the
definition of a new instant form of dynamics \cite{dirac}, the rest-frame
one \cite{lu1}. There is a decoupled point ${\tilde{x}}_{s}^{\mu }(\tau )$
on each Wigner hyperplane describing the ``external'' center of mass of the
isolated system (thus serving as a decoupled point particle clock) with
conjugate momentum $p_{s}^{\mu }.$ $\ {\tilde{x}}_{s}^{\mu }(\tau )$ is a
canonical variable, but it is not covariant like the Newton-Wigner position
operator: it has only covariance under the little group of timelike
Poincar\'{e} orbits.

After the restriction to the Wigner hyperplane only four first class
constraints are left:

i) three of them say that the total 3-momentum of the isolated system
vanishes (rest-frame condition): the natural gauge fixing for these
constraints is the requirement that the ``internal'' 3-center of mass of the
isolated system inside the Wigner hyperplanes coincides with the origin $%
x_{s}^{\mu }(\tau )$ of the coordinates $\vec{\sigma}$ in it (see Refs.\cite
{mate,alp} , where the group-theoretical results of Ref.\cite{pauri} are
used). In this way only ``internal" relative variables describe the isolated
system: they are either Lorentz scalars or Wigner spin 1 3-vectors.

ii) the fourth one identifies the invariant mass of the isolated system as
the Hamiltonian of the evolution in $\tau $ when, with a gauge fixing, $\tau
$ is made to coincide with the Lorentz scalar rest-frame time $%
T_{s}=p_{s}\cdot {\tilde{x}}_{s}/\sqrt{p_{s}^{2}}=p_{s}\cdot x_{s}/\sqrt{%
p_{s}^{2}}$ of the decoupled external center of mass.

In this description the standard manifestly covariant fields like the
Klein-Gordon field $\tilde{\phi}(z^{\mu })$ are replaced by the new fields $%
\phi (\tau ,\vec{\sigma})=\tilde{\phi}(z^{\mu }(\tau ,\vec{\sigma}))$, which
know the embedding and have the non-local information about the equal time
hypersurfaces $\Sigma (\tau )$ built-in. In the case of gauge theories, one
can make a canonical reduction to a canonical basis of Dirac's observables
in the radiation gauge (or Coulomb or generalized Coulomb; the literature is
ambiguous about the terminology to be used), with the only universal
breaking of manifest covariance connected with the external center of mass,
since the relative motions are Wigner covariant.

See Ref.\cite{india} for a complete review of the research program aiming to
give a unified description of the four interactions in terms of
Dirac-Bergmann's observables in the framework of the rest-frame instant form
of dynamics, which is the classical background of the Tomonaga-Schwinger
formulation of quantum field theory.

The description of scalar (or spinning \cite{lu3}) particles on arbitrary
spacelike hypersurfaces requires the choice of the sign of the energy of the
particle. \ This happens because the position of the particle on $\Sigma
(\tau )$ is identified by 3 numbers $\vec{\sigma}=\vec{\eta}(\tau )$ and not
by 4: this implies that the mass-shell first class constraints of the
standard manifestly covariant approach have been solved and that one of the
two disjoint branches of the mass spectrum has been chosen. In this way, one
gets a different Lagrangian for each branch of the mass spectrum of an
isolated system of particles (in the standard manifestly covariant
description the branches are topologically disjoint for free particles):
there is no possibility of crossing of the branches (the classical
background of pair production) when interactions are present inside the
isolated system (for instance charged particles plus the electromagnetic
field) as happens in the manifestly covariant approach. While there is no
problem in the coupling to magnetic fields of these particles with a
definite sign of the energy, the minimal coupling in the Lagrangian will
miss those couplings to the electric fields which are at the basis of the
non-diagonalizability of both the Feshbach-Villars description of the
Klein-Gordon field and of the Dirac equation through the Foldy-Wouthuysen
transformation (in the case of spinning particles these couplings will have
to be extracted from the iterative diagonalization of these theories and
added non-minimally). However, this description of particles with, say,
positive energy (whose quantization requires pseudodifferential operators
\cite{lamme}) seems suited for the description of the asymptotic
Tomonaga-Schwinger states and will be used to introduce a notion of particle
in a future quantization of classical fields on Wigner hyperplanes in the
rest-frame instant form. These asymptotic states will replace the Fock ones
of the standard manifestly covariant theory, which are the main source of
problems in the theory of relativistic quantum bound states (the spurious
solutions of the Bethe-Salpeter equation; see Ref.\cite{india,karui}). This
framework should allow the introduction of bound states among the asymptotic
states.

Coming back to the isolated system formed by $N$ scalar charged particles of
positive energy plus the electromagnetic field, we recall that in Ref.\cite
{lu1}, after canonical reduction to the radiation gauge, the final invariant
mass of the reduced system is a function only of Dirac's observables (gauge
invariant particles dressed with a Coulomb cloud and transverse radiation
field) and contains:

i) the kinetic energy of the radiation field;

ii) the kinetic energy for the particles with minimal coupling to the
radiation field;

iii) the instantaneous action-at-a-distance Coulomb potential among the
charges with the Coulomb self-energies regularized (the $i\not=j$ rule, $%
Q_{i}Q_{j}\not=0$) due to the Grassmann character of the electric charges $%
Q_{i}$ , i.e. $Q_{i}^{2}=0$ [at this semiclassical level we have $%
Q_{i}=e\theta _{i}^{\ast }\theta _{i}$; this does not imply the vanishing of
the fine structure constant $\alpha =e^{2}/4\pi \approx 1/137$, since it
gets contributions from $Q_{i}Q_{j}$; therefore, we are retaining effects of
order $\alpha $ but not of higher order, because, as it will be shown, we do
not have many-body forces].

Then, in Ref.\cite{lu4} there was the evaluation of the retarded
Lienard-Wiechert potentials of the charged particles in the radiation gauge
in the rest-frame instant form. Since the Lienard-Wiechert potentials and
fields are linearly dependent on the charges $Q_i$, the semiclassical
regularization $Q_{i}^{2}=0 $ eliminates the radiation coming from a single
particle (the electromagnetic energy-momentum tensor contains only terms in $%
Q_iQ_j$, $i\not= j$) and the causal problems of the Abraham-Lorentz-Dirac
equation of each charged particle [since the radiation reaction term has the
coefficient $\tau _{o}=2Q^{2}/3mc^{2}$ ($\tau _{o}$ is proportional to the
time needed for light to travel across a classical electron radius), which
vanishes if $Q^{2}=0$, there are neither the Schott term nor the Larmor one
but only the Lorentz forces produced by the other particles]. In Refs \cite
{ror} one may find an extended discussion about this equation and in Ref.
\cite{poisson} a recent review on its derivation and its causal problems
(preacceleration, runaway solutions). See Ref.\cite{agui} for modern
attempts to extract the subset of causal solutions of this equation with the
requirement of selecting only its solutions which admit a smooth limit for $%
\tau _{o}\rightarrow 0$ (the runaway solutions are singular in this limit)
and to find an effective second order equation for this subset of solutions.

Even if at the semiclassical level the ``single charged particle'' has no
acausal behaviour, because, notwithstanding it produces a Grassmann-valued
vector potential, it does not irradiate, we can recover the asymptotic
Larmor formula for a system of charges (considered as external sources of
the electromagnetic field) due to the interference radiation from $Q_{i}Q_{j}
$, $i\not=j$, terms (this result is in accord with macroscopic experimental
facts). See Ref.\cite{lu3} for the extension to spinning particles.

Being in the radiation gauge, at each $\tau $ the retarded Lienard-Wiechert
potentials evaluated in Ref.\cite{lu4} contain also a non-local (in $\vec %
\sigma$)term (coming from the transverse projector; see Eqs.(\ref{V13}), (%
\ref{V22}) in Section V): since this term involves all the points of $\Sigma
(\tau )$, the Lienard-Wiechert potential receives contributions from ``all''
the retarded times before $\tau $, namely from the whole past history of the
particles. In the absence of incoming radiation one could put these retarded
Lienard-Wiechert potentials inside the Lagrangian given in Eq.(74) of Ref.
\cite{lu4} for the description of the isolated system in the rest-frame
radiation gauge on the Wigner hyperplanes: one would get a Fokker-like
action in the radiation gauge, replacing the standard Fokker-Tetrode one in
the Lorentz gauge of the manifestly covariant description, and would have to
face the problem of how to find a Hamiltonian description when there are
integro-differential equations of motion with delay. The existing attempts
are based on the idea of replacing retarded particle coordinates and
velocities with instantaneous coordinates and accelerations of all the
orders (see for instance Refs.\cite{tet1,tet2,tet3,tet4,tet5}). In this way
one replaces integro-defferential equations of motion with an infinite set
of coupled differential ones. Since it is unknown how to formulate the
Cauchy problem for these integro-differential equations (see Ref.\cite{bel};
exceptions are the 1-dimensional case \cite{driver} or the time-asymmetric
case \cite{staru,staru1}), there have been complicated attempts to find
conditions for extracting a set of effective second order differential
equations from the infinite set (see for instance Refs.\cite{bel,tet5}). In
Ref.\cite{tet6} there was an attempt to study the Dirac constraints
originating from actions depending on accelerations of all orders, following
previous attempt of Kerner \cite{kerner} of defining a Hamiltonian approach.
See also the recent approach of Ref.\cite{nik}.

Moreover, one should face a problem similar to the one raised in Ref.\cite
{ander}, that only with symmetric Green's functions like ${\frac{1}{2}}%
(retarded+advanced)$ can the Fokker-Tetrode action corresponding to the
Lorentz gauge give rise to a variational principle whose extremals are
equivalent to the subspace of extremals of the original action defined by
the symmetric Lienard-Wiechert solutions without incoming radiation, i.e.
the adjunct Lienard-Wiechert fields (otherwise there are problems with the
boundary terms). This is compatible with the Feynman-Wheeler \cite{fey}
starting point for their theory of the absorbers (see Ref.\cite{leiter} for
the definition of radiation in this theory). A noncovariant justification of
the results of Ref.\cite{ander} is given in Ref.\cite{tet4}; by ignoring the
self-interactions and assuming a Lagrangian for two charged particles at
equal times in which each particle interacts with the retarded
Lienard-Wiechert potential of the other one, one obtains in the equations of
motion Lorentz forces which correspond to ${\frac{1}{2}}(retarded+advanced)$
interactions, because in this Lagrangian the transition from $retarded$ to ${%
\frac{1}{2}}(retarded+advanced)$ interactions is a total time derivative.
However, self-reaction is ignored in these calculations and it is not clear
how to arrive at a covariant formulation of these results. Let us remark
that from the point of view of quantum field theory its regularization and
renormalization require the use of the complex Feynman Green function (which
does not vanish outside the lightcone; the solutions with the retarded Green
function cannot be regularized at the distributional level): while its
imaginary part is connected to absorption in other channels, its real part
is just the ${\frac{1}{2}}(retarded+advanced)$ Green function like in
Feynman-Wheeler [ see Ref.\cite{halpern} for the extraction of a
Fokker-Tetrode action with ${\frac{1}{2}}(retarded+advanced)$ kernel from
the particle limit of QED (it does not work in QCD)].

However, both in the Dirac derivation \cite{di1} of the
Abraham-Lorentz-Dirac equation through the evaluation of the near zone
self-field (with the same results obtainable with balance equations using
the far zone fields; see Refs.\cite{burko,poisson}) and in the
Feynman-Wheeler approach with the assumption of complete absorbers \cite{ror}
the radiation is determined by the radiative Green function ${\frac{1}{2}}
(retarded-advanced)$ [at the conceptual level this introduces the acausal
advanced Green function and interpretational problems]. Indeed, the
regularization of the self-energy divergence due to radiation reaction is
done by rewriting $retarded={\frac{1}{2}}(retarded+advanced)+{\frac{1}{2}}%
(retarded-advanced)$, by noting that the non radiative Coulomb piece of the
fields (which does not influence the motion of the particle only giving a
divergent electromagnetic contribution to the mass) is in ${\frac{1}{2}}
(retarded+advanced)$ and by discarding this term as a regularization.

However, till now all the calculations have been done in the Lorentz gauge
and it is not clear whether the previous statements are ``gauge invariant".

Since we now have the results of Refs.\cite{lu1,lu4} in the radiation gauge
and a new type of regularization with the semiclassical approximation, it is
interesting to revisit all these problems.

The aim of this paper is to show that, starting from the rest-frame instant
form description of $N$ charged scalar particles plus the electromagnetic
field with Grassmann-valued electric charges $Q_{i}$, the semiclassical
regularization $Q_{i}^{2}=0$ allows one to transform the subspace of
Lienard-Wiechert solutions (without or with incoming radiation) into a
symplectic submanifold of the space of all solutions. \ This takes place
since all the higher accelerations coming from an equal time development of
the delay decouple, being of order $Q_{i}^{2}$ on the solution of the
particle equations of motion. As a consequence, at this semiclassical level
the retarded, advanced and symmetric Lienard-Wiechert solutions coincide by
using the equations of motion, so that there is only one sector of
semiclassical Lienard-Wiechert solutions (modulo the incoming radiation).
The semiclassical Lienard-Wiechert potential and fields can be expressed as
phase space functions and it is possible to eliminate the electromagnetic
degrees of freedom by means of second class constraints added to the reduced
phase space of the radiation gauge in the rest frame. Having gone to Dirac
brackets with respect to these constraints, we get a reduced phase space
containing only particles with mutual instantaneous action-at-a-distance
interactions. We can find new canonical variables for the particles
corresponding to a Darboux basis for these brackets. We can evaluate the
final Hamiltonian, showing that besides the Coulomb potential there are
vector potentials under the particle kinetic energy square roots (coming
from the minimal coupling to the radiation field) and a scalar potential
outside them (coming from the energy of the radiation field): due to $%
Q_{i}^{2}=0$ one can extract the vector potentials from under the square
roots and write a unique effective scalar potential added to the Coulomb
one. This can be done both in the original (no longer) canonical variables
and in the final canonical basis. \ The effective scalar potential is the
complete Darwin potential: at the lowest order in $1/c^{2}$ we obtain the
known form of the Darwin potential.

We find that in the framework of Maxwell (not Feynman-Wheeler) theory, the
semiclassical regularization $Q_{i}^{2}=0,Q_{i}Q_{j}\neq 0$ extracts
automatically the instantaneous action-at-a-distance potential hidden in the
delay, which as mentioned above turns out to be the same in the retarded and
advanced solutions, because the difference is in the $Q_{i}^{2}$ terms which
depend on the higher accelerations (in QFT these effects are hidden in the
radiative corrections coming out from the regularization of the ultraviolet
divergences, and this is possible only with the Feynman Green function).

This means that at this semiclassical level, the elimination of the
electromagnetic degrees of freedom produces a system of particles with
instantaneous action at a distance given by the Coulomb and Darwin
potentials. Since ${\frac{1}{2}} (retarded-advanced)=0$, all the
effects now come from the regularized $retarded = advanced ={\frac{
1}{2}} (retarded+advanced)$ solution and there is no mass
renormalization. Even if the transverse projector implies
contributions from the whole past (or future) history of the
particles, in the semiclassical approximation only the instantaneous
action-at-a-distance effects on $\Sigma_W(\tau )$ survive. Each
particle feels only the action of the other $N-1$ [thus giving us an
effective Abraham-Lorentz-Dirac equation with no self-reaction and
with the Lorentz forces of the other particles replaced by
action-at-a-distance interactions]. Like in the Feynman-Wheeler
approach \cite{leiter}, we can now speak of radiation only as the
effect of the other $N-1$ particles on the one chosen as a detector of
radiation when it is far away from the other particles:\hfill
equations of motion), one has accord with the Larmor formula coming
from the $Q_iQ_j$ interference terms;\hfill(the Larmor formula gives
zero at the semiclassical level due to the particle equations of
motion).

An important and new feature of this formalism is the possibility to find
the final canonical variables for the particles after the introduction of
the Dirac brackets: their use introduce new higher order contributions to
the Darwin potential coming from the kinetic energy square roots. These
contributions lead to a substantial\ cancellation with corresponding terms
coming from what began as the electromagnetic energy integral. Our final
generalized Darwin interaction naturally divides into two portions. \ The
first portion has the same form either as the original lowest order
correction derived originally by Darwin in the retarded case or as the
lowest order $1/c^{2}$ well known form of the Darwin potential in the case
of symmetric [${\frac{1}{2}}(retarded+advanced)$ ] Lienard-Wiechert
potentials but with the masses replaced by the kinetic energies, $%
m_{i}\rightarrow \sqrt{m_{i}^{2}+\kappa _{i}^{2}}$, \ and this is a new
result (strictly speaking this is a higher order correction, but it shows
that we are using the correct relativistic kinematics without $1/c^2$
expansions). The second portion, a double infinite series, is, like that of
the generalization of the $1/c^{2\text{ }}$Darwin potential, is also new. It
is of higher order in $1/c^2$ than the more familiar first portion.
Furthermore, our generalized Darwin interaction for $N$ bodies is equal to
the pairwise sum of two body pieces

For the restricted case of two bodies considerable simplifications result
when one evaluates the series in the center-of-mass rest frame. It then can
be written in closed form.

The Darwin potential we obtain can be regarded as the classical analogue of
the full effects (complete transverse as well as longitudinal) of the single
photon exchange in the Bethe-Salpeter equation since it is the same order in
the coupling constants. \ The effect of the semiclassical regularization $%
Q_{i}^{2}=0$ is to truncate out the classical analogue of the numerous
higher order ladder and cross ladder diagrams. \ To the extent that the
Darwin potential we obtain has a low order ($1/c^{2}$) portion that agrees
with the standard result, it would be expected to contribute correctly to
the spectral results in a quantized formalism. \

For two particles there is the problem of the comparison of the
semiclassical Lienard-Wiechert sector with the 2-particle models defined by
two first class constraints with covariant instantaneous
action-at-a-distance interactions (phenomenological approximations of the
Bethe-Salpeter equation). \ Several authors beginning with \cite{dv,tod,va}
\ and continued by \cite{ll,crater,saz} have developed pairs of \ commuting
generalized mass shell conditions, first class constraints with
instantaneous potentials in the center-of-mass sytem, whose quantization
gives coupled Klein-Gordon equations for two spin zero particles [see Refs.
\cite{crater,saz,crater1,saz1} for similar equations for Dirac particles
deriving from pairs of first class constraints for spinning particles]. Some
of these models were generated as approximations to the Bethe-Salpeter
equation, by reducing it in a covariant instantaneous approximations to a
3-dimensional equation (with the elimination of the spurious abnormal
sectors of relative energy excitations) of the Lippmann-Schwinger type and
then to the equation of the quasipotential approach [see the bibliography of
the quoted references], which Todorov\cite{tod} reformulated as a pair of
first class constraints at the classical level. In Ref.\cite{saz} it is
directly shown how the normal sectors of the Bethe-Salpeter equation are
connected with the quantization of pairs of first class constraints with
instantaneous (in general nonlocal, but approximable with local) potentials
like in Todorov's examples. Ref. \cite{crater} shows how to derive the
Todorov potential for the electromagnetic and world scalar case from
Tetrode-Fokker-Feynman-Wheeler dynamics with scalar and vector potentials
[this theory is connected with ${\frac{1}{2}}(retarded+advanced)$ solutions
with no incoming radiation (adjunct Lienard-Wiechert fields) of Maxwell
equations with particle currents in the Lorentz gauge]; besides the Coulomb
potential, at the order $1/c^{2}$ one gets the standard Darwin potential
(becoming the Breit one at the quantum level when spin is added), which is
known to be phenomenologically correct.

What are the connections between the center-of-mass rest frame \ form of the
two-body interaction Hamiltonian that we develope in this paper to all
orders in $1/c^{2}$ and that obtained in the above references? \ The Darwin
potential becomes a common overlap of the two approaches and thus an
important testing ground for the approach we develope in this paper.
Furthermore, when in future papers pseudoclassical spin is introduced
(extending as in \cite{lu3} to interacting systems of particles and fields)
the types of tests we perform in this paper will be relevant (note, however,
there are difficulties with other categories of Darwin type of interactions
brought on by the introduction of transverse spin-dependent electric field
effects which unlike magnetic fields cannot be diagnonalized by a
Foldy-Wouthuysen transformation \cite{bard,lu3}).

\bigskip

In Section II we give a review of parametrized Minkowski theories on
arbitrary spacelike hypersurfaces.

In Section III we apply this formalism to the isolated system of $N$ charged
scalar particles, with Grassmann-valued electric charges, plus the
electromagnetic field and we arrive at its rest-frame instant form on the
Wigner hyperplanes. We also make the canonical reduction to the radiation
gauge.

In Section IV we study the ``internal" Poincar\'e algebra and the
``internal" center of mass on the Wigner hyperplanes and we derive the
Hamiltonian and Lagrange equations of motion for fields and particles. Also
the energy-momentum tensor is evaluated.

In Section V we evaluate the Lienard-Wiechert potentials, we show that at
the semiclassical level they depend only on particle coordinates and
velocities (since at this level $retarded = advanced$) and we find their
phase space expression. Then we eliminate the electromagnetic degrees of
freedom by means of second class constraints which force them to coincide
with the Lienard-Wiechert solution. We introduce the associated Dirac
brackets and we find their canonical Darboux basis.

In Section VI we derive the physical Hamiltonian and the effective Darwin
potential to all orders in $1/c^2$ in terms of the old (noncanonical) and of
the new (canonical) variables. In the two-particle case we get a closed form
of this potential using the rest-frame condition. Also the final form of the
energy-momentum tensor is given.

In the Conclusions there are some general considerations and hints for
future developments.

Appendix A gives an explicit summation for the vector potential presented in
Section V from the Lienard-Wiechert series in the rest-frame radiation gauge.

In Appendix B we evaluate the field energy and momentum integrals used in
Section VI when the electric and magnetic fields are expressed in terms of
the Lienard-Wiechert series for the vector potential.

We derive in Appendix C a general formula for a certain quantity, which is
important for obtaining the closed form expression of the Darwin potential
of Section VI for $N$=2 in the rest frame.

In Appendix D we use a technique similar to that developed by Kerner ( and
applied in \cite{crater} to obtain the Todorov quasipotential from the
Wheeler-Feynman action) in the transformation of the Lagrangian expression
for the invariant mass to $M$. \ In this proof, (done to all orders) we must
use Dirac brackets since we have used the Lienard-Wiechert constraints as a
strong condition on the dynamical variables. \ This necessitates the
explicit expression for the field momentum integrals developed in the
earlier appendix from the Lienard-Wiechert solutions.

In Appendix E we obtain a special solution of Hamilton's equations for the
two-body problem that is analogous to Schild's solution for circular orbits
\cite{schild}.

\vfill\eject

\section{ Parametrized Minkowski Theory for N Free Scalar Particles.}

In this Section we shall review the description of $N$ scalar free particles
on arbitrary spacelike hypersurfaces \cite{lu1}, leaves of the foliation of
Minkowski spacetime associated with one of its 3+1 splittings following the
suggestions of Refs.\cite{dd,ku}. The scalar parameter $\tau $ labelling the
hypersurfaces $\Sigma (\tau )$ allows one to introduce a covariant concept
of ``equal time'', which will be useful in the description of an isolated
system of interacting particles and fields. It will be shown that in these
parametrized Minkowski theories there are first class constraints implying
the independence of the description of the isolated system from the chosen
3+1 splitting.

As said in the Introduction this requires the addition to the theory of an
infinite number of new configuration variables $z^{\mu }(\tau ,\vec{\sigma})$
describing the spacelike hypersurfaces as an embedding of $R^3$ into
Minkowski spacetime. We use the notation $\sigma^{\check A%
}=(\sigma^{\tau}=\tau ,\sigma^{\check r} )$, i.e. $\check A=(\tau ,\check r)$
[the notation $A=(\tau ,r)$ will be used for the Wigner indices on the
Wigner hyperplane, see Section III].

In the manifestly Lorentz covariant approach the worldlines of scalar
particles are described by 4-vector coordinates $x_{i}^{\mu }(\tau _{i})$,
where the $\tau _{i}$'s are affine parameters (often they are restricted to
be the proper times of the particles). Even if one uses a unique affine
parameter $\tau $ for all the particles, $x_{i}^{\mu }(\tau )$, there is the
problem that the particles times $x_{i}^{0}(\tau )$ are gauge variables due
to the presence of the first class mass-shell constraints $%
p_{i}^{2}-m_{i}^{2}\approx 0$. For each free particle the constraint
manifold is the union of two disjoint submanifolds $p^{0}=\pm \sqrt{%
m_{i}^{2}+{\vec{p}}_{i}^{2}}$. The gauge nature of the $x_{i}^{0}$'s is
connected with: i) the arbitrariness in the choice of the center-of-mass
time; ii) the arbitrariness in the choice of how to trigger the $N$
particles (at equal times or with any conceivable mutual delay; this is the
gauge freedom of relative times). Given the foliation of Minkowski spacetime
with leaves $\Sigma (\tau )$ we can give a covariant description of the
particles at ``equal times'' (covariant zero relative times condition) by
parametrizing the worldlines as $x_{i}^{\mu }(\tau )=z^{\mu }(\eta
_{i}^{A}(\tau ))=z^{\mu }(\tau ,\vec{\eta}_{i}(\tau ))$, with with the
Lorentz scalar coordinates $\eta _{i}^{\check{A}}(\tau )=(\tau ,\eta _{i}^{%
\check{r}}(\tau ))$. Only 3 Lorentz scalar coordinates ${\vec{\eta}}%
_{i}(\tau )$ identify the intersection of the particle worldline with $%
\Sigma (\tau )$. This implies that in this description there are no
mass-shell constraints, namely that we are describing particles with a well
defined sign of the energy: $\eta _{i}=sign\,p_{i}^{0}=\pm 1$. In this paper
we shall consider only positive energy particles, so that $\eta _{i}=1$ for
every $i$. There will be a conjugate momentum ${\vec{\kappa}}_{i}(\tau )$
for each particle and the standard momentum $p_{i}^{\mu }(\tau )$ will be a
derived quantity which satisfies $p_{i}^{2}=m_{i}^{2}$.

The metric induced on $\Sigma (\tau )$ from the Minkowski metric $%
\eta^{\mu\nu}=(+---)$ is

\begin{equation}
g_{\check{A}\check{B}}(z(\tau ,\vec{\sigma})):=\eta _{\mu \nu }z_{\check{A}
}^{\mu }z_{\check{B}}^{\nu },  \label{II1}
\end{equation}

\noindent where we have used the notation

\begin{equation}
z_{\check{A}}^{\mu }(\tau ,\vec{\sigma})={\frac{\partial z^{\mu }}{\partial
\sigma ^{\check{A}}}}:=\partial _{\check{A}}z^{\mu }.  \label{II2}
\end{equation}

If we define the quantity $z^{\check A}_{\mu}(\tau ,\vec \sigma )$ by means
of

\begin{equation}
z_{\mu }^{\check A}z_{\check B}^{\mu }=\delta _{\check B}^{\check A},
\label{II3}
\end{equation}

\noindent they satisfy

\begin{equation}
g_{\check{A}\check{B}}z_{\mu }^{\check{A}}z_{\nu }^{\check{B}}=\eta
_{\lambda \kappa }z_{\check{A}}^{\lambda }z_{\check{B}}^{\kappa }z_{\mu }^{%
\check{A}}z_{\nu }^{\check{B}}=\eta _{\mu \nu }.  \label{II4}
\end{equation}
Therefore, the $z_{\mu }^{\check{A}}(\tau ,\vec{\sigma})$ are a set of
vierbeins, with the $z_{\check{A}}^{\mu }(\tau ,\vec{\sigma})$ the inverse
vierbeins.

Since we require $g_{\tau \tau }>0$ as a condition on the embedding $%
z^{\mu}(\tau ,\vec \sigma )$, $z_{\tau }^{\mu }$ is a time-like 4-vector and
the $z^{\mu}_{\check r}(\tau ,\vec \sigma )$'s are spacelike 4-vectors
tangent to $\Sigma (\tau )$.

The determinant of the metric is

\begin{equation}
g=-det||g_{\check{A}\check{B}}||=(det\,z_{\check{A}}^{\mu })^{2}.
\label{II5}
\end{equation}
The spatial part of the metric has an associated determinant which is
defined by

\begin{equation}
\gamma =-det||g_{\check{r}\check{s}}||;\quad \quad \ \Gamma =\sqrt{\gamma }.
\label{II6}
\end{equation}

We next define the inverse metric $g^{\check{A}\check{B}}$ by

\begin{equation}
g^{\check{A}\check{B}}g_{\check{B}\check{C}}=\delta _{\check{C}}^{\check{A}}.
\label{II7}
\end{equation}
This implies

\begin{equation}
g^{\check A\check B}g_{\check A\check B}=4=g^{\check A\check B}z_{\check A%
}^{\mu }z_{\check B}^{\nu }\eta _{\mu\nu }=\eta^{\mu \nu }\eta _{\mu \nu },
\label{II8}
\end{equation}

\noindent so that

\begin{equation}
\eta ^{\mu \nu }=g^{\check A\check B}z_{\check A}^{\mu }z_{\check B}^{\nu }=
g^{\tau \tau }z_{\tau }^{\mu}z_{\tau }^{\nu }+2g^{\tau \check{r}}z_{\tau
}^{\mu }z_{\check{r}}^{\nu }+g^{\check{r}\check{s}}z_{\check{r}}^{\mu }z_{%
\check{s}}^{\nu }.  \label{II9}
\end{equation}

By definition of the element of an inverse matrix

\begin{equation}
g^{\tau \tau }={\frac{\Gamma ^{2}}{g}=}\frac{\gamma }{g},  \label{II10}
\end{equation}

\noindent while the inverse of the spatial $g_{\check{r}\check{s}}$ is
defined as $\gamma ^{\check{r}\check{s}}$, that is,

\begin{equation}
\gamma ^{\check{r}\check{s}}g_{\check{s}\check{t}}=\delta _{\check{t}}^{%
\check{r}}.  \label{II11}
\end{equation}
To find $g^{\tau \check{u}}$ use the fact that
\begin{equation}
g^{\tau \check{r}}g_{\check{r}\check{s}}+g^{\tau \tau }g_{\tau \check{s}
}=\delta _{\check{s}}^{\tau }=0,  \label{II12}
\end{equation}
and therefore
\begin{equation}
g^{\tau \check{r}}g_{\check{r}\check{s}}=-g^{\tau \tau }g_{\tau \check{s}}.
\label{II13}
\end{equation}
Multiplying by $\gamma^{\check{s}\check{u}}$ leaves us with
\begin{equation}
g^{\tau \check{u}}=-{\frac{\Gamma ^{2}}{g}}g_{\tau \check{s}}\gamma ^{\check{%
s}\check{u}}.  \label{II14}
\end{equation}
Now using this consider
\begin{equation}
g^{\check{r}\check{s}}g_{\check{t}\check{s}}+g^{\check{r}\tau }g_{\check{t}%
\tau }=\delta _{\check{t}}^{\check{r}}.  \label{II15}
\end{equation}
Multiply both sides by $\gamma ^{\check{t}\check{u}}$, use the definition of
$\gamma ^{\check{t}\check{u}}$ and the above expression for $g^{\check{r}%
\tau }$, and we obtain
\begin{equation}
g^{\check{r}\check{u}}=\gamma ^{\check{r}\check{u}}+{\frac{\Gamma ^{2}}{g}}
g_{\tau \check{s}}g_{\tau \check{v}}\gamma^{\check{s}\check{r}} \gamma ^{%
\check{v}\check{u}}.  \label{II16}
\end{equation}

Thus in summary we have expressed the inverse metric in terms of the metric
and the inverse of its spatial parts

\begin{eqnarray}
g^{\tau \tau } &=&{\frac{\Gamma ^{2}}{g}},  \nonumber \\
g^{\tau \check{r}} &=&-{\frac{\Gamma ^{2}}{g}}g_{\tau \check{s}}\gamma ^{%
\check{s}\check{r}},  \nonumber \\
g^{\check{r}\check{s}} &=&\gamma ^{\check{r}\check{s}}+{\frac{\Gamma ^{2}}{g}%
}g_{\tau \check{u}}g_{\tau \check{v}}\gamma ^{ur}\gamma ^{vs}.  \label{II17}
\end{eqnarray}
Moreover, we have

\begin{equation}
\eta ^{\mu \nu }z_{\mu }^{\check A}z_{\nu }^{\check B}=g^{\check C\check D}
z_{\check C}^{\mu }z_{\check D}^{\nu }z_{\mu }^{\check A}z_{\nu }^{\check B%
}=g^{\check A\check B}.  \label{II18}
\end{equation}

The normal to $\Sigma (\tau )$ at the point $z^{\mu }(\tau ,\vec{\sigma})$
is the Lorentz four vector

\begin{equation}
l^{\mu }(\tau ,\vec{\sigma})={\frac{1}{\Gamma (\tau ,\vec \sigma )}}%
\epsilon^{\mu\alpha\beta\gamma }z_{\check{1}\alpha }(\tau ,\vec \sigma ) z_{%
\check{2}\beta }(\tau ,\vec \sigma ) z_{\check{3}\gamma }(\tau ,\vec \sigma
).  \label{II19}
\end{equation}
with the normalization $l^{2}(\tau ,\vec \sigma )=1$. By construction we
have $l_{\mu }(\tau ,\vec \sigma )z_{\check{r}}^{\mu }(\tau , \vec \sigma )=0
$.

The evolution 4-vector $z_{\tau }^{\mu }(\tau ,\vec \sigma )$ can be
decomposed on the mutually orthogonal four vectors $l^{\mu }$ and $%
z_{s}^{\mu }$:

\begin{equation}
z_{\tau }^{\mu }(\tau ,\vec \sigma )=N(\tau ,\vec \sigma ) l^{\mu}(\tau ,%
\vec \sigma )+N^{\check{s}}(\tau ,\vec \sigma ) z_{\check{s}}^{\mu }(\tau ,%
\vec \sigma ).  \label{II20}
\end{equation}

\noindent where $N(\tau ,\vec \sigma )$ is the lapse and $N^{\check{r}}(\tau
,\vec \sigma )$ the shift functions in the terminology of ADM general
relativity. To determine the vector $N^{\check{r}}$ we use the orthogonality
of $z_{\mu \check{r}}$ and $l^{\mu }$. That is multiplying both sides of the
above equation by $z_{\mu \check{r}}$ and using the expression for $g_{%
\check{A}\check{B}}$ and multiplying by $\gamma ^{\check{r}\check{\check{u}}}
$ we obtain.
\begin{equation}
N^{\check{u}}(\tau ,\vec \sigma )=g_{\tau \check{r}}(\tau ,\vec \sigma
)\gamma^{\check{r}\check{u}}(\tau ,\vec \sigma ).  \label{II21}
\end{equation}
Using this expression for $N^{\check{r}}$ we determine the scalar $N$ by\
first multiplying the previous equation by $z_{\mu \tau }$. Then we use the
definition of $l^{\mu }$ and the determinant, multiply the result by $%
g^{\tau \tau }$. \ Then use the definition of $\gamma ^{\check{r}\check{s}}$
, and we obtain $N=\Gamma /\sqrt{\gamma }$. Hence
\begin{equation}
z_{\tau }^{\mu }(\tau ,\vec \sigma )=\Big[ {\frac{\Gamma }{\sqrt{\gamma }}}%
l^{\mu }+g_{\tau \check{r} }\gamma ^{\check{r}\check{s}}z_{\check{s}}^{\mu }%
\Big] (\tau ,\vec \sigma ).  \label{II22}
\end{equation}
Substituting this expression for $z_{\tau }^{\mu }$ into
\begin{equation}
\eta ^{\mu \nu }=g^{\check{A}\check{B}}z_{\check{A}}^{\mu }z_{\check{B}%
}^{\nu }=g^{\tau \tau }z_{\tau }^{\mu }z_{\tau }^{\nu }+g^{\tau \check{r}%
}(z_{\tau }^{\mu }z_{\check{r}}^{\nu }+z_{\tau }^{\nu }z_{\check{r}}^{\mu
})+g^{\check{r}\check{s}}z_{\check{r}}^{\mu }z_{\check{s}}^{\nu },
\label{II23}
\end{equation}
together with those for $g^{\check{A}\check{B}}$, we obtain after some
algebra the following decomposition of the Minkowski metric
\begin{equation}
\eta ^{\mu \nu }=l^{\mu }(\tau ,\vec \sigma ) l^{\nu }(\tau ,\vec \sigma )+
\gamma ^{\check{r}\check{s}}(\tau ,\vec \sigma ) z_{\check{r} }^{\mu }(\tau ,%
\vec \sigma ) z_{\check{s}}^{\nu }(\tau ,\vec \sigma ).  \label{II24}
\end{equation}

Coming back to the scalar particles, the relation between their world line
velocities in the two descriptions is

\begin{equation}
\dot{x_{i}^{\mu }}(\tau )=z_{\tau }^{\mu }(\tau ,\vec{\eta}_{i}(\tau ))+z_{%
\check{r}}^{\mu }(\tau ,\vec{\eta}_{i}(\tau ))\dot{\eta _{i}^{\check{r}}}%
(\tau ).  \label{II25}
\end{equation}
Noting that
\begin{eqnarray}
\dot{x_{i}}^{2} &=&\dot{x_{i}^{\mu }}\dot{x_{i}^{\nu }}\eta _{\mu \nu
}=z_{\tau }^{\mu }z_{\tau }^{\nu }\eta _{\mu \nu }+2z_{\tau }^{\mu }z_{%
\check{r}}^{\nu }\eta _{\mu \nu }\dot{\eta _{i}^{\check{r}}}+z_{\check{s}%
}^{\mu }z_{\check{r}}^{\nu }\eta _{\mu \nu }\dot{\eta _{i}^{\check{r}}}\dot{%
\eta _{i}^{\check{s}}}=  \nonumber \\
&=&g_{\tau \tau }(\tau ,\vec{\eta}_{i}(\tau ))+2g_{\tau \check{r}}(\tau ,%
\vec{\eta}_{i}(\tau ))\dot{\eta _{i}^{\check{r}}}+g_{\check{s}\check{r}%
}(\tau ,\vec{\eta}_{i}(\tau ))\dot{\eta _{i}^{\check{r}}}\dot{\eta _{i}^{%
\check{s}}},  \label{II26}
\end{eqnarray}

\noindent the standard action for $N$ free scalar particles becomes
\begin{equation}
S=\int d\tau \sum_{i}^{N}\Big[-m_{i}\sqrt{\dot{x_{i}}^{2}}\Big]=\int d\tau
L(\tau )=\int d\tau d^{3}\sigma {\cal L}(\tau ,\vec{\sigma}),  \label{II27}
\end{equation}
with the Lagrangian density
\begin{equation}
{\cal L}(\tau ,\vec{\sigma})=-\sum_{i=1}^{N}\delta ^{3}(\vec{\sigma}-{\vec{%
\eta}}_{i}(\tau ))m_{i}\sqrt{g_{\tau \tau }(\tau ,\vec{\sigma})+2g_{\tau {\
\check{r}}}(\tau ,\vec{\sigma}){\dot{\eta}}_{i}^{\check{r}}(\tau )+g_{{\
\check{r}}{\check{s}}}(\tau ,\vec{\sigma}){\dot{\eta}}_{i}^{\check{r}}(\tau )%
{\dot{\eta}}_{i}^{\check{s}}(\tau )},  \label{II28}
\end{equation}
and the Lagrangian
\begin{equation}
L(\tau )=-\sum_{i=1}^{N}m_{i}\sqrt{g_{\tau \tau }(\tau ,{\vec{\eta}}%
_{i}(\tau ))+2g_{\tau {\check{r}}}(\tau ,{\vec{\eta}}_{i}(\tau )){\dot{\eta}}%
_{i}^{\check{r}}(\tau )+g_{{\check{r}}{\check{s}}}(\tau ,{\vec{\eta}}%
_{i}(\tau )){\dot{\eta}}_{i}^{\check{r}}(\tau ){\dot{\eta}}_{i}^{\check{s}%
}(\tau )}.  \label{II29}
\end{equation}
The above action is invariant under separate $\tau $ and $\vec{\sigma}$
reparametrization. This leads naturally to constraints.

The canonical momenta are determined from the\ dependence on the
``velocity'' ${z_{\tau }^{\mu }(\tau ,\vec{\sigma})}$ associated with the
hypersurface and the particle velocities ${{\dot{\eta}}_{i}^{\check{r}}(\tau
).}$

\begin{eqnarray}
\rho _{\mu }(\tau ,\vec{\sigma}) &=&-{\frac{{\partial {\cal L}(\tau ,\vec{%
\sigma})}}{{\partial z_{\tau }^{\mu }(\tau ,\vec{\sigma})}}}%
=\sum_{i=1}^{N}\delta ^{3}(\vec{\sigma}-{\vec{\eta}}_{i}(\tau ))  \nonumber
\\
&&m_{i}{\frac{{z_{\tau \mu }(\tau ,\vec{\sigma})+z_{{\check{r}}\mu }(\tau ,%
\vec{\sigma}){\dot{\eta}}_{i}^{\check{r}}(\tau )}}{\sqrt{g_{\tau \tau }(\tau
,\vec{\sigma})+2g_{\tau {\check{r}}}(\tau ,\vec{\sigma}){\dot{\eta}}_{i}^{%
\check{r}}(\tau )+g_{{\check{r}}{\check{s}}}(\tau ,\vec{\sigma}){\dot{\eta}}%
_{i}^{\check{r}}(\tau ){\ \dot{\eta}}_{i}^{\check{s}}(\tau )}}}=  \nonumber
\\
&=&[(\rho _{\nu }l^{\nu })l_{\mu }+(\rho _{\nu }z_{\check{r}}^{\nu })\gamma
^{{\check{r}}{\check{s}}}z_{{\check{s}}\mu }](\tau ,\vec{\sigma}),  \nonumber
\\
&&{}  \nonumber \\
\kappa _{i{\check{r}}}(\tau ) &=&-{\frac{{\partial L(\tau )}}{{\partial {\
\dot{\eta}}_{i}^{\check{r}}(\tau )}}}=  \nonumber \\
&=&m_{i}{\frac{{g_{\tau {\check{r}}}(\tau ,{\vec{\eta}}_{i}(\tau ))+g_{{\
\check{r}}{\check{s}}}(\tau ,{\vec{\eta}}_{i}(\tau )){\dot{\eta}}_{i}^{%
\check{s}}(\tau )}}{{\ \sqrt{g_{\tau \tau }(\tau ,{\vec{\eta}}_{i}(\tau
))+2g_{\tau {\ \check{r}}}(\tau ,{\vec{\eta}}_{i}(\tau )){\dot{\eta}}_{i}^{%
\check{r}}(\tau )+g_{{\ \check{r}}{\check{s}}}(\tau ,{\vec{\eta}}_{i}(\tau ))%
{\dot{\eta}}_{i}^{\check{r}}(\tau ){\dot{\eta}}_{i}^{\check{s}}(\tau )}}}},
\label{II30}
\end{eqnarray}
Using the above we have

\begin{eqnarray}
&&m_{i}^{2}-\gamma ^{\check{r}\check{s}}\kappa _{i\check{r}}\kappa _{i\check{%
s} }=m_{i}^{2}[1-{\frac{\gamma ^{\check{r}\check{s}}(g_{\tau \check{r}}+g_{%
\check{r}\check{t}} \dot{\eta _{i}^{\check{t}}})(g_{\tau \check{s}}+g_{%
\check{s}\check{u}}\dot{\eta _{i}^{u}})}{g_{\tau \tau } +2g_{\tau \check{r}}%
\dot{\eta _{i}^{\check{r}}}+g_{\check{s}\check{r}} \dot{\eta _{i}^{\check{r}}%
}\dot{\eta _{i}^{\check{s}}}}]}=  \nonumber \\
&&=({\frac{m_{i}}{\sqrt{g_{\tau \tau }+2g_{\tau r}\dot{\eta _{i}^{\check{r}}}%
+g_{\check{s} \check{r}}\dot{\eta _{i}^{\check{r}}}\dot{\eta _{i}^{\check{s}}%
}}}})^{2}(g_{\tau \tau }-\gamma ^{\check{r}\check{s}}g_{\tau \check{r}%
}g_{\tau \check{s}}).  \label{II31}
\end{eqnarray}
Use the following two forms

\begin{eqnarray}
\rho _{\mu }l^{\mu } &=&\sum_{i=1}^{N}{\frac{\delta ^{3}(\vec{\sigma}-\eta
_{i}(\tau ))m_{i}z_{\tau \mu }l^{\mu }}{\sqrt{g_{\tau \tau }(\tau ,\vec{\eta}
_{i}(\tau ))+2g_{\tau \check{r}}(\tau ,\vec{\eta}_{i}(\tau ))\dot{\eta _{i}^{%
\check{r}}}+g_{\check{s}\check{r}}(\tau ,\vec{\eta}_{i}(\tau ))\dot{\eta
_{i}^{\check{r}}}\dot{\eta _{i}^{\check{s}}}}}},  \nonumber \\
z_{\tau }\cdot l &=&{\frac{1}{\gamma }}\epsilon ^{\mu \alpha \beta
\Gamma }z_{\tau \mu }z_{1\alpha }z_{2\beta }z_{3\gamma
}={\frac{\sqrt{g}}{\Gamma }},
\label{II32}
\end{eqnarray}
together with the square root of the above relation and $g^{\tau
\check{s} }=- {\frac{\gamma }{g}}g_{\tau \check{r}}\gamma
^{\check{r}\check{s}}$and

\begin{equation}
g^{\tau \check{s}}g_{\tau \check{s}}=g^{\tau A}g_{\tau A}-g^{\tau \tau
}g_{\tau \tau }=1-{\ \frac{\gamma }{g}}g_{\tau \tau },  \label{II33}
\end{equation}
to obtain
\begin{equation}
\rho _{\mu }l^{\mu }=\sum_{i=1}^{N}\delta ^{3}(\vec{\sigma}-\eta _{i}(\tau
)) \sqrt{m_{i}^{2}-\gamma ^{\check{r}\check{s}}\kappa _{i\check{r}}\kappa
_{i \check{s}}}.  \label{II34}
\end{equation}
Using finally
\begin{eqnarray}
\rho _{\mu }z_{\check{r}}^{\mu }&=&\sum_{i=1}^{N}{\frac{\delta ^{3}(\vec{%
\sigma }-\eta _{i}(\tau ))m_{i}(z_{\check{r}}^{\mu }z_{\tau \mu }+z_{\check{r%
} }^{\mu }z_{\check{s}\mu }\dot{\eta _{i}^{\check{s}}})}{\sqrt{g_{\tau \tau
}(\tau ,\vec{\eta}_{i}(\tau ))+2g_{\tau r}(\tau ,\vec{\eta}_{i}(\tau ))\dot{
\eta _{i}^{\check{r}}}+g_{\check{s}\check{r}}(\tau ,\vec{\eta}_{i}(\tau ))
\dot{\eta _{i}^{\check{r}}}\dot{\eta _{i}^{\check{s}}}}}}=  \nonumber \\
&=&\sum_{i=1}^{N}\delta ^{3}(\vec{\sigma}-\eta _{i}(\tau ))\kappa_{ri},
\label{II35}
\end{eqnarray}
we obtain the form of the four primary first class constraints ${\cal H}%
_{\mu }$ \ following from $\tau $ and $\vec{\sigma}$ reparametrization
invariance:

\begin{eqnarray}
{\cal H}_{\mu }(\tau ,\vec{\sigma}) &=&\rho _{\mu }(\tau ,\vec{\sigma}%
)-l_{\mu }(\tau ,\vec{\sigma})\sum_{i=1}^{N}\delta ^{3}(\vec{\sigma}-{\vec{%
\eta}}_{i}(\tau ))\sqrt{m_{i}^{2}-\gamma ^{{\check{r}}{\check{s}}}(\tau ,%
\vec{\sigma})\kappa _{i{\check{r}}}(\tau )\kappa _{i{\check{s}}}(\tau )}-
\nonumber \\
&-&z_{{\check{r}}\mu }(\tau ,\vec{\sigma})\gamma ^{{\check{r}}{\check{s}}%
}(\tau ,\vec{\sigma})\sum_{i=1}^{N}\delta ^{3}(\vec{\sigma}-{\vec{\eta}}%
_{i}(\tau ))\kappa _{i{\check{s}}}\approx 0.  \label{II36}
\end{eqnarray}
Assuming the following Poisson brackets

\begin{eqnarray}
\{z^{\mu }(\tau ,\vec{\sigma}),\rho _{\nu }(\tau ,{\vec{\sigma}}^{^{\prime
}})\} &=&-\eta _{\nu }^{\mu }\delta ^{3}(\vec{\sigma}-{\vec{\sigma}}%
^{^{\prime }}),  \nonumber \\
\{\eta _{i}^{\check{r}}(\tau ),\kappa _{j{\check{s}}}(\tau )\} &=&-\delta
_{ij}\delta _{\check{s}}^{\check{r}}.  \label{II37}
\end{eqnarray}
one can show the first class nature of the above constraints \cite{lu1}.
Since these constraints are solved in terms of one of the independent
momenta [$\rho _{\mu }(\tau ,\vec{\sigma})$], their Poisson brackets are
exactly zero:

\begin{equation}
\{{\cal H}_{\mu }(\tau ,\vec{\sigma}),{\cal H}_{\nu }(\tau ,{\vec{\sigma}}
^{^{\prime }})\}=0.  \label{II38}
\end{equation}

These constraints imply that the description of the system is independent
from the chosen 3+1 splitting of Minkowski spacetime.

The standard particle momenta $p^{\mu}_i$ are reconstructed as $p^{\mu}_i= (%
\sqrt{m_i^2-\gamma^{\check r\check s}\kappa_{i\check r}\kappa_{i\check s}}; {%
\vec \kappa}_i)$ and satisfy $p_i^2=m_i^2$.

In the next Section we shall add Grassmann-valued electric charges to the
scalar particles, we shall find the new action and constraints, eventually
arriving at the rest-frame instant form.

\vfill\eject

\section{\ N Charged Scalar Particles and the Electromagnetic Field.}

In this Section we will extend the formalism of the previous Section to the
case of an isolated system of $N$ charged scalar particles plus the
electromagnetic field. By using Lorentz scalar electromagnetic potentials
and field strengths, which employ the covariant ``equal time''  concept
associated with the spacelike hypersurfaces $\Sigma (\tau )$, we define the
action for the combined field and particle system, that, like the free
system, is separately invariant under $\tau $ and $\vec{\sigma}$
reparametrizations. As in the free particle case we obtain four primary
first class constraints ${\cal H}^{\mu }(\tau ,\vec{\sigma})\approx 0$ at
each point $\vec{\sigma}$ on each of the space-like surfaces $\Sigma (\tau )$%
, implying the independence from the chosen 3+1 splitting. Moreover, there
are the two additional first class constraints describing the
electromagnetic gauge invariance of the theory. By using a gauge fixing
condition which restricts the $\Sigma (\tau )$ 's to hyperplanes $\Sigma
_{H}(\tau )$ and by using Dirac brackets, the embedding variables $z^{\mu
}(\tau ,\vec{\sigma})$ are reduced to only ten ones. The original
constraints ${\cal H}^{\mu }(\tau ,\vec{\sigma})\approx 0$ are reduced to
just 10 first class global constraints on each of the hyperplanes. We then
specialize to hyperplanes orthogonal to the total timelike four-momentum of
the system with 6 new gauge fixings depending on the standard Wigner boost
for timelike Poincar\'{e} orbits. These hyperplanes, defined by the system
configuration, are called the Wigner hyperplanes $\Sigma _{W}(\tau )$. After
having defined the new Dirac brackets, we remain with a decoupled
``external'' canonical non-covariant center of mass, with Wigner-covariant
particle and field degrees of freedom on the Wigner hyperplane and with only
four first class contraints. One of these four constraints identifies the
invariant mass of the system as the effective Hamiltonian, while the other
three define the rest-frame condition ${\cal \vec{H}}_{p}(\tau )\approx 0$
(vanishing of the total 3-momentum inside the Wigner hyperplane) for the
combined system of particles and fields. Finally, we make the canonical
reduction to eliminate the electromagnetic gauge degrees of freedom, placing
the formalism (including the Poincar\'{e} generators) in the Wigner
covariant rest-frame radiation gauge. Also we give the energy-momentum
tensor of the full isolated system.

\subsection{The Action, The Constraints and the Canonical Reduction.}

Let us now review the isolated system of $N$ charged scalar particles plus
the electromagnetic field following Ref.\cite{lu1}. Just as on the
hypersurface $\Sigma (\tau )$ the positive energy particles are described by
coordinates ${\vec{\eta}}_{i}(\tau )$ such that $x_{i}^{\mu }(\tau )=z^{\mu
}(\tau ,{\vec{\eta}}_{i}(\tau ))$, so the electric charge of each particle
is described in a semiclassical way by means of a pair of complex conjugate
Grassmann variables $\theta _{i}(\tau ),\theta _{i}^{\ast }(\tau )$ \cite
{bar} satisfying [$I_{i}=I_{i}^{\ast }=\theta _{i}^{\ast }\theta _{i}$ is
the generator of the $U_{em}(1)$ group of particle $i$]

\begin{eqnarray}
&&\theta _{i}^{2}=\theta _{i}^{{\ast }2}=0,\quad \quad \theta _{i}\theta
_{i}^{\ast }+\theta _{i}^{\ast }\theta _{i}=0,  \nonumber \\
&&\theta _{i}\theta _{j}=\theta _{j}\theta _{i},\quad \quad \theta
_{i}\theta _{j}^{\ast }=\theta _{j}^{\ast }\theta _{i},\quad \quad \theta
_{i}^{\ast }\theta _{j}^{\ast }=\theta _{j}^{\ast }\theta _{i}^{\ast },\quad
\quad i\not=j.  \label{III1}
\end{eqnarray}

\noindent As said in the Introduction, the formal quantization procedure
sends $\theta _{i}^{\ast }$, $\theta _{i}$ into the Clifford algebra
describing a two-level Fermi oscillator $b_{i}^{\dagger }$, $b_{i}$, and
each Grassmann-valued electric charge $Q_{i}=e\theta _{i}^{\ast }\theta _{i}$
goes into $eb_{i}^{\dagger }b_{i}$ with eigenvalues $\pm e$.

The standard electromagnetic potential $A_{\mu}(z)$ and the field strength $%
F_{\mu\nu}(z)$ are replaced on $\Sigma (\tau )$ by Lorentz-scalar variables $%
A_{\check{A}}(\tau ,\vec{\sigma})$ and $F_{{\check{A}}{\check{B}}}(\tau ,%
\vec{\sigma})$ respectively, defined by

\begin{eqnarray}
&&A_{\check{A}}(\tau ,\vec{\sigma})=z_{\check{A}}^{\mu }(\tau ,\vec{\sigma}%
)A_{\mu }(z(\tau ,\vec{\sigma})),  \nonumber \\
&&F_{{\check{A}}{\check{B}}}(\tau ,\vec{\sigma})={\partial }_{\check{A}}A_{%
\check{B}}(\tau ,\vec{\sigma})-{\partial }_{\check{B}}A_{\check{A}}(\tau ,%
\vec{\sigma})=z_{\check{A}}^{\mu }(\tau ,\vec{\sigma})z_{\check{B}}^{\nu
}(\tau ,\vec{\sigma})F_{\mu \nu }(z(\tau ,\vec{\sigma})).  \label{III2}
\end{eqnarray}
The new potentials $A_{\check{A}}(\tau ,\vec{\sigma})$ have built-in the
covariant concept of ``equal time'' through their implicit dependence on the
embeddings $z^{\mu }(\tau ,\vec{\sigma})$.

With $d^{3}\Sigma ^{\mu }$ the surface element of $\Sigma (\tau )$ we have
the following volume element of Minkowski space-time
\begin{equation}
d^{4}z=z_{\tau }^{\mu }d\tau d^{3}\Sigma _{\mu }=d\tau z_{\tau }^{\mu
}l_{\mu }\Gamma d^{3}\sigma =\sqrt{g}d\tau d^{3}\sigma .  \label{III3}
\end{equation}

The action now depends on the configuration variables $z^{\mu }(\tau ,\vec{%
\sigma})$,$A_{\check{A} }(\tau ,\vec{\sigma})$, ${\vec{\eta}}_{i}(\tau )$, $%
\theta _{i}(\tau )$ and $\theta _{i}^{\ast }(\tau )$,$i=1,..,N$, and
consists of a ``kinetic'' piece for the complex Grassmann charges $\int
\frac{i}{2}[\theta _{i}^{\ast }(\tau ){\dot{\theta}}_{i}(\tau )-{\dot{\theta}%
}_{i}^{\ast }(\tau )\theta _{i}(\tau )]d\tau $, the same particle kinetic
piece as in the previous Section, the kinetic term $\int d^{4}z(-\frac{1}{4}%
F^{\mu \nu }F_{\mu\nu })$ for the electromagnetic field, and the
field-particle interaction term $\int Q_{i}A_{\mu }(x_{i}(\tau ))\dot{x}%
_{i}^{\mu }\left( \tau \right) d\tau $:

\begin{eqnarray}
S &=&\int d\tau d^{3}\sigma \,{\cal L}(\tau ,\vec{\sigma})=\int d\tau L(\tau
),  \nonumber \\
L(\tau ) &=&\int d^{3}\sigma {\cal L}(\tau ,\vec{\sigma}),  \nonumber \\
{\cal L}(\tau ,\vec{\sigma}) &=&{\frac{i}{2}}\sum_{i=1}^{N}\delta ^{3}(\vec{
\sigma}-{\vec{\eta}}_{i}(\tau ))[\theta _{i}^{\ast }(\tau ){\dot{\theta}}
_{i}(\tau )-{\dot{\theta}}_{i}^{\ast }(\tau )\theta _{i}(\tau )]-  \nonumber
\\
&&-\sum_{i=1}^{N}\delta ^{3}(\vec{\sigma}-{\vec{\eta}}_{i}(\tau ))[m_{i}%
\sqrt{g_{\tau \tau }(\tau ,\vec{\sigma})+2g_{\tau {\check{r}}}(\tau ,\vec{%
\sigma}){\dot{\eta}}_{i}^{\check{r}}(\tau )+g_{{\check{r}}{\check{s}} }(\tau
,\vec{\sigma}){\dot{\eta}}_{i}^{\check{r}}(\tau ){\dot{\eta}}_{i}^{\check{s}%
}(\tau )}+  \nonumber \\
&&+Q_i(\tau )(A_{\tau }(\tau ,\vec{\sigma})+A_{\check{r}}(\tau ,\vec{\sigma})%
{\dot{\eta}}_{i}^{\check{r}}(\tau ))]-  \nonumber \\
&&-{\frac{1}{4}}\,\sqrt{g(\tau ,\vec{\sigma})}g^{{\check{A}}{\check{C}}
}(\tau ,\vec{\sigma})g^{{\check{B}}{\check{D}}}(\tau ,\vec{\sigma})F_{{\
\check{A}}{\check{B}}}(\tau ,\vec{\sigma})F_{{\check{C}}{\check{D}}}(\tau ,
\vec{\sigma}),  \nonumber \\
&&{}  \nonumber \\
Q_i(\tau ) &=& e\theta _{i}^{\ast }(\tau )\theta _{i}(\tau ).  \label{III4}
\end{eqnarray}

Since $A_{\tau }(\tau ,\vec{\sigma})$ transforms as a $\tau $-derivative the
action is still invariant under separate $\tau $- and $\vec{\sigma}$%
-reparametrizations as in the free case. In addition it is invariant under
the electromagnetic local gauge transformations and under the odd global
phase transformations $\delta \theta _{i}\mapsto i\alpha \theta _{i}$ ,
generated by the $I_{i}$'s. The $Q_{i}=e I_{i}$ are the constants of motion
associated with this last symmetry [${\frac{d}{{d\tau }}}Q_{i}(\tau )\,%
\stackrel{\circ }{=} \,0$, where '$\stackrel{\circ }{=}$' means evaluated on
the solutions of the Euler-Lagrange equations; from now on we shall write $%
Q_i$ instead of $Q_i(\tau )$].

Since the semiclassical approximation $Q_{i}^{2}=0$ regularizes the Coulomb
self-energy, the criticism of Rohrlich\cite{ror} to this action principle
[that the minimal coupling term and the electromagnetic field term are ill
defined because they diverge on the worldlines of the particles] does not
apply.

The canonical momenta are [$E_{\check{r}}=F_{{\check{r}}\tau }$ and $B_{%
\check{r}}={\frac{1}{2}}\epsilon _{{\check{r}}{\check{s}}{\check{t}} }F_{{%
\check{s}}{\check{t}}}$ ($\epsilon _{{\check{r}}{\check{s}}{\check{t}}
}=\epsilon ^{{\check{r}}{\check{s}}{\check{t}}}$) are the ``electric'' and
``magnetic'' fields respectively; for $g_{\check{A}\check{B}}\rightarrow\eta
_{\check{A}\check{B}}$ one gets $\pi ^{\check{r}}=-E_{\check{r}}=E^{\check{r}%
}$]

\begin{eqnarray}
\rho _{\mu }(\tau ,\vec{\sigma}) &=&-{\frac{{\partial {\cal L}(\tau ,\vec{%
\sigma})}}{{\partial z_{\tau }^{\mu }(\tau ,\vec{\sigma})}}}%
=\sum_{i=1}^{N}\delta ^{3}(\vec{\sigma}-{\vec{\eta}}_{i}(\tau ))m_{i}
\nonumber \\
&&{\frac{{z_{\tau \mu }(\tau ,\vec{\sigma})+z_{{\check{r}}\mu }(\tau ,\vec{%
\sigma}){\dot{\eta}}_{i}^{\check{r}}(\tau )}}{\sqrt{g_{\tau \tau }(\tau ,%
\vec{\sigma})+2g_{\tau {\check{r}}}(\tau ,\vec{\sigma}){\dot{\eta}}_{i}^{%
\check{r}}(\tau )+g_{{\check{r}}{\check{s}}}(\tau ,\vec{\sigma}){\dot{\eta}}%
_{i}^{\check{r}}(\tau ){\dot{\eta}}_{i}^{\check{s}}(\tau )}}}+  \nonumber \\
&&+\frac{\sqrt{g(\tau ,\vec{\sigma})}}{4}[z_{\mu }^{\tau }(\tau ,\vec
\sigma )g^{\breve{A}%
\breve{C}}(\tau ,\vec{\sigma})g^{\breve{B}\breve{D}}(\tau ,\vec{\sigma}%
)-2z_{\mu }^{\check A}(\tau ,\vec \sigma )g^{\breve{B}\breve{D}}(\tau
,\vec{\sigma})g^{\bar{C}%
\tau }(\tau ,\vec{\sigma})-  \nonumber \\
&&-2z_{\mu }^{\check D}(\tau ,\vec \sigma )g^{B\tau }(\tau
,\vec{\sigma})g^{\breve{A}\breve{C} }(\tau
,\vec{\sigma})]F_{{\check{A}}{\check{B}}}(\tau ,\vec{\sigma})F_{{\bar{
C}\bar{D}}}(\tau ,\vec{\sigma})  \nonumber \\
 &=&[(\rho _{\nu }l^{\nu
})l_{\mu }+(\rho _{\nu }z_{\check{r}}^{\nu })\gamma
^{{\check{r}}{\check{s}}}z_{{\check{s}}\mu }](\tau ,\vec{\sigma}),  \nonumber
\\
&&{}  \nonumber \\
\pi ^{\tau }(\tau ,\vec{\sigma}) &=&{\frac{{\partial L}}{{\partial \partial
_{\tau }A_{\tau }(\tau ,\vec{\sigma})}}}=0,  \nonumber \\
\pi ^{\check{r}}(\tau ,\vec{\sigma}) &=&{\frac{{\partial L}}{{\partial
\partial _{\tau }A_{\check{r}}(\tau ,\vec{\sigma})}}}=-{\frac{{\gamma (\tau ,%
\vec{\sigma})}}{\sqrt{g(\tau ,\vec{\sigma})}}}\gamma ^{{\check{r}}{\check{s}}%
}(\tau ,\vec{\sigma})(F_{\tau {\check{s}}}-g_{\tau {\check{v}}}\gamma ^{{%
\check{v}}{\check{u}}}F_{{\check{u}}{\check{s}}})(\tau ,\vec{\sigma})=
\nonumber \\
&=&{\frac{{\gamma (\tau ,\vec{\sigma})}}{\sqrt{g(\tau ,\vec{\sigma})}}}%
\gamma ^{{\check{r}}{\check{s}}}(\tau ,\vec{\sigma})(E_{\check{s}}(\tau ,%
\vec{\sigma})+g_{\tau {\check{v}}}(\tau ,\vec{\sigma})\gamma ^{{\check{v}}{\
\check{u}}}(\tau ,\vec{\sigma})\epsilon _{{\check{u}}{\check{s}}{\check{t}}%
}B_{\check{t}}(\tau ,\vec{\sigma})),  \nonumber \\
&&{}  \nonumber \\
\kappa _{i{\check{r}}}(\tau ) &=&-{\frac{{\partial L(\tau )}}{{\partial {\
\dot{\eta}}_{i}^{\check{r}}(\tau )}}}=  \nonumber \\
&=&m_{i}{\frac{{g_{\tau {\check{r}}}(\tau ,{\vec{\eta}}_{i}(\tau ))+g_{{%
\check{r}}{\check{s}}}(\tau ,{\vec{\eta}}_{i}(\tau )){\dot{\eta}}_{i}^{%
\check{s}}(\tau )}}{{\ \sqrt{g_{\tau \tau }(\tau ,{\vec{\eta}}_{i}(\tau
))+2g_{\tau {\check{r}}}(\tau ,{\vec{\eta}}_{i}(\tau )){\dot{\eta}}_{i}^{%
\check{r}}(\tau )+g_{{\check{r}}{\check{s}}}(\tau ,{\vec{\eta}}_{i}(\tau )){%
\dot{\eta}}_{i}^{\check{r}}(\tau ){\dot{\eta}}_{i}^{\check{s}}(\tau )}}}}+
\nonumber \\
&+&Q_{i}A_{\check{r}}(\tau ,{\ \vec{\eta}}_{i}(\tau )),  \nonumber \\
&&{}  \nonumber \\
\pi _{\theta \,i}(\tau ) &=&{\frac{{\partial L(\tau )}}{{\partial {\dot{%
\theta}}_{i}(\tau )}}}=-{\frac{i}{2}}\theta _{i}^{\ast }(\tau ),  \nonumber
\\
\pi _{\theta ^{\ast }\,i}(\tau ) &=&{\frac{{\partial L(\tau )}}{{\partial {\
\dot{\theta}}_{i}^{\ast }(\tau )}}}=-{\frac{i}{2}}\theta _{i}(\tau ).
\label{III5}
\end{eqnarray}
\noindent The following Poisson brackets are assumed

\begin{eqnarray}
&&\{z^{\mu }(\tau ,\vec{\sigma}),\rho _{\nu }(\tau ,{\vec{\sigma}}^{^{\prime
}}\}=-\eta _{\nu }^{\mu }\delta ^{3}(\vec{\sigma}-{\vec{\sigma}}^{^{\prime
}}),  \nonumber \\
&&\{A_{\check{A}}(\tau ,\vec{\sigma}),\pi ^{\check{B}}(\tau,\vec{\sigma}
^{^{\prime }})\}=\eta _{\check{A}}^{\check{B}}\delta ^{3}(\vec{\sigma}-\vec{
\sigma}^{^{\prime }}),  \nonumber \\
&&\{\eta _{i}^{\check{r}}(\tau ),\kappa _{j{\check{s}}}(\tau )\}=-\delta
_{ij}\delta _{\check{s}}^{\check{r}},  \nonumber \\
&&\{\theta _{i}(\tau ),\pi _{\theta \,j}(\tau )\}=-\delta _{ij},  \nonumber
\\
&&\{\theta _{i}^{\ast }(\tau ),\pi _{\theta ^{\ast }\,j}(\tau
)\}=-\delta_{ij}.  \label{III6}
\end{eqnarray}

The Grassmann momenta give rise to the second class constraints $\pi_{\theta
\, i}+{\frac{i}{2}}\theta^{*}_i\approx 0$, $\pi_{\theta^{*}\, i}+{\frac{i}{2}
} \theta_i\approx 0$ [$\lbrace \pi_{\theta \, i}+{\frac{i}{2}}\theta^{*}_i,
\pi_{\theta^{*}\, j}+{\frac{i}{2}}\theta_j\rbrace =-i\delta_{ij}$]; $\pi
_{\theta \, i}$ and $\pi_{\theta^{*}\, i}$ are then eliminated with the help
of Dirac brackets

\begin{equation}
\lbrace A,B\rbrace {}^{*}=\lbrace A,B\rbrace -i[\lbrace A,\pi_{\theta \, i}+
{\frac{i}{2}}\theta^{*}_i\rbrace \lbrace \pi_{\theta^{*}\, i}+{\frac{i}{2}}
\theta_i,B\rbrace +\lbrace A,\pi_{\theta^{*}\, i}+{\frac{i}{2}} \theta_i
\rbrace \lbrace \pi_{\theta \, i}+{\frac{i}{2}}\theta^{*}_i,B\rbrace ],
\label{III7}
\end{equation}

\noindent so that the remaining Grassmann variables have the fundamental
Dirac brackets [which we will still denote $\lbrace .,.\rbrace$ for the sake
of simplicity]

\begin{eqnarray}
&&\{\theta _{i}(\tau ),\theta _{j}(\tau )\}=\{\theta _{i}^{\ast }(\tau
),\theta _{j}^{\ast }(\tau )\}=0,  \nonumber \\
&&\{\theta _{i}(\tau ),\theta_{j}^{\ast }(\tau )\}=-i\delta _{ij}.
\label{III8}
\end{eqnarray}

As in the free particle case of Section II, we obtain four primary
constraints

\begin{eqnarray}
&{\cal H}_{\mu }&(\tau ,\vec{\sigma})=\rho _{\mu }(\tau ,\vec{\sigma}
)-l_{\mu }(\tau ,\vec{\sigma})[T_{\tau \tau }(\tau ,\vec{\sigma})+
\sum_{i=1}^{N}\delta ^{3}(\vec{\sigma}-{\vec{\eta}}_{i}(\tau ))\times
\nonumber \\
&&\sqrt{m_{i}^{2}-\gamma ^{{\check{r}}{\check{s}}}(\tau ,\vec{\sigma}
)[\kappa _{i{\check{r}}}(\tau )-Q_{i} A_{\check{r}}(\tau ,\vec{\sigma}%
)][\kappa _{i{\check{s}}}(\tau )-Q_{i} A_{\check{s}}(\tau ,\vec{ \sigma})]}%
\,\,]-  \nonumber \\
&-&z_{{\check{r}}\mu }(\tau ,\vec{\sigma})\gamma ^{{\check{r}}{\check{s}}
}(\tau ,\vec{\sigma})\{-T_{\tau \check{s}}(\tau ,\vec{\sigma}
)+\sum_{i=1}^{N}\delta ^{3}(\vec{\sigma}-{\vec{\eta}}_{i}(\tau ))[\kappa _{i{%
\ \check{s}}}-Q_{i} A_{\check{s} }(\tau ,\vec{\sigma})]\}\approx 0,
\label{III9}
\end{eqnarray}

\noindent where

\begin{eqnarray}
T_{\tau \tau }(\tau ,\vec{\sigma}) &=&-{\frac{1}{2}}({\frac{1}{\sqrt{\gamma }
}}\pi ^{\check{r}}g_{{\check{r}}{\check{s}}}\pi ^{\check{s}}-{\frac{\sqrt{
\gamma }}{2}}\gamma ^{{\check{r}}{\check{s}}}\gamma ^{{\check{u}}{\check{v}}
}F_{{\check{r}}{\check{u}}}F_{{\check{s}}{\check{v}}})(\tau ,\vec{\sigma}),
\nonumber \\
T_{\tau {\check{s}}}(\tau ,\vec{\sigma}) &=&-F_{{\check{s}}{\check{t}}}(\tau
,\vec{\sigma})\pi ^{\check{t}}(\tau ,\vec{\sigma})=-\epsilon _{{\check{s}}{\
\check{t}}{\check{u}}}\pi ^{\check{t}}(\tau ,\vec{\sigma})B_{\check{u}}(\tau
,\vec{\sigma})=  \nonumber \\
&=&{[\vec{\pi}(\tau ,\vec{\sigma})\times \vec{B}(\tau ,\vec{\sigma})]}_{%
\check{s}},  \label{III10}
\end{eqnarray}

\noindent are the energy density and the Poynting vector respectively. We
use the notation $(\vec{\pi}\times \vec{B})_{\check{s}}=(\vec{E}\times \vec{%
B })_{\check{s}}$ because it is consistent with $\epsilon _{{\check{s}}{%
\check{ t}}{\check{u}}}\pi ^{\check{t}}B_{\check{u}}$ in the flat metric
limit $g_{{\check{A}}{\check{B}}}\rightarrow \eta _{{\check{A}}{\check{B}}}$%
; in this limit $T_{\tau \tau }\rightarrow {\frac{1}{2}}({\vec{E}}^{2}+{\vec{%
B}}^{2})$.

This form of the constraint displays both the tangential $z_{\check r}^{\mu
}(\tau ,\vec \sigma )$ and normal $l^{\mu }(\tau ,\vec \sigma )$ components
of the momentum $\rho ^{\mu }(\tau,\vec \sigma )$ conjugate to the embedding
variables $z^{\mu }(\tau ,\vec{\sigma})$.

Again, being solved in terms of the momenta $\rho _{\mu }(\tau ,\vec{\sigma}%
) $,these constraints are first class with exactly zero Poisson brackets ($\{%
{\cal H}_{\mu }(\tau ,\vec{\sigma}),{\cal H}_{\nu }(\tau ,{\vec{\sigma}}%
^{^{\prime }})\}=0$) and their existence implies once again that the
description of the system is independent of the choice of foliation.

Moreover, we have the (Lorentz scalar) primary constraints of the
electromagnetic field connected with the gauge invariance of the action

\begin{equation}
\pi^{\tau}(\tau ,\vec \sigma ) \approx 0.  \label{III11}
\end{equation}

From these constraints we construct the Dirac Hamiltonian. But first we must
construct the canonical Hamiltonian $H_{C}$. The canonical Hamiltonian is

\begin{eqnarray}
H_{c} &=&-\sum_{i=1}^{N}\kappa _{i{\check{r}}}(\tau ){\dot{\eta}}_{i}^{%
\check{r}}(\tau )+\int d^{3}\sigma \lbrack \pi ^{\check{A}}(\tau ,\vec{\sigma%
})\partial _{\tau }A_{\check{A}}(\tau ,\vec{\sigma})-\rho _{\mu }(\tau ,\vec{%
\sigma})z_{\tau }^{\mu }(\tau ,\vec{\sigma})-{\cal L}(\tau ,\vec{\sigma})]=
\nonumber \\
&=&\int d^{3}\sigma \lbrack \partial _{\check{r}}(\pi ^{\check{r}}(\tau ,%
\vec{\sigma})A_{\tau }(\tau ,\vec{\sigma}))-A_{\tau }(\tau ,\vec{\sigma}%
)\Gamma (\tau ,\vec{\sigma})]=-\int d^{3}\sigma A_{\tau }(\tau ,\vec{\sigma}%
)\Gamma (\tau ,\vec{\sigma}),  \label{III12}
\end{eqnarray}

\noindent after the elimination of a surface term.

Note that because of the $\tau $ and $\vec{\sigma}$ reparametrization
invariance, $H_{C}$ nearly vanishes, except for the portion involving

\begin{equation}
\Gamma (\tau ,\vec{\sigma})\equiv \partial _{\check{r}}\pi ^{\check{r}}(\tau
,\vec{\sigma})-\sum_{i=1}^{N}Q_{i} \delta ^{3}(\vec{\sigma}-{\vec{\eta}}%
_{i}(\tau )).  \label{III13}
\end{equation}
.

\noindent Thus the Dirac Hamiltonian is ($\lambda ^{\mu }(\tau ,\vec{\sigma})
$ and $\lambda _{\tau }(\tau ,\vec{\sigma})$ are Dirac multipliers)

\begin{equation}
H_{D}=\int d^{3}\sigma \lbrack \lambda ^{\mu }(\tau ,\vec{\sigma}){\cal H}
_{\mu }(\tau ,\vec{\sigma})+\lambda _{\tau }(\tau ,\vec{\sigma})\pi ^{\tau
}(\tau ,\vec{\sigma})-A_{\tau }(\tau ,\vec{\sigma})\Gamma (\tau ,\vec{\sigma}
)].  \label{III14}
\end{equation}

The requirement that the primary constraints be $\tau $ independent ($\{\pi
^{\tau }(\tau ,\vec{\sigma} ),H_{D}\}\approx 0$, $\{ {\cal H}^{\mu}(\tau ,%
\vec \sigma ), H_D \} \approx 0$) leads only to the Gauss's law secondary
constraint
\begin{equation}
\Gamma (\tau ,\vec{\sigma})\approx 0.  \label{III15}
\end{equation}

Since the embedding variables $z^{\mu}(\tau ,\vec \sigma )$ are the only
configuration variables with Lorentz indices, the ten conserved generators
of the Poincar\'{e} transformations are:

\begin{eqnarray}
P^{\mu } &=&p_{s}^{\mu }=\int d^{3}\sigma \rho ^{\mu }(\tau ,\vec{\sigma}),
\nonumber \\
J^{\mu \nu } &=&J_{s}^{\mu \nu }=\int d^{3}\sigma (z^{\mu }\rho ^{\nu
}-z^{\nu }\rho ^{\mu })(\tau ,\vec{\sigma}),  \label{III16}
\end{eqnarray}
(the subscript $s$ stands for hypersurface variable). From the first of
these we obtain

\begin{equation}
\{z^{\mu }(\tau ,\vec{\sigma}),p_{s}^{\nu }\}=-\eta ^{\mu \nu }.
\label{III17}
\end{equation}

We can restrict ourselves to foliations whose leaves are spacelike
hyperplanes $\Sigma _{H}(\tau )$ with constant timelike normal \bigskip $%
b_{\tau}^{\mu }\,\,\,(\partial _{\tau }b_{\tau }^{\mu }=0)$, by imposing the
following gauge fixings [$x^{\mu}_s$ is an arbitrary origin]

\begin{equation}
\zeta ^{\mu }(\tau ,\vec{\sigma})=z^{\mu }(\tau ,\vec{\sigma})-x_{s}^{\mu
}(\tau )-b_{\check{r}}^{\mu }(\tau )\sigma ^{\check{r}}\approx 0.
\label{III18}
\end{equation}
In this expression $b_{\check{r}}^{\mu }(\tau )$, $\check{r}=1,2,3$
are three orthonormal vectors, such that the constant and future
pointing normal to the hyperplane $\Sigma_H(\tau )$ [$b^{\mu}_{\check
A}=(b^{\mu}_{\tau}, b^{\mu}_{\check r})$ are orthonormal tetrads] is
\begin{equation}
l^{\mu }(\tau ,\vec{\sigma})\approx l^{\mu }=b_{\tau }^{\mu }=\varepsilon
_{\alpha \beta \gamma }^{\mu }b_{\check{1}}^{\alpha }(\tau )b_{\check{2}%
}^{\beta }(\tau )b_{\check{3}}^{\gamma }(\tau ).  \label{III19}
\end{equation}
Using the definitions of the vierbeins we obtain from the above gauge fixing
the simplifications

\begin{eqnarray}
z_{\check{r}}^{\mu }(\tau ,\vec{\sigma}) &\approx &b_{\check{r}}^{\mu }(\tau
),  \nonumber \\
z_{\tau }^{\mu }(\tau ,\vec{\sigma}) &\approx &\dot{x}_{s}^{\mu }(\tau )+%
\dot{b}_{\check{r}}^{\mu }(\tau )\sigma ^{\check{r}},  \nonumber \\
g_{\check{r}\check{s}}(\tau ,\vec{\sigma}) &\approx &-\delta _{\check{r}
\check{s}},\quad \gamma ^{\check{r}\check{s}}(\tau ,\vec{\sigma})\approx
-\delta ^{\check{r}\check{s}},\quad \gamma (\tau ,\vec{\sigma})\approx 1,
\label{III20}
\end{eqnarray}
as well as a natural decomposition of the Lorentz generators into orbital
and spin portions

\begin{eqnarray}
J_{s}^{\mu \nu } &=&x_{s}^{\mu }p_{s}^{\nu }-x_{s}^{\nu }p_{s}^{\mu
}+S_{s}^{\mu \nu },  \nonumber \\
&&S_{s}^{\mu \nu }=b_{\check{r}}^{\mu }(\tau )\int d^{3}\sigma \sigma ^{%
\check{r}}\rho ^{\nu }(\tau ,\vec{\sigma})-b_{\check{r}}^{\nu }(\tau )\int
d^{3}\sigma \sigma ^{\check{r}}\rho ^{\mu }(\tau ,\vec{\sigma}).
\label{III21}
\end{eqnarray}
Here $S_{s}^{\mu \nu }$ is the spin part of the Lorentz generators.

These gauge fixings have the following Poisson brackets with the primary
constraint ${\cal H}_{\mu }(\tau ,\vec{\sigma})\approx 0$

\begin{equation}
\{\zeta ^{\mu }(\tau ,\vec{\sigma}),{\cal H}_{\nu }(\tau ,\vec{\sigma}%
)\}=-\eta _{\nu }^{\mu }\delta ^{3}(\vec{\sigma}-\vec{\sigma}^{\prime }).
\label{III22}
\end{equation}
Therefore, we get a continuum set of second class constraints. They can be
eliminated by forming the Dirac brackets
\begin{equation}
\{A,B\}^{\ast }=\{A,B\}-\int [\{A,\zeta ^{\mu }(\tau ,\vec{\sigma})\}\{{\cal %
H}_{\mu }(\tau ,\vec{\sigma}),B\}-\{A,{\cal H}_{\mu }(\tau ,\vec{\sigma}%
)\}\{\zeta ^{\mu }(\tau ,\vec{\sigma}),B\}].  \label{III23}
\end{equation}
For example, one finds that

\begin{equation}
\{x_{s}^{\mu },p_{s}^{\nu }\}^{\ast }=-\eta ^{\mu \nu }.  \label{III24}
\end{equation}
In this way the infinity of continuum hypersurface degrees of freedom
$ z^{\mu }(\tau ,\vec{\sigma})$, $\rho _{\mu }(\tau ,\vec{\sigma})$,
are reduced to 20 : i) 8 are $x_{s}^{\mu }(\tau )$, $p_{s}^{\mu }$;
ii) 12 are the 6 independent pairs of canonical variables hidden in
$b_{\check{A}}^{\mu }$ and $S_{s}^{\mu \nu }=J_{s}^{\mu \nu
}-(x_{s}^{\mu }p_{s}^{\nu }-x_{s}^{\nu }p_{s}^{\mu })$ [they have the
following brackets consistent with the orthonormality of the tetrads
$b^{\mu}_{\check A}$ \cite{lu1}: $\{ b^{\mu}_{\check
A},b^{\nu}_{\check B} \} =0$, $\{ S^{\mu\nu}_s, b^{\rho}_{\check A} \}
= \eta^{\rho\nu}b^{\mu}_{\check A}-\eta^{\rho\mu}b^{\nu}_{\check A}$,
$\{ S^{\mu\nu}_s,S^{\alpha\beta}_s \} =
C^{\mu\nu\alpha\beta}_{\gamma\delta} S^{\gamma\delta}_s$, where
$C^{\mu\nu\alpha\delta}_{\gamma\delta}=\delta^{\nu}_{\gamma}\delta
^{\alpha}_{\delta}\eta^{\mu\beta}+\delta^{\mu}_{\gamma}\delta^{\beta}_{\delta}
\eta^{\nu\alpha}-\delta^{\nu}_{\gamma}\delta^{\beta}_{\delta}\eta^{\mu\alpha}
-\delta^{\mu}_{\gamma}\delta^{\alpha}_{\delta}\eta^{\nu\beta}$ are the structure
constants of the Lorentz algebra].

It can be shown \cite{lu1} that the following 10 first class constraints
survive at the level of the Dirac brackets [they are 10 combinations of the
primary constraints whose gauge freedom is not fixed by the gauge fixings (%
\ref{III18})]

\begin{eqnarray}
{\tilde{{\cal H^{\mu }}}}(\tau )&=&\int d^{3}\sigma {\cal H}^{\mu }(\tau ,%
\vec{\sigma})=  \nonumber \\
&=&p_s^{\mu} - l^{\mu}\Big[ \sum_{i=1}^N \sqrt{m^2_i+[{\vec \kappa}_i(\tau
)-Q_i \vec A(\tau ,{\vec \eta}_i(\tau )) ]^2}+  \nonumber \\
&+& {\frac{1}{2}} \int d^3\sigma [{\vec \pi}^2+{\vec B}^2](\tau ,\vec \sigma
) \Big]+  \nonumber \\
&+& b^{\mu}_{\check r}(\tau ) \Big[ \sum_{i=1}^N[\kappa_{i\check r}(\tau
)-Q_iA_{\check r}(\tau ,{\vec \eta}_i(\tau ))]+\int d^3\sigma [\vec \pi
\times \vec B]_{\check r}(\tau ,\vec \sigma )\Big] \approx 0,  \label{III25}
\end{eqnarray}

\noindent and

\begin{eqnarray}
{\tilde{{\cal H}}}^{\mu \nu }(\tau ) &=&b_{\check{r}}^{\mu }(\tau )\int
d^{3}\sigma \,\sigma ^{\check{r}}{\cal H}^{\nu }(\tau ,\vec{\sigma})-b_{%
\check{r}}^{\nu }(\tau )\int d^{3}\sigma \,\sigma ^{\check{r}}{\cal H}^{\mu
}(\tau ,\vec{\sigma})=  \nonumber \\
&=&S_{s}^{\mu \nu }-[b_{\check{r}}^{\mu }(\tau )b_{\tau }^{\nu }-b_{\check{r}%
}^{\nu }(\tau )b_{\tau }^{\mu }]\Big[\sum_{i=1}^{N}\eta _{i}^{\check{r}%
}(\tau )\sqrt{m_{i}^{2}+[{\vec{\kappa}}_{i}(\tau )-Q_{i}\vec{A}(\tau ,{\vec{%
\eta}}_{i}(\tau ))]^{2}}+  \nonumber \\
&+&{\frac{1}{2}}\int d^{3}\sigma \,\sigma ^{\check{r}}[{\vec{\pi}}^{2}+{\vec{%
B}}^{2}](\tau ,\vec{\sigma})\Big]-  \nonumber \\
&-&[b_{\check{r}}^{\mu }(\tau )b_{\check{s}}^{\nu }(\tau )-b_{\check{r}%
}^{\nu }(\tau )b_{\check{s}}^{\mu }(\tau )]\Big[\sum_{i=1}^{N}\eta _{i}^{%
\check{r}}(\tau )[\kappa _{i}^{\check{s}}(\tau )-Q_{i}A^{\check{s}}(\tau ,{%
\vec{\eta}}_{i}(\tau ))]+  \nonumber \\
&+&\int d^{3}\sigma \,\sigma ^{\check{r}}[\vec{\pi}\times \vec{B}]^{_{%
\check{s}}}(\tau ,\vec{\sigma})\Big]\approx 0.  \label{III26}
\end{eqnarray}

These constraints say that $p^{\mu}_s$ coincides with the total 4-momentum
of the isolated system and that $S_s^{\mu\nu}$ is determined by its spin
tensor.

The Dirac Hamiltonian becomes $H_D={\tilde \lambda}_{\mu}(\tau
){\tilde {\cal H}}^{\mu}(\tau )+{\tilde \lambda}_{\mu\nu}(\tau
){\tilde {\cal H}}^{\mu\nu}(\tau )+\int d^3\sigma [\lambda_{\tau}(\tau
,\vec \sigma )\pi^{\tau}(\tau ,\vec \sigma )-A_{\tau}(\tau ,\vec
\sigma )\Gamma (\tau ,\vec \sigma )]$, where, due to the associated
Hamilton equations, the new Dirac multipliers have the following
interpretation\cite{lu1}: ${\tilde \lambda}^{\mu}(\tau )\, {\buildrel
\circ \over =}\, -{\dot x}^{\mu}_s(\tau )$, ${\tilde \lambda}^{\mu\nu}(\tau )
=-{\tilde \lambda}^{\nu\mu}(\tau )\, {\buildrel \circ \over =}\,
{1\over 2}[{\dot b}^{\mu}_{\check r}(\tau )b^{\nu}_{\check r}(\tau
)-b^{\mu}_{\check r}(\tau ){\dot b}^{\nu}_{\check r}(\tau )]$.

Restricting our considerations to configurations with $p_{s}^{2}>0$, we make
a further canonical reduction to the special foliation whose hyperplanes are
orthogonal to $p_{s}^{\mu }.$ These hyperplanes are intrinsically determined
by the system itself and are called the Wigner hyperplanes $\Sigma _{W}(\tau
)$. They can be identified\cite{lu1} by requiring the gauge fixings $%
b^{\mu}_{\check A}(\tau ) \approx L^{\mu }{}_{\nu =A}(p_s,{\stackrel{\circ }{%
p}}_s)$ for the constraints ${\tilde {{\cal H}}}^{\mu\nu}(\tau )\approx 0$,
where $L^{\mu }{}_{\nu }(p_s,{\stackrel{\circ }{p}}_s)$ is the standard
Wigner boost for timelike Poincar\'e orbits. This implies $l^{\mu }=b_{\tau
}^{\mu }\approx p_{s}^{\mu }/\sqrt{p_{s}^{2}}$.

The rest frame form of a timelike fourvector $p^{\mu }$ is $\stackrel{\circ
}{p}{}^{\mu }=\eta \sqrt{p^{2}}(1;\vec{0})=\eta ^{\mu o}\eta \sqrt{p^{2}}$, $%
\stackrel{\circ }{p}{}^{2}=p^{2}$, where $\eta =sign\,p^{0}$. Since we
restricted ourselves to positive energy particles, $\eta_i=+1$, we shall put
$\eta=1$. The standard Wigner boost transforming $\stackrel{\circ }{p}%
{}^{\mu }$ into $p^{\mu }$ is

\begin{eqnarray}
L^{\mu }{}_{\nu }(p,\stackrel{\circ }{p}) &=&\epsilon _{\nu }^{\mu }(u(p))=
\nonumber \\
&=&\eta _{\nu }^{\mu }+2{\frac{{p^{\mu }{\stackrel{\circ }{p}}_{\nu }}}{{\
p^{2}}}}-{\frac{{(p^{\mu }+{\stackrel{\circ }{p}}^{\mu })(p_{\nu }+{%
\stackrel{\circ }{p}}_{\nu })}}{{p\cdot \stackrel{\circ }{p}+p^{2}}}}=
\nonumber \\
&=&\eta _{\nu }^{\mu }+2u^{\mu }(p)u_{\nu }(\stackrel{\circ }{p})-{\frac{{\
(u^{\mu }(p)+u^{\mu }(\stackrel{\circ }{p}))(u_{\nu }(p)+u_{\nu }(\stackrel{
\circ }{p}))}}{{1+u^{0}(p)}}},  \nonumber \\
&&{}  \nonumber \\
\nu =0 &&\epsilon _{0}^{\mu }(u(p))=u^{\mu }(p)=p^{\mu }/\sqrt{p^{2}},
\nonumber \\
\nu =r &&\epsilon _{r}^{\mu }(u(p))=(-u_{r}(p);\delta _{r}^{i}-{\frac{{\
u^{i}(p)u_{r}(p)}}{{1+u^{0}(p)}}}).  \label{III27}
\end{eqnarray}
The inverse of $L^{\mu }{}_{\nu }(p,\stackrel{\circ }{p})$ is $L^{\mu
}{}_{\nu }(\stackrel{\circ }{p},p)$, the standard boost to the system rest
frame, defined by

\begin{equation}
L^{\mu }{}_{\nu }(\stackrel{\circ }{p},p)=L_{\nu }{}^{\mu }(p,\stackrel{
\circ }{p})=L^{\mu }{}_{\nu }(p,\stackrel{\circ }{p}){|}_{\vec{p}\rightarrow
-\vec{p}}.  \label{III28}
\end{equation}

We also use these boosts to define the following vierbeins [the $\epsilon
_{r}^{\mu }(u(p))$ 's are also called polarization vectors; the indices $r,s$
will be used for $A=$1,2,3 and $\bar{o}$ for $A$=0]

\begin{eqnarray}
&&\epsilon _{A}^{\mu }(u(p))=L^{\mu }{}_{A}(p,\stackrel{\circ }{p}),
\nonumber \\
&&\epsilon _{\mu }^{A}(u(p))=L^{A}{}_{\mu }(\stackrel{\circ }{p},p)=\eta
^{AB}\eta _{\mu \nu }\epsilon _{B}^{\nu }(u(p)),  \nonumber \\
&&{}  \nonumber \\
&&\epsilon _{\mu }^{\bar{o}}(u(p))=\eta _{\mu \nu }\epsilon _{o}^{\nu
}(u(p))=u_{\mu }(p),  \nonumber \\
&&\epsilon _{\mu }^{r}(u(p))=-\delta ^{rs}\eta _{\mu \nu }\epsilon _{r}^{\nu
}(u(p))=(\delta ^{rs}u_{s}(p);\delta _{j}^{r}-\delta ^{rs}\delta _{jh}{\frac{%
{u^{h}(p)u_{s}(p)}}{{1+u^{o}(p)}}}),  \nonumber \\
&&\epsilon_{o}^{A}(u(p))=u_{A}(p),  \label{III29}
\end{eqnarray}

\noindent which satisfy

\begin{eqnarray}
&&\epsilon _{\mu }^{A}(u(p))\epsilon _{A}^{\nu }(u(p))=\eta _{\nu }^{\mu },
\nonumber \\
&&\epsilon _{\mu }^{A}(u(p))\epsilon _{B}^{\mu }(u(p))=\eta _{B}^{A},
\nonumber \\
&&\eta ^{\mu \nu }=\epsilon _{A}^{\mu }(u(p))\eta ^{AB}\epsilon _{B}^{\nu
}(u(p))=u^{\mu }(p)u^{\nu }(p)-\sum_{r=1}^{3}\epsilon _{r}^{\mu
}(u(p))\epsilon _{r}^{\nu }(u(p)),  \nonumber \\
&&\eta _{AB}=\epsilon _{A}^{\mu }(u(p))\eta _{\mu \nu }\epsilon _{B}^{\nu
}(u(p)),  \nonumber \\
&&p_{\alpha }{\frac{{\partial }}{{\partial p_{\alpha }}}}\epsilon _{A}^{\mu
}(u(p))=p_{\alpha }{\frac{{\partial }}{{\partial p_{\alpha }}}}\epsilon_{\mu
}^{A}(u(p))=0.  \label{III30}
\end{eqnarray}
With the Wigner rotation corresponding to the Lorentz transformation $%
\Lambda $ being

\begin{eqnarray}
R^{\mu }{}_{\nu }(\Lambda ,p) &=&{[L(\stackrel{\circ }{p},p)\Lambda
^{-1}L(\Lambda p,\stackrel{\circ }{p})]}^{\mu }{}_{\nu }=\left(
\begin{array}{cc}
1 & 0 \\
0 & R^{i}{}_{j}(\Lambda ,p)
\end{array}
\right) ,  \nonumber \\
{} &&{}  \nonumber \\
R^{i}{}_{j}(\Lambda ,p) &=&{(\Lambda ^{-1})}^{i}{}_{j}-{\frac{{(\Lambda
^{-1})^{i}{}_{o}p_{\beta }(\Lambda ^{-1})^{\beta }{}_{j}}}{{p_{\rho
}(\Lambda ^{-1})^{\rho }{}_{o}+\eta \sqrt{p^{2}}}}}-  \nonumber \\
&-&{\frac{{p^{i}}}{{p^{o}+\eta \sqrt{p^{2}}}}}[(\Lambda ^{-1})^{o}{}_{j}-{\
\frac{{((\Lambda ^{-1})^{o}{}_{o}-1)p_{\beta }(\Lambda ^{-1})^{\beta }{}_{j}}
}{{p_{\rho }(\Lambda ^{-1})^{\rho }{}_{o}+\eta \sqrt{p^{2}}}}}].
\label{III31}
\end{eqnarray}
we have that the polarization vectors transform under the Poincar\'{e}
transformations $(a,\Lambda )$ in the following way:

\begin{equation}
\epsilon _{r}^{\mu }(u(\Lambda p))=(R^{-1})_{r}{}^{s}\,\Lambda ^{\mu
}{}_{\nu }\,\epsilon _{s}^{\nu }(u(p))\text{)} .  \label{III32}
\end{equation}

These boosts can be used to obtain the further canonical reduction referred
to above. \ This takes place in two steps: i) firstly, one boosts to the
rest frame the variables $b_{\check{A}}^{\mu }$, $S_{s}^{\mu \nu }$, with
the standard Wigner boost $L^{\mu }{}_{\nu }(p_{s},\stackrel{\circ }{p}_{s})$
for timelike Poincar\'{e} orbits; ii) then, one adds the gauge-fixings $b_{%
\check{A}}^{\mu }-L^{\mu }{}_{A}(p_{s},\stackrel{\circ }{p}_{s})\approx 0$
and goes to Dirac brackets. It can be shown\cite{lu1} that after this
special gauge fixing the Lorentz scalar 3-indices $\check r$ become
Wigner spin 1 3-indices $r$. Therefore, we get a rest-frame instant
form of dynamics with Wigner covariance.

The Lorentz generators become $J_{s}^{\mu \nu }={\tilde{x}}_{s}^{\mu
}p_{s}^{\nu }-{\tilde{x}}_{s}^{\nu }p_{s}^{\mu }+{\tilde{S}}_{s}^{\mu \nu }$
with ${\tilde{S}}_{s}^{\mu \nu }$ given in Eq.(59) of Ref.\cite{lu1}. We now
get ${\tilde{{\cal H}}}^{\mu \nu }(\tau )\equiv 0$, i.e. $S_{s}^{\mu \nu }$
is forced to coincide with the spin tensor of the isolated system.

If we define the rest-frame spin tensor

\begin{eqnarray}
{\bar{S}}_{s}^{AB} &=&\epsilon _{\mu }^{A}(u(p_{s}))\epsilon _{\nu
}^{B}(u(p_{s}))S_{s}^{\mu \nu }\equiv \lbrack \eta _{\check{r}}^{A}\eta
_{\tau }^{B}-\eta _{\check{r}}^{B}\eta _{\tau }^{A}]\,[{\frac{1}{2}}\int
d^{3}\sigma \sigma ^{\check{r}}\,[{\vec{\pi}}^{2}+{\vec{B}}^{2}](\tau ,\vec{%
\sigma})+  \nonumber \\
&+&\sum_{i=1}^{N}\eta _{i}^{\check{r}}(\tau )\sqrt{m_{i}^{2}+[{\vec{\kappa}}%
_{i}(\tau )-Q_{i}\vec{A}(\tau ,{\vec{\eta}}_{i}(\tau ))]{}^{2}}]-  \nonumber
\\
&-&[\eta _{\check{r}}^{A}\eta _{\check{s}}^{B}-\eta _{\check{r}}^{B}\eta _{%
\check{s}}^{A}]\,[\int d^{3}\sigma \sigma ^{\check{r}}\,{[\vec{\pi}\times
\vec{B}]}_{\check{s}}{(\tau ,\vec{\sigma})}+  \nonumber \\
&+&\sum_{i=1}^{N}\eta _{i}^{\check{r}}(\tau )[\kappa _{i}^{\check{s}}(\tau
)-Q_{i}A^{\check{s}}(\tau ,{\vec{\eta}}_{i}(\tau ))]\,\,],  \nonumber \\
&&{}  \nonumber \\
{\bar{S}}_{s}^{rs} &\equiv &\sum_{i=1}^{N}\Big(\eta _{i}^{r}(\tau )[\kappa
_{i}^{s}(\tau )-Q_{i}A^{s}(\tau ,{\vec{\eta}}_{i}(\tau ))]-  \nonumber \\
&-&\sum_{i=1}^{N}\eta _{i}^{s}(\tau )[\kappa _{i}^{r}(\tau )-Q_{i}A^{s}(\tau
,{\vec{\eta}}_{i}(\tau ))]\Big)+  \nonumber \\
&+&\int d^{3}\sigma \,\Big(\sigma ^{r}[\vec{\pi}\times \vec{B}]^{s}(\tau ,%
\vec{\sigma})-\sigma ^{s}[\vec{\pi}\times \vec{B}]^{r}(\tau ,\vec{\sigma})%
\Big),  \nonumber \\
&&{}  \nonumber \\
{\bar{S}}_{s}^{\bar{0}r} &\equiv &-\sum_{i=1}^{N}\eta _{i}^{r}(\tau )\sqrt{%
m_{i}^{2}+[{\vec{\kappa}}_{i}(\tau )-Q_{i}\vec{A}(\tau ,{\vec{\eta}}%
_{i}(\tau ))]^{2}}-  \nonumber \\
&-&{\frac{1}{2}}\int d^{3}\sigma \,\sigma ^{r}[{\vec{\pi}}^{2}+{\vec{B}}%
^{2}](\tau ,\vec{\sigma}).  \label{III33}
\end{eqnarray}
it can be shown that the form of ${\tilde{S}}_{s}^{\mu \nu }$ implies that
the rest-frame ``external'' Poincar\'{e} generators are \cite{lu1}

\begin{eqnarray}
J_{s}^{ij} &:&={\tilde{x}}_{s}^{i}p_{s}^{j}-{\tilde{x}}_{s}^{j}p_{s}^{i}+%
\delta ^{ir}\delta ^{js}{\bar{S}}_{s}^{rs},  \nonumber \\
J_{s}^{oi} &:&={\tilde{x}}_{s}^{o}p_{s}^{i}-{\tilde{x}}_{s}^{i}p_{s}^{o}-{%
\frac{{\delta ^{ir}{\bar{S}}_{s}^{rs}p_{s}^{s}}}{{\ p_{s}^{o}+\eta _{s}\sqrt{%
p_{s}^{2}}}}}.  \label{III34}
\end{eqnarray}

Only in this special gauge do we get the separation of a decoupled
``external'' canonical non-covariant center of mass described by the 4 pairs
${\tilde{x}}_{s}^{\mu }(\tau )$, $p_{s}^{\mu }$, of canonical variables ($\{%
\tilde{x}_{s}^{\mu },p_{s}^{\nu }\}^{\ast \ast }=-\eta ^{\mu \nu }$)
identifying the Wigner hyperplane $\Sigma _{W}(\tau )$ [see Eq.(59) of Ref.
\cite{lu1} for the expression of ${\tilde{x}}_{s}^{\mu }(\tau )$ in terms of
$x_{s}^{\mu }$ and of the spin tensor] and of the ``internal''
Wigner-covariant canonical variables ${\vec{\eta}}_{i}(\tau )$, ${\vec{\kappa%
}}_{i}(\tau )$, $A_{A}(\tau ,\vec{\sigma})$, $\pi ^{A}(\tau ,\vec{\sigma})$
living inside the Wigner hyperplane and with the Dirac brackets coinciding
with the original Poisson brackets ( $\{\eta _{i}^{\check{r}}(\tau ),\kappa
_{j{\check{s}}}(\tau )\}^{\ast \ast }=-\delta _{ij}\delta _{\check{s}}^{%
\check{r}}$, $\{A_{\check{A}}(\tau ,\vec{\sigma}),\pi ^{\check{B}}(\tau ,%
\vec{\sigma}^{^{\prime }})\}=\eta _{\check{A}}^{\check{B}}\delta ^{3}(\vec{%
\sigma}-\vec{\sigma}^{^{\prime }})$).

As shown in Ref.\cite{lu1} one can replace $\tilde{x}_{s}^{\mu
},p_{s}^{\mu } $ with a new canonical basis $T_{s}=p_{s}\cdot
{\tilde{x}}_{s}/\sqrt{ p_{s}^{2}}=p_s\cdot x_s/\sqrt{p^2_s}$ (it is
the Lorentz-invariant rest frame time), $\varepsilon
_{s}=\sqrt{p_{s}^{2}},\vec{z}_{s}=\sqrt{p_{s}^{2}}(
{\tilde{\vec{x}}}_{s}-{\tilde{x}}_{s}^{o}{\vec{p}}_{s}/p_{s}^{o}),\vec{k}%
_{s}=\vec{u}(p_{s})$ with $\vec{z}_{s}$ having the same covariance of the
Newton-Wigner position operator under the little group $O(3)$ of the
timelike Poincar\'e orbits. The 3-position canonical variable ${\vec
z}_s/\epsilon_s$ is the classical background of this operator and
describes the decoupled ``external" 3-center of mass, whose 4-position
is ${\tilde x}^{\mu}_s$.

In this special gauge there is no restriction on $p_{s}^{\mu }$: the four
velocity $u^{\mu }(p_{s})=p_{s}^{\mu }/\sqrt{p_{s}^{2}}=l^{\mu }$ describes
the orientation of the Wigner hyperplane with respect to an arbitrary
Lorentz frame. \

We obtain the following form for the constraints ${\tilde{{\cal H^{\mu }}}}%
(\tau )\approx 0$:
\begin{eqnarray}
{\tilde{{\cal H^{\mu }}}}(\tau ) &=&\int d^{3}\sigma {\cal H}^{\mu }(\tau ,%
\vec{\sigma})=p_{s}^{\mu }-  \nonumber \\
&&-u^{\mu }(p)\Big({\frac{1}{2}}\int d^{3}\sigma \lbrack {\vec{\pi}}^{2}+{%
\vec{B}}^{2}](\tau ,\vec{\sigma})+  \nonumber \\
&+&\sum_{i=1}^{N}\sqrt{m_{i}^{2}+[{\vec{\kappa}}_{i}(\tau )-Q_{i}\vec{A}%
(\tau ,{\vec{\eta}}_{i}(\tau ))]{}^{2}}\Big)-  \nonumber \\
&-&\epsilon _{r}^{\mu }(u(p))\Big(\int d^{3}\sigma \lbrack {\vec{\pi}}\times
{\vec{B}]}^{r}(\tau ,\vec{\sigma})+  \nonumber \\
&+&\sum_{i=1}^{N}[{\vec{\kappa}}_{i}(\tau )-Q_{i}\vec{A}(\tau ,{\vec{\eta}}%
_{i}(\tau ))]^{r}\Big)\approx 0.  \label{III35}
\end{eqnarray}
Their projections along the normal and the tangents to the Wigner hyperplane
are

\begin{eqnarray}
{\cal H}(\tau ) &=&u^{\mu }(p_{s}){\tilde{{\cal H}}}_{\mu }(\tau )=
\nonumber \\
&=&\sqrt{p_{s}^{2}}-\Big(\sum_{i=1}^{N}\sqrt{m_{i}^{2}+[{\vec{\kappa}}%
_{i}(\tau )-Q_{i}\vec{A}(\tau ,{\vec{\eta}}_{i}(\tau ))]{}^{2}}+  \nonumber
\\
&+&{\frac{1}{2}}\int d^{3}\sigma \lbrack {\vec{\pi}}^{2}+{\vec{B}}^{2}](\tau
,\vec{\sigma})\Big)\approx 0,  \label{III36}
\end{eqnarray}
\begin{equation}
{{\cal \vec{H}}}_{p}(\tau )=\sum_{i=1}^{N}[{\vec{\kappa}}_{i}(\tau )-Q_{i}%
\vec{A}(\tau ,{\vec{\eta}}_{i}(\tau ))]+\int d^{3}\sigma \lbrack {\vec{\pi}}%
\times {\vec{B}]}(\tau ,\vec{\sigma})\approx 0.  \label{III37}
\end{equation}

\ The first one gives the mass spectrum of the isolated field plus particle
system, while the other three say that the total 3-momentum of the $N$
charged particles plus fields vanishes inside the Wigner hyperplane $%
\Sigma_{W}(\tau )$. This condition is the rest-frame condition identifying
the Wigner hyperplane as the rest frame of the isolated system.

The Dirac Hamiltonian is now $H_{D}={\tilde{\lambda}}^{\mu }(\tau ){\tilde{%
{\cal H}}}_{\mu }(\tau )=\lambda (\tau ){\cal H}(\tau )-\vec{\lambda}(\tau
)\cdot {\vec{{\cal H}}}_{p}(\tau )$, with $\lambda (\tau )\approx -{\dot x}%
_{s\mu}(\tau ) u^{\mu}(p_s)$, $\lambda_r(\tau )\approx -{\dot x}_{s\mu}(\tau
) \epsilon^{\mu}_r(u(p_s))$.

The two additional electromagnetic constraints are

\begin{eqnarray}
\pi ^{\tau }(\tau ,\vec{\sigma}) &\approx &0,  \nonumber \\
\Gamma (\tau ,\vec{\sigma}) &\approx &0.  \label{III38}
\end{eqnarray}
In the rest-frame instant form of dynamics on the Wigner hyperplanes they
are Lorentz scalar constraints [$A_{\tau }(\tau ,\vec{\sigma})$ and $\pi
^{\tau }(\tau ,\vec{\sigma})$ are Lorentz scalars, while ${\vec{A}}(\tau ,%
\vec{\sigma})$ and $\vec{\pi}(\tau ,\vec{\sigma})$ are spin-1 Wigner
3-vectors].

We now eliminate the electromagnetic gauge degrees of freedom by decomposing
the above spin-one Wigner 3-vector canonical field variables into their
transverse and longitudinal components\cite{lu1}

\begin{eqnarray}
{\vec{A}}(\tau ,\vec{\sigma}) &=&{\check{\vec{A}}}_{\perp }(\tau ,\vec{%
\sigma })-{\frac{{\vec{\partial}}}{{\triangle }}}\,\vec{\partial}\cdot \vec{A%
}(\tau ,\vec{\sigma}),  \nonumber \\
\vec{\pi}(\tau ,\vec{\sigma}) &=&{\check{\vec{\pi}}}_{\perp }(\tau ,\vec{
\sigma})-{\frac{{\vec{\partial}}}{{\triangle }}}\,\vec{\partial}\cdot \vec{
\pi}(\tau ,\vec{\sigma})\approx  \nonumber \\
&\approx& {\check {\vec \pi}}_{\perp}(\tau ,\vec \sigma )+ {\frac{{\vec %
\partial}}{{\triangle}}} \sum_{i=1}^N Q_i \delta^3(\vec \sigma - {\vec \eta}%
_i(\tau )),  \label{III39}
\end{eqnarray}
with $\Delta =-\vec{\partial}^{2}.$

We re-express everything in terms of the Dirac observables: i) ${\check{\vec{%
A}}}_{\perp }(\tau ,\vec{\sigma})$, ${\ \check{\vec{\pi}}}_{\perp }(\tau ,%
\vec{\sigma})$, $\{{\check{A}}_{\perp }^{r}(\tau ,\vec{\sigma}),{\check{\pi}}%
_{\perp }^{s}(\tau ,{\vec{\sigma}} ^{^{\prime }})\}=-P_{\perp }^{rs}(\vec{%
\sigma})\delta ^{3}(\vec{\sigma}-{\ \vec{\sigma}}^{^{\prime }})$ [$P_{\perp
}^{rs}(\vec{\sigma})=\delta ^{rs}+{\ \frac{{\partial ^{r}\partial ^{s}}}{{%
\triangle }}}$] for the electromagnetic field; ii) ${\vec{\eta}}_{i}(\tau )$%
, ${\check{\vec{\kappa}}}_{i}(\tau )={\vec{\kappa}}_{i}(\tau )+Q_{i}{\frac{{%
\vec{\partial}}}{{\triangle }}}\,\vec{ \partial}\cdot \vec{A}(\tau ,\vec{%
\sigma})$ for the particles [they now become dressed with a Coulomb cloud];
iii) ${\check{\theta}}_{i}^{\ast }(\tau )$, ${\check{\theta}}_{i}(\tau )$,
such that $Q_{i}=e\theta _{i}^{\ast }\theta _{i}=e{\check{\theta}}_{i}^{\ast
}{\check{\theta}}_{i} $.

This is known as the Wigner-covariant rest-frame radiation gauge. Note that
\ ${\vec{\kappa}}_{i}(\tau )-Q_{i}\vec{A}(\tau ,{\vec{\eta}}_{i}(\tau ))=%
\check{\vec{\kappa}}_{i}(\tau )-Q_{i}\vec{A}_{\perp }(\tau ,\vec{\eta}%
_{i}(\tau )).$ Using the Gauss law constraint $\vec \partial \cdot \vec{\pi}%
(\tau ,\vec{\sigma})\approx \Sigma _{i}Q_{i}\delta ^{3}(\vec{\sigma}-\vec{%
\eta}_{i})$ with $\frac{1}{\Delta }\delta ^{3}(\vec{\sigma}-\vec{\eta}_{i})$
$=-1/(4\pi |\vec{\sigma}-\vec{\eta}_{i}|)$ and integrating by parts we
separate out the Coulomb portion of the rest frame energy from the field
energy integral. \ A similar procedure on the field momentum integral
simplifies the rest frame condition (and the expression for the internal
angular momentum in the next section). \ Thus we find that the reduced form
of the 4 constraints is

\begin{eqnarray}
{\cal H}(\tau ) &=&\epsilon _{s}-\{\sum_{i=1}^{N}\sqrt{m_{i}^{2}+(\check{%
\vec{\kappa}}_{i}(\tau )-Q_{i}{\check{\vec{A}}}_{\perp }(\tau ,\vec{\eta}%
_{i}(\tau )))^{2}}+  \nonumber \\
&+&\sum_{i\neq j}\frac{Q_{i}Q_{j}}{4\pi \mid \vec{\eta}_{i}(\tau )-\vec{\eta}%
_{j}(\tau )\mid }+\int d^{3}\sigma {\frac{1}{2}}[\check{\vec{\pi}}_{\perp
}^{2}+\check{\vec{B}}^{2}](\tau ,\vec{\sigma})\}=  \nonumber \\
&=&\epsilon _{s}-M\approx 0,  \nonumber \\
{\vec{{\cal H}}}_{p}(\tau ) &=&\Sigma _{i}\check{\vec{\kappa}}_{i}(\tau
)+\int d^{3}\sigma \lbrack {\check{\vec{\pi}}}_{\perp }\times {\check{\vec{B}%
}}](\tau ,\vec{\sigma})\approx 0,  \label{III40}
\end{eqnarray}
where $\epsilon _{s}=\sqrt{p_{s}^{2}}$ and $M$ is the invariant mass of the
isolated system. The rest-frame condition does not depend any more on the
interaction as it must be in an instant form of dynamics. This procedure not
only extracts the Coulomb potential from field theory but also regularizes
the Coulomb interaction due to the semiclassical property $Q_{i}^{2}=0$.

If we add the gauge fixing $T_s-\tau \approx 0$ we get $\lambda (\tau )=-1$
and the Dirac Hamiltonian for the evolution in the rest-frame time is $H_D=M
-\vec \lambda (\tau )\cdot {\vec {{\cal H}}}_p(\tau )$ [see the more accurate
discussion after Eq.(\ref{IV3})].

The embedding corresponding to the Wigner hyperplanes in the gauge $%
T_s\equiv \tau$ is $z^{\mu}(\tau ,\vec \sigma )=x^{\mu}_s(\tau
)+\epsilon^{\mu}_r(u(p_s)) \sigma^r$ with the origin of the 3-coordinates
given by $x^{\mu}_s(\tau )=x^{\mu}_s(0) +u^{\mu}(p_s) \tau
+\epsilon^{\mu}_r(u(p_s))\int_0^{\tau} d\tau^{^{\prime}}\,
\lambda_r(\tau^{^{\prime}})$ since ${\dot x}^{\mu}_s(\tau )=u^{\mu}(p_s)
+\epsilon^{\mu}_r(u(p_s)) \lambda_r(\tau )$ [instead for the
``external'' center of mass we have ${\dot {\tilde x}} ^{\mu}_s(\tau
)=u^{\mu}(p_s)$]. The final canonical variables are: i) ${\vec z}_s$,
${\vec k}_s$ (the decoupled ``external" 3-center of mass); ii) ${\vec
\eta}_i(\tau )$, ${\check {\vec \kappa}}_i(\tau )$ (the particle variables);
iii) ${\check {\vec A}}_{\perp}(\tau ,\vec \sigma )$, ${\check {\vec
\pi}}_{\perp}(\tau ,\vec \sigma )$ (the transverse radiation field).
They are still restricted by the three rest-frame conditions ${\vec
{\cal H}}_p(\tau )\approx 0$.

\subsection{The Energy-Momentum Tensor.}

The Euler-Lagrange equations from the action (\ref{III4})) are

\begin{eqnarray}
&&({\frac{{\partial {\cal L}}}{{\partial z^{\mu }}}}-\partial _{\check{A}}{%
\frac{{\partial {\cal L}}}{{\partial z_{\check{A}}^{\mu }}}})(\tau ,\vec{%
\sigma})=\eta _{\mu \nu }\partial _{\check{A}}[\sqrt{g}T^{\check{A}\check{B}%
}\,z_{\check{B}}^{\nu }](\tau ,\vec{\sigma})\,\stackrel{\circ }{=}\,0,
\nonumber \\
&&({\frac{{\partial L}}{{\partial \eta _{i}^{r}}}}-{\frac{d}{{d\tau }}}{%
\frac{{\partial L}}{{\partial {\dot{\eta}}_{i}^{r}}}})(\tau )\,\stackrel{%
\circ }{=}\,0,  \nonumber \\
&&({\frac{{\partial {\cal L}}}{{\partial A_{\check{A}}}}}-\partial _{\check{B%
}}{\frac{{\partial {\cal L}}}{{\partial \partial _{\check{B}}A_{\check{A}}}}}%
)(\tau ,\vec{\sigma})\,\stackrel{\circ }{=}\,0,  \label{III41}
\end{eqnarray}

\noindent where we introduced the total energy-momentum tensor [${\dot \eta}^{%
\check A}_i(\tau )=(1; {\dot \eta}_i^{\check r}(\tau ))$]

\begin{eqnarray}
T^{\check A\check B}(\tau ,\vec \sigma )&=&-[ {\frac{2}{\sqrt{g}}} {\frac{{%
\delta S}}{{\delta g_{\check A\check B}}}}](\tau ,\vec \sigma )=  \nonumber
\\
&=&\sum_{i=1}^N \delta^3(\vec \sigma -{\vec \eta}_i(\tau )) {\frac{{m_i {%
\dot \eta}_i^{\check A}(\tau ) {\dot \eta}_i^{\check B}(\tau )}}{{\sqrt{g}
\sqrt{g_{\check C\check D}{\dot \eta}_i^{\check C}(\tau ) {\dot \eta}_i^{%
\check D}(\tau )} }}}+  \nonumber \\
&+& [F^{\check A\check C}F_{\check C}{}^{\check B} +{\frac{1}{4}} g^{\check A%
\check B}F^{\check C\check D}F_{\check C\check D}] (\tau ,\vec \sigma ).
\label{III42}
\end{eqnarray}

When $\partial _{\check{A}}[\sqrt{g}z_{\check{B}}^{\mu }](\tau ,\vec
\sigma )=0$ as happens on the Wigner hyperplanes in the gauge $T_{s}-\tau
\approx 0$, $\vec{\lambda} (\tau )=0$, we get the conservation of the
energy-momentum tensor $T^{\check{ A}\check{B}}(\tau ,\vec{\sigma})$,
i.e. $\partial _{\check{A}}T^{\check{A}%
\check{B}}\,\stackrel{\circ }{=}\,0$. Otherwise there is compensation coming
from the dynamics of the hypersurface.

On the Wigner hyperplanes the energy-momentum tensor becomes

\begin{eqnarray}
T^{\tau\tau}(\tau ,\vec \sigma )&=&\sum_{i=1}^N \delta^3(\vec \sigma - {\vec %
\eta}_i(\tau )) \sqrt{m^2_i+[{\vec \kappa}_i(\tau )-Q_i\vec A(\tau ,{\vec %
\eta}_i(\tau ))]^2} +{\frac{1}{2}}[{\vec \pi}^2+{\vec B}^2](\tau ,\vec \sigma
),  \nonumber \\
T^{r\tau}(\tau ,\vec \sigma )&=&\sum_{i=1}^N\delta^3(\vec \sigma -{\vec \eta}%
_i(\tau )) [\kappa_i^r(\tau )-Q_i A^r(\tau ,{\vec \eta}_i(\tau ))] +[{\vec %
\pi}\times \vec B](\tau ,\vec \sigma ),  \nonumber \\
T^{rs}(\tau ,\vec \sigma )&=& \sum_{i=1}^N \delta^3(\vec \sigma -{\vec \eta}%
_i(\tau )) {\frac{{\ [\kappa_i^r(\tau )-Q_iA^r(\tau ,{\vec \eta}_i(\tau ))]
[\kappa_i^s(\tau )-Q_iA^s(\tau ,{\vec \eta}_i(\tau ))]}}{{\sqrt{m_i^2+[{\vec %
\kappa}_i(\tau )-Q_i\vec A(\tau ,{\vec \eta}_i(\tau ))]^2} }}} -  \nonumber
\\
&-&\Big[ {\frac{1}{2}}\delta^{rs} [{\vec \pi}^2+{\vec B}^2]
-[\pi^r\pi^s+B^rB^s]\Big] (\tau ,\vec \sigma ).  \label{III43}
\end{eqnarray}

Finally, after the canonical reduction, which eliminates the electromagnetic
gauge degrees of freedom and the choice of the gauge $T_s\equiv \tau$, we get

\begin{eqnarray}
T^{\tau\tau}(\tau ,\vec \sigma )&=&\sum_{i=1}^N \delta^3(\vec \sigma - {\vec %
\eta}_i(\tau )) \sqrt{m^2_i+[{\check {\vec \kappa}}_i(\tau ) -Q_i{\check {%
\vec A}}_{\perp}(\tau ,{\vec \eta}_i(\tau ))]^2} +  \nonumber \\
&+&{\frac{1}{2}}[\Big( {\check {\vec \pi}}_{\perp}+\sum_{i=1}^NQ_i{\frac{{%
\vec \partial}}{{\triangle}}}\delta^3(\vec \sigma -{\vec \eta}_i(\tau ))\Big)%
^2 +{\vec B}^2](\tau ,\vec \sigma ),  \nonumber \\
T^{r\tau}(\tau ,\vec \sigma )&=&\sum_{i=1}^N\delta^3(\vec \sigma -{\vec \eta}%
_i(\tau )) [{\check \kappa}_i^r(\tau )-Q_i {\check A}_{\perp}^r(\tau ,{\vec %
\eta}_i(\tau ))] +  \nonumber \\
&+&[\Big( {\check {\vec \pi}}_{\perp}+\sum_{i=1}^NQ_i{\frac{{\vec \partial}}{%
{\triangle}}}\delta^3(\vec \sigma -{\vec \eta}_i(\tau ))\Big)\times \vec B%
](\tau ,\vec \sigma ),  \nonumber \\
T^{rs}(\tau ,\vec \sigma )&=& \sum_{i=1}^N \delta^3(\vec \sigma -{\vec \eta}%
_i(\tau )) {\frac{{\ [{\check \kappa}_i^r(\tau )-Q_i{\check A}%
_{\perp}^r(\tau ,{\vec \eta}_i(\tau ))] [{\check \kappa}_i^s(\tau )-Q_i{%
\check A}_{\perp}^s(\tau ,{\vec \eta}_i(\tau ))]}}{{\sqrt{m_i^2+[{\check {%
\vec \kappa}}_i(\tau )-Q_i{\check {\vec A}}_{\perp}(\tau ,{\vec \eta}_i(\tau
))]^2} }}} -  \nonumber \\
&-&\Big[ {\frac{1}{2}}\delta^{rs} [\Big( {\check {\vec \pi}}%
_{\perp}+\sum_{i=1}^NQ_i{\frac{{\vec \partial}}{{\triangle}}}\delta^3(\vec %
\sigma -{\vec \eta}_i(\tau ))\Big)^2+{\vec B}^2] -  \nonumber \\
&-&[\Big( {\check {\vec \pi}}_{\perp}+\sum_{i=1}^NQ_i{\frac{{\vec \partial}}{%
{\triangle}}}\delta^3(\vec \sigma -{\vec \eta}_i(\tau ))\Big)^r \Big( {%
\check {\vec \pi}}_{\perp}+\sum_{i=1}^NQ_i{\frac{{\vec \partial}}{{\triangle}%
}}\delta^3(\vec \sigma -{\vec \eta}_i(\tau ))\Big)^s +  \nonumber \\
&+&B^rB^s]\Big] (\tau ,\vec \sigma ).  \label{III44}
\end{eqnarray}

\vfill\eject

\section{ Internal Poincar\'e Algebra and Equations of Motion for the
Electromagnetic Field and the N Charged Particles.}

In this Section we first build a realization of the Poincar\'{e} algebra
inside the Wigner hyperplane\ using results of the previous Section. Then,
by identifying the Lorentz scalar rest-frame time $T_{s}$ with the invariant
time $\tau $ labeling the hypersurfaces $\Sigma (\tau )$, we arrive at the
Dirac Hamiltonian $H_{D}=M-\vec{\lambda}(\tau )\cdot {\cal \vec{H}}_{p}(\tau
)$, in which the only gauge freedom left is the one associated with the
rest-frame condition. We then obtain the Hamilton and Lagrange equations for
fields and particles. Then we describe how to find the canonical
``internal'' center of mass ${\vec{q}}_{+}$ for fields and particles on the
Wigner hyperplane. The natural gauge fixing to the rest-frame \ conditions $%
{\cal \vec{H}}_{p}(\tau )\approx 0$ are $\vec{q}_{+}\approx 0$: they imply $%
\vec{\lambda}(\tau )=0$ and the decoupling of the ``internal'' center of
mass from the ``internal'' relative motions. In this way only the
``external'' decoupled 3-center of mass ${\vec z}_s$ remains (the
Newton-Wigner-like 3-position which replaces the 4-center of mass
$\tilde{x}_{s}^{\mu }$ in the gauge $T_s\equiv \tau$). A property of
the particle accelerations of any order, which will be needed in the
next Section, is derived.

\subsection{ Internal Poincar\'{e} Algebra}

In the rest-frame instant form of the dynamics there is another realization
of the Poincar\'{e} algebra besides the ``external'' one given in Eq.(\ref
{III34}). This is the ``internal'' realization built in terms of the
variables living inside each Wigner hyperplane. The associated generators of
this internal Poincar\'{e} group are given by (for positive energies in the
Wigner-covariant rest frame radiation gauge)
\begin{eqnarray}
{\cal P}_{(int)}^{\tau } &=&M=\sum_{i=1}^{N}\sqrt{m_{i}^{2}+(\check{\vec{%
\kappa}}_{i}(\tau )-Q_{i}{\check{\vec{A}}}_{\perp }(\tau ,\vec{\eta}%
_{i}(\tau )))^{2}}+  \nonumber \\
&+&\sum_{i\neq j}\frac{Q_{i}Q_{j}}{4\pi \mid \vec{\eta}_{i}(\tau )-\vec{\eta}%
_{j}(\tau )\mid }+\int d^{3}\sigma {\frac{1}{2}}[\check{\vec{\pi}}_{\perp
}^{2}+\check{\vec{B}}^{2}](\tau ,\vec{\sigma}),  \nonumber \\
&&{}  \nonumber \\
{\cal \vec{P}}_{(int)} &=&{\vec{{\cal H}}}_{p}={\check{\vec{\kappa}}}%
_{+}(\tau )+\int d^{3}\sigma \lbrack {\check{\vec{\pi}}}_{\perp }\times {%
\check{\vec{B}}}](\tau ,\vec{\sigma})\approx 0,  \nonumber \\
&&{}  \nonumber \\
{\cal J}_{(int)}^{r} &=&\varepsilon ^{rst}{\bar{S}}_{s}^{st}=\sum_{i=1}^{N}(%
\vec{\eta}_{i}(\tau )\times {\check{\vec{\kappa}}}_{i}(\tau ))^{r}+\int
d^{3}\sigma \,(\vec{\sigma}\times \,{[{\check{\vec{\pi}}}}_{\perp }{\times {%
\check{\vec{B}}]}}^{r}{(\tau ,\vec{\sigma})},  \nonumber \\
{\cal K}_{(int)}^{r} &=&{\bar{S}}_{s}^{\bar{o}r}=-{\bar{S}}_{s}^{r\bar{o}%
}=-\sum_{i=1}^{N}\vec{\eta}_{i}(\tau )\sqrt{m_{i}^{2}+[{{\check{\vec{\kappa}}%
}}_{i}(\tau )-Q_{i}{\check{\vec{A}}}_{\perp }(\tau ,{\ \vec{\eta}}_{i}(\tau
))]{}^{2}}+  \nonumber \\
&+&\sum_{i=1}^{N}{\LARGE [}\sum_{j\not=i}^{1..N}Q_{i}Q_{j}[{\frac{1}{{%
\triangle _{{\vec{\eta}}_{j}}}}}{\frac{{\partial }}{{\partial \eta _{j}^{r}}}%
}c({\vec{\eta}}_{i}(\tau )-{\vec{\eta}}_{j}(\tau ))-\eta _{j}^{r}(\tau )c({%
\vec{\eta}}_{i}(\tau )-{\vec{\eta}}_{j}(\tau ))]+  \nonumber \\
&+Q_{i}&\int d^{3}\sigma {\check{\pi}}_{\perp }^{r}(\tau ,\vec{\sigma})c(%
\vec{\sigma}-{\ \vec{\eta}}_{i}(\tau )){\LARGE ]}-{\frac{1}{2}}\int
d^{3}\sigma \sigma ^{r}\,({{\check{\vec{\pi}}}}_{\perp }^{2}+{{\check{\vec{B}%
}}}^{2})(\tau ,\vec{\sigma}),  \label{IV1}
\end{eqnarray}
in which ${\check{\vec{\kappa}}}_{+}(\tau )=\Sigma _{i}\check{\vec{\kappa}}%
_{i}(\tau )$ and  $c({\vec{\eta}}_{i}-{\vec{\eta}}_{j}):=-1/(4\pi |\vec{\eta}%
_{j}-\vec{\eta}_{i}|)$. \ The latter two Lorentz generators are determined
as the components of the spin tensor ${\bar{S}}_{s}^{AB}$ defined in Eq.(\ref
{III33}) inside each Wigner hyperplane.

The Dirac Hamiltonian is
\begin{equation}
H_{D}=\lambda (\tau ){\cal H}(\tau )-\vec{\lambda}(\tau )\cdot {\vec{{\cal H}
}}_{p}(\tau ).  \label{IV2}
\end{equation}
As already said, if we add the gauge-fixing

\begin{equation}
\chi =T_{s}-\tau \approx 0, \quad\quad T_{s} \equiv \frac{p_{s}\cdot {\tilde %
x}_{s}}{\sqrt{p_{s}^{2}}}={\frac{{\ p_s\cdot x_s}}{\sqrt{p^2_s}}},
\label{IV3}
\end{equation}

\noindent implying that the Lorentz scalar parameter $\tau$ labelling the
leaves of the foliation of Minkowski spacetime with Wigner hyperplanes
coincides with the rest-frame time $T_s$ of the decoupled point particle
clock (the ``external" center of mass) ${\tilde x}^{\mu}_s$, its
conservation in $\tau $ will imply $\lambda (\tau )=-1$ so that, after
taking the Dirac brackets associated with the second class constraints $%
\epsilon_s-M\approx 0$ and $T_s-\tau \approx 0$ (this eliminates $T_s$ and
$\epsilon_s$), the final Dirac Hamiltonian
in this gauge would be $H_D=-\vec \lambda (\tau ) \cdot {\vec {{\cal H}}}%
_p(\tau )$. However, if we wish to reintroduce the evolution in $\tau \equiv
T_s$ in this frozen phase space [containing the canonical variables
${\vec z}_s$, ${\vec k}_s$, ${\vec \eta}_i(\tau )$, ${\check {\vec
\kappa}}_i(\tau )$, ${\check {\vec A}}_{\perp}(\tau ,\vec \sigma )$,
${\check {\vec \pi}}_{\perp}(\tau ,\vec \sigma )$] we must use the
Hamiltonian

\begin{equation}
H_D=M-\vec{\lambda}(\tau )\cdot \vec{{\cal H}}_{p}(\tau ),  \label{IV4}
\end{equation}
because $M={\cal P}^{\tau}_{(int)}$ is the invariant mass and the
``internal" energy generator of the isolated system [it is like with the
frozen Hamilton-Jacobi theory, in which the time evolution can be
reintroduced by using the energy generator of the Poincar\'e group as
Hamiltonian].

The only remaining first class constraints are the rest-frame conditions ${%
\vec{{\cal H}}}_{p}(\tau )\approx 0$. The Dirac multipliers $\vec{\lambda}%
(\tau )$ describe the remaining gauge freedom on the location of the
``internal'' center of mass on the Wigner hyperplanes. In the next
Subsection we will study the natural gauge fixings for these first class
constraints. After this final canonical reduction the isolated system will
be described by the decoupled ``external'' 3-center-of-mass canonical
variables ${\vec{z}}_{s}$, ${\vec{k}}_{s}$ and by relative Wigner-covariant
degrees of freedom on the Wigner hyperplane, with the invariant mass $M$ as
the Hamiltonian for the evolution in $\tau \equiv T_{s}$.

\subsection{\noindent The equations of motion for particles and fields.}

The Hamilton-Dirac equations associated to the previous Hamiltonian are

\begin{eqnarray}
\dot{\vec{\eta}}_{i}(\tau )\, &\,\stackrel{\circ }{=}&\,\frac{\check{\vec{%
\kappa}}_{i}(\tau )-Q_{i}\check{\vec{A}}_{\perp }(\tau ,\vec{\eta}_{i}(\tau
))}{\sqrt{m_{i}^{2}+(\check{\vec{\kappa}}_{i}(\tau )-Q_{i}\check{\vec{A}}%
_{\perp }(\tau ,\vec{\eta}_{i}(\tau )))^{2}}}-\vec{\lambda}(\tau ),
\nonumber \\
\dot{\check{\vec{\kappa}}}_{i}(\tau )\,\stackrel{\circ }{=} &&\,-\sum_{k\neq
i}\frac{Q_{i}Q_{k}({\vec{\eta}}_{i}(\tau )-{\vec{\eta}}_{k}(\tau ))}{4\pi
\mid \vec{\eta}_{i}(\tau )-\vec{\eta}_{k}(\tau )\mid ^{3}}+  \nonumber \\
&+&Q_{i}(\dot{\eta}_{i}^{u}(\tau )+\lambda ^{u}(\tau )){\frac{{\partial }}{%
\partial {{\vec{\eta}}_{i}}}}{\check{A}}_{\perp }^{u}(\tau ,\vec{\eta}%
_{i}(\tau ))],  \nonumber \\
{\check{\vec{\kappa}}}_{+}(\tau ) &+&\int d^{3}\sigma \lbrack {\check{\vec{%
\pi}}}_{\perp }\times {\check{\vec{B}}}](\tau ,\vec{\sigma})\approx 0.
\label{IV5}
\end{eqnarray}
in which  $\stackrel{\circ }{=}$ means evaluated on the equations of motion.

\noindent The Hamilton-Dirac equations for the fields are

\begin{eqnarray}
&&\dot{{\check A}}_{\perp r}(\tau ,\vec{\sigma})\,\stackrel{\circ }{=}-{%
\check \pi} _{\perp r}(\tau ,\vec{\sigma})-[\vec{\lambda}(\tau )\cdot \vec{%
\partial}]{\check A}_{\perp r}(\tau ,\vec{\sigma}),  \nonumber \\
\dot{{\check \pi}}_{\perp }^{r}(\tau ,\vec{\sigma})\, &\stackrel{\circ }{=}%
&\,\Delta {\check A}_{\perp }^{r}(\tau ,\vec{\sigma})-[\vec{\lambda}(\tau
)\cdot \vec{\partial} ]{\check \pi}_{\perp }^{r}(\tau ,\vec{\sigma})+
\nonumber \\
&-&\sum_{i}Q_{i}P_{\perp }^{rs}(\vec{\sigma})\dot{\eta}_{i}^{s}(\tau )\delta
^{3}(\vec{\sigma}-\vec{\eta}_{i}(\tau )).  \label{IV6}
\end{eqnarray}

\noindent The associated Lagrangian, obtained by means of an inverse
Legendre transformation \cite{lu4}, is

\begin{eqnarray}
L_{R}(\tau ) &=&\dot{\vec{\eta}}_{i}(\tau )\cdot \check{\vec{\kappa}}
_{i}(\tau )-\int d^{3}\sigma {\dot{\check{\vec{A}}}}_{\perp }(\tau ,\vec{
\sigma})\cdot \check{\vec{\pi}}_{\perp }(\tau ,\vec{\sigma})-H_{R}(\tau )=
\nonumber \\
&=&\sum_{i=1}^{N}\Big[ -m_{i}\sqrt{1-(\dot{\vec{\eta}}_{i}(\tau )+\vec{%
\lambda} (\tau ))^{2}}+Q_{i}[\dot{\vec{\eta}}_{i}(\tau )+\vec{\lambda}(\tau
)]\cdot \check{\vec{A}}_{\perp }(\tau ,\vec{\eta}_{i}(\tau ))\Big] +
\nonumber \\
&+&\frac{1}{2}\sum_{i\neq j}\frac{Q_{i}Q_{j}}{4\pi \mid \vec{\eta}_{i}(\tau
)-\vec{\eta}_{j}(\tau )\mid }+  \nonumber \\
&+&\int d^{3}\sigma [\frac{(\dot{\check{\vec{A}}}_{\perp }+[\vec{\lambda}
(\tau )\cdot \vec{\partial}]\check{\vec{A}}_{\perp })^{2}}{2}-\frac{\check{%
\vec{B}}^{2}}{2}](\tau ,\vec{\sigma}).  \label{IV7}
\end{eqnarray}

\noindent Here $\vec{\lambda}(\tau )$ is now interpreted as a non-linear
Lagrange multiplier needed to get the rest-frame conditions ${\vec{{\cal H}}}%
_{p}={\vec{{\cal P}}}_{(int)}\approx 0$. Its Euler-Lagrange equations$\frac{d%
}{dt}\frac{\partial L_{R}}{\partial \dot{\eta}_{ir}}\,\stackrel{\circ }{=}\,%
\frac{\partial L_{R}}{\partial \eta _{ir}};\,\,\,\frac{\partial L_{R}}{%
\partial \vec{\lambda}}\stackrel{\circ }{=}\,0$ \ yield

\begin{eqnarray}
&&\frac{d}{d\tau }[m_{i}\frac{\dot{\vec{\eta}}_{i}(\tau )+\vec{\lambda}(\tau
)}{\sqrt{1-(\dot{\vec{\eta}}_{i}(\tau )+\vec{\lambda}(\tau ))^{2}}}+Q_{i}%
\check{\vec{A}}_{\perp }(\tau ,\vec{\eta}_{i}(\tau ))]\,\stackrel{\circ }{=}
\nonumber \\
&\stackrel{\circ }{=}\,&-\sum_{k\neq i}\frac{Q_{i}Q_{k}({\vec{\eta}}%
_{i}(\tau )-{\vec{\eta}}_{k}(\tau )}{4\pi \mid \vec{\eta}_{i}(\tau )-\vec{%
\eta}_{k}(\tau )\mid ^{3}}+Q_{i}(\dot{\eta}_{i}^{u}(\tau )+\lambda ^{u}(\tau
)){\frac{{\partial }}{\partial {{\vec{\eta}}_{i}}}}{\check{A}}_{\perp
}^{u}(\tau ,\vec{\eta}_{i}(\tau )),  \nonumber \\
&&{}  \nonumber \\
&-&\ddot{\check{A}}_{\perp }^{r}(\tau ,\vec{\sigma})-\frac{d}{d\tau }\{[\vec{%
\lambda}(\tau )\cdot \vec{\partial}]{\check{A}}_{\perp }^{r}(\tau ,\vec{%
\sigma})\}\,\stackrel{\circ }{=}  \nonumber \\
&&\stackrel{\circ }{=}\,\Delta {\check{A}}_{\perp }^{r}(\tau ,\vec{\sigma})+[%
\vec{\lambda}(\tau )\cdot \vec{\partial}]\{\dot{\check{A}}_{\perp }^{r}(\tau
,\vec{\sigma})+[\vec{\lambda}(\tau )\cdot \vec{\partial}]{\check{A}}_{\perp
}^{r}(\tau ,\vec{\sigma})\}-  \nonumber \\
&-&\sum_{i=1}^{N}Q_{i}P_{\perp }^{rs}(\vec{\sigma})[\dot{\eta}_{i}^{s}(\tau
)+\lambda ^{s}(\tau )]\delta ^{3}(\vec{\sigma}-\vec{\eta}_{i}(\tau )),
\label{IV8}
\end{eqnarray}
and [these are the Lagrangian rest-frame conditions]

\begin{eqnarray}
&&\sum_{i=1}^{N}[m_{i}\frac{\dot{\vec{\eta}}_{i}(\tau )+\vec{\lambda}(\tau )
}{\sqrt{1-(\dot{\vec{\eta}}(\tau )+\vec{\lambda}(\tau ))^{2}}}+Q_{i}\check{%
\vec{A}}_{\perp }(\tau ,\vec{\eta}_{i}(\tau ))]+  \nonumber \\
&&+\int d^{3}\sigma \sum_{r}[(\vec{\partial}{\check{A}}_{\perp }^{r})(\dot{%
\check{A}}_{\perp }^{r}+[\vec{\lambda}(\tau )\cdot \vec{\partial}]{\check{A}}%
_{\perp }^{r})](\tau ,\vec{\sigma})\, \stackrel{\circ }{=}\,0.  \label{IV9}
\end{eqnarray}

\noindent The Lagrangian expression for the conserved invariant mass $M=%
{\cal P}^{\tau}_{(int)}$ is

\begin{eqnarray}
E_{rel} &=&\sum_{i=1}^{N}\frac{m_{i}}{\sqrt{1-(\dot{\vec{\eta}}_{i}(\tau )+%
\vec{\lambda}(\tau ))^{2}}}+\sum_{i>j}\frac{Q_{i}Q_{j}}{4\pi \mid \vec{\eta}%
_{i}(\tau )-\vec{\eta}_{j}(\tau )\mid }+  \nonumber \\
&+&\int d^{3}\sigma {\frac{1}{2}}[\check{\vec{E}}_{\perp }^{2}+\check{\vec{B}%
}^{2}](\tau ,\vec{\sigma})=const.  \label{IV10}
\end{eqnarray}

\noindent Eq.(\ref{IV8}) may be rewritten as

\begin{eqnarray}
&&\frac{d}{d\tau }(m_{i}\frac{\dot{\vec{\eta}}_{i}(\tau )+\vec{\lambda}(\tau
)}{\sqrt{1+(\dot{\vec{\eta}}_{i}(\tau )+\vec{\lambda}(\tau ))^{2}}})\,
\nonumber \\
&\stackrel{\circ }{=}&\,-\sum_{k\neq i}\frac{Q_{i}Q_{k}({\vec{\eta}}
_{i}(\tau )-{\vec{\eta}}_{k}(\tau ))}{4\pi \mid \vec{\eta}_{i}(\tau )-\vec{
\eta}_{k}(\tau )\mid ^{3}}+  \nonumber \\
&+&Q_{i}[\check{\vec{E}}_{\perp }(\tau ,\vec{\eta}_{i}(\tau ))+(\dot{\vec{
\eta}}_{i}(\tau )+\vec{\lambda}(\tau ))\times \check{\vec{B}}(\tau ,\vec{%
\eta }_{i}(\tau ))],  \label{IV11}
\end{eqnarray}

\noindent where the notation ${\check E}^{r}_{\perp}=-\check{\dot{A}}^{r}
_{\perp}-[\vec{\lambda}(\tau)\cdot\vec{\partial}]{\check A}^{r}_{\perp}= {\
\check \pi}^r_{\perp}$ has been introduced.

Eqs.(\ref{IV11}) and (\ref{IV8}) are the rest-frame analogues of the usual
equations for charged particles in an external electromagnetic field and of
the electromagnetic field with external particle sources in which both
particles and electromagnetic field are dynamical. Eq.(\ref{IV9}) defines
the rest frame by using the total (Wigner spin 1) 3-momentum of the isolated
system formed by the particles plus the electromagnetic field. Eq.(\ref{IV10}
) gives the constant invariant mass of the isolated system: the
electromagnetic self-energy of the particles has been regularized by the
Grassmann-valued electric charges [$Q_{i}^{2}=0$] so that the invariant mass
is finite.

\subsection{The ``internal" center of mass and the last gauge fixing.}

The rest-frame conditions ${\vec{{\cal H}}}_{p}={\vec{{\cal P}}}%
_{(int)}\approx 0$ show that there are still 3 gauge degrees of freedom
among the reduced canonical variables ${\vec{\eta}}_{i}(\tau )$, ${\check{%
\vec{\kappa}}}_{i}(\tau )$, ${\check{\vec{A}}}_{\perp }(\tau ,\vec{\sigma})$%
, ${\check{\vec{\pi}}}_{\perp }(\tau ,\vec{\sigma})$ on each Wigner
hyperplane. They correspond to our freedom in the choice of the point of the
Wigner hyperplane that locates the ``internal'' 3-center of mass ${\vec{q}}%
_{+}$ of the isolated system. After the gauge fixing ${\vec q}_{+}\approx 0$
only Wigner-covariant relative variables are left on the Wigner hyperplane
and there is no double counting of the center of mass [only the decoupled
canonical non-covariant ``external'' one ${\vec{z}}_{s}$, ${\vec{k}}_{s}$ is
left].

In Refs.\cite{lu1,lu4} there was a naive choice ${\vec \eta}_{+}={\frac{1}{N}%
}\sum_{i=1}^N{\vec \eta}_i$ of the Wigner spin 1 3-vector conjugate to ${%
\vec {{\cal H}}}_p={\vec {{\cal P}}}_{(int)}$. Then, after realizing that ${%
\vec \eta}_{+}\approx 0$ does not imply $\vec \lambda (\tau )=0$, in Ref.
\cite{mate} a different choice ${\vec q}_{+}$ was made by utilizing the
group-theoretical results of Ref.\cite{pauri}: now the time constancy of the
gauge fixings ${\vec q}_{+}\approx 0$ implies $\vec \lambda (\tau )=0$.
Moreover, the nonrelativistic limit of ${\vec q}_{+}$ is now the unique
nonrelativistic center of mass.

In this gauge we get the simplest description of the dynamics on the Wigner
hyperplanes: $\vec{\sigma}={\vec{q}}_{+}\approx 0$ implies that the
``internal'' center of mass is put at the origin $x_{s}^{\mu }(\tau )$ of
the coordinates [$z^{\mu }(\tau ,\vec{\sigma})=x_{s}^{\mu }(\tau )+\epsilon
_{r}^{\mu }(u(p_{s}))\sigma ^{r}$ with $x_{s}^{\mu }(\tau )=x_{s}^{\mu
}(0)+u^{\mu }(p_{s})\tau $] of the Wigner hyperplane. In this gauge the
origin acquires the property ${\dot{x}}_{s}^{\mu }(\tau )=u^{\mu }(p_{s})$
and becomes also the ``external'' Fokker-Pryce center of inertia of the
isolated system \cite{mate} [in a future paper \cite{alp} there will be a
more detailed analysis of these problems].

In order to find ${\vec{q}}_{+}$ one must take advantage of the ``internal"
realization of the Poincar\'{e} algebra inside the Wigner hyperplane. Ref.
\cite{pauri} implies the following definition of the canonical ``internal''
3-center of mass
\begin{eqnarray}
\vec{q}_{+} &=&\frac{-{\cal \vec{K}}_{(int)}}{\sqrt{({\cal P}_{(int)}^{\tau
})^{2}-({\cal \vec{P}}_{(int)})^{2}}}+  \nonumber \\
&+&\frac{{\cal \vec{J}}_{(int)}\times {\cal \vec{P}}_{(int)}}{\sqrt{({\cal P}%
_{(int)}^{\tau })^{2}-({\cal \vec{P}}_{(int)})^{2}}[{\cal P}_{(int)}^{\tau }+%
\sqrt{({\cal P}_{(int)}^{\tau })^{2}-({\cal \vec{P}}_{(int)})^{2}}]}
\nonumber \\
&&+\frac{{\cal \vec{K}}_{(int)}\cdot {\cal \vec{P}}_{(int)}{\cal \vec{P}}%
_{(int)}}{{\cal P}_{(int)}^{\tau }\sqrt{({\cal P}_{(int)}^{\tau })^{2}-(%
{\cal \vec{P}}_{(int)})^{2}}[{\cal P}_{(int)}^{\tau }+\sqrt{({\cal P}%
_{(int)}^{\tau })^{2}-({\cal \vec{P}}_{(int)})^{2}}]}  \nonumber \\
%TCIMACRO{\QATOP{\approx }{{\cal \vec{P}}_{(int)}\approx 0} }%
%BeginExpansion
{\approx  \atop {\cal \vec{P}}_{(int)}\approx 0}%
%EndExpansion
&&\quad \quad -\frac{{\cal \vec{K}}_{(int)}}{{\cal P}_{(int)}^{\tau }}=\vec{R%
}_{+}.  \label{IV12}
\end{eqnarray}

Imposing the rest-frame condition, it is seen that ${\vec{q}}_{+}$
weakly coincides with the noncanonical ``internal'' M\o ller center of
energy ${\vec{ R }}_{+}$. In that same limit it is also equal to the
``internal'' Fokker-Pryce center of inertia defined by
\begin{equation}
\vec{Y}_{+}=\vec{q}_{+}+\frac{\vec{S}_{(int)}\times {\cal \vec{P}}_{(int)}}{%
\sqrt{({\cal P}_{(int)}^{\tau })^{2}-({\cal \vec{P}}_{(int)})^{2}}[{\cal P}%
_{(int)}^{\tau }+\sqrt{({\cal P}_{(int)}^{\tau })^{2}-({\cal \vec{P}}%
_{(int)})^{2}}]},  \label{IV13}
\end{equation}
where
\begin{equation}
\vec{S}_{(int)}\equiv {\cal \vec{J}}_{(int)}-\vec{q}_{+}\times {\cal \vec{P}}%
_{(int)}\approx \vec{S}_{s}.  \label{IV14}
\end{equation}
With the gauge fixing condition $\vec{q}_{+}\approx 0$ and with $T_{S}=\tau $
one finds the following expression for the origin $x_{s}^{\mu }(\tau )$ of
the coordinates on the Wigner hyperplane [$x^{\mu}_s(0)$ is arbitrary]
\begin{equation}
x_{s}^{(\vec{q}_{+})\mu }(T_{s})=x_{s}^{\mu }(0)+u^{\mu }(p_{s})T_{s}.
\label{IV15}
\end{equation}
It can be shown that this coincides with the covariant noncanonical
``external'' Fokker-Pryce center of inertia $Y^{\mu }(\tau )$. \ However, it
is different from both the ``external'' center of mass ${\ \tilde{x}}%
_{s}^{\mu }(\tau )$ and the ``external'' center of energy of M\o ller $R^{\mu
}(\tau )$.

Since ${\frac{d}{{d\tau }}}{\vec{q}}_{+}\,\stackrel{\circ }{=}\,\{{\vec{q}}%
_{+},M-\vec{\lambda}(\tau )\cdot {\vec{{\cal H}}}_{p}\}=-\vec{\lambda}(\tau
)\approx 0$, there is no gauge freedom left and we could eliminate the
variables ${\vec{q}}_{+}$, ${\vec{{\cal P}}}_{(int)}={\vec{{\cal H}}}_{p}$
and look for a canonical basis of (Dirac observable) relative variables on
the Wigner hyperplane.

Instead of doing that [see Ref.\cite{alp}], in this paper we will go on to
work with all the variables ${\vec{\eta}}_{i}(\tau )$, ${\check{\vec{\kappa}}%
}_{i}(\tau )$, ${\check{\vec{A}}}_{\perp }(\tau ,\vec{\sigma})$, ${\check{%
\vec{\pi}}}_{\perp }(\tau ,\vec{\sigma})$, but we shall restrict their
equations of motion to the gauge $\vec{\lambda}(\tau )=0$ without explicitly
introducing ${\vec{q}}_{+}\approx 0$.

The equations of motion for the particles and for the electromagnetic field
then become

\begin{eqnarray}
\frac{d}{d\tau }(m_{i}\frac{\dot{\vec{\eta}}_{i}(\tau )}{\sqrt{1-\dot{\vec{
\eta}}_{i}^{2}(\tau )}})\, &\stackrel{\circ }{=}&\,-\sum_{k\neq i}\frac{
Q_{i}Q_{k}({\vec{\eta}}_{i}(\tau )-{\vec{\eta}}_{k}(\tau ))}{4\pi \mid \vec{%
\eta}_{i}(\tau )-\vec{\eta}_{k}(\tau )\mid ^{3}}+  \nonumber \\
&+&Q_{i}[\check{\vec{E}}_{\perp }(\tau ,\vec{\eta}_{i}(\tau ))+\dot{\vec{%
\eta }}_{i}(\tau )\times \check{\vec{B}}(\tau ,\vec{\eta}_{i}(\tau ))],
\label{IV16}
\end{eqnarray}

\begin{eqnarray}
\Box {\check{A}}_{\perp }^{r}(\tau ,\vec{\sigma}) &=&\ddot{\check{A}}_{\perp
}^{r}(\tau ,\vec{\sigma})+\Delta {\check{A}}_{\perp }^{r}(\tau ,\vec{\sigma}
)\,\stackrel{\circ }{=}\,J_{\perp }^{r}(\tau ,\vec{\sigma})=  \nonumber \\
&=&\sum_{i=1}^{N}Q_{i}P_{\perp }^{rs}(\vec{\sigma})\dot{\eta}^{s}(\tau
)\delta ^{3}(\vec{\sigma}-\vec{\eta}_{i}(\tau ))=  \nonumber \\
&=&\sum_{i=1}^{N}Q_{i}\dot{\eta}^{s}(\tau )(\delta ^{rs}+\frac{\partial
^{r}\partial ^{s}}{\Delta })\delta ^{3}(\vec{\sigma}-\vec{\eta}_{i}(\tau ))=
\nonumber \\
&=&\sum_{i=1}^{N}Q_{i}\dot{\eta}^{s}(\tau )[\delta ^{3}(\vec{\sigma}-\vec{%
\eta}_{i}(\tau ))+  \nonumber \\
&+&\int d^{3}\sigma ^{\prime} \frac{\pi^{rs}(\vec{\sigma}-\vec{\sigma}%
^{\prime })} {\mid \vec{\sigma}-\vec{\sigma}^{\prime }\mid ^{3}}\delta ^{3}(
\vec{\sigma}^{\prime }-\vec{\eta}_{i}(\tau )),  \label{IV17}
\end{eqnarray}

\noindent with

\begin{equation}
\pi ^{rs}(\vec{\sigma}-\vec{\sigma}^{\prime })=\delta ^{rs}-3(\sigma ^{r}-{\
\sigma ^{\prime }}^{r})(\sigma ^{s}-{\sigma ^{\prime }}^{s})/(\vec{\sigma}-{%
\ \vec{\sigma}^{\prime })}^{2}.  \label{IV18}
\end{equation}
\

We point out that defining $\vec{\beta}_{i}(\tau )=\dot{\vec{\eta}}_{i}(\tau
)={\frac{{d{\vec{\eta}}_{i}(\tau )}}{{d\tau }}}={\frac{1}{c}}{\frac{{d{\vec{%
\eta}}_{i}^{^{\prime }}(t)}}{{dt}}}$ [$\tau =ct$, ${\vec{\eta}}%
^{^{\prime}}_{i}(t)={\vec{\eta}}_{i}(\tau )$; even if we use everywhere $c=1$%
, we have momentarily reintroduced it] and $\vec{\beta}_{i}^{(h)}=d^{h}\vec{%
\beta}_{i}/d\tau ^{h}$ and writing the particle equations of motion as (no
sum over $i$)
\begin{equation}
\frac{d}{d\tau }(m_{i}\frac{\vec{\beta}_{i}(\tau )}{\sqrt{1-\vec{\beta}%
_{i}^{2}(\tau )}})\,=\frac{m_{i}}{\sqrt{1-\vec{\beta}_{i}^{2}(\tau )}}[\vec{%
\beta}_{i}^{(1)}+\vec{\beta}_{i}\frac{\vec{\beta}_{i}^{(1)}\cdot \vec{\beta}%
_{i}}{1-\vec{\beta}_{i}^{2}(\tau )}]\stackrel{\circ }{=}{Q}_{i}\vec{F}_{i},
\label{IV19}
\end{equation}
we obtain
\begin{equation}
m_{i}\frac{\vec{\beta}_{i}^{(1)}\cdot \vec{\beta}_{i}}{(1-\vec{\beta}%
_{i}^{2}(\tau ))^{3/2}}\stackrel{\circ }{=}{Q}_{i}\vec{\beta}_{i}\cdot \vec{F%
}_{i},  \label{IV20}
\end{equation}
so that
\begin{equation}
\vec{\beta}_{i}^{(1)}\stackrel{\circ }{=}\frac{\sqrt{1-\vec{\beta}%
_{i}^{2}(\tau )}}{m_{i}}Q_{i}(\vec{F}_{i}-\vec{\beta}_{i}\vec{\beta}%
_{i}\cdot \vec{F}_{i}).  \label{IV21}
\end{equation}
Thus in general we will have for every $h\geq 1$
\begin{equation}
\vec{\beta}_{i}^{(h)}\stackrel{\circ }{=}{Q}_{i}\vec{G}_{i},  \label{IV22}
\end{equation}
so that using the Grassmann property of the charges
\begin{equation}
Q_{i}\vec{\beta}_{i}^{(h)}\stackrel{\circ }{=}{0},\quad \quad h\geq 1.
\label{IV23}
\end{equation}
This will lead to important simplifications later allowing us to drop
acceleration dependent terms in the force.

Due to the projector $P_{\perp }^{rs}(\vec{\sigma})$ required by the
rest-frame radiation gauge, the sources of the transverse (Wigner spin 1)
vector potential becomes non - local and one has a system of
integrodifferential equations (like with the equations generated by
Fokker-Tetrode actions) with the open problem of how to define an initial
value problem.

The Lagrangian equations identifying the rest frame become

\begin{eqnarray}
&&\sum_{i=1}^{N}(\eta _{i}m_{i}\frac{\dot{\vec{\eta}}_{i}(\tau )}{\sqrt{1-{\
\dot{\vec{\eta}}}^{2}(\tau )}}+Q_{i}\check{\vec{A}}_{\perp }(\tau ,\vec{\eta}
_{i}(\tau )))+  \nonumber \\
&&+\int d^{3}\sigma \sum_{r}[(\vec{\partial}{\check{A}}_{\perp }^{r})\dot{%
\check{A}}_{\perp }^{r}](\tau ,\vec{\sigma})\,\stackrel{\circ }{=}\,0.
\label{IV24}
\end{eqnarray}

\vfill\eject

\section{Electromagnetic Lienard-Wiechert Potentials.}

In this Section we will study the Lienard-Wiechert solutions of the previous
radiation gauge field equations in the gauge $\vec{\lambda}(\tau )=0$ in
absence of incoming radiation by using the results of Ref.\cite{lu4}. We
shall study the $retarded$, $advanced$ and ${\frac{1}{2}}(retarded+advanced)$
Lienard-Wiechert potentials. Using the Smart-Winter \cite{sw} expansion for
retarded and advanced time dependence, we obtain an infinite series form of
the retarded and advanced Lienard-Wiechert potentials depending on
instantaneous accelerations of every order. It will be shown that the
results of the previous Section imply that on the solutions of the particle
equations of motion the higher accelerations decouple due to the
semiclassical regularization $Q_{i}^{2}=0$. We show that this implies that
the ${\frac{1}{2}}(retarded-advanced)$ Lienard-Wiechert potential vanishes
at the semiclassical level and that there is only one semiclassical
Lienard-Wiechert potential: $retarded=advanced={\frac{1}{2}}%
(retarded+advanced)$. This allows us to re-express the semiclassical
Lienard-Wiechert potential in terms of particle canonical coordinates and
momenta [the same can be done for the Lienard-Wiechert electric field, as it
will be shown in the next Section]. Therefore, we get a Hamiltonian
description of the Lienard-Wiechert semiclassical solution: this sector of
solutions can be identified as the symplectic submanifold of the space of
solutions of the electromagnetic field equations determined by two pairs of
second class constraints, which force the electromagnetic field to coincide
with the semiclassical Lienard-Wiechert one.. After having gone to Dirac
brackets with respect to them, we get a reduced phase space with only
particles and we find a canonical basis ${\tilde{\vec{\eta}}}_{i}$, ${\tilde{%
\vec{\kappa}}}_{i}$ for these brackets for arbitrary $N$. In the new
variables the rest-frame condition becomes $\sum_{i=1}^{N}{\tilde{\vec{\kappa%
}}}_{i}\approx 0$, as expected in an instant form of dynamics.

\subsection{Grassmann truncated form of the advanced and retarded Lienard
Wiechert Solutions.}

Here we develop the Grassmann truncated forms for the Lienard-Wiechert
vector potential [see the next Section for the transverse electric and
magnetic fields]. \ The ${\frac{1}{2}}( retarded + advanced)$ solutions are
given (for $\vec{\lambda}(\tau )=0$) by [for the sake of notational
simplicity we will use the notation ${\vec{\kappa}}_{i}(\tau )$, ${\vec{A}}%
_{\perp }(\tau ,\vec{\sigma})$, ${\vec{\pi}}_{\perp }(\tau ,\vec{\sigma})$,
instead of ${\check{\vec{\kappa}}}_{i}(\tau )$, ${\check{\vec{A}}}_{\perp
}(\tau ,\vec{\sigma})$, ${\check{\vec{\pi}}}_{\perp }(\tau ,\vec{\sigma})$,
from now on]

\begin{eqnarray}
A_{\perp S}^{r}(\tau ,\vec{\sigma}) &=&\frac{1}{2}[A_{\perp +}^{r}+A_{\perp
-}^{r}](\tau ,\vec{\sigma})=  \nonumber \\
&=&{\frac{1}{2}}{\cal P}_{\perp }^{rs}(\vec{\sigma})\sum_{i=1}^{N}{\frac{{%
Q_{i}}}{{2\pi }}}\int d\tau _{1}d^{3}\sigma _{1}\,[\theta (\tau -\tau
_{1})+\theta (\tau -\tau _{1})]  \nonumber \\
&&\delta \lbrack (\tau -\tau _{1})^{2}-(\vec{\sigma}-{\vec{\sigma}}_{1})^{2}]%
{\dot{\eta}}_{i}^{s}(\tau _{1})\delta ^{3}({\vec{\sigma}}_{1}-{\vec{\eta}}%
_{i}(\tau _{1}))=  \nonumber \\
&=&{\cal P}_{\perp }^{rs}(\vec{\sigma})\sum_{i=1}^{N}\frac{Q_{i}}{2\pi c}%
\int dt_{1}\delta \lbrack (t-t_{1})^{2}-\frac{1}{c^{2}}(\vec{\sigma}-\vec{%
\eta}_{i}(ct_{1}))^{2}]\beta _{i}^{s}(ct_{1}):=  \nonumber \\
&:&=\sum_{i=1}^{N}Q_{i}A_{\perp \,Si}^{r}(\tau ,\vec{\sigma}),  \label{V1}
\end{eqnarray}
in which we have put $\tau =ct$, $\vec{\beta}_{i}(\tau )=\dot{\vec{\eta}}%
_{i}(\tau )\,={\frac{1}{c}}{\frac{{d{\vec{\eta}}_{i}^{^{\prime }}(t)}}{{dt}}}
$ and ${\vec{A}}_{\perp +}={\vec{A}}_{\perp RET}$ (${\vec{A}}_{\perp -}={%
\vec{A}}_{\perp ADV}$) for the retarded (advanced) solution. The equation
for $t_{1}$ is $c^{2}(t-t_{1})^{2}=(\vec{\sigma}-\vec{\eta}_{i}(ct_{1}))^{2}$
with the two solutions being

\begin{eqnarray}
t_{i+}(\tau ,\vec \sigma )&=&{\frac{1}{c}} \tau _{i+}(\tau ,\vec{\sigma}) =t-%
\frac{1}{c}r_{i+}(\tau_{i+}(\tau , \vec{\sigma}),\vec{\sigma})={\frac{{\tau}%
}{c}} -T_{i+}(\tau ,\vec \sigma ),  \nonumber \\
t_{i-}(\tau ,\vec \sigma )&=&{\frac{1}{c}} \tau _{i-}(\tau ,\vec{\sigma}) =t
+\frac{1}{c}r_{i-}(\tau _{i-}(\tau , \vec{\sigma}),\vec{\sigma})={\frac{{\tau%
}}{c}} +T_{i-}(\tau ,\vec \sigma ),  \label{V2}
\end{eqnarray}
for the retarded and for the advanced case respectively. The light cone
delta function is

\begin{eqnarray}
&&\delta [(\tau -\tau_1)^2-(\vec \sigma -{\vec \eta}_i(\tau_1))^2]={\frac{1}{%
{c^2}}} \delta \lbrack (t -t_{1})^{2}-\frac{1}{c^{2}}(\vec{\sigma}-\vec{\eta}
_{i}(ct_{1}))^{2}]=  \nonumber \\
&&=\frac{\delta \lbrack \tau _{1-}-\tau _{i+}(\tau , \vec{\sigma})]}{2|\tau
-\tau _{1}-\vec{\beta}_{i}(\tau _{1})\cdot (\vec{\sigma}-\vec{\eta}_{i}(\tau
_{1}))|}+ \frac{\delta \lbrack \tau _{1-}-\tau _{i-}(\tau ,\vec{\sigma})]}{%
2|\tau -\tau _{1}-\vec{\beta}_{i}(\tau _{1})\cdot (\vec{\sigma}-\vec{\eta }%
_{i}(\tau _{1}))|}.  \label{V3}
\end{eqnarray}
\ The relative space location between the field point and the retarded or
advanced particle position is
\begin{equation}
\vec{\sigma}-\vec{\eta}_{i}(\tau _{i\pm }(\tau ,\vec{\sigma}))=\vec{r}_{i\pm
}(\tau _{i\pm }(\tau ,\vec{\sigma}),\vec{\sigma})=r_{i\pm }(\tau _{i\pm
}(\tau ,\vec{\sigma}),\vec{\sigma})\hat{r}_{i\pm }(\tau _{i\pm }(\tau ,\vec{%
\sigma}),\vec{\sigma}),  \label{V4}
\end{equation}
and its length is related to the time interval by
\begin{eqnarray}
&&r_{i\pm }(\tau _{i\pm }(\tau ,\vec{\sigma}),\vec{\sigma})=|\, \vec \sigma -%
{\vec \eta}_i(\tau_{i\pm}(\tau ,\vec \sigma ))|= cT_{i\pm }(\tau ,\vec{\sigma%
})=|\tau -\tau _{i\pm }(\tau ,\vec{\sigma})| ,  \nonumber \\
&&{}  \nonumber \\
&&\Rightarrow \quad \tau -\tau_{i\pm}(\tau ,\vec \sigma )=\pm cT_{i\pm}
(\tau ,\vec \sigma )=\pm r_{i\pm}(\tau_{i\pm}(\tau ,\vec \sigma ),\vec \sigma
).  \label{V5}
\end{eqnarray}
The effective spatial interval is defined by
\begin{equation}
\rho _{i\pm }(\tau _{i\pm }(\tau ,\vec{\sigma}),\vec{\sigma})=r_{i\pm }(\tau
_{i\pm }(\tau ,\vec{\sigma}),\vec{\sigma})[1\mp \vec{\beta}_{i}(\tau _{i\pm
}(\tau ,\vec{\sigma}))\cdot \hat{r}_{i\pm }(\tau _{i\pm }(\tau ,\vec{\sigma }%
),\vec{\sigma})].  \label{V6}
\end{equation}
In terms of these variables, the retarded, advanced and time symmetric
solutions are
\begin{eqnarray}
A^r_{\perp \pm}(\tau ,\vec \sigma )&=& \sum_{i=1}^N {\frac{{Q_i}}{{4\pi}}}
{\cal P}^{rs}_{\perp}(\vec \sigma ) {\frac{{\ \beta^s_i(\tau_{i\pm}(\tau ,%
\vec \sigma ))}}{{\rho_{i\pm} (\tau_{i\pm}(\tau ,\vec \sigma),\vec \sigma )}}%
},  \nonumber \\
A_{\perp S}^{r}(\tau ,\vec{\sigma})&=&\sum_{i=1}^{N}Q_{i}A_{\perp \,
Si}^{r}(\tau ,\vec{\sigma})=\sum_{i=1}^{N}\frac{Q_{i}}{8\pi }{\cal P}_{\perp
}^{rs}(\vec{\sigma})\left[ \frac{{\beta}_{i}^s(\tau _{i+}(\tau ,\vec{\sigma }%
))}{\rho _{i+}(\tau _{i+}(\tau ,\vec{\sigma}),\vec{\sigma})}+\frac{{\ \beta}%
_{i}^s(\tau _{i-}(\tau ,\vec{\sigma}))}{\rho _{i-}(\tau _{i-}(\tau ,\vec{
\sigma}),\vec{\sigma})}\right] .  \label{V7}
\end{eqnarray}

We use the Smart-Wintner expansion \cite{sw,tet3,tet4}
\begin{eqnarray}
f(\tau _{i\pm }) &=&f(\tau \mp cT_{i\pm }(\tau _{i\pm }((\tau ,\vec{\sigma}),%
\vec{\sigma})))=f(\tau -[\pm r_{i\pm }(\tau _{i\pm }(\tau ,\vec{\sigma}),%
\vec{\sigma})])=  \nonumber \\
&=&f(\tau )+\sum_{k=1}^{\infty }\frac{(-)^{k}}{k!}\frac{d^{k-1}}{d\tau ^{k-1}%
}\left[ \left( \pm r_{i}(\tau ,\vec{\sigma})\right) ^{k}\frac{df(\tau )}{%
d\tau }\right] =  \nonumber \\
&=&\sum_{k=0}^{\infty }\frac{(-)^{k}}{k!}\frac{d^{k}}{d\tau ^{k}}\left[
\left( \pm r_{i}(\tau ,\vec{\sigma})\right) ^{k}[1\mp \vec{\beta}_{i}(\tau
)\cdot \hat{r}_{i}(\tau ,\vec{\sigma})]f(\tau )\right] ,  \label{V8}
\end{eqnarray}
where
\begin{eqnarray}
{\vec{r}}_{i}(\tau ,\vec{\sigma}) &=&r_{i}(\tau ,\vec{\sigma}){\hat{r}}%
_{i}(\tau ,\vec{\sigma})=\vec{\sigma}-{\vec{\eta}}_{i}(\tau )={\vec{r}}%
_{i\pm }(\tau _{\pm }(\tau ,\vec{\sigma}),\vec{\sigma}){|}_{\tau _{i\pm
}(\tau ,\vec{\sigma})=\tau },  \nonumber \\
f(\tau _{i\pm }) &=&\frac{\beta _{i}^{s}(\tau _{i\pm })}{\rho _{i\pm }(\tau
_{i\pm })}=\frac{\beta _{i}^{s}(\tau _{i\pm })}{r_{i\pm }(\tau _{i\pm
})[1\mp \vec{\beta}_{i}(\tau _{i\pm })\cdot \hat{r}_{i\pm }(\tau ,\vec{\sigma%
})]}.  \label{V9}
\end{eqnarray}

\noindent and where the last line in Eq.(\ref{V8}) is identical to the
previous one since ${\frac{{dr_{i}(\tau ,\vec{\sigma})}}{{d\tau }}}=-{\vec{%
\beta}}_{i}(\tau )\cdot {\hat{r}}_{i}(\tau ,\vec{\sigma})$.

Hence we get
\begin{equation}
A_{\perp \pm }^{r}(\tau ,\vec{\sigma})=\sum_{i=1}^{N}\frac{Q_{i}}{4\pi }%
{\cal P}_{\perp }^{rs}(\vec{\sigma})\sum_{k=0}^{\infty }\frac{(\mp )^{k}}{k!}%
\frac{d^{k}}{d\tau ^{k}}[r_{i}^{k-1}(\tau ,\vec{\sigma})\beta _{i}^{s}(\tau
)],\quad A_{\perp S}^{r}={\frac{1}{2}}(A_{\perp +}^{r}+A_{\perp -}^{r}).
\label{V10}
\end{equation}
In order to evaluate the above derivatives we need the Leibnitz formula for
the $k$th derivative of the product $f(\tau )g(\tau )$
\begin{equation}
\frac{d^{k}}{d\tau ^{k}}(fg)=\sum_{m=0}^{k}{\frac{{k!}}{{m!(k-m)!}}}\frac{%
d^{m}f}{d\tau ^{m}}\,\,\frac{d^{k-m}g}{d\tau ^{k-m}},  \label{V11}
\end{equation}
Thus we get
\begin{eqnarray}
\sum_{k=0}^{\infty }\frac{(\mp )^{k}}{k!}\frac{d^{k}}{d\tau ^{k}}%
[r_{i}^{k-1}\beta _{i}^{s}] &=&\sum_{k=0}^{\infty }\frac{(\mp )^{k}}{k!}%
\sum_{m=0}^{k}\frac{k!}{m!(k-m)!}\frac{d^{m}r^{k-1}}{d\tau ^{m}}\frac{%
d^{k-m}\beta ^{s}}{d\tau ^{k-m}}=  \nonumber \\
\sum_{m=0}^{\infty }\sum_{k=m}^{\infty }\frac{(\mp )^{k}}{m!(k-m)!}\frac{%
d^{m}r^{k-1}}{d\tau ^{m}}\frac{d^{k-m}\beta ^{s}}{d\tau ^{k-m}}
&=&\sum_{m=0}^{\infty }\sum_{h=0}^{\infty }\frac{(\mp )^{h+m}}{m!h!}\frac{%
d^{h}\beta ^{s}}{d\tau ^{h}}\frac{d^{m}r^{h+m-1}}{d\tau ^{m}}.  \label{V12}
\end{eqnarray}
Using the notation $\beta ^{(h)s}=\frac{d^{h}\beta ^{s}}{d\tau ^{h}}$ $\,$we
obtain the following expression for the vector potential
\begin{equation}
A_{\perp \pm }^{r}(\tau ,\vec{\sigma})=\sum_{i=1}^{N}\frac{Q_{i}}{4\pi }%
{\cal P}_{\perp }^{rs}(\vec{\sigma})\sum_{h=0}^{\infty }\frac{(\mp )^{h}}{h!}%
\beta _{i}^{(h)s}(\tau )\phi _{i\pm ,h}(\tau ,\vec{\sigma}),  \label{V13}
\end{equation}
in which
\begin{equation}
\phi _{i\pm ,h}(\tau ,\vec{\sigma})=\sum_{m=0}^{\infty }\frac{(\mp )^{m}}{m!}%
\frac{d^{m}r_{i}^{h+m-1}(\tau ,\vec{\sigma})}{d\tau ^{m}}=\sum_{m=0}^{\infty
}\frac{(\mp )^{m}}{m!}\frac{d^{m}}{d\tau ^{m}}[\sqrt{(\vec{\sigma}-\vec{\eta}%
_{i}(\tau ))^{2}}]^{m+h-1}.  \label{V14}
\end{equation}
In order to display the result of the evaluation of the derivative we use
the formula
\begin{equation}
\frac{d^{m}}{d\tau ^{m}}R(f(\tau ))=\sum_{n=0}^{m}\sum_{n_{1}n_{2..}}\frac{m!%
}{n_{1}!n_{2}!..}\frac{d^{n}R(f(\tau ))}{df^{n}}|_{f=f(\tau )}(\frac{1}{1!}%
\frac{df(\tau )}{d\tau })^{n_{1}}(\frac{1}{2!}\frac{d^{2}f(\tau )}{d\tau ^{2}%
})^{n_{2}}...  \label{V15}
\end{equation}
(with the summations restricted so that $\sum_{r}n_{r}=n,\,\,%
\sum_{r}rn_{r}=m $ ) to obtain
\begin{eqnarray}
\phi _{i\pm ,h}(\tau ,\vec{\sigma}) &=&\sum_{m=0}^{\infty }\frac{(\mp )^{m}}{%
m!}\sum_{n=0}^{m}\sum_{n_{1}n_{2..}}\frac{m!}{n_{1}!n_{2}!..}  \nonumber \\
&&\frac{\partial ^{n}r_{i}^{m+h-1}(\tau ,\vec{\sigma})}{\partial \vec{r}%
_{i}^{n}}\circ \left( \frac{-\vec{\beta}_{i}(\tau )}{1!}\right)
^{n_{1}}\left( \frac{-\vec{\beta}_{i}^{(1)}(\tau )}{2!}\right) ^{n_{2}}...
\label{V16}
\end{eqnarray}
In this expression the symbol $\circ $ represents a scalar product between
the tensors to the left and to the right with the summation$\sum_{r}n_{r}=n$
indicating how the indices would be matched. Changing the $m$ summation
index to $k=m-n$ we obtain
\begin{eqnarray}
\phi _{i\pm ,h}(\tau ,\vec{\sigma}) &=&\sum_{n=0}^{\infty
}\sum_{k=0}^{\infty }\sum_{n_{1}n_{2..}}\frac{(\mp )^{k+n}(-)^{\Sigma
_{r}n_{r}=n}}{n_{1}!n_{2}!..}  \nonumber \\
&&\frac{\partial ^{n}r_{i}^{k+n+h-1}(\tau ,\vec{\sigma})}{\partial \vec{r}%
_{i}^{n}}\circ \left( \frac{\vec{\beta}_{i}(\tau )}{1!}\right)
^{n_{1}}\left( \frac{\vec{\beta}_{i}^{(1)}(\tau )}{2!}\right) ^{n_{2}}...
\label{V17}
\end{eqnarray}
(In this latter summation $\sum_{r}n_{r}=n,\,\,\sum_{r}rn_{r}=n+k.$)

Now we can take advantage of the Grassmann charges to significantly simplify
the above multi-summations. \ As we have seen above, with a semiclassical $%
Q_{i}$ there are {\it no accelerations} on shell ($Q_{i}\vec{\beta}_{i}^{(h)}%
\stackrel{\circ }{=}{0}$) in the equations of motion of the particle `$i$',
since both the Coulomb potential and the Lienard-Wiechert Lorentz force on
particle `$i$' produced by the other particles, i.e. $Q_{i}[{\vec{E}}_{\perp
}(\tau ,{\vec{\eta}}_{i}(\tau ))+{\vec{\beta}}_{i}(\tau )\times \vec{B}(\tau
,{\vec{\eta}}_{i}(\tau ))]$, are proportional to $Q_{i}$. Therefore, the
full set of Hamilton equations (\ref{IV5}), (\ref{IV6}) for both fields and
particles imply that at the semiclassical level we have a natural ``order
reduction'' of the final particle equation of motion in the Lienard-Wiechert
sector [only second order differential equations].

One effect of this truncation is the elimination of multi-particle forces;
all the interactions will be pairwise, in both the Lagrangian and
Hamiltonian formalisms. This was to be expected since the rest-frame instant
form is an equal-time description of the $N$ particle system:
(acceleration-independent) 3-body,.. N-body forces appear as soon as we go
to a description with no concept of equal time, like in the standard
approach with $N$ first class constraints \cite{manybody}.

Thus the only contributing indices are $n_{2}=n_{3}=..=0,\,n_{1}=n$ and our
expression for the transverse vector potentials simplify to
\begin{eqnarray}
A_{\perp \pm }^{r}(\tau ,\vec{\sigma}) &\stackrel{\circ }{=}&\,\sum_{i=1}^{N}%
\frac{Q_{i}}{4\pi }{\cal P}_{\perp }^{rs}(\vec{\sigma})\beta _{i}^{s}(\tau
)\phi _{i\pm ,0}(\tau ,\vec{\sigma})\stackrel{\circ }{=}  \nonumber \\
\stackrel{\circ }{=}\, &&\sum_{i=1}^{N}\frac{Q_{i}}{4\pi }{\cal P}_{\perp
}^{rs}(\vec{\sigma})\beta _{i}^{s}(\tau )\sum_{n=0}^{\infty }\frac{(\pm )^{n}%
}{n!}\frac{\partial ^{n}r_{i}^{n-1}(\tau ,\vec{\sigma})}{\partial \vec{r}%
_{i}^{n}}\cdot \left( \frac{\vec{\beta}_{i}(\tau )}{1!}\right) ^{n},
\nonumber \\
A_{\perp S}^{r}(\tau ,\vec{\sigma}) &=&{\frac{1}{2}}(A_{\perp
+}^{r}+A_{\perp -}^{r})(\tau ,\vec{\sigma}).  \label{V18}
\end{eqnarray}
Since $r_{i}=\sqrt{{\vec{r}}_{i}^{2}}$, we see that for odd n=2m+1 we get

\begin{equation}
{\frac{{\partial^{2m+1}}}{{\partial {\vec r}_i^{2m+1}}}} (\sqrt{{\vec r}_i^2}%
)^{2m} ={\frac{{\partial^{2m+1}}}{{\partial {\vec r}_i^{2m+1}}}} ({\vec r}%
_i^2 )^m =0,  \label{V19}
\end{equation}

\noindent and this implies the equality of the retarded, advanced and
symmetric Lienard-Wiechert potentials on-shell

\begin{eqnarray}
A_{\perp S}^{r}(\tau ,\vec{\sigma}) &\stackrel{\circ }{=}&A_{\perp \pm
}^{r}(\tau ,\vec{\sigma})\stackrel{\circ }{=}\,\sum_{i=1,i\neq u}^{N}\frac{%
Q_{i}}{4\pi }{\cal P}_{\perp }^{rs}(\vec{\sigma})\beta _{i}^{s}(\tau )
\nonumber \\
&&\sum_{m=0}^{\infty }\frac{1}{(2m)!}\left( \vec{\beta}_{i}(\tau )\cdot
\frac{\partial ^{2m}}{\partial \vec{r}_{i}^{2m}}\right) r_{i}^{2m-1}(\tau ,%
\vec{\sigma}).  \label{V20}
\end{eqnarray}
Therefore, at the semiclassical level there is only one Lienard-Wiechert
sector with a uniquely determined standard action-at-a-distance interaction.

We use a tensor notation to write the transverse symmetric vector potential
above as

\begin{equation}
\vec{A}_{\perp S}(\tau ,\vec{\sigma})\, {\buildrel \circ \over =}\,
\sum_{i=1}^{N}{\frac{Q_{i}}{4\pi }}{\bf {\cal P}}_{\perp}\cdot \dot{\vec{\eta%
}}_{i}\sum_{m=0}^{\infty }{\frac{\dot{\vec{\eta}}_{ij_{1}}(\tau )..\dot{\vec{%
\eta}}_{ij_{2m}}(\tau ) }{(2m)!}}{\ \frac{\partial ^{2m}|\vec{\sigma}-\vec{%
\eta}_{i}(\tau )|^{2m-1}}{\partial \sigma _{j_{1}}..\partial \sigma _{j_{2m}}%
}}.  \label{V21}
\end{equation}
Using the definition of the Coulomb projection operator
\begin{equation}
{\cal P}(\vec{\sigma})_{\perp hk}F(\vec{\sigma})=\delta _{hk}F(\vec{\sigma})-%
{\frac{1}{4\pi }}\int d^{3}\sigma ^{\prime }{\frac{\partial ^{2}} {\partial
\sigma _{h}\partial \sigma _{k}}}{\frac{1}{|{\vec{\sigma}}^{\prime }-\vec{
\sigma}|}}F({\vec{\sigma}}^{\prime }),  \label{V22}
\end{equation}
and compactifying the notation still further we obtain [${\vec \nabla}%
_{\sigma}=\partial /\partial \vec \sigma$]
\begin{eqnarray}
&&\vec{A}_{\perp S}(\tau ,\vec{\sigma})\, {\buildrel \circ
\over =}\,
\sum_{i=1}^{N}{\frac{Q_{i}}{4\pi }} \sum_{m=0}^{\infty }{\frac{1}{(2m)!}}%
\Big[ \dot{\vec{\eta}}_{i}(\tau )(\dot{\vec{\eta}} _{i}(\tau )\cdot \vec{%
\nabla}_{\sigma })^{2m})\, |\vec{\sigma}-\vec{\eta}_{i}(\tau )|^{2m-1}-
\nonumber \\
&&-{\frac{1}{4\pi }}\int d^{3}\sigma ^{\prime }[\vec{\nabla}_{\sigma }(\dot{%
\vec{\eta}}_{i}(\tau )\cdot \vec{\nabla}_{\sigma }){\frac{1}{|{\vec{\sigma}}%
^{\prime }-\vec{\sigma}|}}](\dot{\vec{\eta}}_{i}(\tau )\cdot \vec{\nabla}%
_{\sigma^{\prime }})^{2m}\, |{\vec{\sigma}}^{\prime }-\vec{\eta}_{i}(\tau
)|^{2m-1}\Big] .  \label{V23}
\end{eqnarray}
Integration by parts and changing from ${\frac{\partial }{\partial {\vec %
\sigma}^{\prime }}}$ to ${\frac{\partial }{\partial \vec \sigma }}$ and
translation gives
\begin{eqnarray}
\vec{A}_{\perp S}(\tau ,\vec{\sigma})\, &{\buildrel \circ \over =}\,&
\sum_{i=1}^{N}{\frac{Q_{i}}{4\pi }} \sum_{m=0}^{\infty }{\frac{1}{(2m)!}} %
\Big[ \dot{\vec{\eta}}_{i}(\tau )(\dot{\vec{\eta}} _{i}(\tau )\cdot \vec{%
\nabla}_{\sigma })^{2m})\, |\vec{\sigma}-\vec{\eta}_{i}(\tau )|^{2m-1}-
\nonumber \\
&&-{\frac{1}{4\pi }}\int d^{3}\sigma ^{\prime }\Big( \vec{\nabla}_{\sigma }(%
\dot{\vec{\eta}}_{i}(\tau )\cdot \vec{\nabla}_{\sigma })^{2m+1}\, {\frac{1}{|%
{\vec{\sigma}}^{\prime }-(\vec{\sigma}-\vec{\eta}_{i}(\tau ))|}}\Big) \, {%
\sigma ^{\prime }}^{2m-1}\Big] .  \label{V24}
\end{eqnarray}
The integral above is finite, and thus we can view it as the ${\Lambda
\rightarrow \infty }$ limit of an integral with a cutoff $\Lambda$ and take
the derivatives out. The integral is thus of the form

\begin{equation}
-{\frac{1}{4\pi }}{\vec{\nabla}}_{\sigma }(\dot{\vec{\eta}}_{i}(\tau )\cdot
\vec{\nabla}_{\sigma })^{2m+1}\int d^{3}\sigma ^{\prime }{\frac{{\sigma
^{\prime }}^{2m-1}}{|{\vec{\sigma}}^{\prime }-(\vec{\sigma}-\vec{\eta}%
_{i}(\tau ))|}},  \label{V25}
\end{equation}
and
\begin{eqnarray}
{\frac{1}{4\pi }}\int_{\Lambda }d^{3}\sigma ^{\prime }{\frac{{\sigma
^{\prime }}^{2m-1}}{|\vec{\sigma}^{\prime }-(\vec{\sigma}-\vec{\eta}_{i})|}}
&=&{{\frac{1}{2}}}\int_{0}^{\Lambda }d\sigma ^{\prime }{\sigma ^{\prime }}%
^{2m+1}\int_{-1}^{1}{\frac{dz}{\sqrt{{\vec{\sigma ^{\prime }}}^{2}+(\vec{%
\sigma}-\vec{\eta}_{i})^{2}-2\sigma ^{\prime }|\vec{\sigma}-\vec{\eta}_{i}|z}%
}}  \nonumber \\
&=&{\frac{1}{2}}\int_{0}^{\Lambda }d\sigma ^{\prime }{\sigma ^{\prime }}%
^{2m+1}{\frac{-1}{\sigma ^{\prime }|\vec{\sigma}-\vec{\eta}_{i}|}}\sqrt{{\
\vec{\sigma ^{\prime }}}^{2}+(\vec{\sigma}-\vec{\eta}_{i})^{2}-2\sigma |\vec{%
\sigma}-\vec{\eta}_{i}|}  \nonumber \\
&=&-{\frac{1}{2|\vec{\sigma}-\vec{\eta}_{i}|}}\int_{0}^{\Lambda }d\sigma
^{\prime }{\sigma ^{\prime }}^{2m}(|{\vec{\sigma}}^{\prime }-|\vec{\sigma}-%
\vec{\eta}_{i}||-|{\vec{\sigma}}^{\prime }+|\vec{\sigma}-\vec{\eta}_{i}||)
\nonumber \\
&=&{\frac{\Lambda ^{2m+1}}{2m+1}}-{\frac{|\vec{\sigma}-\vec{\eta}_{i}|^{2m+1}%
}{(2m+1)(2m+2)}}.  \label{V26}
\end{eqnarray}
Note that the $\Lambda $ cutoff will get killed by the $\sigma $
derivatives. Thus, we obtain
\begin{eqnarray}
\vec{A}_{\perp S}(\tau ,\vec{\sigma})\, &{\buildrel \circ \over =}\,&
\sum_{i=1}^{N}{\frac{Q_{i}}{4\pi }}\sum_{m=0}^{\infty }\Big[{\frac{1}{(2m)!}}%
\dot{\vec{\eta}}_{i}(\tau )(\dot{\vec{\eta}}_{i}(\tau )\cdot {\vec{\nabla}}%
_{\sigma })^{2m}\,|\vec{\sigma}-\vec{\eta}_{i}(\tau )|^{2m-1}-  \nonumber \\
&&-{\frac{1}{(2m+2)!}}{\vec{\nabla}}_{\sigma }(\dot{\vec{\eta}}_{i}(\tau
)\cdot {\vec{\nabla}}_{\sigma })^{2m+1}\,|\vec{\sigma}-\vec{\eta}_{i}(\tau
)|^{2m+1}\Big]:=  \nonumber \\
&:&=\sum_{i=1}^{N}Q_{i}\vec{A}_{\perp Si}(\vec{\sigma}-\vec{\eta}_{i}(\tau ),%
{\dot{\vec{\eta}}}_{i}(\tau )).  \label{V27}
\end{eqnarray}
Using the first half of particle Hamilton equations (\ref{IV5}) [with $\vec{%
\lambda}(\tau )=0$] in the form $\dot{\vec{\eta}_{i}}=\vec{\kappa}_{i}/\sqrt{%
m_{i}^{2}+{\vec{\kappa}_{i}}^{2}}+O(Q_{i})$, we can, as shown in Appendix A,
arrive at the following closed form of the vector potential [${\vec{\eta}}%
_{i}={\vec{\eta}}_{i}(\tau )$, ${\vec{\kappa}}_{i}={\vec{\kappa}}_{i}(\tau )$%
]

\begin{eqnarray}
&&{\vec{A}}_{\perp S}(\tau ,\vec{\sigma})\, {\buildrel \circ
\over =}\,
\sum_{i=1}^{N}Q_{i}{\vec{A}}_{\perp Si}(\vec{\sigma}-{\vec{\eta}}_{i}(\tau ),%
{\vec{\kappa}}_{i}(\tau )),  \nonumber \\
&&{}  \nonumber \\
&&\vec{A}_{\perp Si}(\vec{\sigma}-\vec{\eta}_{i},{\vec{\kappa}}_{i})={\frac{1%
}{4\pi |\vec{\sigma}-\vec{\eta}_{i}|}}\Big[{\frac{\vec{\kappa}_{i}}{\sqrt{%
m_{i}^{2}+(\vec{\kappa}_{i}\cdot {\frac{{\vec{\sigma}-\vec{\eta}_{i}}}{{|%
\vec{\sigma}-\vec{\eta}_{i}|}}})^{2}}}}-  \nonumber \\
&&-\vec{\kappa}_{i}\cdot ({\bf I}-{\frac{(\vec{\sigma}-\vec{\eta}_{i})(\vec{%
\sigma}-\vec{\eta}_{i})}{|\vec{\sigma}-\vec{\eta}_{i}|^{2}}})({\frac{\sqrt{%
m_{i}^{2}+{\ \vec{\kappa}_{i}}^{2}}}{\sqrt{m_{i}^{2}+(\vec{\kappa}_{i}\cdot {%
\frac{{\vec{\sigma}-\vec{\eta}_{i}}}{{|\vec{\sigma}-\vec{\eta}_{i}|}}})^{2}}}%
}-1)\times  \nonumber \\
&&{\frac{\sqrt{{m_{i}^{2}+\vec{\kappa}_{i}}^{2}}}{\vec{\kappa}_{i}^{2}-(\vec{%
\kappa}_{i}\cdot {\frac{{\vec{\sigma}-\vec{\eta}_{i}}}{{|\vec{\sigma}-\vec{%
\eta}_{i}|}}})^{2}}}\Big].  \label{V28}
\end{eqnarray}

\subsection{Lienard-Wiechert Second-Class Constraints, their Dirac Brackets
and the New Canonical Variables.}

\bigskip

Thus far we have the reduced phase space of $N$ charged particles plus the
transverse electromagnetic field. This is a well defined isolated system
with a global Darboux basis [ $\vec{\eta}_{i},\vec{\kappa}_{i},\vec{A}%
_{\perp }(\tau ,\vec{\sigma}),\vec{\pi}_{\perp }(\tau ,\vec{\sigma})$] and a
well defined physical Hamiltonian, the invariant mass $M={\cal P}%
_{(int)}^{\tau }$. All possible configurations of motion take place in this
reduced phase space. \ The space of solutions of Hamilton's equations is a
symplectic space\ in that there is a definition of Poisson brackets on the
space of solutions. The question arises whether one can select a subset of
solutions of the equations of motion which is still a symplectic manifold:
an arbitrarily chosen set of solutions will not form a symplectic manifold.
The method we propose here is to add by hand a set of second class
constraints ``compatible with the equations of motion'' which amounts to the
selection of a symplectic submanifold of the symplectic manifold of
solutions.

\bigskip The above Grassmann truncated semiclassical Lienard Wiechert
solution $\vec{A}_{\perp S}$ for the vector potential with $\vec{\pi}_{\perp
S}={\vec{E}}_{\perp S}=-\frac{\partial }{\partial \tau }\vec{A}_{\perp S}$
for the canonical conjugate field momentum [see Eq.(\ref{VI2}) in the next
Section] and provide us such a set of second class constraints \ by way of

\begin{eqnarray}
\vec{\chi}_{1}(\tau ,\vec{\sigma}) &=&\vec{A}_{\perp }(\tau ,\vec{\sigma}%
)-\sum_{i=1}^{N}Q_{i}\vec{A}_{\perp Si}(\vec{\sigma}-\vec{\eta}_{i}(\tau ),{%
\vec{\kappa}}_{i}(\tau ))\approx 0,  \nonumber \\
\vec{\chi}_{2}(\tau ,\vec{\sigma}) &=&\vec{\pi}_{\perp }(\tau ,\vec{\sigma}%
)-\sum_{i=1}^{N}Q_{i}\vec{\pi}_{\perp Si}(\vec{\sigma}-\vec{\eta}_{i}(\tau ),%
{\vec{\kappa}}_{i}(\tau ))\approx 0.  \label{V29}
\end{eqnarray}

These constraints allow us to eliminate the canonical degrees of freedom of
the radiation field and to get the symmetric Lienard-Wiechert reduced phase
space, in which there are only particle degrees of freedom. This has an
immediate and important consequence: the independent variables $\vec{\eta}%
_{i},\vec{\kappa}_{i}$ will no longer be canonical when one imposes these
constraints by way of modified Dirac brackets.

Now in order to compute the effects of these constraints we must use them in
the construction of Dirac brackets. This requires that we compute the 6x6
matrix of brackets
\begin{equation}
\bigg({%
%TCIMACRO{
%\QATOP{\{\vec{\chi}_{1},\vec{\chi}_{1}\}}{\{\vec{\chi}_{2},\vec{\chi}_{1}\}}}%
%BeginExpansion
{\{\vec{\chi}_{1},\vec{\chi}_{1}\} \atop \{\vec{\chi}_{2},\vec{\chi}_{1}\}}%
%EndExpansion
}{%
%TCIMACRO{
%\QATOP{\{\vec{\chi}_{1},\vec{\chi}_{2}\}}{\{\vec{\chi}_{2},\vec{\chi}_{2}\}}}%
%BeginExpansion
{\{\vec{\chi}_{1},\vec{\chi}_{2}\} \atop \{\vec{\chi}_{2},\vec{\chi}_{2}\}}%
%EndExpansion
}\bigg ).  \label{V30}
\end{equation}
It turns out that this matrix bracket is relatively simple, due to the
Grassmann charges. \ Consider, for example the case of two particles. The
particle or Lienard-Wiechert parts of the matrix bracket vanish since $%
Q_{1}^{2}=0=Q_{2}^{2}$ and cross terms vanish because they involve Poisson
brackets of particle one variables with particle two variables. Thus the
only part of the 6x6 matrix bracket that contributes is from the field
variables. It has the form
\begin{equation}
\{\vec{\chi}_{1}(\tau ,\vec{\sigma}_{1}),\vec{\chi}_{2}(\tau ,\vec{\sigma}%
_{2})\}=({\bf I}-{\frac{\vec{\nabla}\vec{\nabla}}{\vec{\nabla}^{2}}})\delta
^{3}(\vec{\sigma}_{1}-\vec{\sigma}_{2}),  \label{V31}
\end{equation}
and since
\begin{equation}
\{\vec{\chi}_{1},\vec{\chi}_{1}\}=0=\{\vec{\chi}_{2},\vec{\chi}_{2}\},
\label{V32}
\end{equation}
only the 3x3 off diagonal portion contributes.

In order to have a well defined Dirac bracket we need to use a modified form
of the Dirac bracket in which the inverse of the matrix of constraint
Poisson brackets is used. Calling this matrix $C$, we define $\tilde{C}^{-1}$
so that $C\tilde{C}^{-1}=({\bf I}-{\frac{\vec{\nabla}\vec{ \nabla}}{\vec{%
\nabla}^{2}}})\delta ^{3}(\vec{\sigma}_{1}-\vec{\sigma}_{2})$. But the
transverse form of the delta function allows us to use the idempotent
property of the projector to show that the inverse of $C$ in this sense is
just $C$ itself. In that case for two functions $f(\vec{\kappa}_{i},\vec{\eta%
}_{i}),g(\vec{\kappa}_{i},\vec{\eta}_{i})$ of the particle variables the
Dirac bracket becomes

\begin{eqnarray}
\{f,g\}^{\ast } &=&\{f,g\}-  \nonumber \\
&&-[\int d^{3}\sigma \{f,-\sum_{i}Q_{i}\vec{A}_{\perp Si}(\vec{\sigma}-\vec{%
\eta}_{i}(\tau ),{\vec{\kappa}}_{i}{(\tau )})\}\cdot \{-\sum_{j}Q_{j}\vec{\pi%
}_{\perp Sj}(\vec{\sigma}-\vec{\eta}_{j}(\tau ),{\vec{\kappa}}_{j}(\tau
)),g\}-  \nonumber \\
&&-\{f,-\sum_{j}Q_{j}\vec{\pi}_{\perp Sj}(\vec{\sigma}-\vec{\eta}_{j}(\tau ),%
{\vec{\kappa}}_{j}(\tau ))\}\cdot \{-\sum_{i}Q_{i}\vec{A}_{\perp Si}(\vec{%
\sigma}-\vec{\eta}_{i}(\tau ),{\vec{\kappa}}_{i}(\tau )),g\}].  \label{V33}
\end{eqnarray}
This bracket will lead to a new symplectic manifold by altering the basic
commutation relations and providing us with new canonical variables.\ Toward
this end we define the following scalar function.
\begin{equation}
{\cal K}=\sum_{i=1}^{N-1}\sum_{j=i+1}^{N}Q_{i}Q_{j}{\cal K}_{ij}(\vec{\kappa}%
_{i},\vec{\kappa}_{j};\vec{\eta}_{i}-\vec{\eta}_{j}),  \label{V34}
\end{equation}
in which
\begin{eqnarray}
{\cal K}_{ij} &=&\int d^{3}\vec{\sigma}[\vec{A}_{\perp Si}(\vec{\sigma}-\vec{%
\eta}_{i},{\vec{\kappa}}_{i})\cdot \vec{\pi}_{\perp Sj}(\vec{\sigma}-\vec{%
\eta}_{j},{\vec{\kappa}}_{j})-  \nonumber \\
&-&\vec{A}_{\perp Sj}(\vec{\sigma}-\vec{\eta}_{j},{\vec{\kappa}}_{j})\cdot
\vec{\pi}_{\perp Si}(\vec{\sigma}-\vec{\eta}_{i},{\vec{\kappa}}_{i})]=
\nonumber \\
&=&{\cal K}_{ij}(\vec{\kappa}_{i},\vec{\kappa}_{j};\vec{\eta}_{i}-\vec{\eta}%
_{j})=-{\cal K}_{ji}.  \label{V35}
\end{eqnarray}

Let $\widetilde{\vec{\eta}_{i}}=\vec{\eta}_{i}+\vec{\alpha}_{i},$ $%
\widetilde{\vec{\kappa}_{i}}=\vec{\kappa}_{i}+\vec{\beta}_{i},$ $i=1,2,..N$
where
\begin{eqnarray}
\vec{\alpha}_{i} &=&a_{i}\sum_{j=i+1}^{N}Q_{i}Q_{j}\nabla _{\kappa _{i}}%
{\cal K}_{ij}+\bar{a}_{i}\sum_{j=1}^{i-1}Q_{i}Q_{j}\nabla _{\kappa _{i}}%
{\cal K}_{ji},  \nonumber \\
\vec{\beta}_{i} &=&b_{i}\sum_{j=i+1}^{N}Q_{i}Q_{j}\nabla _{\eta _{i}}{\cal K}%
_{ij}+\bar{b}_{i}\sum_{j=1}^{i-1}Q_{i}Q_{j}\nabla _{\eta _{i}}{\cal K}_{ji}.
\label{V36}
\end{eqnarray}
Since they do not appear in these equations, we may choose $\bar{a}_{1}=\bar{%
b }_{1}=a_{N}=b_{N}=0$.

We determine relations between the unknown coefficients by requiring that \
\ $\widetilde{\vec{\eta}_{i}},\widetilde{\vec{\kappa}_{j}}$ be independent
canonical variables. So, for example, (for $k<l$)
\begin{eqnarray}
\{\widetilde{\vec{\eta}_{k}},\widetilde{\vec{\eta}_{l}}\}^{\ast } &=&\{%
\widetilde{\vec{\eta}_{k}},\widetilde{\vec{\eta}_{l}}\}-  \nonumber \\
&-&[\int d^{3}\sigma \{\widetilde{\vec{\eta}_{k}},-\sum_{i}Q_{i}\vec{A}%
_{\perp Si}(\vec{\sigma}-\vec{\eta}_{i},{\vec{\kappa}}_{i})\}\cdot
\{-\sum_{j}Q_{j}\vec{\pi}_{\perp Sj}(\vec{\sigma}-\vec{\eta}_{j},{\vec{\kappa%
}}_{j}),\widetilde{\vec{\eta}_{l}}\}-  \nonumber \\
&&-\{\widetilde{\vec{\eta}_{k}},-\sum_{j}Q_{j}\vec{\pi}_{\perp Sj}(\vec{%
\sigma}-\vec{\eta}_{j},{\vec{\kappa}}_{j})\}\cdot \{-\sum_{i}Q_{i}\vec{A}%
_{\perp Si}(\vec{\sigma}-\vec{\eta}_{i},{\vec{\kappa}}_{i}),\widetilde{\vec{%
\eta}_{l}}\}]=  \nonumber \\
&=&\{\vec{\eta}_{k},\vec{\alpha}_{l}\}+\{\vec{\alpha}_{k},\vec{\eta}%
_{l}\}+Q_{k}Q_{l}\nabla _{\kappa _{k}}\nabla _{\kappa _{l}}{\cal K}_{kl}=0.
\label{V37}
\end{eqnarray}
Then using the expressions for $\vec{\alpha}_{i}$ leads to
\begin{equation}
a_{l}-\bar{a}_{k}=1,\quad k>l;\quad \quad \bar{a}_{l}-a_{k}=-1,\quad l>k.
\label{V38}
\end{equation}
Solving this gives
\begin{eqnarray}
\bar{a}_{2} &=&\bar{a}_{3}=..=\bar{a}_{N}:=\bar{a},  \nonumber \\
a_{1} &=&a_{2}=..a_{N-1}:=a.  \label{V39}
\end{eqnarray}
Similarly, requiring that\ $\widetilde{\{\vec{\kappa}_{k}},\widetilde{\vec{%
\kappa}_{l}}\}^{\ast }=0$ \ leads to
\begin{eqnarray}
\bar{b}_{2} &=&\bar{b}_{3}=..=b_{N}:=\bar{b},  \nonumber \\
b_{1} &=&b_{2}=..=b_{N-1}:=b.  \label{V40}
\end{eqnarray}
Requiring that $\{\widetilde{\vec{\eta}_{i}},$ $\widetilde{\vec{\kappa}_{i}}%
\}^{\ast }={\vec{\vec{1}}}$ \ leads to
\begin{equation}
a_{i}+b_{i}=0,\quad i=1,..,N-1;\quad\quad \bar{a}_{i}+\bar{b}_{i}=0,\quad
i=2,..,N.  \label{V41}
\end{equation}
This condition implies that the first two conditions are equivalent to one
another. The requirement that $\{\widetilde{\vec{\eta}_{k}},$ $\widetilde{%
\vec{\kappa}_{l}}\}^{\ast }=0$ for $k<l$ leads to
\begin{equation}
a_{k}+\bar{b}_{l}=1;\quad \quad k<l,  \label{V42}
\end{equation}
or
\begin{equation}
a_{k}-\bar{a}_{l}=1;\quad\quad k<l,  \label{V43}
\end{equation}
which is also the same as the first condition. \ While for $l<k$ it leads to
\begin{equation}
\bar{a}_{k}+b_{l}=-1;\quad\quad l<k,  \label{V44}
\end{equation}
or
\begin{equation}
\bar{a}_{k}-a_{l}=-1;\quad\quad l<k,  \label{V45}
\end{equation}
which again is the same as the first condition. \ This leaves us with just
two unknowns $a$ and $\bar{a}$.

So in summary we have
\begin{eqnarray}
\widetilde{\vec{\eta}_{i}} &=&\vec{\eta}_{i}+a\sum_{j=i+1}^{N}Q_{i}Q_{j}\vec{%
\nabla}_{\kappa _{i}}{\cal K}_{ij}+\bar{a}\sum_{j=1}^{i-1}Q_{i}Q_{j}\vec{%
\nabla}_{\kappa _{i}}{\cal K}_{ji},  \nonumber \\
\widetilde{\vec{\kappa}_{i}} &=&\vec{\kappa}_{i}-a\sum_{j=i+1}^{N}Q_{i}Q_{j}%
\vec{\nabla}_{\eta _{i}}{\cal K}_{ij}-\bar{a}\sum_{j=1}^{i-1}Q_{i}Q_{j}\vec{%
\nabla}_{\eta _{i}}{\cal K}_{ji}.  \label{V46}
\end{eqnarray}

Let us rewrite the rest frame condition Eq.(\ref{III40})
\begin{eqnarray}
&&{\vec{{\cal H}}}_{p}={\vec{{\cal P}}}_{(int)}=\sum_{i-1}^{N}\vec{\kappa}%
_{i}+\int d^{3}\sigma \lbrack {\vec{\pi}}_{\perp }\times {\vec{B}}](\tau ,%
\vec{\sigma})=  \nonumber \\
&=&\sum_{i-1}^{N}\vec{\kappa}_{i}+\sum_{i<j}Q_{i}Q_{j}\int d^{3}\sigma \Big[%
\vec{\pi}_{\perp Si}(\vec{\sigma}-\vec{\eta}_{i},{\vec{\kappa}}_{i})\times ({%
\vec{\nabla}}_{\sigma }\times \vec{A}_{\perp Sj}(\vec{\sigma}-\vec{\eta}_{j},%
{\vec{\kappa}}_{j}))+  \nonumber \\
&&+\vec{\pi}_{\perp Sj}(\vec{\sigma}-\vec{\eta}_{j},{\vec{\kappa}}%
_{j})\times ({\vec{\nabla}}_{\sigma }\times \vec{A}_{\perp Si}(\vec{\sigma}-%
\vec{\eta}_{i},{\vec{\kappa}}_{i}))\Big]\approx 0.  \label{V47}
\end{eqnarray}
in these new canonical variables. \ Let us expand the cross products,
integrate by parts and use the transverse gauge condition to get
\begin{equation}
{\vec{{\cal H}}}_{p}={\vec{{\cal P}}}_{(int)}=\sum_{i-1}^{N}\vec{\kappa}%
_{i}+\sum_{i<j}Q_{i}Q_{j}\vec{\nabla}_{\eta _{j}}{\cal K}_{ij}=0.
\label{V48}
\end{equation}
If we choose $a=-{\bar{a}=}\frac{1}{2}$ this becomes
\begin{equation}
{\vec{{\cal H}}}_{p}={\vec{{\cal P}}}_{(int)}={\tilde{\vec{\kappa}}}%
_{+}=\sum_{i=1}^{N}\widetilde{\vec{\kappa}_{i}}=0,  \label{V49}
\end{equation}

\noindent like in the case of $N$ either free or interacting particles on
the Wigner hyperplane \cite{lu1} [in an instant form of dynamics the
Poincar\'{e} generators ${\vec{{\cal P}}}_{(int)}$ do not depend on the
interaction].

In other words the rest frame condition is simply that the sum of the $N$
(new) canonical momentum is zero. Note that this same choice gives
\begin{equation}
\sum_{i=1}^{N}\widetilde{\vec{\eta}_{i}}=\sum_{i-1}^{N}\vec{\eta}%
_{i}-\sum_{i<j}Q_{i}Q_{j}\vec{\nabla}_{\kappa _{j}}{\cal K}_{ij}.
\label{V50}
\end{equation}
Therefore, with the choice $a=-\bar{a}=\frac{1}{2},$ Eq.(\ref{V46}) defining
the final canonical variables becomes
\begin{eqnarray}
\widetilde{\vec{\eta}_{i}} &=&\vec{\eta}_{i}+\frac{1}{2}\sum_{j\neq
i}Q_{i}Q_{j}\vec{\nabla}_{\kappa _{j}}{\cal K}_{ij},  \nonumber \\
\widetilde{\vec{\kappa}_{i}} &=&\vec{\kappa}_{i}-\frac{1}{2}\sum_{j\neq
i}Q_{i}Q_{j}\vec{\nabla}_{\eta _{i}}{\cal K}_{ij},  \label{V51}
\end{eqnarray}
with ${\cal K}_{ij}$ given by Eq.(\ref{V35}).

In the next section we will re-express the other internal Poincar\'{e}
generators $M={\cal P}_{(int)}^{\tau }$, ${\cal \vec{J}}_{(int)}$, ${\vec{%
{\cal K}}}_{(int)}$ of Eqs.(\ref{IV1}) and the internal center-of-mass
coordinate ${\vec{q}}_{+}\approx -{\vec{{\cal K}}}_{(int)}/M$ of Eq.(\ref
{IV12}) in these final canonical variables. \vfill\eject

\section{The Exact Darwin Hamiltonian from the Invariant Mass.}

In this Section our aim is to use the explicit semiclassical Lienard
Wiechert solution of Eqs.(\ref{V27}), (\ref{V28}) for the transverse vector
potential to obtain an explicit form of the instantaneous
action-at-a-distance potentials present in the invariant mass $M={\cal P}%
^{\tau}_{(int)}$ of Eq.(\ref{IV1}) after the elimination of the
electromagnetic degrees of freedom. In its phase space form this Hamiltonian
for the $\tau \equiv T_s$-evolution will contain:

i) a vector potential ${\vec{V}}_{i}(\tau )=Q_{i}\sum_{i\not=j}^{1..N}Q_{j}{%
\vec{A}}_{\perp Sj}({\vec{\eta}}_{i}(\tau )-{\vec{\eta}}_{j}(\tau ),{\vec{%
\kappa}}_{j}(\tau ))$, minimally coupled to ${\vec{\kappa}}_{i}(\tau )$,
under the square root  kinetic energy term of each particle `i';

ii) a scalar potential $U(\tau )={\frac{1}{2}}\int d^{3}\sigma \lbrack {\vec{%
\pi}}_{\perp S}^{2}+{\vec{B}}_{S}^{2}](\tau ,\vec{\sigma})$, coming from the
field energy, which adds to the Coulomb potentials.

Due to $Q_{i}^{2}=0$ we can extract the vector potentials from the square
roots: the semiclassical contribution from the vector potentials is a new
effective scalar potential $U_{1}=-\sum_{i=1}^{N}Q_{i}{\frac{{{\vec{\kappa}}%
_{i}\cdot {\vec{V}}_{i}}}{\sqrt{m_{i}^{2}+{\vec{\kappa}}_{i}^{2}}}}$ and we
get the complete Darwin potential $V_{DAR}=U+U_{1}$ added to the Coulomb
one. In this form the invariant mass becomes the exact semiclassical Darwin
Hamiltonian and $V_{DAR}$ is the Darwin potential to all orders of $1/c^{2}$
for every $N$. If we call $V_{LOD}$ the lowest $1/c^{2}$ order historical
Darwin potential [see Eq.(\ref{VI11})], we have $U_{1}=2V_{LOD}+U_{1HOD}$
[see Eq.(\ref{VI10})], $U=-V_{LOD}-U_{1HOD}+U_{HOD}$ [see Eq.(\ref{VI13})], $%
V_{DAR}=V_{LOD}+U_{HOD}$. [``$LOD$'' and ``$HOD$'' mean lowest and higher
order in $1/c^{2}$ respectively]. \ When we re-express the invariant mass in
terms of the final canonical variables, there is an extra contribution $%
U_{HOD}^{\prime }$ coming from the square roots, so that at the end the
final Darwin potential is
\begin{equation}
\tilde{V}_{DAR}=V_{DAR}+U_{HOD}^{\prime
}=V_{LOD}+V_{HOD};\,\,\,V_{HOD}=U_{HOD}+U_{HOD}^{\prime }.  \label{VI1}
\end{equation}

In this Section we shall evaluate the complete (to all orders in $1/c^{2}$)
Darwin potential in the old (no longer canonical) variables and then we
shall re-express it in terms of the new final canonical variables. We begin
by obtaining the contribution of the field energy integrals, expressed in
terms of the canonical particle variables. \ Using the truncation properties
of the Grassmann charges we extract from the kinetic piece the vector
potential portion and combine it \ with the field energy integral. In
addition to the naive kinetic part (expressed in terms of the old canonical
momentum) we obtain the rest frame Coulomb part, a part that generalizes the
standard Darwin interaction and a double infinite series containing all
higher order corrections. \ As an extra check on these generalized Darwin
interactions in $M$ we obtain an independent derivation of this series using
the Lagrangian expression for the invariant mass. \ Then we express the
kinetic portion in terms of the new canonical momentum to obtain the final
form of the complete Darwin Hamiltonian. This interaction Hamiltonian
contains no $N$-body forces and is a sum of \ two-body portions. \ In the
center-of-mass rest frame the double infinite series can be summed exactly
to obtain a\ closed form expression for the special case of two particles.
Then all the generators of the ``internal'' Poincaire` algebra and the
energy-momentum tensor are expressed in the new variables in the $N$%
-particle case. As with the three-momentum we find that the internal angular
momentum does not depend on the interaction. \ Also we obtain the
``internal'' center of mass $\vec{q}_{+}$ and there are some comments on how
to find a collective variable (replacing the center of mass) for a cluster
of particles interacting with the remaining ones of the isolated system.

\subsection{Field Energy and Momentum Integrals.}

Although we have summed exactly to a closed form (\ref{V28}) the
semiclassical Lienard-Wiechert solution, we use the series form (\ref{V27})
for finding the expression for the invariant mass $M$, since the closed form
provides no simplification in obtaining that expression. From the above we
can find expressions for the semiclassical Lienard-Wiechert electric and
magnetic fields. For the electric field we find [${\ddot {\vec \eta}}_i(\tau
)$ does not contribute due to Eq.(\ref{IV23}), and the same is true of ${%
\dot {\vec \kappa}}_i(\tau )$; ${\frac{{\partial}}{{\partial \tau}}} |\vec %
\sigma -{\vec \eta}_i(\tau )|^n=-{\dot {\vec \eta}}_i(\tau )\cdot {\vec %
\nabla}_{\sigma} |\vec \sigma -{\vec \eta}_i(\tau )|^n$]

\begin{eqnarray}
\vec{E}_{\perp S}(\tau ,\vec{\sigma}) &=&{\vec{\pi}}_{\perp S}(\tau ,\vec{%
\sigma})=-{\frac{\partial \vec{A}_{\perp S}(\tau ,\vec{\sigma})}{\partial
\tau }}=  \nonumber \\
&=&\sum_{i=1}^{N}{\frac{Q_{i}}{4\pi }}\sum_{m=0}^{\infty }\Big[{\frac{1}{%
(2m)!}}\dot{\vec{\eta}}_{i}(\tau )(\dot{\vec{\eta}}_{i}(\tau )\cdot {\vec{%
\nabla}}_{\sigma })^{2m+1}\,|\vec{\sigma}-\vec{\eta}_{i}(\tau )|^{2m-1}-
\nonumber \\
&-&{\frac{1}{(2m+2)!}}{\vec{\nabla}}_{\sigma }(\dot{\vec{\eta}}_{i}(\tau
)\cdot {\vec{\nabla}}_{\sigma })^{2m+2}\,|\vec{\sigma}-\vec{\eta}_{i}(\tau
)|^{2m+1}\Big]=  \nonumber \\
&=&\sum_{i=1}^{N}Q_{i}{\vec{E}}_{\perp Si}(\vec{\sigma}-{\vec{\eta}}%
_{i}(\tau ),{\vec{\kappa}}_{i}(\tau ))=  \nonumber \\
&=&\sum_{i=1}^{N}Q_{i}{\frac{{{\vec{\kappa}}_{i}(\tau )\cdot {\vec{\nabla}}%
_{\sigma }}}{\sqrt{m_{i}^{2}+{\vec{\kappa}}_{i}^{2}(\tau )}}}\,{\vec{A}}%
_{\perp Si}(\vec{\sigma}-{\vec{\eta}}_{i}(\tau ),{\vec{\kappa}}_{i}(\tau ))=
\nonumber \\
&=&-\sum_{i=1}^{N}Q_{i}\times  \nonumber \\
&&{\frac{1}{{4\pi |\vec{\sigma}-{\vec{\eta}}_{i}(\tau )|^{2}}}}\Big[{\vec{%
\kappa}}_{i}(\tau )\,{\vec{\kappa}}_{i}(\tau )\cdot {\frac{{\vec{\sigma}-{%
\vec{\eta}}_{i}(\tau )}}{{|\vec{\sigma}-{\vec{\eta}}_{i}(\tau )|}}}{\frac{%
\sqrt{m_{i}^{2}+{\vec{\kappa}}_{i}^{2}(\tau )}}{[m_{i}^{2}+({\vec{\kappa}}%
_{i}(\tau )\cdot {\frac{{\vec{\sigma}-{\vec{\eta}}_{i}(\tau )}}{{|\vec{\sigma%
}-{\vec{\eta}}_{i}(\tau )|}}})^{2}]^{3/2}}}+  \nonumber \\
&&+{\frac{{\vec{\sigma}-{\vec{\eta}}_{i}(\tau )}}{{|\vec{\sigma}-{\vec{\eta}}%
_{i}(\tau )|}}}\Big({\frac{{\ {\vec{\kappa}}_{i}^{2}(\tau )+({\vec{\kappa}}%
_{i}(\tau )\cdot {\frac{{\vec{\sigma}-{\vec{\eta}}_{i}(\tau )}}{{|\vec{\sigma%
}-{\vec{\eta}}_{i}(\tau )|}}})^{2}}}{{{\vec{\kappa}}_{i}^{2}(\tau )-({\vec{%
\kappa}}_{i}(\tau )\cdot {\frac{{\vec{\sigma}-{\vec{\eta}}_{i}(\tau )}}{{|%
\vec{\sigma}-{\vec{\eta}}_{i}(\tau )|}}})^{2}}}}({\frac{\sqrt{m_{i}^{2}+{%
\vec{\kappa}}_{i}^{2}(\tau )}}{\sqrt{m_{i}^{2}+({\vec{\kappa}}_{i}(\tau
)\cdot {\frac{{\vec{\sigma}-{\vec{\eta}}_{i}(\tau )}}{{|\vec{\sigma}-{\vec{%
\eta}}_{i}(\tau )|}}})^{2}}}}-1)+  \nonumber \\
&&+{\frac{{({\vec{\kappa}}_{i}(\tau )\cdot {\frac{{\vec{\sigma}-{\vec{\eta}}%
_{i}(\tau )}}{{|\vec{\sigma}-{\vec{\eta}}_{i}(\tau )|}}})^{2}\sqrt{m_{i}^{2}+%
{\vec{\kappa}}_{i}^{2}(\tau )}}}{{[m_{i}^{2}+({\vec{\kappa}}_{i}(\tau )\cdot
{\frac{{\vec{\sigma}-{\vec{\eta}}_{i}(\tau )}}{{|\vec{\sigma}-{\vec{\eta}}%
_{i}(\tau )|}}})^{2}\,]^{3/2}}}}\Big)\Big].  \label{VI2}
\end{eqnarray}

The magnetic field is [$B_{S}^{r}=\epsilon ^{rsu}{\frac{{\partial }}{{%
\partial \sigma ^{s}}}}A_{\perp S}^{u}$]
\begin{eqnarray}
\vec{B}_{S}(\tau ,\vec{\sigma}) &=&\sum_{i}^{1..N}Q_{i}{\vec{B}}_{Si}(\vec{%
\sigma}-{\vec{\eta}}_{i}(\tau ),{\vec{\kappa}}_{i}(\tau ))=  \nonumber \\
&=&-\sum_{i=1}^{N}{\frac{Q_{i}}{4\pi }}\sum_{m=0}^{\infty }{\frac{1}{(2m)!}}%
\dot{\vec{\eta}}_{i}(\tau )\times {\vec{\nabla}}_{\sigma }(\dot{\vec{\eta}}%
_{i}(\tau )\cdot {\vec{\nabla}}_{\sigma })^{2m})\,|\vec{\sigma}-\vec{\eta}%
_{i}(\tau )|^{2m-1}=  \nonumber \\
&=&\sum_{i=1}^{N}Q_{i}{\frac{1}{{4\pi |\vec{\sigma}-{\vec{\eta}}_{i}(\tau
)|^{2}}}}{\frac{{m_{i}^{2}\,{\vec{\kappa}}_{i}(\tau )\times {\frac{{\vec{%
\sigma}-{\vec{\eta}}_{i}(\tau )}}{{|\vec{\sigma}-{\vec{\eta}}_{i}(\tau )|}}}}%
}{{[m_{i}^{2}+({\vec{\kappa}}_{i}(\tau )\cdot {\frac{{\vec{\sigma}-{\vec{\eta%
}}_{i}(\tau )}}{{|\vec{\sigma}-{\vec{\eta}}_{i}(\tau )|}}})^{2}\,]^{3/2}}}}.
\label{VI3}
\end{eqnarray}
In these expressions ${\dot{\vec{\eta}}}_{i}(\tau )$ may be replaced with ${%
\vec{\kappa}}_{i}(\tau )/\sqrt{m_{i}^{2}+{\vec{\kappa}}_{i}^{2}(\tau )}$. We
have also given the closed form of the fields.

Let us remark that the Feynman-Wheeler complete absorber assumption is
violated in the Maxwell theory, since it would imply (see for instance Ref.
\cite{leiter}, where there is the definition of radiation in the
Feynman-Wheeler theory) ${\vec E}_{\perp S}(\tau ,\vec \sigma )={\vec E}%
_{\perp \pm}(\tau ,\vec \sigma )=0$ and ${\vec B}_S(\tau ,\vec \sigma )={%
\vec B}_{\pm}(\tau ,\vec \sigma )=0$ everywhere (inside and outside the
absorbers).

For the energy we need ${\vec{E}}_{\perp S}^{2}+{\vec{B}}_{S}^{2}$ and for
the momentum we need ${\vec{E}}_{\perp S}\times {\vec{B}}_{S}$. In Appendix
B we evaluate these and show that [${\vec{\eta}}_{ij}(\tau )=\eta _{ij}(\tau
){\hat{\eta}}_{ij}(\tau )={\vec{\eta}}_{i}(\tau )-{\vec{\eta}}_{j}(\tau )$; $%
{\vec{\nabla}}_{ij}=\partial /\partial {\vec{\eta}}_{ij}$]
\begin{eqnarray}
&&U(\tau )={\frac{1}{2}}\int d^{3}\sigma ({\vec{E}}_{\perp S}^{2}+{\vec{B}}%
_{S}^{2})(\tau ,\vec{\sigma}):=\sum_{i<j}^{1..N}{\frac{{Q_{i}Q_{j}}}{{4\pi }}%
}h_{1}(\dot{\vec{\eta}_{i}},\dot{\vec{\eta}_{j}},\vec{\eta}_{ij})=  \nonumber
\\
&&  \nonumber \\
&=&\sum_{i<j}^{1..N}{\frac{Q_{i}Q_{j}}{4\pi }}\sum_{m=0}^{\infty
}\sum_{n=0}^{\infty }\Big[\dot{\vec{\eta}_{i}}\cdot \dot{\vec{\eta}_{j}}{%
\frac{(\dot{\vec{\eta}_{i}}\cdot \vec{\nabla}_{ij})^{2m+1}(\dot{\vec{\eta}%
_{j}}\cdot \vec{\nabla}_{ij})^{2n+1}\eta _{ij}^{2n+2m+1}}{(2n+2m+2)!}}-
\nonumber \\
&-&{\frac{(\dot{\vec{\eta}_{i}}\cdot \vec{\nabla}_{ij})^{2m+2}(\dot{\vec{\eta%
}_{j}}\cdot \vec{\nabla}_{ij})^{2n+2}\eta ^{2n+2m+3}}{(2n+2m+4)!}}+
\nonumber \\
&+&\dot{\vec{\eta}_{i}}\cdot \dot{\vec{\eta}_{j}}{\frac{(\dot{\vec{\eta}_{i}}%
\cdot \vec{\nabla}_{ij})^{2m}(\dot{\vec{\eta}_{j}}\cdot \vec{\nabla}%
_{ij})^{2n}\vec{\nabla}_{ij}^{2}\eta _{ij}^{2n+2m+1}}{(2n+2m+2)!}}-
\nonumber \\
&-&{\frac{(\dot{\vec{\eta}_{i}}\cdot \vec{\nabla}_{ij})^{2m+1}(\dot{\vec{\eta%
}_{j}}\cdot \vec{\nabla}_{ij})^{2n+1}\eta _{ij}^{2n+2m+1}}{(2n+2m+2)!}}\Big],
\label{VI4}
\end{eqnarray}
and
\begin{eqnarray}
&&\int d^{3}\sigma ({\vec{E}}_{\perp S}\times {\vec{B}}_{S})(\tau ,\vec{%
\sigma}):=\sum_{i<j}^{1..N}Q_{i}Q_{j}\vec{h}_{1}(\dot{\vec{\eta}_{i}},\dot{%
\vec{\eta}_{j}},\vec{\eta}_{ij})=  \nonumber \\
&&  \nonumber \\
&=&\sum_{i<j}^{1..N}{\frac{Q_{i}Q_{j}}{4\pi }}\Big[\sum_{m=0}^{\infty
}\sum_{n=0}^{\infty }\Big({\vec{\nabla}}_{ij}[\dot{\vec{\eta}_{i}}\cdot \dot{%
\vec{\eta}_{j}}{\frac{(\dot{\vec{\eta}_{i}}\cdot \vec{\nabla}_{ij})^{2m+1}(%
\dot{\vec{\eta}_{j}}\cdot \vec{\nabla}_{ij})^{2n}\eta _{ij}^{2n+2m+1}}{%
(2n+2m+2)!}}-  \nonumber \\
&-&{\frac{(\dot{\vec{\eta}_{i}}\cdot \vec{\nabla}_{ij})^{2m+2}(\dot{\vec{\eta%
}_{j}}\cdot \vec{\nabla}_{ij})^{2n+1}\eta _{ij}^{2n+2m+3}}{(2n+2m+4)!}}]\Big)%
+(i\longleftrightarrow j)\Big].  \label{VI5}
\end{eqnarray}

\subsection{The complete Darwin Hamiltonian in terms of the old canonical
variables}

Our first aim is to express the invariant mass $M$, i.e. the Hamiltonian for
the $\tau \equiv T_s$-evolution, in terms of the original canonical
variables. Later we will obtain this Hamiltonian in terms of the new
canonical variables.\ The Hamiltonian $M$ and the internal 3-momentum are

\begin{eqnarray}
M &=&{\cal P}_{(int)}^{\tau }=\sum_{i=1}^{N}\sqrt{m_{i}^{2}+(\vec{\kappa}%
_{i}(\tau )-Q_{i}\vec{A}_{\perp S}(\tau ,\vec{\eta}_{i}(\tau )))^{2}}+
\nonumber \\
&+&\sum_{i\neq j}^{1..N}\frac{Q_{i}Q_{j}}{4\pi \mid \vec{\eta}_{i}(\tau )-%
\vec{\eta}_{j}(\tau )\mid }+\int d^{3}\sigma {\frac{1}{2}}[\vec{\pi}_{\perp
S}^{2}(\tau ,\vec{\sigma})+\vec{B}_{S}^{2}(\tau ,\vec{\sigma})]=  \nonumber
\\
&=&\sum_{i=1}^{N}\sqrt{m_{i}^{2}+[{\vec{\kappa}}_{i}(\tau )-{\vec{V}}%
_{i}(\tau )]^{2}}+\sum_{i\not=j}^{1..N}{\frac{{Q_{i}Q_{j}}}{{4\pi |{\vec{\eta%
}}_{i}(\tau )-{\vec{\eta}}_{j}(\tau )|}}}+U(\tau )=  \nonumber \\
&=&\sum_{i=1}^{N}\sqrt{m_{i}^{2}+{\vec{\kappa}}_{i}^{2}}+\sum_{i\not=%
j}^{1..N}{\frac{{Q_{i}Q_{j}}}{{4\pi |{\vec{\eta}}_{i}(\tau )-{\vec{\eta}}%
_{j}(\tau )|}}}+V_{DAR}(\tau ),  \nonumber \\
&&{}  \nonumber \\
V_{DAR}(\tau )\, &{:=}\,&V_{LOD}(\tau )+U_{HOD}(\tau ),  \nonumber \\
&&{}  \nonumber \\
{\vec{{\cal P}}}_{(int)} &=&{\vec{{\cal H}}}_{p}(\tau )={\vec{\kappa}}%
_{+}(\tau )+\int d^{3}\sigma \lbrack {\vec{\pi}}_{\perp S}\times {\vec{B}}%
_{S}](\tau ,\vec{\sigma})\approx 0,  \label{VI6}
\end{eqnarray}
The first line of $M$ is
\begin{eqnarray}
&&\sum_{i=1}^{N}\sqrt{m_{i}^{2}+(\vec{\kappa}_{i}(\tau )-Q_{i}\vec{A}_{\perp
S}(\tau ,\vec{\eta}_{i}(\tau )))^{2}}=\sum_{i=1}^{N}\left( \sqrt{m_{i}^{2}+%
\vec{\kappa}_{i}(\tau )^{2}}-\frac{\vec{\kappa}_{i}(\tau )\cdot Q_{i}\vec{A}%
_{\perp S}(\tau ,\vec{\eta}_{i}(\tau ))}{\sqrt{m_{i}^{2}+\vec{\kappa}%
_{i}(\tau )^{2}}}\right) =  \nonumber \\
&=&\sum_{i=1}^{N}\sqrt{m_{i}^{2}+{\vec{\kappa}}_{i}^{2}}+U_{1},  \nonumber \\
&&{}  \nonumber \\
&&U_{1}=-\sum_{i=1}^{N}{\frac{{{\vec{\kappa}}_{i}\cdot {\vec{V}}_{i}}}{\sqrt{
m_{i}^{2}+{\vec{\kappa}}_{i}^{2}}}},\quad \Rightarrow \quad V_{DAR}(\tau
)=U_{1}(\tau )+U(\tau )=V_{LOD}(\tau )+U_{HOD}(\tau ),  \label{VI7}
\end{eqnarray}
in which the vector potential is given by the semiclassical Lienard-Wiechert
transverse potential Eq.(\ref{V27}). By re-expressing this transverse vector
potential in terms of the momenta [one may use in that expression either the
new or old canonical variables because of the Grassmann truncation] we obtain

\begin{eqnarray}
\vec{A}_{\perp S}(\tau ,\vec{\sigma}) &=&\sum_{i=1}^{N}{\frac{Q_{i}}{4\pi }}%
\sum_{m=0}^{\infty }\Big[{\frac{1}{(2m)!}\frac{\vec{\kappa}_{i}}{\sqrt{%
m_{i}^{2}+{\vec{\kappa}_{i}}^{2}}}}({\frac{\vec{\kappa}_{i}}{\sqrt{m_{i}^{2}+%
{\vec{\kappa}_{i}}^{2}}}}\cdot \vec{\nabla}_{\sigma })^{2m}\,|\vec{\sigma}-%
\vec{\eta}_{i}|^{2m-1}-  \nonumber \\
&&-{\frac{1}{(2m+2)!}}\vec{\nabla}_{\sigma }({\frac{\vec{\kappa}_{i}}{\sqrt{%
m_{i}^{2}+{\vec{\kappa}_{i}}^{2}}}}\cdot \vec{\nabla}_{\sigma })^{2m+1}\,|%
\vec{\sigma}-\vec{\eta}_{i}|^{2m+1}\Big],  \label{VI8}
\end{eqnarray}
so that for the scalar potential $U_{1}$ we get
\begin{eqnarray}
U_{1} &=&-\sum_{i=1}^{N}{\frac{{{\vec{\kappa}}_{i}\cdot {\vec{V}}_{i}}}{%
\sqrt{m_{i}^{2}+{\vec{\kappa}}_{i}^{2}}}}=\sum_{i=1}^{N}\left( -\frac{\vec{%
\kappa}_{i}\cdot Q_{i}\vec{A}_{\perp S}(\tau ,\vec{\eta}_{i}(\tau ))}{\sqrt{%
m_{i}^{2}+\vec{\kappa}_{i}(\tau )^{2}}}\right) =  \nonumber \\
&=&-\sum_{i<j}^{1..N}{\frac{Q_{i}Q_{j}}{4\pi }}\sum_{m=0}^{\infty }\Big[({%
\frac{\vec{\kappa}_{i}}{\sqrt{m_{i}^{2}+{\vec{\kappa}_{i}}^{2}}}\cdot \frac{%
\vec{\kappa}_{j}}{\sqrt{m_{j}^{2}+{\vec{\kappa}_{j}}^{2}}})}  \nonumber \\
&&\Big(({\frac{\vec{\kappa}_{i}}{\sqrt{m_{i}^{2}+{\vec{\kappa}_{i}}^{2}}}}%
\cdot \vec{\nabla}_{ij})^{2m}+({\frac{\vec{\kappa}_{j}}{\sqrt{m_{j}^{2}+{%
\vec{\kappa}_{j}}^{2}}}}\cdot \vec{\nabla}_{ij})^{2m}\Big){\frac{\eta
_{ij}^{2m-1}}{(2m)!}}-  \nonumber \\
&&-\Big(({\frac{\vec{\kappa}_{j}}{\sqrt{m_{j}^{2}+{\vec{\kappa}_{j}}^{2}}}}%
\cdot \vec{\nabla}_{ij})({\frac{\vec{\kappa}_{i}}{\sqrt{m_{i}^{2}+{\vec{%
\kappa}_{i}}^{2}}}}\cdot \vec{\nabla}_{ij})^{2m+1}+  \nonumber \\
&&+({\frac{\vec{\kappa}_{i}}{\sqrt{m_{i}^{2}+{\vec{\kappa}_{i}}^{2}}}}\cdot
\vec{\nabla}_{ij})({\frac{\vec{\kappa}_{j}}{\sqrt{m_{j}^{2}+{\vec{\kappa}j}%
^{2}}}}\cdot \vec{\nabla}_{ij})^{2m+1}\Big){\frac{\eta _{ij}^{2m+1}}{(2m+2)!}%
}\Big].  \label{VI9}
\end{eqnarray}
The lowest order part of this kinetic contribution (the $m=0$ term) is twice
the familiar Darwin interaction (but with $m_{i}\rightarrow \sqrt{m_{i}^{2}+{%
\vec{\kappa}}_{i}^{2}}$; strictly speaking this is a higher order
correction). Thus we have
\begin{equation}
U_{1}=-\sum_{i=1}^{N}{\frac{{{\vec{\kappa}}_{i}\cdot {\vec{V}}_{i}}}{\sqrt{%
m_{i}^{2}+{\vec{\kappa}}_{i}^{2}}}}=2V_{LOD}(\tau )+U_{1HOD},  \label{VI10}
\end{equation}
with
\begin{eqnarray}
&&V_{LOD}=-\sum_{i<j}^{1..N}{\frac{Q_{i}Q_{j}}{4\pi }}\Big({\frac{\vec{\kappa%
}_{i}}{\sqrt{m_{i}^{2}+{\vec{\kappa}_{i}}^{2}}}}\cdot {\frac{\vec{\kappa}_{j}%
}{\sqrt{m_{j}^{2}+{\vec{\kappa}_{j}}^{2}}}\frac{1}{\eta _{ij}}}-  \nonumber
\\
&&-({\frac{\vec{\kappa}_{i}}{\sqrt{m_{i}^{2}+{\vec{\kappa}_{i}}^{2}}}}\cdot
\vec{\nabla}_{ij})({\frac{\vec{\kappa}_{j}}{\sqrt{m_{j}^{2}+{\vec{\kappa}_{j}%
}^{2}}}}\cdot \vec{\nabla}_{ij})\frac{\eta _{ij}}{2}\Big)=  \nonumber \\
&=&-\sum_{i<j}^{1..N}{\frac{Q_{i}Q_{j}}{8\pi \eta _{ij}}}\Big({\frac{\vec{%
\kappa}_{i}}{\sqrt{m_{i}^{2}+{\vec{\kappa}_{i}}^{2}}}}\cdot {\frac{\vec{%
\kappa}_{j}}{\sqrt{m_{j}^{2}+{\vec{\kappa}_{j}}^{2}}}}+({\frac{\vec{\kappa}%
_{i}}{\sqrt{m_{i}^{2}+{\vec{\kappa}_{i}}^{2}}}}\cdot \hat{\eta}_{ij})({\frac{%
\vec{\kappa}_{j}}{\sqrt{m_{j}^{2}+{\vec{\kappa}_{j}}^{2}}}}\cdot \hat{\eta}%
_{ij})\Big).  \label{VI11}
\end{eqnarray}
The standard form for the historical Darwin term is the above but with $%
\sqrt{m_{i}^{2}+\kappa _{i}^{2}}\rightarrow m_{i}$.

The remaining compensating part of the familiar Darwin interaction plus all
higher order parts come from the field energy $U(\tau ).$ In terms of
momentum variables, using Grassmann truncations, the expression Eq.(\ref{VI4}
 ) for the field energy integral simply
) becomes

\begin{eqnarray}
&&U(\tau )={\frac{1}{2}}\int d^3\sigma ({\vec E}_{\perp S}^2+{\vec B}%
_S^2)(\tau ,\vec \sigma )=\sum_{i<j}^{1..N}{\frac{Q_{i}Q_{j}}{4\pi }}h_{1}({%
\frac{\vec{\kappa}_{i}}{\sqrt{m_i^2+ {\vec{\kappa}_{i}}^{2}}}},{\frac{\vec{%
\kappa}_{j}}{\sqrt{m_j^2+{\vec{ \kappa}_{j}}^{2}}}},\vec{\eta}_{ij})=
\nonumber \\
&&=\sum_{i<j}^{N}{\frac{Q_{i}Q_{j}}{4\pi }}\sum_{m=0}^{\infty
}\sum_{n=0}^{\infty}\Big[ {\frac{\vec{\kappa}_{i}}{\sqrt{m_i^2+{\vec{\kappa}%
_{i}} ^{2}}}}\cdot {\frac{\vec{\kappa}_{j}}{\sqrt{m_j^2+{\vec{\kappa}_{j}}
^{2}}}}\times  \nonumber \\
&&{\frac{({\frac{\vec{\kappa}_{i}}{\sqrt{m_i^2+{\vec{\kappa}_{i}} ^{2}}}}%
\cdot \vec{\nabla}_{ij})^{2m+1}({\frac{\vec{\kappa}_{j}}{\sqrt{m_j^2+{\vec{%
\kappa}_{j}}^{2}}}}\cdot \vec{\nabla}_{ij})^{2n+1}\eta _{ij}^{2n+2m+1}}{%
(2n+2m+2)!}}-  \nonumber \\
&&-{\frac{({\frac{\vec{\kappa}_{i}}{\sqrt{m_i^2+{\vec{\kappa}_{i}}^{2}}}}
\cdot \vec{\nabla}_{ij})^{2m+2}({\frac{\vec{\kappa}_{j}}{\sqrt{m_j^2+{\vec{%
\kappa} _{j}}^{2}}}}\cdot \vec{\nabla}_{ij})^{2n+2}\eta _{ij}^{2n+2m+3}}{
(2n+2m+4)!}}+  \nonumber \\
&&+{\frac{\vec{\kappa}_{i}}{\sqrt{m_i^2+{\vec{\kappa}_{i}}^{2}}}}\cdot {%
\frac{\vec{\kappa}_{j}}{\sqrt{m_j^2+{\vec{\kappa}_{j}}^{2}}}\frac{({\ \frac{%
\vec{\kappa}_{i}}{\sqrt{m_i^2+{\vec{\kappa}_{i}}^{2}}}}\cdot \vec{ \nabla}%
_{ij})^{2m}({\frac{\vec{\kappa}_{j}}{\sqrt{m_j^2+{\vec{\kappa}_{j}} ^{2}}}}%
\cdot \vec{\nabla}_{ij})^{2n}\eta _{ij}^{2n+2m-1}}{(2n+2m)!}}  \nonumber \\
&&-{\frac{({\frac{\vec{\kappa}_{i}}{\sqrt{m_i^2+{\vec{\kappa}_{i}}^{2}}}}
\cdot \vec{\nabla}_{ij})^{2m+1}({\frac{\vec{\kappa}_{j}}{\sqrt{m_j^2+{\vec{%
\kappa} _{j}}^{2}}}}\cdot \vec{\nabla}_{ij})^{2n+1}\eta _{ij}^{2n+2m+1}}{
(2n+2m+2)!}}\Big] .  \label{VI12}
\end{eqnarray}

Single infinite sum pieces can be split off from the double infinite sum in
the last two lines of the above expression for the field energy integral.
This naturally separates out the compensating portion of \ the familiar
lowest order Darwin parts plus all remaining higher order Darwin parts,
including a piece that cancels exactly $U_{1HOD}.$
\begin{eqnarray}
&&U(\tau )=\sum_{i<j}^{1..N}{\frac{Q_{i}Q_{j}}{4\pi }}h_{1}({\frac{\vec{%
\kappa}_{i}}{\sqrt{m_{i}^{2}+{\vec{\kappa}_{i}}^{2}}}},{\frac{\vec{\kappa}%
_{j}}{\sqrt{m_{j}^{2}+{\vec{\kappa}_{j}}^{2}}}},\vec{\eta}_{ij})=  \nonumber
\\
&=&\sum_{i<j}^{1..N}{\frac{Q_{i}Q_{j}}{4\pi }}\left( {\frac{\vec{\kappa}_{i}%
}{\sqrt{m_{i}^{2}+{\vec{\kappa}_{i}}^{2}}}}\cdot {\frac{\vec{\kappa}_{j}}{%
\sqrt{m_{j}^{2}+{\vec{\kappa}_{j}}^{2}}}\frac{1}{\eta _{ij}}-}({\frac{\vec{%
\kappa}_{i}}{\sqrt{m_{i}^{2}+{\vec{\kappa}_{i}}^{2}}}}\cdot \vec{\nabla}%
_{ij})({\frac{\vec{\kappa}_{j}}{\sqrt{m_{j}^{2}+{\vec{\kappa}_{j}}^{2}}}}%
\cdot \vec{\nabla}_{ij})\frac{\eta _{ij}}{2}\right) +  \nonumber \\
&&+\sum_{i<j}^{N}{\frac{Q_{i}Q_{j}}{4\pi }}\sum_{m=0}^{\infty
}\sum_{n=0}^{\infty }  \nonumber \\
&&\Big[{\frac{\vec{\kappa}_{i}}{\sqrt{m_{i}^{2}+{\vec{\kappa}_{i}}^{2}}}}%
\cdot {\frac{\vec{\kappa}_{j}}{\sqrt{m_{j}^{2}+{\vec{\kappa}_{j}}^{2}}}\frac{%
({\frac{\vec{\kappa}_{i}}{\sqrt{m_{i}^{2}+{\vec{\kappa}_{i}}^{2}}}}\cdot
\vec{\nabla}_{ij})^{2m+1}({\frac{\vec{\kappa}_{j}}{\sqrt{m_{j}^{2}+{\vec{%
\kappa}_{j}}^{2}}}}\cdot \vec{\nabla}_{ij})^{2n+1}\eta _{ij}^{2n+2m+1}}{%
(2n+2m+2)!}}-  \nonumber \\
&&-{\frac{({\frac{\vec{\kappa}_{i}}{\sqrt{m_{i}^{2}+{\vec{\kappa}_{i}}^{2}}}}%
\cdot \vec{\nabla}_{ij})^{2m+2}({\frac{\vec{\kappa}_{j}}{\sqrt{m_{j}^{2}+{%
\vec{\kappa}_{j}}^{2}}}}\cdot \vec{\nabla}_{ij})^{2n+2}\eta _{ij}^{2n+2m+3}}{%
(2n+2m+4)!}}\Big]+  \nonumber \\
&&+\sum_{i<j}^{1..N}{\frac{Q_{i}Q_{j}}{4\pi }}\sum_{m=1}^{\infty }{\frac{1}{%
(2m)!}}\Big[{\frac{\vec{\kappa}_{j}}{\sqrt{m_{j}^{2}+{\vec{\kappa}_{j}}^{2}}}%
\cdot \frac{\vec{\kappa}_{i}}{\sqrt{m_{i}^{2}+{\vec{\kappa}_{i}}^{2}}}}
\nonumber \\
&&\left( ({\frac{\vec{\kappa}_{i}}{\sqrt{m_{i}^{2}+{\vec{\kappa}_{i}}^{2}}}}%
\cdot \vec{\nabla}_{ij})^{2m}+({\frac{\vec{\kappa}_{j}}{\sqrt{m_{j}^{2}+{%
\vec{\kappa}_{j}}^{2}}}}\cdot \vec{\nabla}_{ij})^{2m}\right) )\eta
_{ij}^{2m-1}-  \nonumber \\
&&-{\frac{1}{(2m+2)!}}\Big({\frac{\vec{\kappa}_{j}}{\sqrt{m_{j}^{2}+{\vec{%
\kappa}_{j}}^{2}}}}\cdot \vec{\nabla}_{ij}({\frac{\vec{\kappa}_{i}}{\sqrt{%
m_{i}^{2}+{\vec{\kappa}_{i}}^{2}}}}\cdot \vec{\nabla}_{ij})^{2m+1}+
\nonumber \\
&+&{\frac{\vec{\kappa}_{i}}{\sqrt{m_{i}^{2}+{\vec{\kappa}_{i}}^{2}}}}\cdot
\vec{\nabla}_{ij}({\frac{\vec{\kappa}_{j}}{\sqrt{m_{j}^{2}+{\vec{\kappa}_{j}}%
^{2}}}}\cdot \vec{\nabla}_{ij})^{2m+1}\Big)\eta _{ij}^{2m+1}\Big]+  \nonumber
\\
&&+\sum_{i<j}^{1..N}{\frac{Q_{i}Q_{j}}{4\pi }}\sum_{m=0}^{\infty
}\sum_{n=0}^{\infty }\Big[{\frac{\vec{\kappa}_{i}}{\sqrt{m_{i}^{2}+{\vec{%
\kappa}_{i}}^{2}}}}\cdot {\frac{\vec{\kappa}_{j}}{\sqrt{m_{j}^{2}+{\vec{%
\kappa}_{j}}^{2}}}}\times  \nonumber \\
&&\frac{{{({\frac{\vec{\kappa}_{i}}{\sqrt{m_{i}^{2}+{\vec{\kappa}_{i}}^{2}}}}%
\cdot \vec{\nabla}_{ij})^{2m+2}({\frac{\vec{\kappa}_{j}}{\sqrt{m_{j}^{2}+{%
\vec{\kappa}_{j}}^{2}}}}\cdot \vec{\nabla}_{ij})^{2n+2}\eta _{ij}^{2n+2m+3}}}%
}{{(2n+2m+4)!}}-  \nonumber \\
&&-{\frac{({\frac{\vec{\kappa}_{i}}{\sqrt{m_{i}^{2}+{\vec{\kappa}_{i}}^{2}}}}%
\cdot \vec{\nabla}_{ij})^{2m+3}({\frac{\vec{\kappa}_{j}}{\sqrt{m_{j}^{2}+{%
\vec{\kappa}_{j}}^{2}}}}\cdot \vec{\nabla}_{ij})^{2n+3}\eta _{ij}^{2n+2m+5}}{%
(2n+2m+6)!}}\Big]=  \nonumber \\
&=&-V_{LOD}-U_{1HOD}+U_{HOD},  \nonumber \\
&&{}  \nonumber \\
&&\Rightarrow \quad V_{DAR}(\tau )=U_1(\tau )+U(\tau )=V_{LOD}(\tau
)+U_{HOD}(\tau ).  \label{VI13}
\end{eqnarray}
In this form we see that the first summation \ is $-V_{LOD}$ \ and combines
with the gauge part of the kinetic piece [$2V_{LOD}(\tau )$] to give the
familiar lowest order Darwin piece $V_{LOD}(\tau )$. \ The third set of
summations is $-U_{1HOD}$ and exactly cancels with the corresponding term in
the kinetic piece. \ The second and fourth set of summations ( the double
sums which begin at higher order in $(1/c^{2})$) we define as $U_{HOD}$, \
coming only from $U(\tau )$. \ Altogether we obtain [$\eta _{ij}=|{\vec{\eta}%
}_{ij}|=|{\vec{\eta}}_{i}-{\vec{\eta}}_{j}|$]

\begin{eqnarray}
M &=&\sum_{i=1}^{N}\sqrt{m_{i}^{2}+(\vec{\kappa}_{i}(\tau )-Q_{i}\vec{A}%
_{\perp S}(\tau ,\vec{\eta}_{i}(\tau )))^{2}}+  \nonumber \\
&+&\sum_{i\neq j}^{1..N}\frac{Q_{i}Q_{j}}{4\pi \mid \vec{\eta}_{i}(\tau )-%
\vec{\eta}_{j}(\tau )\mid }+\int d^{3}\sigma {\frac{1}{2}}[\vec{\pi}_{\perp
S}^{2}+\vec{B}_{S}^{2}](\tau ,\vec{\sigma})=  \nonumber \\
&=&\sum_{i=1}^{N}\sqrt{m_{i}^{2}+[{\vec{\kappa}}_{i}(\tau )-{\vec{V}}%
_{i}(\tau )]^{2}}+\sum_{i\not=j}^{1..N}{\frac{{Q_{i}Q_{j}}}{{4\pi |{\vec{\eta%
}}_{i}(\tau )-{\vec{\eta}}_{j})(\tau )|}}}+U(\tau )=  \nonumber \\
&=&\sum_{i=1}^{N}\sqrt{m_{i}^{2}+{\vec{\kappa}}_{i}^{2}(\tau )}+\sum_{i\not=%
j}^{1..N}{\frac{{Q_{i}Q_{j}}}{{|{\vec{\eta}}_{i}(\tau )-{\vec{\eta}}%
_{j}(\tau )|}}}+V_{LOD}(\tau )+V_{HOD}(\tau )=  \nonumber \\
&&{}  \nonumber \\
&=&\sum_{i=1}^{N}\sqrt{m_{i}^{2}+\vec{\kappa}_{i}(\tau )^{2}}+\sum_{i\neq j}%
\frac{Q_{i}Q_{j}}{4\pi \mid \vec{\eta}_{i}(\tau )-\vec{\eta}_{j}(\tau )\mid }%
-  \nonumber \\
&-&\sum_{i<j}^{1..N}{\frac{Q_{i}Q_{j}}{4\pi }}\left( {\frac{\vec{\kappa}_{i}%
}{\sqrt{m_{i}^{2}+{\vec{\kappa}_{i}}^{2}}}}\cdot {\frac{\vec{\kappa}_{j}}{%
\sqrt{m_{j}^{2}+{\vec{\kappa}_{j}}^{2}}}\frac{1}{\eta _{ij}}-}({\frac{\vec{%
\kappa}_{i}}{\sqrt{m_{i}^{2}+{\vec{\kappa}_{i}}^{2}}}}\cdot \vec{\nabla}%
_{ij})({\frac{\vec{\kappa}_{j}}{\sqrt{m_{j}^{2}+{\vec{\kappa}_{j}}^{2}}}}%
\cdot \vec{\nabla}_{ij})\frac{\eta _{ij}}{2}\right) +  \nonumber \\
&+&\sum_{i<j}^{1..N}{\frac{Q_{i}Q_{j}}{4\pi }}\sum_{m=0}^{\infty
}\sum_{n=0}^{\infty }\Big[{\frac{\vec{\kappa}_{i}}{\sqrt{m_{i}^{2}+{\vec{%
\kappa}_{i}}^{2}}}}\cdot {\frac{\vec{\kappa}_{j}}{\sqrt{m_{j}^{2}+{\vec{%
\kappa}_{j}}^{2}}}}\times  \nonumber \\
&&{\frac{({\frac{\vec{\kappa}_{i}}{\sqrt{m_{i}^{2}+{\vec{\kappa}_{i}}^{2}}}}%
\cdot \vec{\nabla}_{ij})^{2m+1}({\frac{\vec{\kappa}_{j}}{\sqrt{m_{j}^{2}+{%
\vec{\kappa}_{j}}^{2}}}}\cdot \vec{\nabla}_{ij})^{2n+1}\eta _{ij}^{2n+2m+1}}{%
(2n+2m+2)!}}-  \nonumber \\
&&-{\frac{({\frac{\vec{\kappa}_{i}}{\sqrt{m_{i}^{2}+{\vec{\kappa}_{i}}^{2}}}}%
\cdot \vec{\nabla}_{ij})^{2m+2}({\frac{\vec{\kappa}_{j}}{\sqrt{m_{j}^{2}+{%
\vec{\kappa}_{j}}^{2}}}}\cdot \vec{\nabla}_{ij})^{2n+2}\eta ^{2n+2m+3}}{%
(2n+2m+4)!}}+  \nonumber \\
&&+{\frac{\vec{\kappa}_{i}}{\sqrt{m_{i}^{2}+{\vec{\kappa}_{i}}^{2}}}}\cdot {%
\frac{\vec{\kappa}_{j}}{\sqrt{m_{j}^{2}+{\vec{\kappa}_{j}}^{2}}}\frac{({%
\frac{\vec{\kappa}_{i}}{\sqrt{m_{i}^{2}+{\vec{\kappa}_{i}}^{2}}}}\cdot \vec{%
\nabla}_{ij})^{2m+2}({\frac{\vec{\kappa}_{j}}{\sqrt{m_{j}^{2}+{\vec{\kappa}%
_{j}}^{2}}}}\cdot \vec{\nabla}_{ij})^{2n+2}\eta _{ij}^{2n+2m+3}}{(2n+2m+4)!}}%
-  \nonumber \\
&&-{\frac{({\frac{\vec{\kappa}_{i}}{\sqrt{m_{i}^{2}+{\vec{\kappa}_{i}}^{2}}}}%
\cdot \vec{\nabla}_{ij})^{2m+3}({\frac{\vec{\kappa}_{j}}{\sqrt{m_{j}^{2}+{%
\vec{\kappa}_{j}}^{2}}}}\cdot \vec{\nabla}_{ij})^{2n+3}\eta _{ij}^{2n+2m+5}}{%
(2n+2m+6)!}}\Big].  \nonumber \\
&&{}  \label{VI14}
\end{eqnarray}

In Appendix C we show how the multiple directional derivatives in the
generalized higher order Darwin interactions can be evaluated in general,
thereby obtaining a more readily usable form.\ (The expression for $M$ and
for $V_{DAR}$ for arbitrary $N$ is an immediate generalization of the case
for $N=2$ and for simplicity of notation the details in the appendices will
be limited to the two-body case.) However, because of the significant
complexity of the above form it will be of value to first obtain an
alternative derivation of this series. For $N$=2 we have the following
Lagrangian expression for the invariant mass [see Eq.(\ref{IV10}) with $\vec{%
\lambda}(\tau )=0$; $h_{1}$ is defined in Eq.(\ref{VI4})]

\begin{eqnarray}
E_{rel} &=&h(\dot{\vec{\eta}_{1}},\dot{\vec{\eta}_{2}},\vec{\eta})={\frac{%
m_{1}}{\sqrt{1-\dot{\vec{\eta}_{1}}^{2}}}}+{\frac{m_{2}}{\sqrt{1-\dot{\vec{%
\eta}_{2}}^{2}}}}+{\frac{Q_{1}Q_{2}}{4\pi |\vec{\eta}|}}+{\frac{{Q_{1}Q_{2}}%
}{{4\pi }}}h_{1}(\dot{\vec{\eta}_{1}},\dot{\vec{\eta}_{2}},\vec{\eta}):=
\nonumber \\
&:&=h_{0}(\dot{\vec{\eta}_{1}},\dot{\vec{\eta}_{2}},\vec{\eta})+{\frac{{%
Q_{1}Q_{2}}}{{4\pi }}}h_{1}(\dot{\vec{\eta}_{1}},\dot{\vec{\eta}_{2}},\vec{%
\eta}).  \label{VI15}
\end{eqnarray}
In order to find the Hamiltonian $H(\vec{\kappa}_{1},\vec{\kappa}_{2};\vec{%
\eta})$ from $h_{1}$ we must demand that Hamilton's equation be satisfied .\
We use the Dirac bracket since we have used the constraint as a strong
condition on the dynamical variables. \ This will lead to a set of
differential equations for $H(\vec{\kappa}_{1},\vec{\kappa}_{2};\vec{\eta})$
and to a result that agrees exactly with the above expression for $M$. \ The
details of this analysis are given in Appendix D.

\bigskip Here we mention a\ comparison of these cross checked results (valid
to all order of $1/c^{2}$) with approximate results obtained elsewhere. \ In
\cite{yang} one obtains a single time Lagrangian by expanding the symmetric
Green function in the Fokker action, used by Wheeler and Feynman, to all
orders in $1/c^{2}$. \ From this Lagrangian one obtains the Legendre
Hamiltonian $\tilde{h}(\dot{\vec{\eta}_{1}},\dot{\vec{\eta}_{2}},\vec{\eta}%
)+ $ terms involving the acceleration and all higher order derivatives. \
Ignoring those higher order accelerations one finds the same Legendre
Hamiltonian as above. From that Hamiltonian the authors obtain a final
Hamiltonian that although agreeing through order $1/c^{2}$ with the results
above (including the standard Darwin interaction to that order) they differ
from our common results (above and in \ Appendix D) at order $1/c^{4}$ (no
terms of higher order are computed in \cite{yang}). The failure to obtain
results that agree with our result here is the neglect there of using the
proper Dirac brackets in the Kerner reduction. \ The reason that those Dirac
brackets were not used is that the authors took as a starting point the
Fokker action, not taking into account that this action itself \ is a result
of imposing constraints on the solutions of the electromagnetic field
equations.

By using Eqs.(\ref{V51}) we get
\begin{eqnarray}
\sqrt{m_{i}^{2}+\vec{\kappa}_{i}{}^{2}} &=&\sqrt{m_{i}^{2}+\widetilde{\vec{%
\kappa}_{i}}^{2}+\widetilde{\vec{\kappa}_{i}}\cdot \sum_{j\neq i}Q_{i}Q_{j}%
\vec{\nabla}_{\eta _{i}}{\cal K}_{ij}}=  \nonumber \\
&=&\sqrt{m_{i}^{2}+\widetilde{\vec{\kappa}_{i}}^{2}}+\frac{\widetilde{\vec{%
\kappa}_{i}}\cdot \sum_{j\neq i}Q_{i}Q_{j}\vec{\nabla}_{\tilde{\eta}_{i}}%
\widetilde{{\cal K}}_{ij}}{2\sqrt{m_{i}^{2}+\widetilde{\vec{\kappa}_{i}}^{2}}%
},  \nonumber \\
&&\text{with \ }  \nonumber \\
Q_{i}Q_{j}{\cal K}_{ij}(\vec{\kappa}_{i,}\vec{\kappa}_{j,}\vec{\eta}_{i}-%
\vec{\eta}_{j}) &=&Q_{i}Q_{j}\widetilde{{\cal K}}_{ij}(\widetilde{\vec{\kappa%
}_{i}},\widetilde{\vec{\kappa}_{j}},\widetilde{\vec{\eta}_{i}}-\widetilde{%
\vec{\eta}_{j}}),  \nonumber \\
\widetilde{\vec{\kappa}_{i}}\cdot \nabla _{\tilde{\eta}_{i}}\widetilde{{\cal %
K}}_{ij} &=&\int d^{3}\sigma \lbrack {\large (}\widetilde{\vec{\kappa}_{i}}%
\cdot \vec{\nabla}_{\tilde{\eta}_{i}}\vec{A}_{\perp Si}(\widetilde{\vec{%
\kappa}_{i}},\widetilde{\vec{\kappa}_{j}},\widetilde{\vec{\eta}_{i}}-%
\widetilde{\vec{\eta}_{j}}){\large )}\cdot \vec{\pi}_{\perp Si}(\widetilde{%
\vec{\kappa}_{i}},\widetilde{\vec{\kappa}_{j}},\widetilde{\vec{\eta}_{i}}-%
\widetilde{\vec{\eta}_{j}})  \nonumber \\
&&-\vec{A}_{\perp Si}(\widetilde{\vec{\kappa}_{i}},\widetilde{\vec{\kappa}%
_{j}},\widetilde{\vec{\eta}_{i}}-\widetilde{\vec{\eta}_{j}})\cdot {\large (}%
\widetilde{\vec{\kappa}_{i}}\cdot \vec{\nabla}_{\tilde{\eta}_{i}}\vec{\pi}%
_{\perp Si}(\widetilde{\vec{\kappa}_{i}},\widetilde{\vec{\kappa}_{j}},%
\widetilde{\vec{\eta}_{i}}-\widetilde{\vec{\eta}_{j}}){\large )}],
\label{VI16}
\end{eqnarray}
so that $M={\cal P}_{(int)}^{\tau }$ becomes (due to Grassmann truncation we
can replace the old variables by the \ new variables in the interaction
terms)
\[
M={\cal P}_{(int)}^{\tau }=\sum_{i=1}^{N}\sqrt{m_{i}^{2}+\widetilde{\vec{%
\kappa}_{i}}^{2}}+\sum_{i=1}^{N}\frac{\widetilde{\vec{\kappa}_{i}}\cdot
\sum_{j\neq i}Q_{i}Q_{j}\vec{\nabla}_{\tilde{\eta}_{i}}{\cal K}_{ij}}{2\sqrt{%
m_{i}^{2}+\widetilde{\vec{\kappa}_{i}}^{2}}}+\sum_{i\neq j}\frac{Q_{i}Q_{j}}{%
4\pi \mid \widetilde{\vec{\eta}_{i}}-\widetilde{\vec{\eta}_{j}}\mid }-
\]
\begin{eqnarray}
&-&\sum_{i<j}^{1..N}{\frac{Q_{i}Q_{j}}{4\pi }}\left( {\frac{\widetilde{\vec{%
\kappa}}_{i}}{\sqrt{m_{i}^{2}+\widetilde{\vec{\kappa}}_{i}^{2}}}}\cdot {%
\frac{\widetilde{\vec{\kappa}}_{j}}{\sqrt{m_{j}^{2}+\widetilde{\vec{\kappa}}%
_{j}^{2}}}\frac{1}{\tilde{\eta}_{ij}}-}({\frac{\widetilde{\vec{\kappa}}_{i}}{%
\sqrt{m_{i}^{2}+\widetilde{\vec{\kappa}}_{i}^{2}}}}\cdot \vec{\nabla}_{ij})({%
\frac{\widetilde{\vec{\kappa}}_{j}}{\sqrt{m_{j}^{2}+\widetilde{\vec{\kappa}}%
_{j}^{2}}}}\cdot \vec{\nabla}_{ij})\frac{\tilde{\eta}_{ij}}{2}\right) +
\nonumber \\
&+&\sum_{i<j}^{1..N}{\frac{Q_{i}Q_{j}}{4\pi }}\sum_{m=0}^{\infty
}\sum_{n=0}^{\infty }\Big[{\frac{\widetilde{\vec{\kappa}}_{i}}{\sqrt{%
m_{i}^{2}+\widetilde{(}{\vec{\kappa}}_{i}}^{2}}}\cdot {\frac{\widetilde{\vec{%
\kappa}}_{j}}{\sqrt{m_{j}^{2}+{\vec{\kappa}_{j}}^{2}}}}\times  \nonumber \\
&&{\frac{({\frac{\widetilde{\vec{\kappa}}_{i}}{\sqrt{m_{i}^{2}+{\widetilde{%
\vec{\kappa}}_{i}}^{2}}}}\cdot \vec{\nabla}_{ij})^{2m+1}({\frac{\widetilde{%
\vec{\kappa}}_{j}}{\sqrt{m_{j}^{2}+{\widetilde{\vec{\kappa}}_{j}}^{2}}}}%
\cdot \vec{\nabla}_{ij})^{2n+1}\tilde{\eta}_{ij}^{2n+2m+1}}{(2n+2m+2)!}}-
\nonumber \\
&&-{\frac{({\frac{\widetilde{\vec{\kappa}}_{i}}{\sqrt{m_{i}^{2}+{\widetilde{%
\vec{\kappa}}_{i}}^{2}}}}\cdot \vec{\nabla}_{ij})^{2m+2}({\frac{\widetilde{%
\vec{\kappa}}_{j}}{\sqrt{m_{j}^{2}+{\widetilde{\vec{\kappa}}_{j}}^{2}}}}%
\cdot \vec{\nabla}_{ij})^{2n+2}\,\tilde{\eta}_{ij}^{2n+2m+3}}{(2n+2m+4)!}}+
\nonumber \\
&&+{\frac{\widetilde{\vec{\kappa}}_{i}}{\sqrt{m_{i}^{2}+{\widetilde{\vec{%
\kappa}}_{i}}^{2}}}}\cdot {\frac{\widetilde{\vec{\kappa}}_{j}}{\sqrt{%
m_{j}^{2}+{\widetilde{\vec{\kappa}}_{j}}^{2}}}\frac{({\frac{\widetilde{\vec{%
\kappa}}_{i}}{\sqrt{m_{i}^{2}+{\widetilde{\vec{\kappa}}_{i}}^{2}}}}\cdot
\vec{\nabla}_{ij})^{2m+2}({\frac{\widetilde{\vec{\kappa}}_{j}}{\sqrt{%
m_{j}^{2}+{\widetilde{\vec{\kappa}}_{j}}^{2}}}}\cdot \vec{\nabla}%
_{ij})^{2n+2}\,\tilde{\eta}_{ij}^{2n+2m+3}}{(2n+2m+4)!}}-  \nonumber \\
&&-{\frac{({\frac{\widetilde{\vec{\kappa}}_{i}}{\sqrt{m_{i}^{2}+{\widetilde{%
\vec{\kappa}}_{i}}^{2}}}}\cdot \vec{\nabla}_{ij})^{2m+3}({\frac{\widetilde{%
\vec{\kappa}}_{j}}{\sqrt{m_{j}^{2}+{\widetilde{\vec{\kappa}}_{j}}^{2}}}}%
\cdot \vec{\nabla}_{ij})^{2n+3}\,\tilde{\eta}_{ij}^{2n+2m+5}}{(2n+2m+6)!}}%
\Big].  \label{VI17}
\end{eqnarray}
where $\vec{\nabla}_{ij}=\partial /\partial \tilde{\eta}_{ij}$. \ We
emphasize at this point that the interaction terms are all two-body forces;
there are no $N$ body forces in our semiclassical treatment. \

\subsection{\protect\bigskip Complete Darwin Hamiltonian in terms of the
Final Canonical Variables for the N-Body Problem. \protect\bigskip}

In this subsection we will present the final expression for $M={\cal P}%
_{(int)}^{\tau }$ in the final canonical variables for arbitrary $N$ . Now \
from Eq.(\ref{V48})above we can replace $\sum_{i<j}\nabla _{\eta
_{i}}Q_{i}Q_{j}{\cal K}_{ij}$ by $-\int d^{3}\sigma \lbrack {\vec{\pi}}%
_{\perp S}\times {\vec{B}}_{S}](\tau ,\vec{\sigma})$. We have already
performed that integral (see Eq.(\ref{d14})) for the two-body problem and
the result for the $N$ body problem is an immediate generalization. Using
that complete expression we obtain [$\eta _{ij}=|{\vec{\eta}}_{ij}|=|{\vec{%
\eta}}_{i}-{\vec{\eta}}_{j}|$]
\begin{eqnarray}
&&\sum_{i=1}^{N}\sqrt{m_{i}^{2}+\vec{\kappa}_{i}(\tau )^{2}}=\sqrt{m_{i}^{2}+%
\widetilde{\vec{\kappa}_{i}}(\tau )^{2}}+U_{HOD}^{\prime }(\tau )  \nonumber
\\
&&  \nonumber \\
&&U_{HOD}^{\prime }(\tau )=-\sum_{i<j}{\frac{Q_{i}Q_{j}}{8\pi }}%
\sum_{m=0}^{\infty }\sum_{n=0}^{\infty }\Big[{\frac{\widetilde{\vec{\kappa}}%
_{i}}{\sqrt{m_{i}^{2}+{\widetilde{\vec{\kappa}}_{i}}^{2}}}}\cdot {\frac{%
\widetilde{\vec{\kappa}}_{j}}{\sqrt{m_{j}^{2}+{\widetilde{\vec{\kappa}}_{j}}%
^{2}}}}\frac{1}{(2n+2m+2)!}  \nonumber \\
&&\times \Big(({\frac{\widetilde{\vec{\kappa}}_{i}}{\sqrt{m_{i}^{2}+{%
\widetilde{\vec{\kappa}}_{i}}^{2}}}}\cdot \vec{\nabla _{ij}})^{2m+2}({\frac{%
\widetilde{\vec{\kappa}}_{j}}{\sqrt{m_{j}^{2}+{\widetilde{\vec{\kappa}}_{j}}%
^{2}}}}\cdot \vec{\nabla _{ij}})^{2n}+  \nonumber \\
&&+2({\frac{\widetilde{\vec{\kappa}}_{i}}{\sqrt{m_{i}^{2}+{\widetilde{\vec{%
\kappa}}_{i}}^{2}}}}\cdot \vec{\nabla _{ij}})^{2m+1}({\frac{\widetilde{\vec{%
\kappa}}_{j}}{\sqrt{m_{j}^{2}+{\vec{\kappa}_{j}}^{2}}}}\cdot \vec{\nabla
_{ij}})^{2n+1}+  \nonumber \\
&&({\frac{\widetilde{\vec{\kappa}}_{i}}{\sqrt{m_{i}^{2}+{\widetilde{\vec{%
\kappa}}_{i}}^{2}}}}\cdot \vec{\nabla _{ij}})^{2m}({\frac{\widetilde{\vec{%
\kappa}}_{j}}{\sqrt{m_{j}^{2}+{\vec{\kappa}_{j}}^{2}}}}\cdot \vec{\nabla
_{ij}})^{2n+2}\Big)\tilde{\eta}_{ij}^{2n+2m+1}-  \nonumber \\
&&-\frac{1}{(2n+2m+4)!}\Big(({\frac{\widetilde{\vec{\kappa}}_{i}}{\sqrt{%
m_{i}^{2}+{\vec{\kappa}_{i}}^{2}}}}\cdot \vec{\nabla _{ij}})^{2m+3}({\frac{%
\widetilde{\vec{\kappa}}_{j}}{\sqrt{m_{j}^{2}+{\widetilde{\vec{\kappa}}_{j}}%
^{2}}}}\cdot \vec{\nabla _{ij}})^{2n+1}+  \nonumber \\
&&+2({\frac{\widetilde{\vec{\kappa}}_{i}}{\sqrt{m_{i}^{2}+{\widetilde{\vec{%
\kappa}}_{i}}^{2}}}}\cdot \vec{\nabla _{ij}})^{2m+2}({\frac{\widetilde{\vec{%
\kappa}}_{j}}{\sqrt{m_{j}^{2}+{\vec{\kappa}_{j}}^{2}}}}\cdot \vec{\nabla
_{ij}})^{2n+2}+  \nonumber \\
&+&({\frac{\widetilde{\vec{\kappa}}_{i}}{\sqrt{m_{i}^{2}+{\widetilde{\vec{%
\kappa}}_{i}}^{2}}}}\cdot \vec{\nabla _{ij}})^{2m+1}({\frac{\widetilde{\vec{%
\kappa}}_{j}}{\sqrt{m_{j}^{2}+{\vec{\kappa}_{j}}^{2}}}}\cdot \vec{\nabla
_{ij}})^{2n+3}\Big)\tilde{\eta}_{ij}^{2n+2m+3}\Big].  \label{VI18}
\end{eqnarray}
We now combine this portion of the complete Darwin Hamiltonian coming from
the vector potentials ${\vec{V}}_{i}(\tau )$ in the kinetic terms with the
potential $U(\tau )$ arising from the field energy integral [which contains $%
V_{LOH}(\tau )$] to obtain the final form of the invariant mass in the rest
frame
\begin{eqnarray}
M &=&{\cal P}_{(int)}^{\tau }=\sum_{i=1}^{M}\sqrt{m_{i}^{2}+(\vec{\kappa}%
_{i}(\tau )-Q_{i}\vec{A}_{\perp S}(\tau ,\vec{\eta}_{i}(\tau )))^{2}}+
\nonumber \\
&+&\sum_{i<j}\frac{Q_{i}Q_{j}}{4\pi \eta _{ij}}+\int d^{3}\sigma {\frac{1}{2}%
}[\vec{\pi}_{\perp S}^{2}+\vec{B}_{S}^{2}](\tau ,\vec{\sigma})=  \nonumber \\
&=&\sum_{i=1}^{N}\sqrt{m_{i}^{2}+[{\vec{\kappa}}_{i}^{2}(\tau )-{\vec{V}}%
_{i}(\tau )]^{2}}+\sum_{i<j}\frac{Q_{i}Q_{j}}{4\pi \eta _{ij}}+U(\tau )=
\nonumber \\
&=&\sum_{i=1}^{N}\sqrt{m_{i}^{2}+{\vec{\kappa}}_{i}^{2}(\tau )}+\sum_{i<j}%
\frac{Q_{i}Q_{j}}{4\pi \eta _{ij}}+V_{DAR}(\tau )=\quad \quad \lbrack
V_{DAR}=V_{LOD}+V_{HOD}]  \nonumber \\
&&{}  \nonumber \\
&=&\sum_{i=1}^{N}\sqrt{m_{i}^{2}+\widetilde{\vec{\kappa}_{i}}^{2}}+\sum_{i<j}%
\frac{Q_{i}Q_{j}}{4\pi \eta _{ij}}+\tilde{V}_{DAR}(\tau ),\,\,\,\,\,\,\,\,[%
\tilde{V}_{DAR}=V_{DAR}+U_{HOD}^{\prime }]  \nonumber \\
&&  \nonumber \\
&\tilde{V}_{DAR}=-&\sum_{i<j}\frac{Q_{i}Q_{j}}{4\pi }\left( {\frac{%
\widetilde{\vec{\kappa}}_{i}}{\sqrt{m_{i}^{2}+{\widetilde{\vec{\kappa}}_{i}}%
^{2}}}}\cdot {\frac{\widetilde{\vec{\kappa}}_{j}}{\sqrt{m_{j}^{2}+{%
\widetilde{\vec{\kappa}}_{j}}^{2}}}\frac{1}{\tilde{\eta}}-}({\frac{%
\widetilde{\vec{\kappa}}_{i}}{\sqrt{m_{i}^{2}+{\widetilde{\vec{\kappa}}_{i}}%
^{2}}}}\cdot \vec{\nabla}_{ij})({\frac{\widetilde{\vec{\kappa}}_{j}}{\sqrt{%
m_{j}^{2}+{\widetilde{\vec{\kappa}}_{j}}^{2}}}}\cdot \vec{\nabla}_{ij})\frac{%
\tilde{\eta}_{ij}}{2}\right) -  \nonumber \\
&-&{\frac{Q_{i}Q_{j}}{8\pi }}\sum_{m=0}^{\infty }\sum_{n=0}^{\infty }\Big[{%
\frac{\widetilde{\vec{\kappa}}_{i}}{\sqrt{m_{i}^{2}+{\widetilde{\vec{\kappa}}%
_{i}}^{2}}}}\cdot {\frac{\widetilde{\vec{\kappa}}_{j}}{\sqrt{m_{j}^{2}+{%
\widetilde{\vec{\kappa}}_{j}}^{2}}}}\frac{1}{(2n+2m+2)!}  \nonumber \\
&&\times \Big(({\frac{\widetilde{\vec{\kappa}}_{i}}{\sqrt{m_{i}^{2}+{%
\widetilde{\vec{\kappa}}_{i}}^{2}}}}\cdot \vec{\nabla}_{ij})^{2m+2}({\frac{%
\widetilde{\vec{\kappa}}_{j}}{\sqrt{m_{j}^{2}+{\widetilde{\vec{\kappa}}_{j}}%
^{2}}}}\cdot \vec{\nabla}_{ij})^{2n}+  \nonumber \\
&+&({\frac{\widetilde{\vec{\kappa}}_{i}}{\sqrt{m_{i}^{2}+{\widetilde{\vec{%
\kappa}}_{i}}^{2}}}}\cdot \vec{\nabla}_{ij})^{2m}({\frac{\widetilde{\vec{%
\kappa}}_{j}}{\sqrt{m_{j}^{2}+{\widetilde{\vec{\kappa}}_{j}}^{2}}}}\cdot
\vec{\nabla}_{ij})^{2n+2}\Big)\tilde{\eta}_{ij}^{2n+2m+1}-  \nonumber \\
&-&\frac{1}{(2n+2m+4)!}\Big(({\frac{\widetilde{\vec{\kappa}}_{i}}{\sqrt{%
m_{i}^{2}+{\vec{\kappa}_{i}}^{2}}}}\cdot \vec{\nabla}_{ij})^{2m+3}({\frac{%
\widetilde{\vec{\kappa}}_{j}}{\sqrt{m_{j}^{2}+{\widetilde{\vec{\kappa}}_{j}}%
^{2}}}}\cdot \vec{\nabla}_{ij})^{2n+1}+  \nonumber \\
&+&({\frac{\widetilde{\vec{\kappa}}_{i}}{\sqrt{m_{i}^{2}+{\widetilde{\vec{%
\kappa}}_{i}}^{2}}}}\cdot \vec{\nabla _{ij}})^{2m+1}({\frac{\widetilde{\vec{%
\kappa}}_{j}}{\sqrt{m_{j}^{2}+{\widetilde{\vec{\kappa}}_{j}}^{2}}}}\cdot
\vec{\nabla _{ij}})^{2n+3}\Big)\tilde{\eta}_{ij}^{2n+2m+3}]-  \nonumber \\
&-&2{\frac{\widetilde{\vec{\kappa}}_{i}}{\sqrt{m_{i}^{2}+{\widetilde{\vec{%
\kappa}}_{i}}^{2}}}}\cdot {\frac{\widetilde{\vec{\kappa}}_{j}}{\sqrt{%
m_{j}^{2}+{\widetilde{\vec{\kappa}}_{j}}^{2}}}}{\frac{({\ \frac{\widetilde{%
\vec{\kappa}}_{i}}{\sqrt{m_{i}^{2}+{\widetilde{\vec{\kappa}}_{i}}^{2}}}}%
\cdot \vec{\nabla}_{ij})^{2m+2}({\frac{\widetilde{\vec{\kappa}}_{j}}{\sqrt{%
m_{j}^{2}+{\widetilde{\vec{\kappa}}_{j}}^{2}}}}\cdot \vec{\nabla}%
_{ij})^{2n+2}\tilde{\eta}_{ij}^{2n+2m+3}}{(2n+2m+4)!}}+  \nonumber \\
&+&2{\frac{({\frac{\widetilde{\vec{\kappa}}_{i}}{\sqrt{m_{i}^{2}+{\widetilde{%
\vec{\kappa}}_{i}}^{2}}}}\cdot \vec{\nabla}_{ij})^{2m+3}({\frac{\widetilde{%
\vec{\kappa}}_{j}}{\sqrt{m_{j}^{2}+{\vec{\kappa}_{j}}^{2}}}}\cdot \vec{\nabla%
}_{ij})^{2n+3}\tilde{\eta}_{ij}^{2n+2m+5}}{(2n+2m+6)!}}\Big].  \label{VI19}
\end{eqnarray}
Notice that the $N$ body interaction Hamiltonian is a sum of individual 2
body Hamiltonians.(Grassmann truncation eliminates $N$-body forces). \ In
this sense these interactions correspond to a sum of disconnected and
spectator Feynman diagrams.

\subsection{Further Reductions of the N-Body Darwin Hamiltonian and Closed
Form Solutions for the Two-Body Problem.}

\bigskip

It is of interest to see if we can make further simplifications by deriving
an expression for the\ multiple derivatives. We show in Appendix C that the
problematic expression $(\widetilde{\vec{\kappa}}_{i}\cdot \vec{\nabla}%
_{ij})^{a}(\vec{\kappa}_{j}\cdot \vec{\nabla}_{ij})^{b}\tilde{\eta}^{a+b-1}$
is obtained in terms of powers of \ [$\tilde{\eta}_{ij}=|\widetilde{\vec{\eta%
}_{i}}-\widetilde{\vec{\eta}_{j}}|;\,\hat{\eta}_{ij}=(\widetilde{\vec{\eta}%
_{i}}-\widetilde{\vec{\eta}_{j}})/\tilde{\eta}_{ij}$]
\begin{equation}
\cos ^{2}\phi _{ij}:=\frac{(\widetilde{\vec{\kappa}}_{i}\cdot \widetilde{%
\vec{\kappa}}_{j}-\vec{\kappa}_{i}\cdot \hat{\eta}_{ij}\widetilde{\vec{\kappa%
}}_{j}\cdot \hat{\eta}_{ij})^{2}}{(\vec{\kappa}_{i}^{2}-(\widetilde{\vec{%
\kappa}}_{i}\cdot \hat{\eta}_{ij})^{2})(\widetilde{\vec{\kappa}}_{j}^{2}-(%
\widetilde{\vec{\kappa}}_{j}\cdot \hat{\eta}_{ij})^{2})}.  \label{VI20}
\end{equation}
For $N=2$ this expression reduces to 1 if $\widetilde{\vec{\kappa}}_{1}=-%
\widetilde{\vec{\kappa}}_{2}$ (the center of mass rest frame condition for
the two body problem) . The result for $a=2m+1,\,b=2n+1$ is
\begin{eqnarray}
&&(\widetilde{\vec{\kappa}}_{i}\cdot \vec{\nabla}_{ij})^{2m+1}(\widetilde{%
\vec{\kappa}}_{j}\cdot \vec{\nabla}_{ij})^{2n+1}\tilde{\eta}_{ij}^{2(m+n)+1}=
\nonumber \\
&&{}  \nonumber \\
&=&2\frac{[(2m+2n+1)!!]^{2}[(n+m+1)!]^{2}}{(2(m+n+1))!}  \nonumber \\
&&\frac{(\widetilde{\vec{\kappa}}_{i}\cdot \widetilde{\vec{\kappa}}_{j}-%
\widetilde{\vec{\kappa}}_{i}\cdot \hat{\eta}_{ij}\widetilde{\vec{\kappa}}%
_{j}\cdot \hat{\eta}_{ij})(\widetilde{\vec{\kappa}}_{i}^{2}-(\widetilde{\vec{%
\kappa}}_{i}\cdot \hat{\eta}_{ij})^{2})^{m}(\widetilde{\vec{\kappa}}%
_{j}^{2}-(\widetilde{\vec{\kappa}}_{j}\cdot \hat{\eta}_{ij})^{2})^{n}}{%
\tilde{\eta}_{ij}}\times   \nonumber \\
&&\sum_{l=0}^{n}\sum_{k=0}^{n-l}\sum_{h=0}^{k}%
%TCIMACRO{\binom{2n+1}{l}}%
%BeginExpansion
{2n+1 \choose l}%
%EndExpansion
%TCIMACRO{\binom{2m+1}{l+m-n}}%
%BeginExpansion
{2m+1 \choose l+m-n}%
%EndExpansion
%TCIMACRO{\binom{2(n-l)+1}{2k}}%
%BeginExpansion
{2(n-l)+1 \choose 2k}%
%EndExpansion
%TCIMACRO{\binom{k}{h}}%
%BeginExpansion
{k \choose h}%
%EndExpansion
(-1)^{k+h}(\cos ^{2}\phi_{ij})^{n-l+h-k}.  \label{VI21}
\end{eqnarray}

For $\ $the two body case in the center-of-mass rest frame $\widetilde{\vec{%
\kappa}}_{1}=-\widetilde{\vec{\kappa}}_{2}$ $:=\widetilde{\vec{\kappa}}$
this expression reduces to
\begin{equation}
-\frac{[(2m+2n+1)!!]^{2}}{\tilde{\eta}}(\widetilde{\vec{\kappa}}^{2}-(%
\widetilde{\vec{\kappa}}\cdot \hat{\eta})^{2})^{m+n+1}.  \label{VI22}
\end{equation}
In the general case this expression can be more simply written in terms of $%
\cos (2k+1)\phi _{ij}$ [see Eq.(\ref{c55})]. Using the identity
\begin{equation}
\frac{\lbrack (2m+2n+1)!!][(n+m+1)!]}{(2(m+n+1))!}=\frac{1}{2^{m+n+1}},
\label{VI23}
\end{equation}
one obtains for $m\geq n$ the expression
\begin{eqnarray}
&&(\widetilde{\vec{\kappa}}_{i}\cdot \vec{\nabla}_{ij})^{2m+1}(\widetilde{%
\vec{\kappa}}_{j}\cdot \vec{\nabla}_{ij})^{2n+1}\tilde{\eta}_{ij}^{2(m+n)+1}=
\nonumber \\
&&{}  \nonumber \\
&=&\frac{2}{\tilde{\eta}_{ij}}\frac{(2(m+n+1))!(\widetilde{\vec{\kappa}}%
_{i}^{2}-(\widetilde{\vec{\kappa}}_{i}\cdot \hat{\eta}_{ij})^{2})^{m+1/2}(%
\widetilde{\vec{\kappa}}_{j}^{2}-(\widetilde{\vec{\kappa}}_{j}\cdot \hat{\eta%
}_{ij})^{2})^{n+1/2}}{2^{m+n+2}}  \nonumber \\
&&\times \sum_{k=0}^{n}\cos (2k+1)\phi _{ij}%
%TCIMACRO{\binom{2n+1}{n-k}}%
%BeginExpansion
{2n+1 \choose n-k}%
%EndExpansion
%TCIMACRO{\binom{2m+1}{m-k} }%
%BeginExpansion
{2m+1 \choose m-k}%
%EndExpansion
\label{VI24}
\end{eqnarray}
Hence
\begin{eqnarray}
&&(\widetilde{\vec{\kappa}}_{i}\cdot \vec{\nabla}_{ij})^{2m+3}(\widetilde{%
\vec{\kappa}}_{j}\cdot \vec{\nabla}_{ij})^{2n+1}\tilde{\eta}_{ij}^{2(m+n)+1}=
\nonumber \\
&&{}  \nonumber \\
&=&\frac{2}{\tilde{\eta}_{ij}}\frac{(2(m+n+2))!(\widetilde{\vec{\kappa}}%
_{i}^{2}-(\widetilde{\vec{\kappa}}_{i}\cdot \hat{\eta}_{ij})^{2})^{m+3/2}(%
\widetilde{\vec{\kappa}}_{j}^{2}-(\widetilde{\vec{\kappa}}_{j}\cdot \hat{\eta%
}_{ij})^{2})^{n+1/2}}{2^{m+n+3}}  \nonumber \\
&&\times \sum_{k=0}^{n}\cos (2k+1)\phi
%TCIMACRO{\binom{2n+1}{n-k}}%
%BeginExpansion
{2n+1 \choose n-k}%
%EndExpansion
%TCIMACRO{\binom{2m+3}{m+1-k} }%
%BeginExpansion
{2m+3 \choose m+1-k}%
%EndExpansion
\label{VI25}
\end{eqnarray}

By similar methods one finds that
\begin{eqnarray}
&&(\widetilde{\vec{\kappa}}_{i}\cdot \vec{\nabla}_{ij})^{2m+2}(\widetilde{%
\vec{\kappa}}_{j}\cdot \vec{\nabla}_{ij})^{2n+2}\tilde{\eta}_{ij}^{2(m+n)+3}=
\nonumber \\
&&{}  \nonumber \\
&=&\frac{2}{\tilde{\eta}_{ij}}\frac{(2(m+n+2))!(\widetilde{\vec{\kappa}}%
_{i}^{2}-(\widetilde{\vec{\kappa}}_{i}\cdot \hat{\eta}_{ij})^{2})^{m+1}(%
\widetilde{\vec{\kappa}}_{j}^{2}-(\widetilde{\vec{\kappa}}_{j}\cdot \hat{\eta%
}_{ij})^{2})^{n+1}}{2^{m+n+4}}\times  \nonumber \\
&&(\sum_{k=0}^{n}\cos (2k+1)\phi _{ij}%
%TCIMACRO{\binom{2n+1}{n-k}}%
%BeginExpansion
{2n+1 \choose n-k}%
%EndExpansion
%TCIMACRO{\binom{2m+1}{m-k}}%
%BeginExpansion
{2m+1 \choose m-k}%
%EndExpansion
+%
%TCIMACRO{\binom{2m+2}{m+1}}%
%BeginExpansion
{2m+2 \choose m+1}%
%EndExpansion
%TCIMACRO{\binom{2n+2}{n+1} }%
%BeginExpansion
{2n+2 \choose n+1}%
%EndExpansion
\label{VI26}
\end{eqnarray}
To determine the other combination note
\begin{eqnarray}
&&(\widetilde{\vec{\kappa}}_{i}\cdot \vec{\nabla}_{ij})^{2m}(\widetilde{\vec{%
\kappa}}_{j}\cdot \vec{\nabla}_{ij})^{2n}\tilde{\eta}_{ij}^{2(m+n)-1}=
\nonumber \\
&&{}  \nonumber \\
&=&\frac{2}{\tilde{\eta}_{ij}}\frac{(2(m+n))!(\widetilde{\vec{\kappa}}%
_{i}^{2}-(\widetilde{\vec{\kappa}}_{i}\cdot \hat{\eta}_{ij})^{2})^{m}(%
\widetilde{\vec{\kappa}}_{j}^{2}-(\widetilde{\vec{\kappa}}_{j}\cdot \hat{\eta%
}_{ij})^{2})^{n}}{2^{m+n+2}}\times  \nonumber \\
&&(\sum_{k=0}^{n-1}\cos (2k+1)\phi _{ij}%
%TCIMACRO{\binom{2n-1}{n-1-k}}%
%BeginExpansion
{2n-1 \choose n-1-k}%
%EndExpansion
%TCIMACRO{\binom{2m-1}{m-1-k}}%
%BeginExpansion
{2m-1 \choose m-1-k}%
%EndExpansion
+%
%TCIMACRO{\binom{2m}{m}}%
%BeginExpansion
{2m \choose m}%
%EndExpansion
%TCIMACRO{\binom{2n}{n} }%
%BeginExpansion
{2n \choose n}%
%EndExpansion
\label{VI27}
\end{eqnarray}
Thus
\begin{eqnarray}
&&(\widetilde{\vec{\kappa}}_{i}\cdot \vec{\nabla}_{ij})^{2m+2}(\widetilde{%
\vec{\kappa}}_{j}\cdot \vec{\nabla}_{ij})^{2n}\tilde{\eta}_{ij}^{2(m+n)+1}=
\nonumber \\
&&{}  \nonumber \\
&=&\frac{2}{\tilde{\eta}_{ij}}\frac{(2(m+n+1))!(\widetilde{\vec{\kappa}}%
_{i}^{2}-(\widetilde{\vec{\kappa}}_{i}\cdot \hat{\eta}_{ij})^{2})^{m+1}(%
\widetilde{\vec{\kappa}}_{j}^{2}-(\widetilde{\vec{\kappa}}_{j}\cdot \hat{\eta%
}_{ij})^{2})^{n}}{2^{m+n+3}}\times  \nonumber \\
&&\sum_{k=0}^{n-1}\cos (2k+1)\phi _{ij}%
%TCIMACRO{\binom{2n-1}{n-1-k}}%
%BeginExpansion
{2n-1 \choose n-1-k}%
%EndExpansion
%TCIMACRO{\binom{2m+1}{m+1-k}}%
%BeginExpansion
{2m+1 \choose m+1-k}%
%EndExpansion
+%
%TCIMACRO{\binom{2m+2}{m+1}}%
%BeginExpansion
{2m+2 \choose m+1}%
%EndExpansion
%TCIMACRO{\binom{2n}{n} }%
%BeginExpansion
{2n \choose n}%
%EndExpansion
\label{VI28}
\end{eqnarray}
Unfortunately, the $k$ sums cannot be performed analytically to known closed
forms for the $N$-body problem.

\ In the special case of the two body system we \ can obtain a closed form
if we use the rest frame condition $\widetilde{\vec{\kappa}}_{1}+\widetilde{%
\vec{\kappa}}_{2}=0$ . \ The expression we get in this way may be used with
the Dirac brackets associated with $\widetilde{\vec{\kappa}}_{1}+\widetilde{%
\vec{\kappa}}_{2}\approx 0,\,\,\,\vec{q}_{+}\approx 0$, \ \ \ so that the
final reduced phase contains only $\tilde{\eta}=|\widetilde{\vec{\eta}_{1}}-%
\widetilde{\vec{\eta}_{2}}|$ and $\widetilde{\vec{\kappa}}:=\widetilde{\vec{%
\kappa}}_{1}=-\widetilde{\vec{\kappa}}_{2}.$ Using the identity in Eq.(\ref
{VI22}) the higher order Darwin part $V_{HOD}$ $=U_{HOD}+U_{HOD\text{ }%
}^{\prime }$ becomes

\begin{eqnarray}
V_{HOD} &=&-{\frac{Q_{1}Q_{2}}{8\pi \tilde{\eta}}}\sum_{m=0}^{\infty
}\sum_{n=0}^{\infty }\Big[-\widetilde{\vec{\kappa}}^{2}\frac{%
[(2m+2n+1)!!]^{2}}{(2n+2m+2)!}[\widetilde{\vec{\kappa}}^{2}-(\widetilde{\vec{%
\kappa}}\cdot \hat{\eta})^{2}]^{n+m+1}  \nonumber \\
&&\times ({\frac{1}{\sqrt{m_{1}^{2}+{\widetilde{\vec{\kappa}}}^{2}}}}%
)^{2m+1}({\frac{1}{\sqrt{m_{2}^{2}+{\widetilde{\vec{\kappa}}}^{2}}}})^{2n+1}(%
{\frac{1}{m_{1}^{2}+{\vec{\kappa}}^{2}}}+{\frac{1}{m_{2}^{2}+{\widetilde{%
\vec{\kappa}}}^{2}}})+  \nonumber \\
&&+\frac{[(2m+2n+3)!!]^{2}}{(2n+2m+4)!}  \nonumber \\
&&[\widetilde{\vec{\kappa}}^{2}-(\widetilde{\vec{\kappa}}\cdot \hat{\eta}%
)^{2}]^{n+m+2}(({\frac{1}{\sqrt{m_{1}^{2}+{\widetilde{\vec{\kappa}}}^{2}}}}%
)^{2m+1}({\frac{1}{\sqrt{m_{2}^{2}+{\widetilde{\vec{\kappa}}}^{2}}}})^{2n+1}(%
{\frac{1}{m_{1}^{2}+{\widetilde{\vec{\kappa}}}^{2}}}+{\frac{1}{m_{2}^{2}+{%
\widetilde{\vec{\kappa}}}^{2}}})+  \nonumber \\
&&+2\widetilde{\vec{\kappa}}^{2}\frac{[(2m+2n+3)!!]^{2}}{(2n+2m+4)!}[%
\widetilde{\vec{\kappa}}^{2}-(\widetilde{\vec{\kappa}}\cdot \hat{\eta}%
]^{2})^{n+m+2}({\frac{1}{\sqrt{m_{1}^{2}+{\vec{\kappa}}^{2}}}})^{2m+3}({%
\frac{1}{\sqrt{m_{2}^{2}+{\widetilde{\vec{\kappa}}}^{2}}}})^{2n+3}-
\nonumber \\
&&-2\frac{[(2m+2n+5)!!]^{2}}{(2n+2m+6)!}[\widetilde{\vec{\kappa}}^{2}-(%
\widetilde{\vec{\kappa}}\cdot \hat{\eta})^{2}]^{n+m+3}({\frac{1}{\sqrt{%
m_{1}^{2}+{\widetilde{\vec{\kappa}}}^{2}}}})^{2m+3}({\frac{1}{\sqrt{%
m_{2}^{2}+{\widetilde{\vec{\kappa}}}^{2}}}})^{2n+3}\big].  \nonumber \\
&&{}  \label{VI29}
\end{eqnarray}

We use
\begin{equation}
\frac{\lbrack (2m+2n+1)!!]^{2}}{(2n+2m+2)!}=\frac{(-)^{n+m}}{2(n+m+1)}%
%TCIMACRO{\binom{-3/2}{n+m}}%
%BeginExpansion
{-3/2 \choose n+m}%
%EndExpansion
,  \label{VI30}
\end{equation}
and let $m+n=l$ \ so that $0\leq m\leq l$ and $0\leq l<\infty $. Then we
perform the $m$ sum using
\begin{equation}
\sum_{m=0}^{l}\left( \frac{x}{y}\right) ^{m}=\frac{y^{l+1}-x^{l+1}}{%
y^{l}(y-x)},  \label{VI31}
\end{equation}
and obtain
\begin{eqnarray}
V_{HOD} &=&-{\frac{Q_{1}Q_{2}}{8\pi \tilde{\eta}}}\sum_{l=0}^{\infty }\Big[{-%
}\widetilde{\vec{\kappa}}^{2}\frac{(-)^{l}}{2(l+1)}%
%TCIMACRO{\binom{-3/2}{l}}%
%BeginExpansion
{-3/2 \choose l}%
%EndExpansion
[\widetilde{\vec{\kappa}}^{2}-(\widetilde{\vec{\kappa}}\cdot \hat{\eta}%
)^{2}]^{l+1}[\frac{({\frac{1}{\sqrt{m_{2}^{2}+{\widetilde{\vec{\kappa}}}^{2}}%
}})^{2l+2}-({\frac{1}{\sqrt{m_{1}^{2}+{\widetilde{\vec{\kappa}}}^{2}}}}%
)^{2l+2}}{m_{1}^{2}-m_{2}^{2}}]  \nonumber \\
&&\times \sqrt{m_{1}^{2}+{\widetilde{\vec{\kappa}}}^{2}}\sqrt{m_{2}^{2}+{%
\widetilde{\vec{\kappa}}}^{2}}({\frac{1}{m_{1}^{2}+{\widetilde{\vec{\kappa}}}%
^{2}}}+{\frac{1}{m_{2}^{2}+{\widetilde{\vec{\kappa}}}^{2}}})+  \nonumber \\
&&+\frac{(-)^{l+1}}{2(l+2)}%
%TCIMACRO{\binom{-3/2}{l+1}}%
%BeginExpansion
{-3/2 \choose l+1}%
%EndExpansion
[\widetilde{\vec{\kappa}}^{2}-(\widetilde{\vec{\kappa}}\cdot \hat{\eta}%
]^{2})^{l+2}[\frac{({\frac{1}{\sqrt{m_{2}^{2}+{\widetilde{\vec{\kappa}}}^{2}}%
}})^{2l+2}-({\frac{1}{\sqrt{m_{1}^{2}+{\widetilde{\vec{\kappa}}}^{2}}}}%
)^{2l+2}}{m_{1}^{2}-m_{2}^{2}}]  \nonumber \\
&&\times \sqrt{m_{1}^{2}+{\widetilde{\vec{\kappa}}}^{2}}\sqrt{m_{2}^{2}+{%
\widetilde{\vec{\kappa}}}^{2}}({\frac{1}{m_{1}^{2}+{\widetilde{\vec{\kappa}}}%
^{2}}}+{\frac{1}{m_{2}^{2}+{\widetilde{\vec{\kappa}}}^{2}}})+  \nonumber \\
&&+\widetilde{\vec{\kappa}}^{2}\frac{(-)^{l+1}}{(l+2)}%
%TCIMACRO{\binom{-3/2}{l+1}}%
%BeginExpansion
{-3/2 \choose l+1}%
%EndExpansion
[\widetilde{\vec{\kappa}}^{2}-(\widetilde{\vec{\kappa}}\cdot \hat{\eta}%
)^{2}]^{l+2}\times  \nonumber \\
&&[\frac{({\frac{1}{\sqrt{m_{2}^{2}+{\widetilde{\vec{\kappa}}}^{2}}}}%
)^{2l+2}-({\frac{1}{\sqrt{m_{1}^{2}+{\widetilde{\vec{\kappa}}}^{2}}}})^{2l+2}%
}{m_{1}^{2}-m_{2}^{2}}]({\frac{1}{\sqrt{m_{1}^{2}+{\widetilde{\vec{\kappa}}}%
^{2}}}})({\frac{1}{\sqrt{m_{2}^{2}+{\widetilde{\vec{\kappa}}}^{2}}}})-
\nonumber \\
&&-\frac{(-)^{l}}{(l+3)}%
%TCIMACRO{\binom{-3/2}{l+2}}%
%BeginExpansion
{-3/2 \choose l+2}%
%EndExpansion
[\widetilde{\vec{\kappa}}^{2}-(\widetilde{\vec{\kappa}}\cdot \hat{\eta}%
)^{2}]^{l+3}\times  \nonumber \\
&&[\frac{({\frac{1}{\sqrt{m_{2}^{2}+{\widetilde{\vec{\kappa}}}^{2}}}}%
)^{2l+2}-({\frac{1}{\sqrt{m_{1}^{2}+{\widetilde{\vec{\kappa}}}^{2}}}})^{2l+2}%
}{m_{1}^{2}-m_{2}^{2}}]({\frac{1}{\sqrt{m_{1}^{2}+{\widetilde{\vec{\kappa}}}%
^{2}}})}({\frac{1}{\sqrt{m_{2}^{2}+{\widetilde{\vec{\kappa}}}^{2}}})}\Big].
\label{VI32}
\end{eqnarray}
We use

\begin{eqnarray}
\sum_{l=0}^{\infty }\frac{(-)^{l}}{2(l+1)}
%TCIMACRO{\binom{-3/2}{l}}%
%BeginExpansion
{-3/2 \choose l}%
%EndExpansion
x^{l} &=&\frac{1}{x}((1-x)^{-1/2}-1),  \nonumber \\
\sum_{l=0}^{\infty }\frac{(-)^{l+1}}{2(l+2)}
%TCIMACRO{\binom{-3/2}{l+1}}%
%BeginExpansion
{-3/2 \choose l+1}%
%EndExpansion
x^{l} &=&\frac{1}{x^{2}}(2(1-x)^{-1/2}-2-x)=\frac{1}{x^{2}}
(2(1-x)^{-1/2}+(1-x)-3) ,  \nonumber \\
\sum_{l=0}^{\infty }\frac{(-)^{l}}{2(l+3)}
%TCIMACRO{\binom{-3/2}{l+2}}%
%BeginExpansion
{-3/2 \choose l+2}%
%EndExpansion
x^{l} &=&\frac{1}{x^{3}}((1-x)^{-1/2}-\frac{3}{8}(1-x)^{2}+\frac{5}{4}(1-x)-
\frac{15}{8}),  \label{VI33}
\end{eqnarray}
and finally determine

\begin{eqnarray}
V_{HOD} &=&-{\frac{Q_{1}Q_{2}}{8\pi \tilde{\eta}(m_{1}^{2}-m_{2}^{2})}}\times
\nonumber \\
&&\Big({-}\widetilde{\vec{\kappa}}^{2}\Big[\sqrt{\frac{m_{2}^{2}+{\widetilde{%
\vec{\kappa}}}^{2}}{m_{2}^{2}+(\widetilde{\vec{\kappa}}\cdot \hat{\eta})^{2}}%
}-\sqrt{\frac{m_{1}^{2}+{\vec{\kappa}}^{2}}{m_{1}^{2}+(\widetilde{\vec{\kappa%
}}\cdot \hat{\eta})^{2}}}\Big](\sqrt{\frac{m_{1}^{2}+{\widetilde{\vec{\kappa}%
}}^{2}}{m_{2}^{2}+{\widetilde{\vec{\kappa}}}^{2}}}+\sqrt{\frac{m_{2}^{2}+{%
\widetilde{\vec{\kappa}}}^{2}}{m_{1}^{2}+{\widetilde{\vec{\kappa}}}^{2}}})+
\nonumber \\
&&+\Big[(m_{2}^{2}+{\widetilde{\vec{\kappa}}}^{2})[2\sqrt{\frac{m_{2}^{2}+{%
\widetilde{\vec{\kappa}}}^{2}}{m_{2}^{2}+(\widetilde{\vec{\kappa}}\cdot \hat{%
\eta})^{2}}}+\frac{m_{2}^{2}+(\vec{\kappa}\cdot \hat{\eta})^{2}}{m_{2}^{2}+{%
\widetilde{\vec{\kappa}}}^{2}}-3]-  \nonumber \\
&&-(m_{1}^{2}+{\widetilde{\vec{\kappa}}}^{2})[2\sqrt{\frac{m_{1}^{2}+{%
\widetilde{\vec{\kappa}}}^{2}}{m_{1}^{2}+(\widetilde{\vec{\kappa}}\cdot \hat{%
\eta})^{2}}}+\frac{m_{1}^{2}+(\widetilde{\vec{\kappa}}\cdot \hat{\eta})^{2}}{%
m_{1}^{2}+{\widetilde{\vec{\kappa}}}^{2}}-3]\Big]\times  \nonumber \\
&&(\sqrt{\frac{m_{1}^{2}+{\widetilde{\vec{\kappa}}}^{2}}{m_{2}^{2}+{%
\widetilde{\vec{\kappa}}}^{2}}}+\sqrt{\frac{m_{2}^{2}+{\widetilde{\vec{\kappa%
}}}^{2}}{m_{1}^{2}+{\widetilde{\vec{\kappa}}}^{2}}})+  \nonumber \\
&&+2\kappa ^{2}\Big[(m_{2}^{2}+{\widetilde{\vec{\kappa}}}^{2})[2\sqrt{\frac{%
m_{2}^{2}+{\widetilde{\vec{\kappa}}}^{2}}{m_{2}^{2}+(\widetilde{\vec{\kappa}}%
\cdot \hat{\eta})^{2}}}+\frac{m_{2}^{2}+(\widetilde{\vec{\kappa}}\cdot \hat{%
\eta})^{2}}{m_{2}^{2}+{\widetilde{\vec{\kappa}}}^{2}}-3]-  \nonumber \\
&&-(m_{1}^{2}+{\widetilde{\vec{\kappa}}}^{2})[2\sqrt{\frac{m_{1}^{2}+{%
\widetilde{\vec{\kappa}}}^{2}}{m_{1}^{2}+(\widetilde{\vec{\kappa}}\cdot \hat{%
\eta})^{2}}}+\frac{m_{1}^{2}+(\widetilde{\vec{\kappa}}\cdot \hat{\eta})^{2}}{%
m_{1}^{2}+{\widetilde{\vec{\kappa}}}^{2}}-3]\Big]\times  \nonumber \\
&&({\frac{1}{\sqrt{m_{1}^{2}+{\widetilde{\vec{\kappa}}}^{2}}}})({\frac{1}{%
m_{2}^{2}+\sqrt{{\widetilde{\vec{\kappa}}}^{2}}}})-  \nonumber \\
&&-2\Big[(m_{2}^{2}+{\widetilde{\vec{\kappa}}}^{2})^{2}\Big(\sqrt{\frac{%
m_{2}^{2}+{\vec{\kappa}}^{2}}{m_{2}^{2}+(\widetilde{\vec{\kappa}}\cdot \hat{%
\eta})^{2}}}-\frac{3}{8}\left( \frac{m_{2}^{2}+(\widetilde{\vec{\kappa}}%
\cdot \hat{\eta})^{2}}{m_{2}^{2}+{\vec{\kappa}}^{2}}\right) ^{2}+  \nonumber
\\
&+&\frac{5}{4}\left( \frac{m_{2}^{2}+(\widetilde{\vec{\kappa}}\cdot \hat{\eta%
})^{2}}{m_{2}^{2}+{\widetilde{\vec{\kappa}}}^{2}}\right) -\frac{15}{8}\Big)-
\nonumber \\
&&-(m_{1}^{2}+{\widetilde{\vec{\kappa}}}^{2})^{2}\Big(\sqrt{\frac{m_{1}^{2}+{%
\widetilde{\vec{\kappa}}}^{2}}{m_{1}^{2}+(\widetilde{\vec{\kappa}}\cdot \hat{%
\eta})^{2}}}-\frac{3}{8}\left( \frac{m_{1}^{2}+(\widetilde{\vec{\kappa}}%
\cdot \hat{\eta})^{2}}{m_{1}^{2}+{\widetilde{\vec{\kappa}}}^{2}}\right) ^{2}+
\nonumber \\
&+&\frac{5}{4}\left( \frac{m_{1}^{2}+(\widetilde{\vec{\kappa}}\cdot \hat{\eta%
})^{2}}{m_{1}^{2}+{\widetilde{\vec{\kappa}}}^{2}}\right) -\frac{15}{8}\Big)%
\Big]\times  \nonumber \\
&&({\frac{1}{\sqrt{m_{1}^{2}+{\widetilde{\vec{\kappa}}}^{2}}})}({\frac{1}{%
\sqrt{m_{2}^{2}+{\widetilde{\vec{\kappa}}}^{2}}})}\Big).  \label{VI34}
\end{eqnarray}
So our final two-body expression is
\begin{eqnarray}
M &=&\sqrt{m_{1}^{2}+\widetilde{\vec{\kappa}}^{2}}+\sqrt{m_{2}^{2}+%
\widetilde{\vec{\kappa}}^{2}}+\frac{Q_{1}Q_{2}}{4\pi \tilde{\eta}}+\tilde{V}%
_{DAR},  \nonumber \\
&&\tilde{V}_{DAR}=V_{LOD}+V_{HOD},\,\,\,\,\,\,V_{HOD}=U_{HOD}+U_{HOD}^{%
\prime }  \nonumber \\
&&{}  \nonumber \\
V_{LOD} &=&{\frac{Q_{1}Q_{2}}{8\pi \eta }[}\widetilde{\vec{\kappa}}^{2}-(%
\vec{\kappa}\cdot \hat{\eta})^{2}]({\frac{1}{\sqrt{m_{1}^{2}+{\widetilde{%
\vec{\kappa}}}^{2}}})}({\frac{1}{\sqrt{m_{2}^{2}+{\widetilde{\vec{\kappa}}}%
^{2}}})}.  \label{VI35}
\end{eqnarray}
in which we have used the rest-frame condition $\widetilde{\vec{\kappa}}%
_{1}=-\widetilde{\vec{\kappa}}_{2}:=\widetilde{\vec{\kappa}}$. \ Note that $%
Q_{1}Q_{2}\vec{\kappa}=Q_{1}Q_{2}\widetilde{\vec{\kappa}}.$

\ In the equal mass limit $m_{1}\rightarrow m_{2}:=m$
\begin{eqnarray}
&&V_{HOD}=+{\frac{Q_{1}Q_{2}}{8\pi \tilde{\eta}}}\times   \nonumber \\
&&\frac{m^{2}[3\widetilde{\vec{\kappa}}^{2}+(\widetilde{\vec{\kappa}}\cdot
\hat{\eta})^{2}]-2\vec{\kappa}^{2}[\widetilde{\vec{\kappa}}^{2}-3(\widetilde{%
\vec{\kappa}}\cdot \hat{\eta})^{2}]\sqrt{\frac{m^{2}+\kappa ^{2}}{m^{2}+(%
\widetilde{\vec{\kappa}}\cdot \hat{\eta})^{2}}}-[3\widetilde{\vec{\kappa}}%
^{2}+(\widetilde{\vec{\kappa}}\cdot \hat{\eta})^{2}][m^{2}+(\widetilde{\vec{%
\kappa}}\cdot \hat{\eta})^{2}]}{(m^{2}+\kappa ^{2})[m^{2}+(\widetilde{\vec{%
\kappa}}\cdot \hat{\eta})^{2}]},  \nonumber \\
&&{}  \label{VI36}
\end{eqnarray}
so that
\begin{eqnarray}
&&M={\cal P}_{(int)}^{\tau }=2\sqrt{m^{2}+\widetilde{\vec{\kappa}}^{2}}+%
\frac{Q_{1}Q_{2}}{4\pi \tilde{\eta}}+{\frac{Q_{1}Q_{2}}{8\pi \tilde{\eta}}}%
\times   \nonumber \\
&&(\frac{m^{2}[3\widetilde{\vec{\kappa}}^{2}+(\widetilde{\vec{\kappa}}\cdot
\hat{\eta})^{2}]-2\widetilde{\vec{\kappa}}^{2}[\widetilde{\vec{\kappa}}%
^{2}-3(\widetilde{\vec{\kappa}}\cdot \hat{\eta})^{2}]\sqrt{\frac{m^{2}+%
\widetilde{\vec{\kappa}}^{2}}{m^{2}+(\widetilde{\vec{\kappa}}\cdot \hat{\eta}%
)^{2}}}-2[\widetilde{\vec{\kappa}}^{2}+(\widetilde{\vec{\kappa}}\cdot \hat{%
\eta})^{2}][m^{2}+(\widetilde{\vec{\kappa}}\cdot \hat{\eta})^{2}]}{(m^{2}+%
\widetilde{\vec{\kappa}}^{2})[m^{2}+(\widetilde{\vec{\kappa}}\cdot \hat{\eta}%
)^{2}]}).  \nonumber \\
&&{}  \label{VI37}
\end{eqnarray}
Our result differs from the appropriate terms at order $1/c^{4}$ with those
in \cite{yang}, \cite{mlna} and \cite{scot} although it does agree with
their Darwin interaction at order $1/c^{2}$. \ Since each of these latter
three sources took as a starting point the Fokker particle Lagrangian they
have not used, as we have done here, the canonical variables which would
come from using the pair of secondary constraints arising from the reduction
of the field plus particle Lagrangian to the Fokker action. We point out
that \ Molina et al in \cite{mlna}, like earlier work by Golubenkov and
Smorodinski \cite{golub}obtain order $1/c^{4}$ corrections to the
Hamiltonian. \ However, unlike them these authors not only use the Coulomb
force law \ to replace acceleration dependent terms on the Lagrangian but
also include the effects of that substitution on the choice of canonical
variables by viewing it as a constraint (thus including Dirac brackets), in
the reduction to Hamiltonian forms. (Note that since these three approaches,
unlike ours, do not use Grassmann charges, they contain acceleration driven
terms not contained in our approach. \ The comparisons we are talking about
here with our approach refer to the terms not driven by acceleration
dependent Lagrangian potentials).

It is of interest that Hamilton's equations has a solution for circular
orbits just as does the Schild solution for Feynman-Wheeler electrodynamics
\cite{schild}. In Appendix E we show how this comes about for equal masses.
\ The case of unequal masses is similar.

\subsection{${\cal J}_{(int)}^{r}$, ${\cal K}_{(int)}^{r}$, ${\vec{q}}_{+}$
and the Energy-Momentum Tensor in the Final Canonical Variables.}

\bigskip Eqs.(\ref{V49}) and (Eq.(\ref{VI19}) [(Eq.(\ref{VI35})) for $N=2$]
give ${\cal \vec{P}}_{(int)}$ and $M={\cal P}_{(int)}^{\tau }$ in terms of
the final canonical variables. For the internal angular momentum we get
\begin{eqnarray}
{\cal J}_{(int)}^{r} &=&\varepsilon ^{rst}{\bar{S}}_{s}^{st}=\sum_{i=1}^{N}(%
\vec{\eta}_{i}(\tau )\times {\vec{\kappa}}_{i}(\tau ))^{r}+\int d^{3}\sigma
\,(\vec{\sigma}\times \,{[\vec{\pi}}_{\perp }{\times }\vec{B}{{]}(\tau ,\vec{%
\sigma})})^{r}=  \nonumber \\
&=&\sum_{i=1}^{N}[(\widetilde{\vec{\eta}_{i}}-\vec{\alpha}_{i})\times (%
\widetilde{\vec{\kappa}_{i}}-\vec{\beta}_{i})]^{r}+  \nonumber \\
&+&\frac{1}{2}\sum_{i=1}^{N}\sum_{j\neq i}^{N}Q_{i}Q_{j}[\int d^{3}\sigma \,(%
\vec{\sigma}\times \,{[{\vec{\pi}}_{\perp Si}\times (\vec{\nabla}}_{\sigma
}\times {\vec{A}}_{\perp Sj}){{]}+i\leftrightarrow j)}.  \label{VI38}
\end{eqnarray}
Using the forms for $\vec{\alpha}_{i}$ and $\vec{\beta}_{i}$ given in Eqs.(%
\ref{V51}) together with the expression for ${\cal K}_{ij}$ and expanding
the cross products in the integral we obtain
\begin{eqnarray}
{\cal J}_{(int)}^{r} &=&\sum_{i=1}^{N}(\widetilde{\vec{\eta}_{i}}\times
\widetilde{\vec{\kappa}_{i}})^{r}+  \nonumber \\
&&+\frac{1}{2}\sum_{i=1}^{N}\sum_{j\neq i}^{N}Q_{i}Q_{j}\{(\vec{\eta}%
_{i}\times \nabla _{\eta _{i}}+\vec{\kappa}_{i}\times \nabla _{\kappa
_{i}})^{r}\int d^{3}\sigma ({\vec{A}_{\perp Si}\cdot {\vec{\pi}}_{\perp Sj}-%
\vec{A}_{\perp Sj}\cdot {\vec{\pi}}_{\perp Si})\}}  \nonumber \\
&&+\frac{1}{2}\sum_{i=1}^{N}\sum_{j\neq i}^{N}Q_{i}Q_{j}[\int d^{3}\sigma
\varepsilon ^{ruv}\sigma ^{u}[\,{{\vec{\pi}}_{\perp Si}\cdot \partial }%
_{\sigma }^{v}{\vec{A}_{\perp Sj}+{\vec{\pi}}_{\perp Sj}\cdot \partial }%
_{\sigma }^{v}{\vec{A}_{\perp Si}-}  \nonumber \\
&&{-}\,{{\vec{\pi}}_{\perp Si}\cdot \vec{\nabla}}_{\sigma }A^{v}{_{\perp Sj}-%
{\vec{\pi}}_{\perp Sj}\cdot \vec{\nabla}}_{\sigma }A^{v}{_{\perp Si}]}.
\label{VI39}
\end{eqnarray}
Using the transverse nature of the field together with vanishing surface
terms we find that
\begin{equation}
{\cal \vec{J}}_{(int)}=\sum_{i=1}^{N}\widetilde{\vec{\eta}_{i}}\times
\widetilde{\vec{\kappa}_{i}},  \label{VI40}
\end{equation}
if
\begin{eqnarray}
&&(\vec{\eta}_{i}\times \vec{\nabla}_{\eta _{i}}+\vec{\kappa}_{i}\times \vec{%
\nabla}_{\kappa _{i}})\int d^{3}\sigma ({\vec{A}_{\perp Si}\cdot {\vec{\pi}}%
_{\perp Sj}-\vec{A}_{\perp Sj}\cdot {\vec{\pi}}_{\perp Si})}  \nonumber \\
&=&\int d^{3}\sigma \lbrack {\vec{A}_{\perp Si}\times {\vec{\pi}}_{\perp Sj}-%
\vec{\sigma}\times (\vec{\nabla}}_{\sigma }A^{k}{_{\perp Si})\pi _{\perp
Sj}^{k}+\vec{A}_{\perp Sj}\times {\vec{\pi}}_{\perp Si}+\vec{\sigma}\times (}%
A^{k}{_{\perp Si}{\vec{\nabla}}_{\sigma }\pi _{\perp Sj}^{k})]}.
\label{VI41}
\end{eqnarray}
We have given explicit forms forms for ${\vec{A}_{\perp Si}}$ and ${{\vec{\pi%
}}_{\perp Si}}$ in Eqs.(\ref{V28}) and (\ref{VI2}) respectively. Their
general forms are
\begin{eqnarray}
{\vec{A}_{\perp Si}} &=&\frac{1}{4\pi |\vec{\sigma}-\vec{\eta}_{i}|}[\vec{%
\kappa}_{i}f_{i}(\kappa _{i}^{2},\vec{\kappa}_{i}\cdot \hat{\rho}_{i})+\hat{%
\rho}_{i}g_{i}(\kappa _{i}^{2},\vec{\kappa}_{i}\cdot \hat{\rho}_{i})],
\nonumber \\
{{\vec{\pi}}_{\perp Si}} &=&\frac{1}{4\pi |\vec{\sigma}-\vec{\eta}_{i}|^{2}}[%
\vec{\kappa}_{i}h_{i}(\kappa _{i}^{2},\vec{\kappa}_{i}\cdot \hat{\rho}_{i})+%
\hat{\rho}_{i}c_{i}(\kappa _{i}^{2},\vec{\kappa}_{i}\cdot \hat{\rho}_{i})].
\label{VI42}
\end{eqnarray}
in which $\hat{\rho}_{i}:=\frac{(\vec{\sigma}-\vec{\eta}_{i})}{|\vec{\sigma}-%
\vec{\eta}_{i}|}$ . Using
\begin{equation}
\vec{\nabla}_{\sigma }\frac{1}{|\vec{\sigma}-\vec{\eta}_{i}|^{n}}=-\frac{n%
\hat{\rho}_{i}}{|\vec{\sigma}-\vec{\eta}_{i}|^{n+1}},\quad \quad \vec{\nabla}%
_{\sigma }\hat{\rho}_{i}=\frac{-1}{|\vec{\sigma}-\vec{\eta}_{i}|}({\bf I-}%
\hat{\rho}_{i}\hat{\rho}_{i}),  \label{VI43}
\end{equation}
we find that
\begin{equation}
\lbrack (\vec{\eta}_{i}\times \vec{\nabla}_{\eta _{i}}+\vec{\kappa}%
_{i}\times \vec{\nabla}_{\kappa _{i}}){\vec{A}_{\perp Si}]\cdot {\vec{\pi}}%
_{\perp Sj}=-\vec{\sigma}\times (\vec{\nabla}}_{\sigma }A^{k}{_{\perp
Si})\pi _{\perp Sj}^{k}+\vec{A}_{\perp Si}\times {\vec{\pi}}_{\perp Sj}},
\label{VI44}
\end{equation}
while
\begin{equation}
-A^{k}{_{\perp Sj}}[(\vec{\eta}_{i}\times \vec{\nabla}_{\eta _{i}}+\vec{%
\kappa}_{i}\times \vec{\nabla}_{\kappa _{i}}){\pi _{\perp Si}^{k}=A^{k}{%
_{\perp Sj}}\vec{\sigma}\times \vec{\nabla}}_{\sigma }\pi ^{k}{_{\perp Si}+%
\vec{A}_{\perp Sj}\times {\vec{\pi}}_{\perp Si}},  \label{VI45}
\end{equation}
thus verifying Eq.(\ref{VI40}). \

Thus as expected, in an instant form of dynamics, the internal angular
momentum does not depend upon the interaction. \ On the other hand, the
interaction-dependent internal boosts have the form of Eq.(\ref{IV1}). \
Using the above forms for the final dynamical variables we obtain
\begin{eqnarray}
{\cal \vec{K}}_{(int)} &=&-\sum_{i=1}^{N}\widetilde{\vec{\eta}_{i}}{\large [}%
\sqrt{m_{i}^{2}+\widetilde{\vec{\kappa}_{i}}^{2}}+  \nonumber \\
&&+\frac{\widetilde{\vec{\kappa}_{i}}\cdot \sum_{j\neq i}Q_{i}Q_{j}[\vec{%
\nabla}_{\tilde{\eta}_{i}}\frac{1}{2}\widetilde{{\cal K}}_{ij}(\widetilde{%
\vec{\kappa}_{i}},\widetilde{\vec{\kappa}_{j}},\widetilde{\vec{\eta}_{i}}-%
\widetilde{\vec{\eta}_{j}})-2\vec{A}{{_{\perp Sj}(}}\widetilde{\vec{\kappa}%
_{j}},\widetilde{\vec{\eta}_{i}}-\widetilde{\vec{\eta}_{j}})]}{2\sqrt{%
m_{i}^{2}+\widetilde{\vec{\kappa}_{i}}^{2}}}{\large ]}-  \nonumber \\
&&-\frac{1}{2}\sum_{i=1}^{N}\sum_{j\neq i}Q_{i}Q_{j}\sqrt{m_{i}^{2}+%
\widetilde{\vec{\kappa}_{i}}^{2}}\vec{\nabla}_{\tilde{\kappa}_{i}}\widetilde{%
{\cal K}}_{ij}(\widetilde{\vec{\kappa}_{i}},\widetilde{\vec{\kappa}_{j}},%
\widetilde{\vec{\eta}_{i}}-\widetilde{\vec{\eta}_{j}})+  \nonumber \\
&+&\sum_{i=1}^{N}\sum_{j\not=i}\frac{Q_{i}Q_{j}}{8\pi }\frac{\widetilde{\vec{%
\eta}_{i}}-\widetilde{\vec{\eta}_{j}}}{|\widetilde{\vec{\eta}_{i}}-%
\widetilde{\vec{\eta}_{j}}|}-\sum_{i=1}^{N}\sum_{j\not=i}\frac{Q_{i}Q_{j}}{%
4\pi }\int d^{3}\sigma \frac{{\check{\pi}}_{\perp Sj}^{r}(\vec{\sigma}-%
\widetilde{\vec{\eta}_{j}},\widetilde{\vec{\kappa}_{j}})}{|\vec{\sigma}-%
\widetilde{\vec{\eta}_{i}}|}-  \nonumber \\
&&-{\frac{1}{2}}\sum_{i=1}^{N}\sum_{j\neq i}Q_{i}Q_{j}\int d^{3}\sigma \vec{%
\sigma}[{\vec{\pi}_{\perp Si}}(\vec{\sigma}-\widetilde{\vec{\eta}_{i}},%
\widetilde{\vec{\kappa}_{i}})\cdot {\vec{\pi}}_{\perp Sj}(\vec{\sigma}-%
\widetilde{\vec{\eta}_{j}},\widetilde{\vec{\kappa}_{j}})+  \nonumber \\
&&+\vec{B}{_{Si}}(\vec{\sigma}-\widetilde{\vec{\eta}_{i}},\widetilde{\vec{%
\kappa}_{i}})\cdot {\vec{B}}_{Sj}(\vec{\sigma}-\widetilde{\vec{\eta}_{j}},%
\widetilde{\vec{\kappa}_{j}})]=  \nonumber \\
&=&-{\cal P}_{(int)}^{\tau }\vec{R}_{+}.  \label{VI46}
\end{eqnarray}
In the last line the internal M\o ller center of energy (\ref{IV12})
is shown explicitly. \ This equation allows us to express the internal
canonical center of mass $\vec{q}_{+}\approx \vec{R}_{+}$ defined in
Eq.(\ref{IV12}) in terms of the final canonical variables. \ The
natural gauge fixing to the rest frame constraints ${\cal
\vec{P}}_{(int)}=\sum_{i=1}^{N}\widetilde{\vec{%
\kappa}_{i}}\approx 0$ is $\vec{q}_{+}\approx 0.$

From Eq.(\ref{III44}) we get the following expression for the conserved
energy-momentum tensor:

\begin{eqnarray}
T^{\tau \tau }(T_{s},\vec{\sigma}) &=&\sum_{i=1}^{N}\delta ^{3}(\vec{\sigma}-%
{\vec{\eta}}_{i}(T_{s}))\sqrt{m_{i}^{2}+[{\vec{\kappa}}_{i}(T_{s})-Q_{i}%
\sum_{j\not=i}Q_{j}{\vec{A}}_{\perp Sj}(T_{s},{\vec{\eta}}_{i}(T_{s}))]^{2}}+
\nonumber \\
&+&{\frac{1}{2}}\sum_{i\not= j}^{1..N}Q_{i}Q_{j}[\Big({\vec{\pi}}_{\perp Si}+%
{\frac{{\vec{\partial}}}{{\triangle }}}\delta ^{3}(\vec{\sigma}-{\vec{\eta}}%
_{i}(\tau ))\Big)\cdot \Big({\vec{\pi}}_{\perp Sj}+{\frac{{\vec{\partial}}}{{%
\triangle }}}\delta ^{3}(\vec{\sigma}-{\vec{\eta}}_{j}(\tau ))\Big)+
\nonumber \\
&+&{\vec{B}}_{Si}\cdot {\vec{B}}_{Sj}](T_{s},\vec{\sigma}),  \nonumber \\
T^{r\tau }(T_{s},\vec{\sigma}) &=&\sum_{i=1}^{N}\delta ^{3}(\vec{\sigma}-{%
\vec{\eta}}_{i}(T_{s}))[{\kappa }_{i}^{r}(T_{s})-Q_{i}\sum_{j\not=i}Q_{j}{A}%
_{\perp Sj}^{r}(T_{s},{\vec{\eta}}_{i}(T_{s}))]+  \nonumber \\
&+&Q_i\sum_{i\not=j}^{1..N}Q_j[\Big({\vec{\pi}}_{\perp Si}+{\frac{{\vec{%
\partial}} }{{\triangle }}}\delta ^{3}(\vec{\sigma}-{\vec{\eta}}_{i}(T_{s}))%
\Big)\times {\vec{B}}_{Sj}+ (i \leftrightarrow j) ](T_{s},\vec{\sigma}),
\nonumber \\
T^{rs}(T_{s},\vec{\sigma}) &=&\sum_{i=1}^{N}\delta ^{3}(\vec{\sigma}-{\vec{%
\eta}}_{i}(T_{s}))  \nonumber \\
&&{\frac{{\ [{\kappa }_{i}^{r}(T_{s})-Q_{i}\sum_{j\not=i}Q_{j}{A}_{\perp
Sj}^{r}(T_{s},{\vec{\eta}}_{i}(T_{s}))][{\kappa }_{i}^{s}(T_{s})-Q_{i}\sum_{j%
\not=i}Q_{j}{A}_{\perp Sj}^{s}(T_{s},{\vec{\eta}}_{i}(T_{s}))]}}{\sqrt{%
m_{i}^{2}+[{\vec{\kappa}}_{i}(T_{s})-Q_{i}\sum_{j\not=i}Q_{j}{\vec{A}}%
_{\perp Sj}(T_{s},{\vec{\eta}}_{i}(T_{s}))]^{2}}}}-  \nonumber \\
&-&\sum_{i\not=j}^{1..N}Q_{i}Q_{j}\Big[{\frac{1}{2}}\delta ^{rs}[\Big({\vec{%
\pi}}_{\perp Si}+{\frac{{\vec{\partial}}}{{\triangle }}}\delta ^{3}(\vec{%
\sigma}-{\vec{\eta}}_{i}(T_{s}))\Big)\cdot \Big({\vec{\pi}}_{\perp Sj}+{%
\frac{{\vec{\partial}}}{{\triangle }}}\delta ^{3}(\vec{\sigma}-{\vec{\eta}}%
_{j}(T_{s}))\Big)+  \nonumber \\
&+&{\vec{B}}_{Si}\cdot {\vec{B}}_{Sj}]-  \nonumber \\
&-&[\Big(\pi _{\perp Si}^{r}+{\frac{{\ \partial ^{r}}}{{\triangle }}}\delta
^{3}(\vec{\sigma}-{\vec{\eta}}_{i}(T_{s}))\Big)\Big(\pi _{\perp Sj}^{s}+{%
\frac{{\ \partial ^{s}}}{{\triangle }}}\delta ^{3}(\vec{\sigma}-{\vec{\eta}}%
_{j}(T_{s}))\Big)+  \nonumber \\
&+&B_{Si}^{r}B_{Sj}^{s}]\Big](T_{s},\vec{\sigma}).  \label{VI47}
\end{eqnarray}

By using Eqs.(\ref{V51}) we get [${\tilde {\vec \eta}}_i={\tilde {\vec \eta}}%
_i(T_s)$, ${\tilde {\vec \kappa}}_i={\tilde {\vec \kappa}}_i(T_s)$; ${\vec A}%
_{\perp Si}(\vec \sigma -{\tilde {\vec \eta}}_i,{\tilde {\vec \kappa}}_i)$
is given in Eq.(\ref{V28}); ${\tilde {{\cal K}}}_{ij}={\tilde {{\cal K}}}%
_{ij}({\tilde {\vec \kappa}}_i,{\tilde {\vec \kappa}}_j, {\tilde {\vec \eta}}%
_i-{\tilde {\vec \eta}}_j)$ is given in Eqs.(\ref{V34}), (\ref{V35})]

\begin{eqnarray}
T^{\tau\tau}(T_s,\vec \sigma )&=&\sum_{i=1}^N \delta^3(\vec \sigma -{\tilde {%
\vec \eta}}_i+{\frac{1}{2}}\sum_{j\not= i}^{1..N}Q_iQ_j {\vec \nabla}_{{%
\tilde \kappa}_j} {\tilde {{\cal K}}}_{ij})  \nonumber \\
&&\sqrt{m_i^2+\Big[ {\tilde {\vec \kappa}}_i-Q_i\sum_{j\not= i}^{1..N}Q_j({%
\vec A}_{\perp Sj}({\tilde {\vec \eta}}_i-{\tilde {\vec \eta}}_j,{\tilde {%
\vec \kappa}}_j) +{\frac{1}{2}}{\vec \nabla}_{{\tilde \eta}_i}{\tilde {{\cal %
K}}}_{ij})\Big]^2}+  \nonumber \\
&+&{\frac{1}{2}} Q_i\sum_{j\not= i}^{1..N} \Big[ \Big( {\vec E}_{\perp Si}(%
\vec \sigma -{\tilde {\vec \eta}}_i, {\tilde {\vec \kappa}}_i)+{\frac{{\vec %
\partial}}{{\triangle}}}\delta^3(\vec \sigma -{\tilde {\vec \eta}}_i\Big) %
\cdot  \nonumber \\
&&\cdot \Big( {\vec E}_{\perp Sj}(\vec \sigma -{\tilde {\vec \eta}}_j, {%
\tilde {\vec \kappa}}_j)+{\frac{{\vec \partial}}{{\triangle}}}\delta^3(\vec %
\sigma -{\tilde {\vec \eta}}_j\Big) +  \nonumber \\
&+&{\vec B}_{Si}(\vec \sigma -{\tilde {\vec \eta}}_i,{\tilde {\vec \kappa}}%
_i) \cdot {\vec B}_{Sj}(\vec \sigma - {\tilde {\vec \eta}}_j,{\tilde {\vec %
\kappa}}_j) \Big] =  \nonumber \\
&=& \sum_{i=1}^N \sqrt{m_i^2+{\tilde {\vec \kappa}}_i^2} \Big[ \delta^3(\vec %
\sigma -{\tilde {\vec \eta}}_i)  \nonumber \\
&&\Big( 1- {\frac{{Q_i\sum_{j\not= i} Q_j {\tilde {\vec \kappa}}_i\cdot [{%
\vec A}_{\perp Sj}({\tilde {\vec \eta}}_i-{\tilde {\vec \eta}}_j,{\tilde {%
\vec \kappa}}_j)+{\frac{1}{2}}{\vec \nabla}_{{\tilde \eta}_i}{\tilde {{\cal K%
}}}_{ij}]}}{{m_i^2+ {\tilde {\vec \kappa}}_i^2}}}\Big) +  \nonumber \\
&+&{\frac{1}{2}} {\vec \partial}_{\sigma} \delta^3(\vec \sigma -{\tilde {%
\vec \eta}}_i) \cdot Q_i \sum_{j\not= i}Q_j {\vec \nabla}_{{\tilde \kappa}%
_j} {\tilde {{\cal K}}}_{ij}\Big] +  \nonumber \\
&+&{\frac{1}{2}} Q_i\sum_{j\not= i}^{1..N} \Big[ \Big( {\vec E}_{\perp Si}(%
\vec \sigma -{\tilde {\vec \eta}}_i, {\tilde {\vec \kappa}}_i)+{\frac{{\vec %
\partial}}{{\triangle}}}\delta^3(\vec \sigma -{\tilde {\vec \eta}}_i\Big) %
\cdot  \nonumber \\
&&\cdot \Big( {\vec E}_{\perp Sj}(\vec \sigma -{\tilde {\vec \eta}}_j, {%
\tilde {\vec \kappa}}_j)+{\frac{{\vec \partial}}{{\triangle}}}\delta^3(\vec %
\sigma -{\tilde {\vec \eta}}_j\Big) +  \nonumber \\
&+&{\vec B}_{Si}(\vec \sigma -{\tilde {\vec \eta}}_i,{\tilde {\vec \kappa}}%
_i) \cdot {\vec B}_{Sj}(\vec \sigma - {\tilde {\vec \eta}}_j,{\tilde {\vec %
\kappa}}_j) \Big] ,  \nonumber \\
&&{}  \nonumber \\
T^{r\tau}(T_s,\vec \sigma )&=& \sum_{i=1}^N \delta^3(\vec \sigma -{\tilde {%
\vec \eta}}_i+{\frac{1}{2}}\sum_{j\not= i}^{1..N}Q_iQ_j {\vec \nabla}_{{%
\tilde \kappa}_j} {\tilde {{\cal K}}}_{ij})  \nonumber \\
&&\Big[ {\tilde \kappa}_i^r-Q_i\sum_{j\not= i}Q_j \Big( A^r_{\perp Sj}({%
\tilde {\vec \eta}}_i-{\tilde {\vec \eta}}_j,{\tilde {\vec \kappa}}_j) +{%
\frac{1}{2}} \nabla^r_{{\tilde \eta}_i} {\tilde {{\cal K}}}_{ij}\Big) \Big] +
\nonumber \\
&+&Q_i\sum_{j\not= i}^{1..N}Q_j \Big[ \Big( {\vec E}_{\perp Si}(\vec \sigma -%
{\tilde {\vec \eta}}_i,{\tilde {\vec \kappa}}_i)+{\frac{{\vec \partial}}{{%
\triangle}}} \delta^3(\vec \sigma -{\tilde {\vec \eta}}_i)\Big) \times
\nonumber \\
&&\times {\vec B}_{Sj}(\vec \sigma -{\tilde {\vec \eta}}_j,{\tilde {\vec %
\kappa}}_j) + (i \leftrightarrow j) \Big] =  \nonumber \\
&=&\sum_{i=1}^N \Big[ \delta^3(\vec \sigma -{\tilde {\vec \eta}}_i)
\nonumber \\
&&[{\tilde \kappa}_i^r-Q_i\sum_{j\not= i}Q_j \Big( A^r_{\perp Sj}({\tilde {%
\vec \eta}}_i-{\tilde {\vec \eta}}_j,{\tilde {\vec \kappa}}_j) +{\frac{1}{2}}
\nabla^r_{{\tilde \eta}_i} {\tilde {{\cal K}}}_{ij}\Big)]+  \nonumber \\
&+&{\frac{1}{2}} {\vec \partial}_{\sigma} \delta^3(\vec \sigma -{\tilde {%
\vec \eta}}_i) \cdot {\tilde \kappa}_i^r Q_i \sum_{j\not= i}Q_j {\vec \nabla}%
_{{\tilde \kappa}_j} {\tilde {{\cal K}}}_{ij}\Big]+  \nonumber \\
&+&Q_i\sum_{j\not= i}^{1..N}Q_j \Big[ \Big( {\vec E}_{\perp Si}(\vec \sigma -%
{\tilde {\vec \eta}}_i,{\tilde {\vec \kappa}}_i)+{\frac{{\vec \partial}}{{%
\triangle}}} \delta^3(\vec \sigma -{\tilde {\vec \eta}}_i)\Big) \times
\nonumber \\
&&\times {\vec B}_{Sj}(\vec \sigma -{\tilde {\vec \eta}}_j,{\tilde {\vec %
\kappa}}_j) + (i \leftrightarrow j) \Big] ,  \nonumber \\
&&{}  \nonumber \\
T^{rs}(T_s,\vec \sigma )&=&\sum_{i=1}^N \delta^3(\vec \sigma -{\tilde {\vec %
\eta}}_i+{\frac{1}{2}}\sum_{j\not= i}^{1..N}Q_iQ_j {\vec \nabla}_{{\tilde %
\kappa}_j} {\tilde {{\cal K}}}_{ij})  \nonumber \\
&& {\frac{1}{\sqrt{m_i^2+{\tilde {\vec \kappa}}_i^2}}}\Big( 1+ {\frac{{%
Q_i\sum_{j\not= i}^{1..N}Q_j {\tilde {\vec \kappa}}_i\cdot [{\vec A} _{\perp
Sj}({\tilde {\vec \eta}}_i-{\tilde {\vec \eta}}_j,{\tilde {\vec \kappa}}_j)+{%
\frac{1}{2}} {\vec \nabla}_{{\tilde \eta}_i}{\tilde {{\cal K}}}_{ij}]}}{{%
m_i^2+{\tilde {\vec \kappa}}_i^2}}} \Big)  \nonumber \\
&&\Big[ {\tilde \kappa}_i^r-Q_i\sum_{j\not= i}Q_j \Big( A^r_{\perp Sj}({%
\tilde {\vec \eta}}_i-{\tilde {\vec \eta}}_j,{\tilde {\vec \kappa}}_j) +{%
\frac{1}{2}} \nabla^r_{{\tilde \eta}_i} {\tilde {{\cal K}}}_{ij}\Big) \Big]
\nonumber \\
&&\Big[ {\tilde \kappa}_i^s-Q_i\sum_{j\not= i}Q_j \Big( A^s_{\perp Sj}({%
\tilde {\vec \eta}}_i-{\tilde {\vec \eta}}_j,{\tilde {\vec \kappa}}_j) +{%
\frac{1}{2}} \nabla^s_{{\tilde \eta}_i} {\tilde {{\cal K}}}_{ij}\Big) \Big] -
\nonumber \\
&-&Q_i\sum_{j\not= i}^{1..N}Q_j \Big[ {\frac{1}{2}} \delta^{rs} \Big( [{\vec %
E}_{\perp Si}(\vec \sigma -{\tilde {\vec \eta}}_i, {\tilde {\vec \kappa}}_i)+%
{\frac{{\vec \partial}}{{\triangle}}} \delta^3(\vec \sigma -{\tilde {\vec %
\eta}}_i)]\cdot  \nonumber \\
&&\cdot [{\vec E}_{\perp Sj}(\vec \sigma -{\tilde {\vec \eta}}_j, {\tilde {%
\vec \kappa}}_j)+{\frac{{\vec \partial}}{{\triangle}}} \delta^3(\vec \sigma -%
{\tilde {\vec \eta}}_j)] +  \nonumber \\
&+&{\vec B}_{Si}(\vec \sigma -{\tilde {\vec \eta}}_i,{\tilde {\vec \kappa}}%
_i)\cdot {\vec B}_{Sj}(\vec \sigma -{\tilde {\vec \eta}}_j,{\tilde {\vec %
\kappa}}_j) \Big) -  \nonumber \\
&-&[E^r_{\perp Si}(\vec \sigma -{\tilde {\vec \eta}}_i, {\tilde {\vec \kappa}%
}_i)+{\frac{{\ \partial^r}}{{\triangle}}} \delta^3(\vec \sigma -{\tilde {%
\vec \eta}}_i)] [E^s_{\perp Sj}(\vec \sigma -{\tilde {\vec \eta}}_j, {%
\tilde {\vec \kappa}}_j)+{\frac{{\partial^s}}{{\triangle}}} \delta^3(\vec %
\sigma -{\tilde {\vec \eta}}_j)] -  \nonumber \\
&-&B^r_{Si}(\vec \sigma -{\tilde {\vec \eta}}_i,{\tilde {\vec \kappa}}_i)
B^s_{Sj}(\vec \sigma -{\tilde {\vec \eta}}_j,{\tilde {\vec \kappa}}_j) \Big] %
=  \nonumber \\
&=&\sum_{i=1}^N {\frac{1}{\sqrt{m_i^2+{\tilde {\vec \kappa}}_i^2}}} \Big[ %
\delta^3(\vec \sigma -{\tilde {\vec \kappa}}_i)  \nonumber \\
&&\Big( {\tilde \kappa}^r_i{\tilde \kappa}_i^s [ 1+ {\frac{{Q_i\sum_{j\not=
i}^{1..N}Q_j {\tilde {\vec \kappa}}_i\cdot [{\vec A} _{\perp Sj}({\tilde {%
\vec \eta}}_i-{\tilde {\vec \eta}}_j,{\tilde {\vec \kappa}}_j)+{\frac{1}{2}}
{\vec \nabla}_{{\tilde \eta}_i}{\tilde {{\cal K}}}_{ij}]}}{{m_i^2+{\tilde {%
\vec \kappa}}_i^2}}} ]-  \nonumber \\
&-&Q_i \sum_{j\not= i}^{1..N} Q_j \Big[ {\tilde \kappa}^r_i (A^s_{\perp Sj}({%
\tilde {\vec \eta}}_i-{\tilde {\vec \eta}}_j,{\tilde {\vec \kappa}}_j)+{%
\frac{1}{2}} \nabla^s_{{\tilde \eta}_i} {\tilde {{\cal K}}}_{ij})+  \nonumber
\\
&+&{\tilde \kappa}^s_i (A^r_{\perp Sj}({\tilde {\vec \eta}}_i-{\tilde {\vec %
\eta}}_j,{\tilde {\vec \kappa}}_j)+{\frac{1}{2}} \nabla^r_{{\tilde \eta}_i} {%
\tilde {{\cal K}}}_{ij}) \Big] \Big) +  \nonumber \\
&+& {\frac{1}{2}} {\vec \partial}_{\sigma} \delta^3(\vec \sigma -{\tilde {%
\vec \eta}}_i) \cdot {\tilde \kappa}_i^r{\tilde \kappa}_i^s Q_i \sum_{j\not=
i}Q_j {\vec \nabla}_{{\tilde \kappa}_j} {\tilde {{\cal K}}}_{ij}\Big] -
\nonumber \\
&-&Q_i\sum_{j\not= i}^{1..N}Q_j \Big[ {\frac{1}{2}} \delta^{rs} \Big( [{\vec %
E}_{\perp Si}(\vec \sigma -{\tilde {\vec \eta}}_i, {\tilde {\vec \kappa}}_i)+%
{\frac{{\vec \partial}}{{\triangle}}} \delta^3(\vec \sigma -{\tilde {\vec %
\eta}}_i)]\cdot  \nonumber \\
&&\cdot [{\vec E}_{\perp Sj}(\vec \sigma -{\tilde {\vec \eta}}_j, {\tilde {%
\vec \kappa}}_j)+{\frac{{\vec \partial}}{{\triangle}}} \delta^3(\vec \sigma -%
{\tilde {\vec \eta}}_j)] +  \nonumber \\
&+&{\vec B}_{Si}(\vec \sigma -{\tilde {\vec \eta}}_i,{\tilde {\vec \kappa}}%
_i)\cdot {\vec B}_{Sj}(\vec \sigma -{\tilde {\vec \eta}}_j,{\tilde {\vec %
\kappa}}_j) \Big) -  \nonumber \\
&-&[E^r_{\perp Si}(\vec \sigma -{\tilde {\vec \eta}}_i, {\tilde {\vec \kappa}%
}_i)+{\frac{{\ \partial^r}}{{\triangle}}} \delta^3(\vec \sigma -{\tilde {%
\vec \eta}}_i)] [E^s_{\perp Sj}(\vec \sigma -{\tilde {\vec \eta}}_j, {%
\tilde {\vec \kappa}}_j)+{\frac{{\partial^s}}{{\triangle}}} \delta^3(\vec %
\sigma -{\tilde {\vec \eta}}_j)] -  \nonumber \\
&-&B^r_{Si}(\vec \sigma -{\tilde {\vec \eta}}_i,{\tilde {\vec \kappa}}_i)
B^s_{Sj}(\vec \sigma -{\tilde {\vec \eta}}_j,{\tilde {\vec \kappa}}_j) \Big] %
.  \label{VI48}
\end{eqnarray}

In the final canonical variables there is a dipole term [gradient of delta
function] for each particle like it happens with spinning particles\cite{lu2}%
: the role of the spin is taken by ${\frac{1}{2}} Q_i \sum_{j\not= i}^{1..N}
Q_j {\vec \partial}_{{\tilde \kappa}_j} {\tilde {{\cal K}}}_{ij}$.

Let us remark that following \cite{mate,alp} we can define Dixon's
multipolar expansion \cite{dixon} of the energy momentum tensor about an
arbitrary point on the Wigner hyperplane. There, the requirement of having
the mass dipole vanishing could be shown to be equivalent to the
identification of the point with $\vec{R}_{+}\approx \vec{q}_{+}$, namely
the internal center of mass.

For a cluster of $n$ particles inside the isolated $N-$body system, we can
now define a non-conserved energy-momentum tensor $T^{AB}_{(n)}(T_s,\vec %
\sigma )$ by collecting in the previous equation all the terms depending on
the canonical coordinates $\widetilde{\vec{\eta}_{i}},$ $\widetilde{\vec{%
\kappa}_{i}}$ of the $n$ particles of the cluster. \ This cluster
energy-momentum tensor depends also on the canonical coordinates of
the other $N-n$ particles: for the cluster these are ``external
fields''. \ If we make a Dixon multipole expansion of this cluster
energy-momentum tensor with respect to an arbitrary point on the
Wigner hyperplane and we require that the cluster mass dipole vanish,
then we can identify a M\o ller center of energy $\vec{R}_{+}^{(n)}$
for the cluster. \ This is the only collective configuration variable
which can be defined for a non-isolated cluster, which has no internal
conserved Poincar\'{e} algebra associated with it, besides the
nonconserved cluster 3-momentum ${\cal \vec{P}}%
_{(int)}^{(n)}=\sum_{i\varepsilon \{n\}}\widetilde{\vec{\kappa}_{i}}$.
Moreover, while $T^{\tau\tau}_{(n)}(T_s,\vec \sigma )$ is the
(non-conserved) energy density of the cluster, by analogy with the theory of
dissipative fluids \cite{israel} we can say that: i) $q^{\mu}(T_s,\vec \sigma
)=\epsilon^{\mu}_r(u(p_s)) T^{r\tau}_{(n)}(T_s,\vec \sigma )$ is the heat
flow; ii) ${\cal P}(T_s,\vec \sigma )={\frac{1}{3}} \sum_r T^{rr}_{(n)}(T_s,%
\vec \sigma )$ is the pressure; iii) $T^{rs}_{(n)(an)}(T_s,\vec \sigma
)=T^{rs}_{(n)}(T_s,\vec \sigma )-{\frac{1}{3}} \sum_u T^{uu}_{(n)}(T_s,\vec %
\sigma )$ is the shear (or anisotropic) stress tensor.

\vfill\eject

\section{Conclusions.}

\bigskip In this paper we analyzed how to extract the action-at-a-distance
interparticle potential hidden in the semiclassical Lienard-Wiechert
solution of the electromagnetic field equations, a subset of the solutions
of the equations of motion for the isolated system formed by $N$ scalar
charged particles plus the electromagnetic field. \ The problem is
formulated in the Wigner-covariant rest-frame instant form of dynamics,
which is defined on the Wigner hyperplanes orthognal to the total time-like
four-momentum of the isolated system and which requires the choice of the
sign of the energy of the particles (in this paper we considered only
positive energies).

This was possible due to the semiclassical approximation of using
Grassmann-valued electric charges ($Q^2_i=0$, $Q_iQ_j\not= 0$ for $i\not= j$%
) as an alternative to the extended electron models used for the
regularization of the Coulomb self energies. How this happens was shown in
Ref.\cite{lu1}, where the Coulomb potential was extracted from the
electromagnetic potential by making the canonical reduction of the
electromagnetic gauge freedom via the Shanmugadhasan canonical
transformation. This is equivalent to the use of a Wigner-covariant
radiation (or Coulomb) gauge in the rest-frame instant form.

Ref.\cite{lu4} presented the retarded Lienard-Wiechert solution for the
transverse electromagnetic field in the rest frame instant radiation gauge:
\ in this gauge, due to the transversality, the retarded Lienard-Wiechert
potential associated with each charged particle depends on the whole
past history of the other particles. At the semiclassical level a
single charged particle with Grassmann-valued electric charge does not
radiate even if it has a non-trivial Lienard-Wiechert potential,
avoiding therefore the acausal features of the Abraham-Lorentz-Dirac
equations, and has no mass renormalization. \ However, a system of $N$
charged particles produces, by virtue of the interference terms from
the various retarded Lienard-Wiechert potentials of the particles, a
radiation which reproduces the standard Larmor expression for
radiation in the wave zone, when the particles are considered as
external sources of the electromagnetic field and their equations of
motion are not used.

If instead the particles are considered dynamical, the use of their
equations of motion and of the semiclassical approximation lead to a drastic
simplification of the Lienard-Wiechert potentials and fileds. Indeed, if we
make an equal time expansion of the delay by expressing these potentials and
fields in terms of particle coordinates, velocities and accelerations of
every order, it turns out that all the accelerations decouple at the
semiclassical level due to the particle equations of motion.

Therefore, at the semiclassical level the retarded, advanced and symmetric
Lienard-Wiechert potentials and the electric and magnetic fields coincide
and depend only on the positions and velocities of the particles, so that we
can find their phase space expression in terms of particle positions and
momenta.

In this way the semiclassical Lienard-Wiechert potential and fields can be
reinterpreted as scalar and vector interparticle instantaneous
action-at-a-distance potentials. It is then possible to identify a
semiclassical reduced phase space containing only particles by eliminating
the electromagnetic field by adding by hand second class contraints which
force the transverse potential and electric field canonical variables to
coincide with the semiclassical Lienard-Wiechert ones in the absence of
incoming radiation: $\vec{A}_{\perp }(\tau ,\vec{\sigma})-\vec{A}_{\perp
LW}(\tau ,\vec{\sigma})\approx 0$, $\vec{\pi}_{\perp }(\tau ,\vec{\sigma})-%
\vec{\pi}_{\perp LW}(\tau ,\vec{\sigma})\approx 0$. Let us remark that this
could be done also in presence of an arbitrary incoming radiation ${\vec{A}}%
_{\perp (rad)}(\tau ,\vec{\sigma})$, ${\vec{\pi}}_{\perp (rad)}(\tau ,\vec{%
\sigma})=-{\frac{{\partial }}{{\partial \tau }}}{\vec{A}}_{\perp (rad)}(\tau
,\vec{\sigma})$ [it is an arbitrary solution of the homogeneous wave
equation and must not be interpreted as a pair of canonical variables] by
modifying the constraints to the form $\vec{A}_{\perp }(\tau ,\vec{\sigma})-%
\vec{A}_{\perp LW}(\tau ,\vec{\sigma})-{\vec{A}}_{\perp (rad)}(\tau ,\vec{%
\sigma})\approx 0$, $\vec{\pi}_{\perp }(\tau ,\vec{\sigma})-\vec{\pi}_{\perp
LW}(\tau ,\vec{\sigma})-{\vec{\pi}}_{\perp (rad)}(\tau ,\vec{\sigma})\approx
0$.

The reduced phase space is obtained by means of the introduction of the
Dirac brackets associated with these second class constraints. Since the old
particle positions and momenta are no longer canonical in this reduced phase
space, we had to find the new (Darboux) basis of particle canonical
variables. The generators of the ``internal'' Poincaire` group inside the
Wigner hyperplanes in the rest-frame instant form of dynamics can be
reexpressed in terms of these new variables: the 3-momentum ${\cal \vec{P}}%
_{(int)}$ and the angular momentum ${\cal \vec{J}}_{(int)}$ \ become equal
to those for $N$ free scalar particles (as expected in an instant form). \
The interaction dependent boosts ${\vec{{\cal K}}}_{(int)}$ are proportional
to the ``internal'' canonical center of mass $\vec{q}_{+}$ inside the Wigner
hyperplane: $\vec{q}_{+}\approx 0$ are the gauge-fixings to be be added to
the rest-frame conditions ${\cal \vec{P}}_{(int)}\approx 0$, if one wishes
to re-express the dynamics only in terms of particle ``internal'' relative
variables. Also the energy-momentum tensor has been evalulated in the new
canonical variables and there is a suggestion on how to find the M\o ller
center of energy of a cluster of $n$ particles contained in the N particle
isolated system.

The Hamiltonian in the rest frame frame instant form, generating the
evolution in the rest-frame time of the decoupled ``external'' canonical
center of mass, is the ``internal'' energy generator $M={\cal P}%
_{(int)}^{\tau }$ (the invariant mass of the isolated $N$ particle system).
\ The semiclassical Lienard-Wiechert solution implies the existence of
interparticle action-at-a-distance potentials of two types: vector
potentials minimally coupled to the Wigner spin 1 particle three-momentum
under the square root associated with the kinetic energies; ii) a scalar
potential (including the Coulomb potential) outside the square roots. \ In
the semiclassical approximation all these potentials can be replaced by a
unique scalar potential, which is the sum of the Coulomb potential and of a
generalized Darwin one for arbitrary $N$. It is the (semiclassical) static
and non-static complete potential corresponding to the \ one photon exchange
tree Feynman diagrams of scalar electrodynamics and is a completely new
result. The expression we find contains no $N$-body forces, being simply a
sum of two particle interactions. \ This is a consequence of our use of
Grassmann charges.

In the $N=2$ case we obtain a closed form of the solution by evaluating it
in the rest frame after the gauge fixing ${\vec q}_{+}\approx 0$: the lowest
order in $1/c^{2}$ contribution of the generalized Darwin potential agrees
with the expression of the standard Darwin potential. We then show that in a
\ semiclassical sense a special solution of the Hamilton equations is the
Schild solution \cite{schild} in which the two particles move in concentric
circular orbits. We evaluate the frequency for equal masses

Future work will proceed along three parallel courses.

The first is the extention of this work to include semiclassical spinning
particles. \ This will not only build upon the work here but also that of
Ref.\cite{lu4}. Of particular concern there will be the issue of reproducing
the correct spin-orbit and Darwin terms of the appropriate order (the
two-body extentions of such terms in the one-body Dirac equation).

The second line is that of the quantization of our general Hamiltonian. Let
us remark that quantization of the closed form Hamiltonian for $N=2$ in
configuration space would involve not only the usual nonlocal operators for
the kinetic energy but also non-local operators for the Darwin portions of
the potential. For free scalar particles the positive energy wave equation $%
i \partial_{\tau} \phi (\tau ,\vec \sigma )=\sqrt{m^2+\triangle} \phi (\tau ,%
\vec \sigma )$ has been studied in Ref.\cite{lamme} by using
pseudo-differential operators. Instead see Refs.\cite{lucha} for the
difficulties in quantizing the equal mass two-body problem ($H=2\sqrt{m^2+{%
\vec \kappa}^2} +{\frac{{\alpha}}{{\eta}}}$) with only the Coulomb potential
outside the square root. Note that the so-called spinless Salpeter equation
would correspond to the quantization of our Hamiltonian with just the
Coulomb interaction and at most the lowest order Darwin interaction. For the
quantization when there are only scalar potentials inside the square root ($%
\sqrt{m^2+V(|\vec \eta |)+{\vec \kappa}^2}$) see Ref.\cite{ll}.

In our semiclassical approximation we have two options for the
action-at-a-distance potentials: i) the vector ones ${\vec{V}}_{i}$ under
the square roots and a scalar one $U$ outside; ii) a unique effective scalar
potential sum of the Coulomb and Darwin ($V_{DAR}=V_{LOD}+V_{HOD}$) ones
outside the square roots. It can be expected that the results of the
quantization of the two options would produce inequivalent theories.

The third line is the development of the quantization of scalar
electrodynamics on the Wigner hyperplanes in the rest-frame instant
form. This would be a special Wigner-covariant instance of
Tomonaga-Schwinger quantum field theory with a well defined covariant
concept of ``equal times''. To introduce a particle concept in such a
quantum formulation, one will have to define Tomonaga-Schwinger
asymptotic states and a reduction formalism. The natural candidates
for the $N$-particle wave functions in such asymptotic states in the
case of the Klein-Gordon field would be the wave functions
corresponding to the quantization of $N$ positive energy scalar
particles on the Wigner hyperplanes in the rest-frame instant form.

Moreover, in the rest-frame quantum field theory there will be new
covariant ``equal time'' Green functions. This should allow the
definition of a relativistic Schr\"{o}dinger equation for bound states
(replacing the Bethe-Salpeter equation for $N$=2 and avoiding by
construction its problems with the spurious solutions, which are a
byproduct of the use of asymptotic Fock states, since in a tensor
product one cannot eliminate the possibility that an ``in'' particle
be in the absolute future of another one). In the rest-frame quantum
field theory it should also be possible to include bound states among
the asymptotic Tomonaga-Schwinger states: they should be described by
the quantization of isolated $N$ particle systems like the one studied
in this paper.

ACKNOWLEDGMENT: L.Lusanna wishes to thank Prof. K.Kuchar for helpful
discussions about the need of compatibility with the equations of motion of
the second class constraints to be added by hand to select a symplectic
subspace of solutions in the space of solutions of a given isolated system.

\vfill\eject

\newpage \appendix

\section{Exact Summation of the Transverse Vector Potential Series}

\setcounter{section}{1}

\bigskip

In this Appendix we perform explicitly the summation of the vector potential
below [see \ Eq.(\ref{V27}\} with ${\dot {\vec \eta}}_i={\vec \beta}_i$
replaced by ${\vec \kappa}_i/\sqrt{m_i^2+{\vec \kappa}_i^2}$]

\begin{eqnarray}
\vec{A}_{\perp }(\tau ,\vec{\sigma}) &=&\sum_{i=1}^{N}{\frac{Q_{i}}{4\pi }}%
\sum_{m=0}^{\infty }\Big[{\frac{1}{(2m)!}\frac{\vec{\kappa}_{i}}{\sqrt{%
m_{i}^{2}+{\vec{\kappa}_{i}}^{2}}}}({\frac{\vec{\kappa}_{i}}{\sqrt{m_{i}^{2}+%
{\vec{\kappa}_{i}}^{2}}}}\cdot {\vec{\nabla}}_{\sigma })^{2m})\,|\vec{\sigma}%
-\vec{\eta}_{i}|^{2m-1}-  \nonumber \\
&&-{\frac{1}{(2m+2)!}}{\vec{\nabla}}_{\sigma }({\frac{\vec{\kappa}_{i}}{%
\sqrt{m_{i}^{2}+{\vec{\kappa}_{i}}^{2}}}}\cdot {\vec{\nabla}}_{\sigma
})^{2m+1}\,|\vec{\sigma}-\vec{\eta}_{i}|^{2m+1}\Big]:=  \nonumber \\
&:&=\vec{A}_{\perp 1}(\vec{\sigma},\tau )+\vec{A}_{\perp 2}(\vec{\sigma}%
,\tau ).  \label{a1}
\end{eqnarray}
Using the result of Appendix C that $(\vec{\kappa}_{i}\cdot {\vec{\nabla}}%
_{\eta })^{2m}|\vec{\eta}|^{2m-1}=[(2m-1)!!]^{2}{\frac{1}{|\vec{\eta}|}}[%
\vec{\kappa}_{i}^{2}-(\vec{\kappa}_{i}\cdot {\frac{{\vec{\eta}}}{{|\vec{\eta}%
|}}})^{2}]^{m}$, we get [${\vec \nabla}_{\eta}=\partial /\partial \vec \eta$%
]
\begin{equation}
\vec{A}_{\perp 1}(\tau ,\vec{\sigma})=\sum_{i=1}^{N}{\frac{Q_{i}}{4\pi }}%
\sum_{m=0}^{\infty }\Big[{\frac{[(2m-1)!!]^{2}}{(2m)!|\vec{\sigma}-\vec{\eta}%
_{i}|}\frac{\vec{\kappa}_{i}}{(m_{i}^{2}+{\vec{\kappa}_{i}}^{2})^{m+1/2}}}(%
\vec{\kappa}_{i}^{2}-(\vec{\kappa}_{i}\cdot \frac{\vec{\sigma}-\vec{\eta}_{i}%
}{|\vec{\sigma}-\vec{\eta}_{i}|})^{2})^{m}\Big].  \label{a2}
\end{equation}
By using
\begin{eqnarray}
\frac{\lbrack (2m-1)!!]^{2}}{(2m)!} &=&\frac{(2m)!}{(m!)^{2}2^{2m}}=\frac{%
(m-1/2)!}{\sqrt{\pi }m!}=\frac{\sqrt{\pi }(-)^{m-1}(m-1/2)}{(1/2-m)!m!}=
\nonumber \\
&=&\frac{\sqrt{\pi }(-)^{m}}{(-1/2-m)!m!}=(-)^{m}%
%TCIMACRO{\binom{-1/2}{m} }%
%BeginExpansion
{-1/2 \choose m}%
%EndExpansion
\label{a3}
\end{eqnarray}
we find that
\begin{eqnarray}
\vec{A}_{\perp 1}(\tau ,\vec{\sigma}) &=&\sum_{i=1}^{N}{\frac{Q_{i}}{4\pi }%
\Big[\frac{\vec{\kappa}_{i}}{\sqrt{m_{i}^{2}+{\vec{\kappa}_{i}}^{2}}\,|\vec{%
\sigma}-\vec{\eta}_{i}|}}\sum_{m=0}^{\infty }(-)^{m}%
%TCIMACRO{\binom{-1/2}{m}}%
%BeginExpansion
{-1/2 \choose m}%
%EndExpansion
\left( \frac{\vec{\kappa}_{i}^{2}-(\vec{\kappa}_{i}\cdot \frac{\vec{\sigma}-%
\vec{\eta}_{i}}{|\vec{\sigma}-\vec{\eta}_{i}|})^{2}}{m_{i}^{2}+{\vec{\kappa}%
_{i}}^{2}}\right) ^{m}]=  \nonumber \\
&=&\sum_{i=1}^{N}{\frac{Q_{i}}{4\pi }\frac{\vec{\kappa}_{i}}{|\vec{\sigma}-%
\vec{\eta}_{i}|}}\frac{1}{\sqrt{m_{i}^{2}+(\vec{\kappa}_{i}\cdot \frac{\vec{%
\sigma}-\vec{\eta}_{i}}{|\vec{\sigma}-\vec{\eta}_{i}|})^{2}}}.  \label{a4}
\end{eqnarray}

For $\vec{A}_{\perp 2}(\tau ,\vec{\sigma})$ we need an expression for $(\vec{%
\kappa}_{i}\cdot {\vec \nabla}_{\eta} )^{2m+1} |\vec \eta |^{2m+1}$. One can
show by an induction procedure that
\begin{equation}
({\vec{\kappa}}_{i}\cdot {\vec \nabla}_{\sigma })^{2m+1}|\vec{\sigma}-\vec{%
\eta} _{i}|^{2m+1}=[(2m+1)!!]^{2}\sum_{l=0}^{m}(-)^{l}
%TCIMACRO{\binom{m}{l}}%
%BeginExpansion
{m \choose l}%
%EndExpansion
(\vec{\kappa}^{2})^{m-l}\frac{(\vec{\kappa}_{i}\cdot \frac{\vec{\sigma}-
\vec{\eta}_{i}}{|\vec{\sigma}-\vec{\eta}_{i}|})^{2l+1}}{2l+1},  \label{a5}
\end{equation}
and hence
\begin{eqnarray}
\vec{A}_{\perp 2}(\tau ,\vec{\sigma})&=&-\sum_{i=1}^{N}{\frac{Q_{i}}{4\pi }}
\frac{{\vec \nabla}_{\sigma }}{\sqrt{m_i^2+{\vec{\kappa}_{i}}^{2}}}
\sum_{m=0}^{\infty }\frac{[(2m+1)!!]^{2}}{(2m+2)!}\left( \frac{\vec{\kappa}
_{i}^{2}}{m_i^2+{\vec{\kappa}_{i}}^{2}}\right) ^{m}\times  \nonumber \\
&&\sum_{l=0}^{m}(-)^{l}
%TCIMACRO{\binom{m}{l}}%
%BeginExpansion
{m \choose l}%
%EndExpansion
(\vec{\kappa}_{i}^{2})^{-l}\frac{(\vec{\kappa}_{i}\cdot \frac{\vec{\sigma}-
\vec{\eta}_{i}}{|\vec{\sigma}-\vec{\eta}_{i}|})^{2l+1}}{2l+1}.  \label{a6}
\end{eqnarray}
Now
\begin{equation}
\frac{\lbrack (2m+1)!!]^{2}}{(2m+2)!}=(-)^{m+1}
%TCIMACRO{\binom{-1/2}{m+1} }%
%BeginExpansion
{-1/2 \choose m+1}%
%EndExpansion
\label{a7}
\end{equation}
so that
\begin{eqnarray}
\vec{A}_{\perp 2}(\tau ,\vec{\sigma}) &=&-\sum_{i=1}^{N}{\frac{Q_{i}}{4\pi }}
\frac{{\vec \nabla}_{\sigma }}{\sqrt{m_i^2+{\vec{\kappa}_{i}}^{2}}}
\sum_{m=0}^{\infty }(-)^{m+1}
%TCIMACRO{\binom{-1/2}{m+1}}%
%BeginExpansion
{-1/2 \choose m+1}%
%EndExpansion
\left( \frac{\vec{\kappa}_{i}^{2}}{m_i^2+{\vec{\kappa}_{i}}^{2}}\right)
^{m}\times  \nonumber \\
&&\int_{0}^{(\vec{\kappa}_{i}\cdot \frac{(\vec{\sigma}-\vec{\eta}_{i})}{|
\vec{\sigma}-\vec{\eta}_{i}|})}\sum_{l=0}^{m}(-)^{l}
%TCIMACRO{\binom{m}{l}}%
%BeginExpansion
{m \choose l}%
%EndExpansion
\left( \frac{w^{2}}{\vec{\kappa}_{i}^{2}}\right) ^{l}dw =  \nonumber \\
&=&-\sum_{i=1}^{N}{\frac{Q_{i}}{4\pi }}\frac{{\vec \nabla}_{\sigma }}{\sqrt{%
m_i^2+{\vec{\kappa}_{i}}^{2}}}\sum_{m=0}^{\infty }(-)^{m+1}
%TCIMACRO{\binom{-1/2}{m+1}}%
%BeginExpansion
{-1/2 \choose m+1}%
%EndExpansion
\left( \frac{\vec{\kappa}_{i}^{2}}{m_i^2+{\vec{\kappa}_{i}}^{2}}\right)
^{m}\times  \nonumber \\
&&\int_{0}^{(\vec{\kappa}_{i}\cdot \frac{(\vec{\sigma}-\vec{\eta}_{i})}{|
\vec{\sigma}-\vec{\eta}_{i}|})}\left( 1-\frac{w^{2}}{\vec{\kappa}_{i}^{2}}
\right) ^{m}dw =  \nonumber \\
&=&-\sum_{i=1}^{N}{\frac{Q_{i}}{4\pi }}\frac{{\vec \nabla}_{\sigma }}{\sqrt{%
m_i^2+{\vec{\kappa}_{i}}^{2}}}\int_{0}^{(\vec{\kappa}_{i}\cdot \frac{(\vec{%
\sigma}-\vec{\eta}_{i})}{|\vec{\sigma}-\vec{\eta}_{i}|})} \left( \frac{\vec{%
\kappa}_{i}^{2}-w^{2}}{m_i^2+{\vec{\kappa}_{i}}^{2}}\right) ^{-1}\times
\nonumber \\
&&\sum_{m=0}^{\infty }(-)^{m+1}
%TCIMACRO{\binom{-1/2}{m+1}}%
%BeginExpansion
{-1/2 \choose m+1}%
%EndExpansion
\left( \frac{\vec{\kappa}_{i}^{2}-w^{2}}{m_i^2+{\vec{\kappa}_{i}}^{2}}
\right) ^{m+1}dw =  \nonumber \\
&=&-\sum_{i=1}^{N}{\frac{Q_{i}}{4\pi }}{\vec \nabla}_{\sigma }\int_{0}^{(%
\vec{\kappa}_{i}\cdot \frac{(\vec{\sigma}-\vec{\eta}_{i})}{|\vec{\sigma}-%
\vec{\eta }_{i}|})}\left( \frac{\vec{\kappa}_{i}^{2}-w^{2}}{m_i^2+{\vec{%
\kappa}_{i}} ^{2}}\right) ^{-1}\Big[ \frac{1}{\sqrt{m_i^2+w^{2}}}-\frac{1}{%
\sqrt{ m_i^2+{\vec{\kappa}_{i}}^{2}}}\Big] dw.  \label{a8}
\end{eqnarray}
Now
\begin{equation}
{\vec \nabla}_{\sigma }\Big( \vec{\kappa}_{i}\cdot \frac{\vec{\sigma}-\vec{%
\eta}_{i}}{ |\vec{\sigma}-\vec{\eta}_{i}|}\Big) =\frac{\vec{\kappa}_{i}}{|%
\vec{\sigma}-\vec{ \eta}_{i}|}\cdot \Big( {\bf I}-\frac{\vec{\sigma}-\vec{%
\eta}_{i}}{|\vec{\sigma}- \vec{\eta}_{i}|}\frac{\vec{\sigma}-\vec{\eta}_{i}}{%
|\vec{\sigma}-\vec{\eta} _{i}|}\Big) .  \label{a9}
\end{equation}
So we obtain
\begin{eqnarray}
\vec{A}_{\perp 2}(\tau ,\vec{\sigma}) &=&-\sum_{i=1}^{N}{\frac{Q_{i}}{4\pi }}
\frac{\vec{\kappa}_{i}}{|\vec{\sigma}-\vec{\eta}_{i}|}\cdot \Big( {\bf I}-%
\frac{\vec{\sigma}-\vec{\eta}_{i}}{|\vec{\sigma}-\vec{\eta}_{i}|}\frac{\vec{
\sigma}-\vec{\eta}_{i}}{|\vec{\sigma}-\vec{\eta}_{i}|}\Big)\times  \nonumber
\\
&&({\frac{\sqrt{m_i^2+{\vec{\kappa}_{i}}^{2}}}{\sqrt{m_{i}^{2}+(\vec{ \kappa}%
_{i}\cdot {\frac{{\vec{\sigma}-\vec{\eta}_{i}}}{{|\vec{\sigma}-\vec{\eta}
_{i}|}}})^{2}}}}-1){\frac{\sqrt{m_i^2+{\vec{\kappa}_{i}}^{2}}}{\vec{\kappa}
_{i}^{2}-(\vec{\kappa}_{i}\cdot {\frac{{\vec{\sigma}-\vec{\eta}_{i}}}{{|\vec{%
\sigma}- \vec{\eta}_{i}|}}})^{2}}} ,  \label{a10}
\end{eqnarray}
which when combined with the expression for $\vec{A}_{\perp 1}(\tau ,\vec{%
\sigma})$ yields the results given by Eq.(\ref{V28}).

\vfill\eject

\section{ Computation of Field Energy and Momentum Integrals}

Here we carry out the details in the computation of the field energy and
momentum for the case $N=2$. \ The general $N$ results obtained in the text
are an immediate generalization. From Eq.(\ref{VI2}) and Eq.(\ref{VI3}) we
find that

\begin{eqnarray}
{\vec{E}}^{2}_{\perp S}(\tau ,\sigma)&=&{\frac{Q_{1}Q_{2}}{8\pi ^{2}}}
\sum_{m=0}^{\infty }\sum_{n=0}^{\infty }{\frac{1}{(2n)!(2m)!}}\, \dot{\vec{%
\eta} }_{1}\cdot \dot{\vec{\eta}}_{2} \times  \nonumber \\
&&[(\dot{\vec{\eta}}_{1}\cdot \vec{\nabla} _{\sigma })^{2m+1}|\vec{\sigma}-%
\vec{\eta}_{1}|^{2m-1}][(\dot{\vec{\eta}} _{2}\cdot \vec{\nabla}_{\sigma
})^{2n+1}|\vec{\sigma}-\vec{\eta} _{2}|^{2n-1}]-  \nonumber \\
&&-{\frac{Q_{1}Q_{2}}{8\pi ^{2}}}\sum_{m=0}^{\infty }\sum_{n=0}^{\infty }{\
\frac{1}{(2n+2)!(2m)!}}\times  \nonumber \\
&&[(\dot{\vec{\eta}}_{1}\cdot \vec{\nabla}_{\sigma })^{2m+1}|\vec{\sigma}-%
\vec{\eta}_{1}|^{2m-1}][(\dot{\vec{\eta}}_{1}\cdot \vec{\nabla}_{\sigma })(%
\dot{\vec{\eta}}_{2}\cdot \vec{\nabla}_{\sigma })^{2n+2}|\vec{\sigma}-\vec{%
\eta}_{2}|^{2n+1}]-  \nonumber \\
&&-{\frac{Q_{1}Q_{2}}{8\pi ^{2}}}\sum_{m=0}^{\infty }\sum_{n=0}^{\infty }{\
\frac{1}{(2m+2)!(2n)!}}\times  \nonumber \\
&&[(\dot{\vec{\eta}}_{2}\cdot \vec{\nabla}_{\sigma })^{2n+1})|\vec{\sigma}-%
\vec{\eta}_{2}|^{2n-1}][(\dot{\vec{\eta}}_{2}\cdot \vec{\nabla}_{\sigma })(%
\dot{\vec{\eta}}_{1}\cdot \vec{\nabla}_{\sigma })^{2m+2})|\vec{\sigma}-\vec{%
\eta}_{1}|^{2m+1}]+  \nonumber \\
&&+{\frac{Q_{1}Q_{2}}{8\pi ^{2}}}\sum_{m=0}^{\infty }\sum_{n=0}^{\infty }{%
\frac{1}{(2m+2)!(2n+2)!}}\times  \nonumber \\
&&[\vec{\nabla}_{\sigma }(\dot{\vec{\eta}}_{2}\cdot \vec{\nabla}_{\sigma
})^{2n+2})|\vec{\sigma}-\vec{\eta}_{2}|^{2n+1}]\cdot [ \vec{\nabla}_{\sigma
}(\dot{\vec{\eta}}_{1}\cdot \vec{\nabla}_{\sigma })^{2m+2})|\vec{\sigma}-%
\vec{\eta}_{1}|^{2m+1}] ,  \label{b1}
\end{eqnarray}
\begin{eqnarray}
{\vec{B}}^{2}_S(\tau ,\vec \sigma )&=&{\frac{Q_{1}Q_{2}}{8\pi ^{2}}}
\sum_{m=0}^{\infty }\sum_{n=0}^{\infty }{\frac{1}{(2n)!(2m)!}}\dot{\vec{\eta}
}_{1}\cdot \dot{\vec{\eta}}_{2}  \nonumber \\
&&[(\vec{\nabla}_{\sigma}(\dot{\vec{\eta}} _{1}\cdot \vec{\nabla}_{\sigma
})^{2m})|\vec{\sigma}-\vec{\eta} _{1}|^{2m-1}]\cdot [(\vec{\nabla}_{\sigma }(%
\dot{\vec{\eta}}_{2}\cdot \vec{ \nabla}_{\sigma })^{2n})|\vec{\sigma}-\vec{%
\eta}_{2}|^{2n-1}]-  \nonumber \\
&&-{\frac{Q_{1}Q_{2}}{8\pi ^{2}}}\sum_{m=0}^{\infty }\sum_{n=0}^{\infty }{%
\frac{1}{(2n)!(2m)!}}  \nonumber \\
&&[(\dot{\vec{\eta}}_{2}\cdot \vec{\nabla}_{\sigma })( \dot{\vec{\eta}}%
_{1}\cdot \vec{\nabla}_{\sigma })^{2m})|\vec{\sigma}-\vec{ \eta}%
_{1}|^{2m-1}]\cdot [(\dot{\vec{\eta}}_{1}\cdot \vec{\nabla}_{\sigma })( \dot{%
\vec{\eta}}_{2}\cdot \vec{\nabla}_{\sigma })^{2n})|\vec{\sigma}-\vec{ \eta}%
_{2}|^{2n-1}] ,  \label{b2}
\end{eqnarray}

\begin{eqnarray}
&&(\vec{E}_{\perp S}(\tau ,\vec{\sigma})\times \vec{B}_{S}(\tau ,\vec{\sigma}%
))_{k}=  \nonumber \\
&=&{\frac{Q_{1}Q_{2}}{16\pi ^{2}}}\sum_{m=0}^{\infty }\sum_{n=0}^{\infty }{%
\frac{1}{(2n)!(2m)!}}\,\dot{\vec{\eta}}_{1}\cdot \dot{\vec{\eta}}_{2}\times
\nonumber \\
&&[(\dot{\vec{\eta}}_{1}\cdot {\vec{\nabla}}_{\sigma })^{2m+1}|\vec{\sigma}-%
\vec{\eta}_{1}|^{2m-1}][{\vec{\nabla}}_{\sigma }(\dot{\vec{\eta}}_{2}\cdot
\nabla _{\sigma })^{2n}|\vec{\sigma}-\vec{\eta}_{2}|^{2n-1}]-  \nonumber \\
&&-{\frac{Q_{1}Q_{2}}{16\pi ^{2}}}\sum_{m=0}^{\infty }\sum_{n=0}^{\infty }{%
\frac{1}{(2n)!(2m)!}}\,\dot{\vec{\eta}}_{2}\times  \nonumber \\
&&[(\dot{\vec{\eta}}_{1}\cdot {\vec{\nabla}}_{\sigma })(\dot{\vec{\eta}}%
_{2}\cdot {\vec{\nabla}}_{\sigma })^{2n}|\vec{\sigma}-\vec{\eta}%
_{2}|^{2n-1}][(\dot{\vec{\eta}}_{1}\cdot {\vec{\nabla}}_{\sigma })^{2m+1}|%
\vec{\sigma}-\vec{\eta}_{1}|^{2m-1}]-  \nonumber \\
&&-{\frac{Q_{1}Q_{2}}{16\pi ^{2}}}\sum_{m=0}^{\infty }\sum_{n=0}^{\infty }{%
\frac{1}{(2n)!(2m+2)!}}\times  \nonumber \\
&&{\vec{\nabla}}_{\sigma }(\dot{\vec{\eta}}_{2}\cdot {\vec{\nabla}}_{\sigma
})^{2n}|\vec{\sigma}-\vec{\eta}_{2}|^{2n-1}][(\dot{\vec{\eta}}_{2}\cdot {%
\vec{\nabla}}_{\sigma })(\dot{\vec{\eta}}_{1}\cdot {\vec{\nabla}}_{\sigma
})^{2m+2}|\vec{\sigma}-\vec{\eta}_{1}|^{2m+1}]+  \nonumber \\
&&+{\frac{Q_{1}Q_{2}}{16\pi ^{2}}}\sum_{m=0}^{\infty }\sum_{n=0}^{\infty }{%
\frac{1}{(2n)!(2m+2)!}}\,\dot{\vec{\eta}}_{2}\times  \nonumber \\
&&[{\vec{\nabla}}_{\sigma }(\dot{\vec{\eta}}_{1}\cdot {\vec{\nabla}}_{\sigma
})^{2m+2}|\vec{\sigma}-\vec{\eta}_{1}|^{2m+1}]\cdot \lbrack {\vec{\nabla}}%
_{\sigma }(\dot{\vec{\eta}}_{2}\cdot {\vec{\nabla}}_{\sigma })^{2n}|\vec{%
\sigma}-\vec{\eta}_{2}|^{2n-1}+  \nonumber \\
&&+(1\longleftrightarrow 2).  \label{b3}
\end{eqnarray}
Our aim here is to compute
\begin{eqnarray}
{{\frac{1}{2}}}\int d^{3}\sigma (\vec{E}_{\perp S}^{2}+\vec{B}_{S}^{2})(\tau
,\vec{\sigma}) &:&=\frac{Q_{1}Q_{2}}{16\pi ^{2}}\sum_{m=0}^{\infty
}\sum_{n=0}^{\infty }{\frac{1}{(2n)!(2m)!}}\Big[(\dot{\vec{\eta}}_{1}\cdot
\dot{\vec{\eta}}_{2})I_{1mn}-  \nonumber \\
&&-\frac{1}{(2n+1)(2n+2)}I_{2mn}-\frac{1}{(2m+1)(2m+2)}I_{3mn}+  \nonumber \\
&&+\frac{1}{(2n+1)(2n+2)(2m+1)(2m+2)}I_{4mn}+\dot{\vec{\eta}}_{1}\cdot \dot{%
\vec{\eta}}_{2}I_{5mn}-I_{6mn}\Big],  \nonumber \\
&&{}  \label{b4}
\end{eqnarray}
\begin{eqnarray}
\ \int d^{3}\sigma (\vec{E}_{\perp S}\times \vec{B}_{S})(\tau ,\vec{\sigma})
&:&={\frac{Q_{1}Q_{2}}{16\pi ^{2}}}\sum_{m=0}^{\infty }\sum_{n=0}^{\infty }{%
\frac{1}{(2n)!(2m)!}}\Big[(\dot{\vec{\eta}}_{1}\cdot \dot{\vec{\eta}}_{2}%
\vec{I}_{7mn}-\dot{\vec{\eta}}_{2}I_{8mn}-  \nonumber \\
&&-\frac{1}{(2m+1)(2m+2)}\vec{I}_{9mn}+\frac{\dot{\vec{\eta}}_{2}}{%
(2m+1)(2m+2)}I_{10mn}\Big]+  \nonumber \\
&&+(1\rightarrow 2),  \label{b5}
\end{eqnarray}
which involve ten different integrals defined by
\begin{eqnarray}
I_{1mn} &=&\int d^{3}\sigma \Big[(\dot{\vec{\eta}}_{1}\cdot {\vec{\nabla}}%
_{\sigma })^{2m+1}|\vec{\sigma}-\vec{\eta}_{1}|^{2m-1}\Big]\Big[(\dot{\vec{%
\eta}}_{2}\cdot {\vec{\nabla}}_{\sigma })^{2n+1}|\vec{\sigma}-\vec{\eta}%
_{2}|^{2n-1}\Big],  \nonumber \\
I_{2mn} &=&\int d^{3}\sigma \Big[(\dot{\vec{\eta}}_{1}\cdot {\vec{\nabla}}%
_{\sigma })^{2m+1}|\vec{\sigma}-\vec{\eta}_{1}|^{2m-1}\Big]\Big[(\dot{\vec{%
\eta}}_{1}\cdot {\vec{\nabla}}_{\sigma })(\dot{\vec{\eta}}_{2}\cdot {\vec{%
\nabla}}_{\sigma })^{2n+2}|\vec{\sigma}-\vec{\eta}_{2}|^{2n+1}\Big],
\nonumber \\
I_{3mn} &=&\int d^{3}\sigma \Big[(\dot{\vec{\eta}}_{2}\cdot {\vec{\nabla}}%
_{\sigma })^{2n+1}|\vec{\sigma}-\vec{\eta}_{2}|^{2n-1}\Big]\Big[(\dot{\vec{%
\eta}}_{2}\cdot {\vec{\nabla}}_{\sigma })(\dot{\vec{\eta}}_{1}\cdot {\vec{%
\nabla}}_{\sigma })^{2m+2}|\vec{\sigma}-\vec{\eta}_{1}|^{2m+1}\Big],
\nonumber \\
I_{4mn} &=&\int d^{3}(\nabla _{\sigma }(\dot{\vec{\eta}}_{1}\cdot {\vec{%
\nabla}}_{\sigma })^{2m+2}|\vec{\sigma}-\vec{\eta}_{1}|^{2m+1}\Big]\cdot %
\Big[{\vec{\nabla}}_{\sigma }(\dot{\vec{\eta}}_{2}\cdot {\vec{\nabla}}%
_{\sigma })^{2n+2}|\vec{\sigma}-\vec{\eta}_{2}|^{2n+1}\Big],  \nonumber \\
I_{5mn} &=&\int d^{3}\Big[{\vec{\nabla}}_{\sigma }(\dot{\vec{\eta}}_{1}\cdot
{\vec{\nabla}}_{\sigma })^{2m}|\vec{\sigma}-\vec{\eta}_{1}|^{2m-1}\Big]\cdot %
\Big[{\vec{\nabla}}_{\sigma }(\dot{\vec{\eta}}_{2}\cdot {\vec{\nabla}}%
_{\sigma })^{2n}|\vec{\sigma}-\vec{\eta}_{2}|^{2n-1}\Big],  \nonumber \\
I_{6mn} &=&\int d^{3}\sigma \Big[(\dot{\vec{\eta}}_{2}\cdot {\vec{\nabla}}%
_{\sigma })(\dot{\vec{\eta}}_{1}\cdot {\vec{\nabla}}_{\sigma })^{2m}|\vec{%
\sigma}-\vec{\eta}_{1}|^{2m-1}\Big]\Big[(\dot{\vec{\eta}}_{1}\cdot {\vec{%
\nabla}}_{\sigma })(\dot{\vec{\eta}}_{2}\cdot {\vec{\nabla}}_{\sigma })^{2n}|%
\vec{\sigma}-\vec{\eta}_{2}|^{2n-1}\Big],  \nonumber \\
\vec{I}_{7mn} &=&\int d^{3}\sigma \Big[(\dot{\vec{\eta}}_{1}\cdot {\vec{%
\nabla}}_{\sigma })^{2m+1}|\vec{\sigma}-\vec{\eta}_{1}|^{2m-1}\Big]\Big[{%
\vec{\nabla}}_{\sigma }(\dot{\vec{\eta}}_{2}\cdot {\vec{\nabla}}_{\sigma
})^{2n}|\vec{\sigma}-\vec{\eta}_{2}|^{2n-1}\Big],  \nonumber \\
I_{8mn} &=&\int d^{3}\sigma \Big[(\dot{\vec{\eta}}_{1}\cdot {\vec{\nabla}}%
_{\sigma })(\dot{\vec{\eta}}_{2}\cdot {\vec{\nabla}}_{\sigma })^{2n}|\vec{%
\sigma}-\vec{\eta}_{2}|^{2n-1}\Big]\Big[(\dot{\vec{\eta}}_{1}\cdot {\vec{%
\nabla}}_{\sigma })^{2m+1}|\vec{\sigma}-\vec{\eta}_{1}|^{2m-1}\Big],
\nonumber \\
\vec{I}_{9mn} &=&\int d^{3}\sigma \Big[{\vec{\nabla}}_{\sigma }(\dot{\vec{%
\eta}}_{2}\cdot {\vec{\nabla}}_{\sigma })^{2n}|\vec{\sigma}-\vec{\eta}%
_{2}|^{2n-1}\Big]\Big[(\dot{\vec{\eta}}_{2}\cdot {\vec{\nabla}}_{\sigma })(%
\dot{\vec{\eta}}_{1}\cdot {\vec{\nabla}}_{\sigma })^{2m+2}|\vec{\sigma}-\vec{%
\eta}_{1}|^{2m+1}\Big],  \nonumber \\
I_{10mn} &=&\int d^{3}\sigma \Big[{\vec{\nabla}}_{\sigma }(\dot{\vec{\eta}}%
_{1}\cdot {\vec{\nabla}}_{\sigma })^{2m+2}|\vec{\sigma}-\vec{\eta}%
_{1}|^{2m+1}\Big]\cdot \Big[{\vec{\nabla}}_{\sigma }(\dot{\vec{\eta}}%
_{2}\cdot {\vec{\nabla}}_{\sigma })^{2n}|\vec{\sigma}-\vec{\eta}_{2}|^{2n-1}%
\Big].  \label{b6}
\end{eqnarray}
The powers of $\sigma $ in each of the ten integrands is $\sigma
^{2m-1-2m-1}\sigma ^{2n-1-2n-1}\sim \sigma ^{-4}$ Therefore the integrals
converge. Thus
\begin{equation}
I_{i}={%
%TCIMACRO{\QATOP{lim}{\Lambda \rightarrow \infty }}%
%BeginExpansion
{lim \atop \Lambda \rightarrow \infty }%
%EndExpansion
}\int_{0}^{\Lambda }\sigma ^{2}d\sigma \int d\hat{\Omega}_{\sigma }(\ \ ):={%
%TCIMACRO{\QATOP{lim}{\Lambda \rightarrow \infty }}%
%BeginExpansion
{lim \atop \Lambda \rightarrow \infty }%
%EndExpansion
}\int_{\Lambda }d^{3}\sigma ()={%
%TCIMACRO{\QATOP{lim}{\Lambda \rightarrow \infty }}%
%BeginExpansion
{lim \atop \Lambda \rightarrow \infty }%
%EndExpansion
}I_{i}(\Lambda ).  \label{b7}
\end{equation}
Since $I_{i}(\Lambda )$ is finite for finite $\Lambda $, we can bring out
the derivatives. Now perform integrations by parts, change to $\partial
/\partial \eta $ from $\partial /\partial \sigma $, translate, and then use
the fact that the integrals are finite so that they can be replaced by $%
\int_{\Lambda }$. This gives ($\vec{\eta}:={\vec{\eta}}_{12}=\vec{\eta}_{1}-%
\vec{\eta}_{2}$)

\begin{eqnarray}
I_{1mn} &=&-(\dot{\vec{\eta}}_{1}\cdot {\vec{\nabla}}_{\eta })^{2m+1}(\dot{%
\vec{\eta}}_{2}\cdot \nabla _{\eta })^{2n+1}\int_{\Lambda }d^{3}\sigma |\vec{%
\sigma}-(\vec{\eta}_{1}-\vec{\eta}_{2})|^{2m-1}\sigma ^{2n-1},  \nonumber \\
I_{2mn} &=&-(\dot{\vec{\eta}}_{1}\cdot {\vec{\nabla}}_{\eta })^{2m+2}(\dot{%
\vec{\eta}}_{2}\cdot {\vec{\nabla}}_{\eta })^{2n+2}\int_{\Lambda
}d^{3}\sigma |\vec{\sigma}-(\vec{\eta}_{1}-\vec{\eta}_{2})|^{2m-1}\sigma
^{2n+1},  \nonumber \\
I_{3mn} &=&-(\dot{\vec{\eta}}_{1}\cdot {\vec{\nabla}}_{\eta })^{2m+2}(\dot{%
\vec{\eta}}_{2}\cdot {\vec{\nabla}}_{\eta })^{2n+2}\int_{\Lambda
}d^{3}\sigma |\vec{\sigma}-(\vec{\eta}_{1}-\vec{\eta}_{2})|^{2m+1}\sigma
^{2n-1},  \nonumber \\
I_{4mn} &=&-{\vec{\nabla}}_{\eta }^{2}(\dot{\vec{\eta}}_{1}\cdot {\vec{\nabla%
}}_{\eta })^{2m+2}(\dot{\vec{\eta}}_{2}\cdot {\vec{\nabla}}_{\eta
})^{2n+2}\int_{\Lambda }d^{3}\sigma |\vec{\sigma}-(\vec{\eta}_{1}-\vec{\eta}%
_{2})|^{2m+1}\sigma ^{2n+1},  \nonumber \\
I_{5mn} &=&-{\vec{\nabla}}_{\eta }^{2}(\dot{\vec{\eta}}_{1}\cdot {\vec{\nabla%
}}_{\eta })^{2m}(\dot{\vec{\eta}}_{2}\cdot {\vec{\nabla}}_{\eta
})^{2n}\int_{\Lambda }d^{3}\sigma |\vec{\sigma}-(\vec{\eta}_{1}-\vec{\eta}%
_{2})|^{2m-1}\sigma ^{2n-1},  \nonumber \\
I_{6mn} &=&-(\dot{\vec{\eta}}_{1}\cdot {\vec{\nabla}}_{\eta })^{2m+1}(\dot{%
\vec{\eta}}_{2}\cdot {\vec{\nabla}}_{\eta })^{2n+1}\int_{\Lambda
}d^{3}\sigma |\vec{\sigma}-(\vec{\eta}_{1}-\vec{\eta}_{2})|^{2m-1}\sigma
^{2n-1},  \nonumber \\
\vec{I}_{7mn} &=&-\vec{\nabla}_{\eta }(\dot{\vec{\eta}}_{1}\cdot {\vec{\nabla%
}}_{\eta })^{2m+1}(\dot{\vec{\eta}}_{2}\cdot {\vec{\nabla}}_{\eta
})^{2n}\int_{\Lambda }d^{3}\sigma |\vec{\sigma}-(\vec{\eta}_{1}-\vec{\eta}%
_{2})|^{2m-1}\sigma ^{2n-1},  \nonumber \\
I_{8mn} &=&-(\dot{\vec{\eta}}_{1}\cdot {\vec{\nabla}}_{\eta })^{2m+2}(\dot{%
\vec{\eta}}_{2}\cdot {\vec{\nabla}}_{\eta })^{2n}\int_{\Lambda }d^{3}\sigma |%
\vec{\sigma}-(\vec{\eta}_{1}-\vec{\eta}_{2})|^{2m-1}\sigma ^{2n-1},
\nonumber \\
\vec{I}_{9mn} &=&-\vec{\nabla}_{\eta }(\dot{\vec{\eta}}_{1}\cdot {\vec{\nabla%
}}_{\eta })^{2m+2}(\dot{\vec{\eta}}_{2}\cdot {\vec{\nabla}}_{\eta
})^{2n+1}\int_{\Lambda }d^{3}\sigma |\vec{\sigma}+(\vec{\eta}_{1}-\vec{\eta}%
_{2})|^{2m+1}\sigma ^{2n-1},  \nonumber \\
I_{10}mn &=&-{\vec{\nabla}}_{\eta }^{2}(\dot{\vec{\eta}}_{1}\cdot {\vec{%
\nabla}}_{\eta })^{2m+2}(\dot{\vec{\eta}}_{2}\cdot {\vec{\nabla}}_{\eta
})^{2n}\int_{\Lambda }d^{3}\sigma |\vec{\sigma}-(\vec{\eta}_{1}-\vec{\eta}%
_{2})|^{2m+1}\sigma ^{2n-1}.  \label{b8}
\end{eqnarray}
The integrals that remain to be evaluated are each of the form
\begin{eqnarray}
&&\int_{\Lambda }d^{3}\sigma \sigma ^{2n-1}|\vec{\sigma}-(\vec{\eta}_{1}-%
\vec{\eta}_{2})|^{2m-1}=2\pi \int_{0}^{\Lambda }d\sigma \sigma
^{2n+1}\int_{-1}^{1}dz(\sigma ^{2}+\eta ^{2}-2\eta \sigma z)^{m-1/2}=
\nonumber \\
&=&-{\frac{2\pi }{\eta (2m+1)}}\int_{0}^{\Lambda }d\sigma \sigma ^{2n}\Big(
\nonumber \\
&=&-{\frac{2\pi \eta ^{2n+2m+1}}{(2m+1)}}\sum_{k=0}^{2m+1}\Big[{\frac{%
(\Lambda /\eta )^{2n+k+1}}{2n+k+1}}({%
%TCIMACRO{\QATOP{2m+1}{k}}%
%BeginExpansion
{2m+1 \atop k}%
%EndExpansion
})[(-)^{1-k}-1]-2{\frac{(-)^{1-k}}{2n+k+1}}({%
%TCIMACRO{\QATOP{2m+1}{k}}%
%BeginExpansion
{2m+1 \atop k}%
%EndExpansion
)}\Big].  \label{b9}
\end{eqnarray}
It is of interest to show how the $\Lambda $ dependent terms vanish after
the $\eta $ derivatives have acted. First note that only the even $k$ terms
of the $\Lambda $ dependent sum survive. Of the ones that remain, the power
of $\eta $ is $\eta ^{2n+2m+1-2n-2k-1}=\eta ^{2m-2k}=\eta ^{even\ power}$.
But since the power is even and since the number of $\eta $ derivatives
always exceeds that even power, the derivative vanishes,. e.g.
\begin{eqnarray}
&&{\frac{\partial ^{3}}{\partial \eta _{i}\partial \eta _{j}\partial \eta
_{k}}}\eta ^{2}={\frac{\partial ^{2}}{\partial \eta _{i}\partial \eta _{j}}}%
2\eta _{k}=0,  \nonumber \\
&&{\frac{\partial ^{5}}{\partial \eta _{i}\partial \eta _{j}\partial \eta
_{k}\partial \eta _{l}\partial \eta _{m}}}\eta ^{4}=4{\frac{\partial ^{4}}{%
\partial \eta _{i}\partial \eta _{j}\partial \eta _{k}\partial \eta _{l}}}%
\eta ^{2}\eta _{m}=4{\frac{\partial ^{3}}{\partial \eta _{i}\partial \eta
_{j}\partial \eta _{k}}}(2\eta _{l}\eta _{m}+\eta ^{2}\delta _{ml})=0.
\label{b10}
\end{eqnarray}
Note further that the last sum is
\begin{eqnarray}
&&2\sum_{k=0}^{2m+1}(-)^{k}({%
%TCIMACRO{\QATOP{2m+1}{k}}%
%BeginExpansion
{2m+1 \atop k}%
%EndExpansion
})\int_{0}^{1}dx\ x^{2n+k}=2\int_{0}^{1}dx\ x^{2n}(1-x)^{2m+1}=2B(2n+1,2m+2).
\nonumber \\
&&{}  \label{b11}
\end{eqnarray}
Thus the portion of the integral that ``survives'' is
\begin{equation}
``\int_{\Lambda }d^{3}\sigma \sigma ^{2n-1}|\vec{\sigma}-\vec{\eta}%
|^{2m-1}"=-{\frac{4\pi (2n)!(2m)!}{(2n+2m+2)!}}\eta ^{2n+2m+1},  \label{b12}
\end{equation}
and our ten integrals appear as (using $\vec{\nabla}^{2}\eta ^{l}=l\left(
l-1\right) \eta ^{l-2}$)
\begin{eqnarray}
&&{\frac{I_{1mn}}{16\pi ^{2}(2n)!(2m)!}}={\frac{1}{4\pi (2n+2m+2)!}}(\dot{%
\vec{\eta}}_{1}\cdot {\vec{\nabla}}_{\eta })^{2m+1}(\dot{\vec{\eta}}%
_{2}\cdot {\vec{\nabla}}_{\eta })^{2n+1}\eta ^{2n+2m+1},  \nonumber \\
&&-{\frac{I_{2mn}}{16\pi ^{2}(2n+2)!(2m)!}}=-{\frac{1}{4\pi (2n+2m+4)!}}(%
\dot{\vec{\eta}}_{1}\cdot {\vec{\nabla}}_{\eta })^{2m+2}(\dot{\vec{\eta}}%
_{2}\cdot {\vec{\nabla}}_{\eta })^{2n+2}\eta ^{2n+2m+3},  \nonumber \\
&&-{\frac{I_{3mn}}{16\pi ^{2}(2n)!(2m+2)!}}=-{\frac{1}{4\pi (2n+2m+4)!}}(%
\dot{\vec{\eta}}_{1}\cdot {\vec{\nabla}}_{\eta })^{2m+2}(\dot{\vec{\eta}}%
_{2}\cdot {\vec{\nabla}}_{\eta })^{2n+2}\eta ^{2n+2m+3},  \nonumber \\
&&{\frac{I_{4mn}}{16\pi ^{2}(2n+2)!(2m+2)!}}={\frac{1}{4\pi (2n+2m+6)!}}(%
\dot{\vec{\eta}}_{1}\cdot {\vec{\nabla}}_{\eta })^{2m+2}(\dot{\vec{\eta}}%
_{2}\cdot {\vec{\nabla}}_{\eta })^{2n+2}{\vec{\nabla}}_{\eta }^{2}\eta
^{2n+2m+5}=  \nonumber \\
&=&\frac{I_{3mn}}{16\pi ^{2}(2n)!(2m+2)!},  \nonumber \\
&&{\frac{I_{5mn}}{16\pi ^{2}(2n)!(2m)!}}={\frac{1}{4\pi (2n+2m+2)!}}{\vec{%
\nabla}}_{\eta }^{2}(\dot{\vec{\eta}}_{1}\cdot {\vec{\nabla}}_{\eta })^{2m}(%
\dot{\vec{\eta}}_{2}\cdot {\vec{\nabla}}_{\eta })^{2n}\eta ^{2n+2m+1},
\nonumber \\
&&-{\frac{I_{6mn}}{16\pi ^{2}(2n)!(2m)!}}=-{\frac{1}{4\pi (2n+2m+2)!}}(\dot{%
\vec{\eta}}_{1}\cdot {\vec{\nabla}}_{\eta })^{2m+1}(\dot{\vec{\eta}}%
_{2}\cdot {\vec{\nabla}}_{\eta })^{2n+1}\eta ^{2n+2m+1},  \nonumber \\
&&{\frac{\vec{I}_{7mn}}{16\pi ^{2}(2n)!(2m)!}}={\frac{1}{4\pi (2n+2m+2)!}}%
\vec{\nabla}_{\eta }(\dot{\vec{\eta}}_{1}\cdot {\vec{\nabla}}_{\eta
})^{2m+1}(\dot{\vec{\eta}}_{2}\cdot {\vec{\nabla}}_{\eta })^{2n}\eta
^{2n+2m+1},  \nonumber \\
&&-{\frac{I_{8mn}}{16\pi ^{2}(2n)!(2m)!}}=-{\frac{1}{4\pi (2n+2m+2)!}}(\dot{%
\vec{\eta}}_{1}\cdot {\vec{\nabla}}_{\eta })^{2m+2}(\dot{\vec{\eta}}%
_{2}\cdot {\vec{\nabla}}_{\eta })^{2n}\eta ^{2n+2m+1},  \nonumber \\
&&-{\frac{\vec{I}_{9mn}}{16\pi ^{2}(2n)!(2m+2)!}}=-{\frac{1}{4\pi (2n+2m+4)!}%
}\vec{\nabla}_{\eta }(\dot{\vec{\eta}}_{1}\cdot {\vec{\nabla}}_{\eta
})^{2m+2}(\dot{\vec{\eta}}_{2}\cdot {\vec{\nabla}}_{\eta })^{2n+1}\eta
^{2n+2m+3},  \nonumber \\
&&{\frac{I_{10mn}}{16\pi ^{2}(2n)!(2m+2)!}}={\frac{1}{4\pi (2n+2m+4)!}}\vec{%
\nabla}_{\eta }^{2}(\dot{\vec{\eta}}_{1}\cdot {\vec{\nabla}}_{\eta })^{2m+2}(%
\dot{\vec{\eta}}_{2}\cdot {\vec{\nabla}}_{\eta })^{2n}\eta ^{2n+2m+3}=
\nonumber \\
&=&{\frac{I_{8mn}}{16\pi ^{2}(2n)!(2m)!}}.  \label{b13}
\end{eqnarray}

Substitution into Eq.(\ref{b4}) and Eq.(\ref{b5}) leads to Eq.(\ref{VI4})
and Eq.(\ref{VI5})

\vfill\eject

\section{The Evaluation of $(\vec{\protect\kappa}_{1}\cdot \vec{\partial}
)^{a}(\vec{\protect\kappa}_{2}\cdot \vec{\partial})^{b}\protect\eta ^{a+b-1}$%
}

Consider the case with $a=2m+1,\ \ b=2n+1$,

\begin{equation}
(\vec{\kappa}_{1}\cdot \vec{\partial})^{2m+1}(\vec{\kappa}_{2}\cdot \vec{%
\partial})^{2n+1}\eta ^{2m+2n+1}.  \label{c1}
\end{equation}
Let
\begin{equation}
\vec{\kappa}_{1}=\kappa _{1}\hat{\kappa}_{1};\quad \vec{\kappa}_{2}=\kappa
_{2}\hat{\kappa}_{2}\ ;\quad \hat{\kappa}_{1}^{2}=1=\hat{\kappa}_{2}^{2},
\label{c2}
\end{equation}
and
\begin{eqnarray}
\hat{\kappa}_{2} &=&\alpha \hat{\kappa}_{1}+\beta \hat{\kappa _{1}}_{\perp
}\ ;\quad \hat{\kappa}_{1}\cdot \hat{\kappa _{1}}_{\perp }=0;\quad \hat{%
\kappa}_{1}\cdot \hat{\kappa}_{2}=\alpha \ ;  \nonumber \\
\ \alpha ^{2}+\beta ^{2} &=&1;\quad \quad \beta =(1-(\hat{\kappa}_{1}\cdot
\hat{\kappa}_{2})^{2})^{{\frac{1}{2}}},  \label{c3}
\end{eqnarray}
so
\begin{eqnarray}
&&Eq.(C1)=\kappa _{1}^{2m+1}\kappa _{2}^{2n+1}\sum_{j=0}^{2n+1}{%
%TCIMACRO{\binom{2n+1}{j}}%
%BeginExpansion
{2n+1 \choose j}%
%EndExpansion
}\alpha ^{j}\beta ^{2n+1-j}(\hat{\kappa}_{1}\cdot \vec{\partial})^{2m+1+j}(%
\hat{\kappa _{1}}_{\perp }\cdot \vec{\partial})^{2n+1-j}\eta ^{2m+n+1}.
\nonumber \\
&&{}  \label{c4}
\end{eqnarray}
Orient our axes so that
\begin{equation}
\vec{\eta}=x\hat{\imath}+y\hat{\jmath}+Z\hat{k},  \label{c5}
\end{equation}
with
\begin{equation}
Z=\vec{\eta}\cdot (\hat{\kappa}_{1}\times \hat{\kappa}_{2}),  \label{c6}
\end{equation}
so that, the $Z$ direction is perpendicular to the plane containing $\vec{%
\kappa}_{1}$ and $\vec{\kappa}_{2}$. Further orient axes so that
\begin{equation}
\vec{\eta}=x\hat{\kappa}_{1}+y\hat{\kappa}_{2}+Z\hat{\kappa}_{1}\times \hat{%
\kappa}_{2}.  \label{c7}
\end{equation}
Thus
\begin{eqnarray}
&&Eq.(C4)=\kappa _{1}^{2m+1}\kappa _{2}^{2m+1}\sum_{j=0}^{2n+1}{%
%TCIMACRO{\binom{2n+1}{j}}%
%BeginExpansion
{2n+1 \choose j}%
%EndExpansion
}\alpha ^{j}\beta ^{2n+1-j}(\partial _{x})^{2m+1+j}(\partial
_{y})^{2n+1-j}\eta ^{2m+n+1}.  \nonumber \\
&&{}  \label{c8}
\end{eqnarray}
Consider just the portion
\begin{equation}
\partial _{x}^{2m+1+j}\partial _{y}^{2n+1-j}\eta ^{2m+2n+1}.  \label{c9}
\end{equation}
Let
\begin{equation}
\eta =(\rho ^{2}+Z^{2})^{{\frac{1}{2}}}\ ;\ \rho ^{2}=x^{2}+y^{2}=zz^{\ast
}\ ;\ z=x+iy,z^{\ast }=x-iy.  \label{c10}
\end{equation}
Thus
\begin{equation}
\partial _{x}={\frac{\partial }{\partial z}}+{\frac{\partial }{\partial
z^{\ast }}}={\frac{\partial z}{\partial x}}{\frac{\partial }{\partial z}}+{%
\frac{\partial z^{\ast }}{\partial x}}{\frac{\partial }{\partial z^{\ast }}}%
;\quad \partial _{y}=i(\partial _{z}-\partial _{z^{\ast }}),\   \label{c11}
\end{equation}
and
\begin{eqnarray}
Eq.(C9) &=&i^{2n+1-j}\sum_{h=0}^{2m+1+j}{%
%TCIMACRO{\binom{2m+1+j}{h}}%
%BeginExpansion
{2m+1+j \choose h}%
%EndExpansion
}\sum_{k=0}^{2n+1+j}{%
%TCIMACRO{\binom{2n+1-j}{k}}%
%BeginExpansion
{2n+1-j \choose k}%
%EndExpansion
}  \nonumber \\
&&\times \partial _{z}^{h+k}(-)^{2n+1-y-k}\partial _{z^{\ast
}}^{2m+2n+2-h-k}(zz^{\ast }+Z^{2})^{m+n+{\frac{1}{2}}}.  \label{c12}
\end{eqnarray}
Since $\partial _{z}z^{\ast }=0=\partial _{z{^{\ast }}}z$, the derivatives
become relatively simple
\begin{equation}
\partial _{z}^{h+k}(zz^{\ast }+Z^{2})^{m+m+{\frac{1}{2}}}={\frac{(m+n+{\frac{%
1}{2}})!}{(m+n+{\frac{1}{2}}-h-k)!}}z^{\ast h+k}(zz^{\ast }+Z^{2})^{m+n+{{%
\frac{1}{2}}}-h-k},  \label{c13}
\end{equation}
and
\begin{eqnarray}
&&\partial _{z^{\ast }}^{2m+2n+2-h-k}z^{\ast h+k}(zz^{\ast }+Z^{2})^{m+n+{%
\frac{1}{2}}-h-k}=  \nonumber \\
&=&\sum_{i=0}^{2m+2n+2-h-k}{%
%TCIMACRO{\binom{2m+2n+2-h-k}{i}}%
%BeginExpansion
{2m+2n+2-h-k \choose i}%
%EndExpansion
}\partial _{z^{\ast }}^{i}z^{\ast (h+k)}\partial _{z^{\ast
}}^{2m+2n+2-h-k-i}(zz^{\ast }+z^{2})^{l+{\frac{1}{2}}-h-k}.  \nonumber \\
&&{}  \label{c14}
\end{eqnarray}
Now
\begin{equation}
\partial _{z^{\ast }}^{i}z^{\ast h+h}={\frac{(h+k)!}{(h+k-i)!}}z^{\ast
h+k-i},  \label{c15}
\end{equation}
Note that the factorial in denominator takes care of cutoff, so we can avoid
having to worry about upper limit on sum.
\begin{eqnarray}
&&\partial _{z^{\ast }}^{2(m+n+1)-h-k-i}(zz^{\ast }+Z^{2})^{l+{\frac{1}{2}}%
-h-k)}=  \nonumber \\
&=&{\frac{(m+n+{\frac{1}{2}}-h-k)!}{(i-m-n-{\frac{3}{2})}!}}%
z^{2(m+n+1)-h-k-i}(zz^{\ast }+Z^{2})^{i-m-n-{\frac{3}{2}}}.  \label{c16}
\end{eqnarray}
Thus combining factors
\begin{eqnarray}
&&Eq.(C8)=\kappa _{1}^{2m+1}\kappa _{2}^{2n+1}(m+n+{\frac{1}{2}})!\eta
^{-2m-2n3}z^{2m+2n+1}(i)\beta ^{2n+1}(-)^{n+1}\times   \nonumber \\
&&\sum_{j=0}^{2n+1}{%
%TCIMACRO{\binom{2n+1}{j}}%
%BeginExpansion
{2n+1 \choose j}%
%EndExpansion
}({\frac{i\alpha }{\beta }})^{j}\sum_{h=0}^{2(m+n+1)+j}\sum_{k=0}^{2n+1-j}%
\sum_{i=0}^{(2(m+n+1-h-k}\bigg({\frac{\eta ^{2}}{zz^{\ast }}}\bigg)^{i}\bigg(%
{\frac{z^{\ast }}{z}}\bigg)^{h+k}(-)^{k}\times   \nonumber \\
&&{%
%TCIMACRO{\binom{2m+1+j}{h}}%
%BeginExpansion
{2m+1+j \choose h}%
%EndExpansion
}{%
%TCIMACRO{\binom{2n+1-j}{k}}%
%BeginExpansion
{2n+1-j \choose k}%
%EndExpansion
}{%
%TCIMACRO{\binom{2m+2n+2-h-k}{i}}%
%BeginExpansion
{2m+2n+2-h-k \choose i}%
%EndExpansion
}{\frac{1}{(i-m-n-{\frac{3}{2}})}}{\frac{(n+k)!}{(h+k-i)!}}.  \nonumber \\
&&{}  \label{c17}
\end{eqnarray}
Consider second two lines and let upper limits be $\infty $ and let the
factorials set the limits. That becomes of form
\begin{equation}
\sum_{j=0}^{\infty }\sum_{i=0}^{\infty }\sum_{h=0}^{\infty
}\sum_{k=0}^{\infty }f(j,h,k,h+k,i).  \label{c18}
\end{equation}
Let $q=2n+1-j,\ p=h+k.$ \ Then
\begin{eqnarray}
Eq.(C18) &=&(2n+1)!\bigg({\frac{i\alpha }{\beta }}\bigg)^{2n+1}\sum_{q=0}^{%
\infty }\sum_{l=0}^{\infty }\sum_{p=0}^{\infty }\sum_{k=0}^{\rho }\bigg({%
\frac{-i\beta }{\alpha }}\bigg)^{q}\bigg({\frac{n^{2}}{zz^{\ast }}}\bigg)^{i}%
\bigg({\frac{z^{\ast }}{z}}\bigg)^{p}(-)^{k}  \nonumber \\
&&\times \ {\frac{1}{q!}}{\frac{1}{(2n+1-q)!}}{\frac{q!}{k!(q-k)!}}{\frac{%
(2(n+m+1)-q)!(2(m+m+1)-p)!}{(p-k)!(2(m+m+1)-q-p+k)!}}  \nonumber \\
&&\times {\frac{p!}{i!(2(m+n+1)-p-i)!}}{\frac{i}{(p-i)!}}{\frac{1}{(i-m-n-{%
\frac{3}{2}})}}.  \label{c19}
\end{eqnarray}

Use factorials to place upper bound on summation limits. Consider $q$ first;
$q\geq 0,\ q\geq k,\ q\leq 2(m+n+1)-p+k;q\geq 2n+1.$ Let $s=q-k\geq 0$.
Upper bound appears ambiguous. One can make the replacement
\begin{equation}
{\frac{(2(n+m+1)-q)!}{(2n+1-q)!}}=\bigg({\frac{d}{dw}}\bigg) %
^{2m+1}w^{2(n+m+1)-q}\mid _{w=1}=\bigg({\frac{d}{dw}}^{2m+1}\bigg) %
w^{2(n+m+1)-k-s}\mid _{w=1},  \label{c20}
\end{equation}
and so our $s$ sum is restricted to
\begin{equation}
0\leq s\leq 2(m+n+1)-p\ .  \label{c21}
\end{equation}
Note that the right hand side must be positive. So we can perform the $q$(or
$s$) sum involved in (\ref{c19}). It is (using ${{1={\frac{(2(m+n+1)-p)!}{
(2(m+n+1)-p)!}}}}$ )
\begin{eqnarray}
&&\sum_{s=0}^{2(m+n+1)-p}w^{2(n+m+1)-k-s}\bigg({\frac{-i\beta }{\alpha }}%
\bigg)^{k+s}{\frac{1}{s!}}{\frac{1}{(2(n+m+1)-p-s)!}}=  \nonumber \\
&=&w^{2(n+n+1)-k}\bigg({\frac{-i\beta }{\alpha }}\bigg)^{k}\bigg(1-{\frac{
i\beta }{\alpha w}}\bigg)^{2(m+n+1)-p}{\frac{1}{(2(m+n+1)-p)!}}=  \nonumber
\\
&=&w^{p-k}\bigg({\frac{-i\beta }{\alpha }}\bigg)^{k}\bigg(w-{\frac{i\beta }{
\alpha }}\bigg)^{2(m+n+1)-p}{\frac{1}{(2(m+n+1)-p)!}}.  \label{c22}
\end{eqnarray}
So our expression (\ref{c19}) reduces to (canceling the last factor)
\begin{eqnarray}
&&Eq.(C19)=(2n+1)!\left( \frac{i\alpha }{\beta }\right) ^{2n+1}\left( {\frac{%
d }{dw}}\right) ^{2n+1}\sum_{i=0}^{\infty }\sum_{p=0}^{\infty
}\sum_{k=0}^{p}w^{p-k}\bigg({\frac{+i\beta }{\alpha }}\bigg)^{k}\bigg({\frac{
w-i\beta }{\alpha }}\bigg)^{2(m+m+1)-p}  \nonumber \\
&&\bigg({\frac{\eta ^{2}}{zz^{\ast }}}\bigg)^{2}\bigg({\frac{z^{\ast }}{z}}%
\bigg)^{p}{\frac{1}{k!}}{\frac{p!}{(p-k)!}}{\frac{1}{i!}}{\frac{1}{(p-i)!}}{%
\ \frac{1}{(2(m+n+1)-p-i)!}}{\frac{1}{(i-m-n-{\frac{3}{2}})!}}\mid _{w=1}.
\label{c23}
\end{eqnarray}
We can perform the $k$ sum and note that the $p$ sum now has well defined
limits since $2(m+n+1)-p-1\geq 0$ and therefore $i\leq p\leq 2(m+n+1)-i$.

Thus
\begin{eqnarray}
&&Eq.(C23) =(2n+1)!\left( \frac{i\alpha }{\beta }\right) ^{2n+1}\left( {%
\frac{d}{dw}}\right) ^{2m+1}\sum_{i=0}^{\infty
}\sum_{p=i}^{2(m+n+1)-i}w^{p}\left( 1+{\frac{i\beta }{\alpha \omega }}
\right) ^{p}  \nonumber \\
&&\times \bigg(w-{\frac{i\beta }{\alpha }}\bigg)^{2(m+n+1)}\bigg({\frac{\eta
^{2}}{zz^{\ast }}}\bigg)^{i}\bigg({\frac{z^{\ast }}{z}}\bigg)^{p}{\frac{1}{
(2(m+n+1)-p-i)!}\frac{1}{i!}}{\frac{1}{(i-m-n-{\frac{3}{2}})!}}.  \label{c24}
\end{eqnarray}
Next, noting that $w^{p}\bigg(1+{\frac{i\beta }{\alpha w}}\bigg)^{p}=(w+{%
\frac{i\beta }{\alpha }})^{p}$ and that the $p$ sum produces ${\frac{1}{
(2(m+n+1)-2i)!}}$ we see that $i$ is restricted to $0\leq i\leq m+n+1$.

The $p$ sum is best performed by changing variables to $r=p-i$. Then $0\leq
r\leq 2(m+n+1-i)$. Thus we obtain
\begin{eqnarray}
&&Eq.(C24)=(2n+1)!\bigg({\frac{i\alpha }{\beta }}\bigg)^{2n+1}\bigg({\frac{d%
}{dw}}\bigg)^{2m+1}\sum_{i=0}^{m+n+1}{\frac{1}{i!}}{\frac{1}{(2(m+n+1-i)!}}{%
\ \frac{1}{(i-m-n-{\frac{3}{2}})!}}\times  \nonumber \\
&&\left( {\frac{w+i\beta /\alpha )}{(w-i\beta /\alpha )}}\right) ^{i}\left( {%
\frac{\eta ^{2}}{zz^{\ast }}}\right) ^{i}\bigg({\frac{z^{\ast }}{z}}\bigg)%
^{i}\bigg[1+{\frac{z^{\ast }(w+{\frac{i\beta }{\alpha }})}{z(w-i{\frac{\beta
}{\alpha }})}}\bigg]^{2(m+n+1-i)}\bigg(w-{\frac{i\beta }{\alpha }}\bigg)%
^{2(m+n+1)}.  \label{c25}
\end{eqnarray}
To perform the $i$ sum, notice that
\begin{eqnarray}
&&{\frac{1}{(i-m-n-3/2)!}}{\frac{1}{(2(m+n+1-i))!}}=  \nonumber \\
&=&{\frac{\pi ^{\frac{1}{2}}}{2^{(m+n+1-i)}(m+n+1-i)!(m+n+{\frac{1}{2}}%
-i)!(1-m-n-3/2)!}}=  \nonumber \\
&=&{\frac{\pi ^{\frac{1}{2}}}{2^{2(m+n+1-1)}}}{\frac{sin(\pi (i-m-n-{\frac{1%
}{2}})}{(m+n+1-i)!}}={\frac{(-)^{i-m-n+1}}{\pi ^{\frac{1}{2}%
}2^{(m+n+1-i)}(m+n+1-i)!}}.  \label{c26}
\end{eqnarray}
Further notice that the 2nd line of (\ref{c25}) can be simplified to
\begin{equation}
\bigg(w^{2}+\beta ^{2}/\alpha ^{2}\bigg)^{i}(\eta ^{2})^{i}{\frac{%
[z(w-i\beta /\alpha )+z^{\ast }(w+\beta /\alpha )]^{2(m+n+1)-1}}{z^{2(m+n+1)}%
}}.  \label{c27}
\end{equation}
So
\begin{eqnarray}
Eq.(C25) &=&(2n+1)!\bigg({\frac{i\alpha }{\beta }})^{2n+1}\bigg({\frac{d}{dw}%
}\bigg)^{2m+1}{\frac{(-)^{m+n+1}}{\pi ^{\frac{1}{2}}2^{2(m+n+1)}}}(z(w-{\
\frac{1}{2}}\beta )+z^{\ast }(w+{\frac{i\beta }{\alpha }}))^{(2m+n+1)}
\nonumber \\
&&\times \sum_{i=0}^{m+n+1}{\frac{1}{i!}}{\frac{2^{2i}}{(m+n+1-i)!}}{\frac{%
\eta ^{2i}(w^{2}+\beta ^{2}/\alpha ^{2})^{i}(-)^{i}}{[z(w-i\beta /\alpha
)+z^{\ast }(w+i\beta /\alpha )]^{2i}}}.  \label{c28}
\end{eqnarray}
The sum produces
\begin{equation}
{\frac{1}{(m+n+1)!}}\bigg[1-{\frac{4\eta ^{2}(w^{2}+\beta ^{2}/\alpha ^{2})}{%
[z(w-i\beta \alpha )+z^{\ast }(w+i\beta /\alpha )]^{2}}}\bigg]^{m+n+1}.
\label{c29}
\end{equation}
Combine this with (\ref{c28}) and then (\ref{c17}) gives
\begin{eqnarray}
&&Eq.(C17)=\kappa _{1}^{2m+1}\kappa _{2}^{2n+1}{\frac{(m+n+{\frac{1}{2}})!}{%
\pi ^{{\frac{1}{2}}}2^{2(m+n+1)}}}\eta ^{-2m-2n-3}(-)i^{2}\alpha
^{2n+1}(2n+1)!(-)^{m+n+1}\times  \nonumber \\
&&\big({\frac{d}{dw}}\big)^{2m+1}\bigg[{\frac{(z(w-i\beta /\alpha )+z^{\ast
}(w+i\beta /\alpha ))^{2}-4\eta ^{2}(w^{2}+\beta ^{2}/\alpha ^{2})}{(m+n+1)!}%
}\bigg]^{m+n+1}.  \label{c30}
\end{eqnarray}

Before expanding or simplifying consider case when
\[
\vec{\kappa}_{2}=-\vec{\kappa}_{1}=-\vec{\kappa}\Rightarrow \beta =0.
\]
Then
\begin{eqnarray}
Eq.(C30) &=&(-)^{m+n}{\frac{\kappa ^{2m+2n+2}}{\pi ^{\frac{1}{2}%
}2^{2(m+n+1)} }}(m+n+{\frac{1}{2}})!\eta ^{-2m-2n-3}(2n+1)!\times  \nonumber
\\
&&\bigg({\frac{d}{dw}}\bigg)^{2m+1}{\frac{(w^{2}(4x^{2}-\eta ^{2}))^{m+n+1}}{
(m+n+1)!}}.  \label{c31}
\end{eqnarray}
Use
\begin{equation}
{\frac{d^{2m+1}}{dw^{2m+1}}}w^{2m+2n+1}\mid _{w=1}={\frac{(2(m+n+1))!}{
(2n+1)!}}.  \label{c32}
\end{equation}
Also $(m+n+{\frac{1}{2}})!={\frac{(2(m+n+1))!}{(m+n+1)!}}{\frac{\pi ^{\frac{%
1 }{2}}}{2^{2m+2n+1}}}$. Thus
\begin{eqnarray}
Eq.(C30) &=&-\kappa ^{2m+2n+2}{\frac{[(2(m+n+1))!]^{2}}{
[(m+n+1)!]^{2}2^{2m+2n+1}}}\bigg(1-{\frac{x^{2}}{\eta ^{2}}}\bigg)^{m+n+1}{%
\frac{1}{\eta }}=  \nonumber \\
&=&-[(2m+2n+1)!!]^{2}{\frac{1}{\eta }}(\kappa ^{2}-(\kappa \cdot \hat{\eta}
)^{2})^{m+n+1}.  \label{c33}
\end{eqnarray}
Return to (\ref{c30}) and consider $z(w-i\beta /\alpha )+z^{\ast }(w+i\beta
/\alpha )=2(xw+y\beta /\alpha ).$ \ In (\ref{c30}) this factor becomes
\begin{eqnarray}
\lbrack \ \ \ \ \ ]^{m+n+1} &=&2^{(m+n+1)}(-)^{m+n+1}\bigg[\eta
^{2}(w^{2}+\beta ^{2}/\alpha ^{2})-x^{2}w^{2}-2xyw\beta /\alpha -y^{2}\beta
^{2}/\alpha ^{2}]^{m+n+1}=  \nonumber \\
&=&2^{2(m+n+1)}(-)^{m+n+1}[w^{2}(\eta ^{2}-x^{2})+\beta ^{2}/\alpha
^{2}(\eta ^{2}-y^{2})-2xyw\beta /\alpha ]^{m+n+1}=  \nonumber \\
&=&2^{2(m+n+1)}(-)^{m+n+1}[\eta ^{2}(w^{2}+\beta ^{2}/\alpha
^{2})-(xw+y\beta /\alpha )^{2}]^{m+n+1}.  \label{c34}
\end{eqnarray}
Thus
\begin{eqnarray}
Eq.(C30) &=&\kappa _{1}^{2m+1}\kappa _{2}^{2n+1}{\frac{[2(m+n+1)]!}{\
2^{2m+2n+2}[(n+m+1)!]^{2}}}{\frac{\alpha ^{2n+1}(2n+1)!}{\eta }}\times
\nonumber \\
&&\bigg({\frac{d}{dw}}\bigg)^{2m+1}\bigg((w^{2}+\beta ^{2}/\alpha ^{2})-({%
\frac{xw}{\eta }}+{\frac{y}{\eta }}{\frac{\beta }{\alpha }})^{2}\bigg) %
^{m+n+1}\bigg|_{w=1}.  \label{c35}
\end{eqnarray}
The derivative is of the form
\begin{equation}
\left( \frac{d}{dw}\right) ^{2m+1}\bigg(aw^{2}+bw+c\bigg)^{m+n+1}=\bigg({\
\frac{d}{dw}}\bigg)^{2m+1}F(Q(w))=\bigg({\frac{d}{dw}}\bigg)^{2m+1}f(w).
\label{c36}
\end{equation}
This radical is
\begin{equation}
\omega ^{2}\bigg(1-{\frac{x^{2}}{\eta ^{2}}}\bigg)-{\frac{2xy\beta w}{\eta
^{2}\alpha }}+{\frac{\beta ^{2}}{d^{2}}}\bigg(1-{\frac{y^{2}}{\eta ^{2}}} %
\bigg),  \label{c37}
\end{equation}
with roots
\begin{equation}
w={\frac{{\frac{xy\beta }{\eta ^{2}\alpha }}\pm \sqrt{{\frac{x^{2}y^{2}\beta
^{2}}{\eta ^{4}\alpha ^{2}}}-{\frac{\beta ^{2}}{\alpha ^{2}}}\bigg(1-{\frac{
y^{2}}{\eta ^{2}}}\bigg)\bigg(1-{\frac{x^{2}}{\eta ^{2}}}\bigg)}}{\bigg(1-{%
\frac{x^{2}}{\eta ^{2}}}\bigg)}}.  \label{c38}
\end{equation}
Since the argument of the root is
\begin{equation}
{\frac{-\beta ^{2}}{\alpha ^{2}}}\bigg(1-{\frac{x^{2}}{\eta ^{2}}}-{\frac{
z^{2}}{\eta ^{2}}}\bigg)<0,  \label{c39}
\end{equation}
the roots are complex. Thus
\[
aw^{2}+bw+c=a(w-\gamma -i\delta )(w-\gamma +i\delta ),
\]
where
\begin{equation}
\gamma ={\frac{xy\beta }{\alpha (\eta ^{2}-x^{2})}};\ \delta ={\frac{\beta }{
\alpha }}{\frac{Z\eta }{(\eta ^{2}-x^{2})}}\ .  \label{c40}
\end{equation}

Consider the derivative:
\begin{eqnarray}
&&{\frac{d^{2m+1}}{dw^{2m+1}}}(w-\gamma -i\delta )^{m+n+1}(w-\gamma +i\delta
)^{m+n+1}=  \nonumber \\
&=&\sum_{j=0}^{2m+1}{\
%TCIMACRO{\binom{2m+1}{j}}%
%BeginExpansion
{2m+1 \choose j}%
%EndExpansion
}\bigg({\frac{d}{dw}}\bigg)^{j}(w-\gamma -i\delta )^{m+n+1}\bigg({\frac{d}{%
dw }}\bigg)^{2m+1-j}(w-\gamma +i\delta )^{m+n+1}=  \nonumber \\
&=&\sum_{j=0}^{2m+1}{\
%TCIMACRO{\binom{2m+1}{j}}%
%BeginExpansion
{2m+1 \choose j}%
%EndExpansion
}{\frac{[(m+n+1)!]^{2}}{(m+n+1-j)!(n+j-m)!}}  \nonumber \\
&&(w-\gamma -i\delta )^{m+n+1-j}(w-\gamma +i\delta )^{n+j-m} .  \label{c41}
\end{eqnarray}

The factorial further restricts the sum to
\[
\sum_{j=max(0,n-m)}^{min(2m+1,m+n+1)}.
\]
Consider the case when $\beta =0=\delta =\gamma .$ Then as $w=1$
\begin{eqnarray}
Eq.(C41) &=&\sum_{j=max(0,n-m)}^{min(2m+1,n+m+1)}{\
%TCIMACRO{\binom{2m+1}{j}}%
%BeginExpansion
{2m+1 \choose j}%
%EndExpansion
}{\frac{(m+n+1)!}{(m+n+1)-j)!}}{\frac{(m+n+1)!}{(n+j-m)!}}=  \nonumber \\
&=&{\frac{[(m+n+1)!]^{2}}{(2n+1)!}}\sum_{j=max(0,n-m)}^{min(2m+1,n+m+1)}{\
%TCIMACRO{\binom{2m+1}{j}}%
%BeginExpansion
{2m+1 \choose j}%
%EndExpansion
}{%
%TCIMACRO{\binom{2n+1}{n+j-m}}%
%BeginExpansion
{2n+1 \choose n+j-m}%
%EndExpansion
} .  \label{c42}
\end{eqnarray}

In order to perform this sum, consider the related product (for $s=1)$
\begin{eqnarray}
&&\sum_{j=0}^{2m+1}{\
%TCIMACRO{\binom{2m+1}{j}}%
%BeginExpansion
{2m+1 \choose j}%
%EndExpansion
}s^{j}\sum_{i=m-n}^{n+m+1}{%
%TCIMACRO{\binom{2n+1}{n+j-m}}%
%BeginExpansion
{2n+1 \choose n+j-m}%
%EndExpansion
}s^{i}=  \nonumber \\
&=&({\rm let}\ l=i-m-n)=\sum_{j=0}^{2m+1}{\
%TCIMACRO{\binom{2m+1}{j}}%
%BeginExpansion
{2m+1 \choose j}%
%EndExpansion
}s^{j}\sum_{l=0}^{2n+1}{%
%TCIMACRO{\binom{2n+1}{l}}%
%BeginExpansion
{2n+1 \choose l}%
%EndExpansion
}s^{\ell }s^{2n+m}=  \nonumber \\
&=&s^{(n+m)}(1+s)^{2(m+n+1)}=\sum_{h=0}^{2(m+n+1)}{\
%TCIMACRO{\binom{2(m+n+1)}{h}}%
%BeginExpansion
{2(m+n+1) \choose h}%
%EndExpansion
}s^{h}s^{n+m}.  \label{c43}
\end{eqnarray}
But consider the product of the two sums. Let $h=j+\ell $ so that $0\leq
h\leq 2(m+n+1)$. Let $l=h-j$. Thus, since $l\geq 0=>h\geq j$ and since $%
2m+1\geq j$, we have $j\leq min(2m+1,h)$. Now we also have $2n+1\geq l$ or $%
j\geq h-2n-1$; thus $j\geq max(0,h-2n-1)$ so
\begin{equation}
\sum_{h=0}^{2(m+n+1)}{\
%TCIMACRO{\binom{2(m+n+1)}{h}}%
%BeginExpansion
{2(m+n+1) \choose h}%
%EndExpansion
}s^{h}=\sum_{h=0}^{2(m+n+1)}s^{h}\sum_{j=max(0,h-2n-1)}^{min(2m+1,h)}{%
%TCIMACRO{\binom{2m+1}{j}}%
%BeginExpansion
{2m+1 \choose j}%
%EndExpansion
}{%
%TCIMACRO{\binom{2n+1}{h-j}}%
%BeginExpansion
{2n+1 \choose h-j}%
%EndExpansion
}.  \label{c44}
\end{equation}
Consider the term $h=m+n+1$ in the sum. Then
\begin{equation}
{\frac{(2(m+n+1)!}{[(m+n+1)!]^{2}}}=\sum_{j=max(0,m-n)}^{min(2m+1,m+n+1}{\
%TCIMACRO{\binom{2m+1}{j}}%
%BeginExpansion
{2m+1 \choose j}%
%EndExpansion
}{%
%TCIMACRO{\binom{2n+1}{n+j-m}}%
%BeginExpansion
{2n+1 \choose n+j-m}%
%EndExpansion
},  \label{c45}
\end{equation}
where we have used ${\
%TCIMACRO{\binom{2n+1}{h-j}}%
%BeginExpansion
{2n+1 \choose h-j}%
%EndExpansion
}={%
%TCIMACRO{\binom{2n+1}{n+m+1-j}}%
%BeginExpansion
{2n+1 \choose n+m+1-j}%
%EndExpansion
}={%
%TCIMACRO{\binom{2n+1}{n+j-m}}%
%BeginExpansion
{2n+1 \choose n+j-m}%
%EndExpansion
}$. Thus
\[
Eq.(C42)={\frac{(2(m+n+1))!}{(2n+1)!}}.
\]
Note that
\begin{eqnarray}
Eq.(C35)|_{\beta =0} &=&-\kappa ^{2(m+n+1)}{\frac{(2(m+n+1)!)^{2}}{%
2^{2(m+n+1)}[(n+m+1)!]^{2}}}{\frac{1}{\eta }}\bigg(1-{\frac{x^{2}}{\eta ^{2}}%
}\bigg)^{m+n+1}=  \nonumber \\
&=&-([2m+2n+1)!!]^{2}{\frac{1}{\eta }}(\kappa ^{2}-(\kappa \cdot \eta
)^{2})^{m+n+1},  \label{c46}
\end{eqnarray}
which checks again with Eq.(C33). Thus, with this we write again
\begin{eqnarray}
Eq.(C35) &=&\kappa _{1}^{2m+1}\kappa _{2}^{2n+1}{\frac{(2(m+n+1))!}{%
2^{2m+2+2}[(n+m+1)!]^{2}}}{\frac{\alpha ^{2n+1}}{\eta }}\bigg(1-{\frac{x^{2}%
}{\eta ^{2}}}\bigg)^{m+n+1}\times   \nonumber \\
&&\sum_{j=max(0,m-n)}^{min(2m+1,n+m+1)}{\
%TCIMACRO{\binom{2m+1}{j}}%
%BeginExpansion
{2m+1 \choose j}%
%EndExpansion
}{%
%TCIMACRO{\binom{2n+1}{n+j-m}}%
%BeginExpansion
{2n+1 \choose n+j-m}%
%EndExpansion
}  \nonumber \\
&&(w-\gamma -i\delta )^{m+n+1-j}(w-\gamma +i\delta )^{n+j-m}|_{w=1}.
\label{c47}
\end{eqnarray}
To simplify, let $j=k-n$. Also, note
\begin{eqnarray}
&&(w-\gamma -i\delta )^{m+n+1-j}(w-\gamma +i\delta )^{n+j-m}|_{w=1}=
\nonumber \\
&=&\bigg[{\frac{1}{\alpha (\eta ^{2}-x^{2})}}\bigg]^{2n+1}[\alpha (\eta
^{2}-x^{2})-\beta (xy+i\eta Z)]^{m+2n+1-k}  \nonumber \\
&&[\alpha (\eta ^{2}-x^{2})-\beta (xy-i\eta z)]^{k-m}.  \label{c48}
\end{eqnarray}
Thus Eq.(\ref{c47}) becomes
\begin{eqnarray}
&&{\frac{[(2m+2n+1)!!]^{2}}{(2(m+n+1))!}}{\frac{\kappa _{1}^{2m+1}\kappa
_{2}^{2n+1}}{\eta }}{\frac{\bigg(1-{\frac{x^{2}}{\eta ^{2}}}\bigg)}{\eta
^{4n+2}}}[(m+n+1))!]^{2}\sum_{k=max(m,n)}^{min(2m+n+1,2n+m+1)}\times
\nonumber \\
&&{%
%TCIMACRO{\binom{2m+1}{k-n}}%
%BeginExpansion
{2m+1 \choose k-n}%
%EndExpansion
}{%
%TCIMACRO{\binom{2n+1}{k-m}}%
%BeginExpansion
{2n+1 \choose k-m}%
%EndExpansion
}(\sigma ^{2}\alpha -\beta (xy+i\eta Z))^{m+2n+1-k}(\sigma ^{2}\alpha -\beta
(xy-i\eta z))^{k-m}.  \label{c49}
\end{eqnarray}
where
\begin{equation}
\sigma ^{2}=\eta ^{2}-x^{2}.  \label{c50}
\end{equation}
If $m\geq n$ then sum becomes $(l:=k-m)$
\begin{equation}
\sum =\sum_{l=0}^{2m+1}{\
%TCIMACRO{\binom{2n+1}{l}}%
%BeginExpansion
{2n+1 \choose l}%
%EndExpansion
}{%
%TCIMACRO{\binom{2m+1}{\ell +m+n}}%
%BeginExpansion
{2m+1 \choose \ell +m+n}%
%EndExpansion
}(\sigma ^{2}\alpha -(xy+i\eta Z)\beta )^{2n+1-l}(\sigma ^{2}\alpha
-(xy-i\eta Z)\beta )^{l}.  \label{c51}
\end{equation}
Because of the symmetry we have
\begin{equation}
\sum =\sum_{l=0}^{\eta }{\
%TCIMACRO{\binom{2n+1}{l}}%
%BeginExpansion
{2n+1 \choose l}%
%EndExpansion
}{%
%TCIMACRO{\binom{2m+1}{\ell +m-n}}%
%BeginExpansion
{2m+1 \choose \ell +m-n}%
%EndExpansion
}\bigg[(\sigma ^{2}\alpha -(xy+i\eta Z)\beta )^{2n+1-l}(\sigma ^{2}\alpha
-(xy-i\eta Z)\beta )^{l}+c.c.\bigg].  \label{c52}
\end{equation}

The bracket $[\ ]$ in (\ref{c52}) above is of the form
\begin{eqnarray}
\lbrack \ \ ] &=&(u-iv)^{2n+1-l}(u+iv)^{l}+c.c.=({\rm let\,\,\,}
u+iv=re^{i\phi })=  \nonumber \\
&=&r^{2n+1}(e^{i(2(l-n)-1)\phi }+c.c.)\ =2\gamma ^{2n+1}{\rm cos}
(2(n-l)+1)\phi ,  \label{c53}
\end{eqnarray}
in which
\begin{eqnarray}
r &=&\sqrt{(\sigma ^{2}\alpha -xy\beta )^{2}+\eta ^{2}Z^{2}\beta ^{2}},
\nonumber \\
\phi &=&tan^{-1}{\frac{\beta \eta z}{\sigma ^{2}\alpha -xy\beta }}.
\label{c54}
\end{eqnarray}
We further need
\begin{eqnarray}
cos((2(n-l)+1)\phi ) &=&\sum_{k=0}^{n-l}{\
%TCIMACRO{\binom{2(n-l)+1}{2k}}%
%BeginExpansion
{2(n-l)+1 \choose 2k}%
%EndExpansion
}(-)^{k}(sin\phi )^{2k}(cos\phi )^{2(n-l-k)+1}=  \nonumber \\
&=&\sum_{k=0}^{n-l}{\
%TCIMACRO{\binom{2(n-l)+1)}{2k}}%
%BeginExpansion
{2(n-l)+1) \choose 2k}%
%EndExpansion
}(-)^{k}\sum_{j=0}^{k}{%
%TCIMACRO{\binom{k}{j}}%
%BeginExpansion
{k \choose j}%
%EndExpansion
}(-)^{j}(cos\phi )^{2(n-l-k+j)+1},  \label{c55}
\end{eqnarray}
where
\begin{equation}
cos\phi ={\frac{\sigma ^{2}\alpha -xy\beta }{r}}.  \label{c56}
\end{equation}

Recall the original variables
\begin{eqnarray}
\eta &=&|\vec{\eta}|;\ \ \eta ^{2}=\rho ^{2}+Z^{2}=\sigma
^{2}+x^{2}=x^{2}+y^{2}+Z^{2},  \nonumber \\
\alpha &=&\hat{\kappa}_{1}\cdot \hat{\kappa}_{2};\ \ \beta =\sqrt{1-(\hat{
\kappa}_{1}\cdot \hat{\kappa}_{2})^{2}},  \nonumber \\
x &=&\hat{\kappa}_{1}\cdot \vec{\eta};\ \ \beta y=\hat{\kappa}_{1\perp
}\cdot \vec{\eta}=(\hat{\kappa}_{2}-\alpha \hat{\kappa}_{1})\cdot \vec{\eta}.
\label{c57}
\end{eqnarray}
Thus
\begin{eqnarray}
\sigma ^{2}\alpha -xy\beta &=&\alpha (\eta ^{2}-(\vec{\eta}\cdot \hat{\kappa}
_{1})^{2})-\vec{\eta}\hat{\cdot}\kappa _{1}(\hat{\kappa}_{2}-\alpha \hat{
\kappa}_{1})\cdot \vec{\eta}=  \nonumber \\
&=&\eta ^{2}(\hat{\kappa}_{1}\cdot \hat{\kappa}_{2}-\hat{\kappa}_{1}\cdot
\hat{\eta}\hat{\kappa}_{2}\cdot \hat{\eta}),  \nonumber \\
\eta ^{2}Z^{2}\beta ^{2} &=&\eta ^{4}(1-(\hat{\eta}\cdot \hat{\kappa}
_{1})^{2}-(\hat{\eta}\cdot \hat{\kappa}_{2})^{2}-(\hat{\kappa}_{1}\cdot \hat{
\kappa}_{2})^{2}-2\hat{\kappa}_{1}\cdot \hat{\kappa}_{2}\hat{\kappa}\cdot
\hat{\eta}\hat{\kappa}_{2}\hat{\eta})),  \label{c58}
\end{eqnarray}
and
\begin{equation}
r^{2}=\eta ^{4}(1-(\hat{\kappa}_{1}\cdot \hat{\eta})^{2})(1-(\hat{\kappa}
_{2}\cdot \hat{\eta})^{2}).  \label{c59}
\end{equation}
Thus
\begin{equation}
cos\phi ={\frac{\vec{\kappa}_{1}\cdot \vec{\kappa}_{2}-\vec{\kappa}
_{1}\cdot \hat{\eta}\vec{\kappa}_{2}\cdot \vec{\eta}}{\sqrt{\kappa _{1}^{2}-(%
\hat{\kappa}_{1}\cdot \vec{\eta})^{2}}\sqrt{\vec{\kappa} _{2}^{2}-(\kappa
_{2}\cdot \hat{\eta})^{2}}}},  \label{c60}
\end{equation}
and
\begin{eqnarray}
&&{\frac{r^{2n+1}}{\eta ^{4n+2}}}\bigg(1-{\frac{x^{2}}{\eta ^{2}}}\bigg)%
^{m-n}\kappa _{1}^{2m+1}\kappa _{2}^{2n+1}=  \nonumber \\
&=&(1-(\hat{\kappa}_{1}\cdot \hat{\eta})^{2})^{n}(1-(\hat{\kappa}_{2}\cdot
\hat{\eta})^{2})^{n}(\kappa _{1}^{2}-(\kappa _{1}\hat{\eta}
)^{2})^{1/2}(\kappa _{2}^{2}-(\hat{\kappa}_{2}\cdot \hat{\eta})^{2})^{1/2},
\label{c61}
\end{eqnarray}
\begin{eqnarray}
&&(\kappa _{1}^{2}-(\kappa _{1}\cdot \hat{\eta})^{2})^{m-n}\kappa
_{1}^{2n}\kappa _{2}^{2n}=  \nonumber \\
&=&(\kappa _{1}^{2}-(\kappa _{1}\cdot \vec{\eta})^{2})^{\frac{1}{2}}(\kappa
_{2}^{2}-(\vec{\kappa}_{2}\cdot \hat{\eta})^{2})^{1/2}(\kappa _{2}^{2}-(\vec{
\kappa}_{2}\cdot \hat{\eta})^{2})^{n}(\kappa _{1}^{2}-\vec{\kappa}_{1}\cdot
\eta )^{2})^{m}.  \label{c62}
\end{eqnarray}

So finally factoring out $cos\phi $ we obtain
\begin{eqnarray}
&&(\vec{\kappa}_{1}\cdot \vec{\partial})^{2m+1}(\vec{\kappa}_{2}\cdot \vec{%
\partial})^{2n+1}\eta ^{2m+2n+1}=  \nonumber \\
&&{}  \nonumber \\
&=&2{\frac{[(2m+2n+1)!!]^{2}}{(2(m+n+1))!}}[(n+m+1))!]^{2}  \nonumber \\
&&\times {\frac{(\vec{\kappa}_{1}\cdot \vec{\kappa}_{2}-\vec{\kappa}%
_{1}\cdot \hat{\eta}\vec{\kappa}_{2}\cdot \hat{\eta})(\kappa _{1}^{2}-(\vec{%
\kappa}_{1}\cdot \hat{\eta}))^{2})(\kappa _{2}^{2}-\vec{\kappa}_{2}\cdot
\hat{\eta}))^{2})^{n}}{\eta }}  \nonumber \\
&&\times \sum_{l=0}^{n}\sum_{k=0}^{n-l}\sum_{j=0}^{2k}{\
%TCIMACRO{\binom{2n+1}{l}}%
%BeginExpansion
{2n+1 \choose l}%
%EndExpansion
}{%
%TCIMACRO{\binom{2m+1}{l+m-n}}%
%BeginExpansion
{2m+1 \choose l+m-n}%
%EndExpansion
}{%
%TCIMACRO{\binom{2(n-\ell )+1}{2k}}%
%BeginExpansion
{2(n-\ell )+1 \choose 2k}%
%EndExpansion
}{%
%TCIMACRO{\binom{2k}{j}}%
%BeginExpansion
{2k \choose j}%
%EndExpansion
}(-)^{k+j}(cos^{2}\phi )^{n-l+j-k}.  \label{c63}
\end{eqnarray}

Check once again the limit $\vec{\kappa}_{1}=-\vec{\kappa}_{2}$ where
Eq.(63) becomes
\begin{equation}
Eq.(C63)=2{\frac{[(2m+2n+1)!!]^{2}}{(2(m+n+1))!}}[(n+m+1)!]^{2}{\frac{%
(\kappa ^{2}-(\vec{\kappa}\cdot \hat{\eta})^{2})}{\eta }}^{m+n+1}\sum ,
\label{c64}
\end{equation}
where the triple sum $\sum $ collapses to a single sum:
\[
\sum (cos\phi =1)\Rightarrow j=k=0,
\]
so that
\begin{equation}
\sum =\sum_{l=0}^{n}{\
%TCIMACRO{\binom{2n+1}{l}}%
%BeginExpansion
{2n+1 \choose l}%
%EndExpansion
}{%
%TCIMACRO{\binom{2m+1}{l+m-n}}%
%BeginExpansion
{2m+1 \choose l+m-n}%
%EndExpansion
}={\frac{1}{2}}{\frac{(2(m+n+1))!}{(n+m+1)!]^{2}}}.  \label{c65}
\end{equation}

\vfill\eject

\section{Derivation of the invariant mass $M$ using the new Dirac brackets}

Kerner has shown\cite{kerner} \ that it is possible to develope a
single-time Hamiltonian formulation of Wheeler-Feynman dynamics. His idea,
basically is to replace the infinitude of ``field'' coordinates by an
infinity of mechnical ones and then with the high order of the equations of
motion replaced by higher powers of the momentum in the interaction. \ In
Ref.\cite{yang}, Crater and Yang give a modification of his approach to
obtain a Hamiltonian expression for both scalar and vector interactions
through order $1/c^{4}$. \ \ The approach taken in this Appendix is similar
to that given in \cite{yang} with two important distinctions: 1) terms of
all order in $1/c^{2}$ are included and 2) the effects of the new Dirac
brackets are included. \ The net result is an expression for $M$ that agrees
exactly with the results obtained in Eq.(\ref{VI14}). \ This result would
not be obtained without the use of these brackets in working our Hamilton's
equations.

\bigskip We begin with the Lagrangian expression for the invariant mass [see
Eq.(\ref{IV10}) with $\vec \lambda (\tau )=0$; $h_1$ is defined in Eq.(\ref
{VI4})]

\begin{eqnarray}
E_{rel} &=&h(\dot{\vec{\eta}_{1}},\dot{\vec{\eta}_{2}},\vec{\eta})={\frac{%
m_{1}}{\sqrt{1-\dot{\vec{\eta}_{1}}^{2}}}}+{\frac{m_{2}}{\sqrt{1-\dot{\vec{%
\eta}_{2}}^{2}}}}+{\frac{Q_{1}Q_{2}}{4\pi |\vec{\eta}|}}+Q_{1}Q_{2}h_{1}(%
\dot{\vec{\eta}_{1}},\dot{\vec{\eta}_{2}},\vec{\eta}):=  \nonumber \\
&:&=h_{0}(\dot{\vec{\eta}_{1}},\dot{\vec{\eta}_{2}},\vec{\eta}%
)+Q_{1}Q_{2}h_{1}(\dot{\vec{\eta}_{1}},\dot{\vec{\eta}_{2}},\vec{\eta}).
\label{d1}
\end{eqnarray}
In order to find the Hamiltonian $H(\vec{\kappa}_{1},\vec{\kappa}_{2};\vec{%
\eta})$ from $h_{1}(\dot{\vec{\eta}_{1}},\dot{\vec{\eta}_{2}},\vec{\eta})$
we must demand that Hamilton's equation be satisfied .\ We use the Dirac
bracket since we have used the constraint as a strong condition on the
dynamical variables. \ Thus we begin with

\begin{eqnarray}
\dot{\vec{\eta}_{i}} &=&\{\vec{\eta}_{i},H\}^{\ast }=\{\vec{\eta}_{i},H\}-
\nonumber \\
&&-[\int d^{3}\sigma \{\vec{\eta}_{i},-\sum_{i}Q_{i}\vec{A}_{\perp Si}(\tau ,%
\vec{\sigma}-\vec{\eta}_{i},{\vec{\kappa}}_{i})\}\cdot \{-\sum_{j}Q_{j}\vec{%
\Pi}_{\perp Sj}(\tau ,\vec{\sigma}-\vec{\eta}_{j},{\vec{\kappa}}_{j}),H\}-
\nonumber \\
&&-\{\vec{\eta}_{i},-\sum_{j}Q_{j}\vec{\Pi}_{\perp Sj}(\tau ,\vec{\sigma}-%
\vec{\eta}_{j},{\vec{\kappa}}_{j})\}\cdot \{-\sum_{i}Q_{i}\vec{A}_{\perp
Si}(\tau ,\vec{\sigma}-\vec{\eta}_{i},{\vec{\kappa}}_{i}),H\}],  \label{d2}
\end{eqnarray}
in which
\begin{equation}
H=H_{0}+Q_{1}Q_{2}H_{1}(\vec{\kappa}_{1},\vec{\kappa}_{2};\vec{\eta}),
\label{d3}
\end{equation}
and
\begin{equation}
H_{0}=\sqrt{m_{1}^{2}+{\vec{\kappa}_{1}}^{2}}+\sqrt{m_{2}^{2}+{\vec{\kappa}%
_{2}}^{2}}.  \label{d4}
\end{equation}
Substituting this Hamiltonian into the above bracket and using Grassmann
truncation yields
\begin{eqnarray}
\dot{\vec{\eta}_{1}} &=&{\frac{\vec{\kappa}_{1}}{\sqrt{m_{1}^{2}+{\vec{\kappa%
}_{1}}^{2}}}}+Q_{1}Q_{2}{\frac{\partial H_{1}}{\partial \vec{\kappa}_{1}}}-
\nonumber \\
&&-[\int d^{3}\sigma \{\vec{\eta}_{1},-Q_{1}\vec{A}_{\perp S1}(\tau ,\vec{%
\sigma}-\vec{\eta}_{1},{\vec{\kappa}}_{1})\}\cdot \{-Q_{2}\vec{\Pi}_{\perp
S2}(\tau ,\vec{\sigma}-\vec{\eta}_{2},{\vec{\kappa}}_{2}),\sqrt{m_{2}^{2}+{%
\vec{\kappa}_{2}}^{2}}\}-  \nonumber \\
&&-\{\vec{\eta}_{1},-Q_{1}\vec{\Pi}_{\perp S1}(\tau ,\vec{\sigma}-\vec{\eta}%
_{1},{\vec{\kappa}}_{1})\}\cdot \{-Q_{2}\vec{A}_{\perp S2}(\tau ,\vec{\sigma}%
-\vec{\eta}_{2},{\vec{\kappa}}_{2}),\sqrt{m_{2}^{2}+{\vec{\kappa}_{2}}^{2}}%
\}=  \nonumber \\
&=&{\frac{\vec{\kappa}_{1}}{\sqrt{m_{1}^{2}+{\vec{\kappa}_{1}}^{2}}}}%
+Q_{1}Q_{2}{\frac{\partial H_{1}}{\partial \vec{\kappa}_{1}}-}\frac{\partial
}{\partial \vec{\kappa}_{1}}\left( \frac{\vec{\kappa}_{2}}{\sqrt{m_{2}^{2}+{%
\vec{\kappa}_{2}}^{2}}}\cdot \frac{\partial }{\partial \vec{\eta}_{2}}%
\right) {\cal K}_{12},  \label{d5}
\end{eqnarray}
\begin{eqnarray}
\dot{\vec{\eta}_{2}} &=&{\frac{\vec{\kappa}_{2}}{\sqrt{m_{2}^{2}+{\vec{\kappa%
}_{2}}^{2}}}}+Q_{1}Q_{2}{\frac{\partial H_{1}}{\partial \vec{\kappa}_{2}}}-
\nonumber \\
&&-[\int d^{3}\sigma \{\vec{\eta}_{2},-Q_{2}\vec{A}_{\perp S2}(\tau ,\vec{%
\sigma}-\vec{\eta}_{2},{\vec{\kappa}}_{2})\}\cdot \{-Q_{1}\vec{\Pi}_{\perp
S1}(\tau ,\vec{\sigma}-\vec{\eta}_{1},{\vec{\kappa}}_{1}),\sqrt{m_{1}^{2}+{%
\vec{\kappa}_{1}}^{2}}\}-  \nonumber \\
&&-\{\vec{\eta}_{2},-Q_{2}\vec{\Pi}_{\perp S2}(\tau ,\vec{\sigma}-\vec{\eta}%
_{2},{\vec{\kappa}}_{2})\}\cdot \{-Q_{1}\vec{A}_{\perp S1}(\tau ,\vec{\sigma}%
-\vec{\eta}_{1},{\vec{\kappa}}_{1}),\sqrt{m_{1}^{2}+{\vec{\kappa}_{1}}^{2}}%
\}=  \nonumber \\
&=&{\frac{\vec{\kappa}_{2}}{\sqrt{m_{2}^{2}+{\vec{\kappa}_{2}}^{2}}}}%
+Q_{1}Q_{2}{\frac{\partial H_{1}}{\partial \vec{\kappa}_{2}}+}\frac{\partial
}{\partial \vec{\kappa}_{2}}\left( \frac{\vec{\kappa}_{1}}{\sqrt{m_{1}^{2}+{%
\vec{\kappa}_{1}}^{2}}}\cdot \frac{\partial }{\partial \vec{\eta}_{2}}%
\right) {\cal K}_{12}.  \label{d6}
\end{eqnarray}
But
\begin{equation}
\int d^{3}\sigma (\vec{E}_{\perp S}\times \vec{B}_{S})(\tau ,\vec{\sigma})=%
\frac{\partial }{\partial \vec{\eta}_{2}}{\cal K}_{12}=-\frac{\partial }{%
\partial \vec{\eta}_{1}}{\cal K}_{12}:=-\frac{\partial }{\partial \vec{\eta}}%
{\cal K}_{12}.  \label{d7}
\end{equation}
Hence

\begin{eqnarray}
\dot{\vec{\eta}_{1}} &=&{\frac{\vec{\kappa}_{1}}{\sqrt{m_1^2+{\vec{\kappa}%
_{1}} ^{2}}}}+Q_{1}Q_{2}{\frac{\partial H_{1}}{\partial \vec{\kappa}_{1}} -}%
\frac{\partial }{\partial \vec{\kappa}_{1}}\left( \frac{\vec{\kappa}_{2}}{%
\sqrt{m_2^2+{\vec{\kappa}_{2}}^{2}}}\cdot \int d^{3}\sigma (\vec{E}_{\perp
S}\times \vec{B}_S)(\tau ,\vec \sigma )\right) ,  \nonumber \\
\dot{\vec{\eta}_{2}} &=&{\frac{\vec{\kappa}_{2}}{\sqrt{m_2^2+{\vec{\kappa}%
_{2}} ^{2}}}}+Q_{1}Q_{2}{\frac{\partial H_{1}}{\partial \vec{\kappa}_{2}} -}%
\frac{\partial }{\partial \vec{\kappa}_{2}}\left( \frac{\vec{\kappa}_{1}}{%
\sqrt{m_1^2+{\vec{\kappa}_{1}}^{2}}}\cdot \int d^{3}\sigma (\vec{E}_{\perp
S}\times \vec{B}_S)(\tau ,\vec \sigma )\right) .  \label{d8}
\end{eqnarray}
Substituting this into $h_{0}$ we find that

\begin{eqnarray}
h_{0} &=&H_{0}+{\frac{Q_{1}Q_{2}}{4\pi \eta }}+  \nonumber \\
&&+{\frac{(m_1^2+\vec{\kappa} _{1}^{2})}{m_{1}^{2}}}\vec{\kappa}_{1}\cdot
\frac{\partial }{ \partial \vec{\kappa}_{1}}(Q_{1}Q_{2}H_{1}-\frac{\vec{%
\kappa}_{2}}{\sqrt{m_2^2+{\ \vec{\kappa}_{2}}^{2}}}\cdot \int d^{3}\sigma (%
\vec{E}_{\perp S}\times \vec{B}_S)(\tau ,\vec \sigma )+  \nonumber \\
&&+{\frac{(m_2^2+\vec{\kappa}_{2}^{2})}{m_{2}^{2}}}\vec{\kappa}_{2}\cdot
\frac{\partial }{\partial \vec{\kappa}_{2}}(Q_{1}Q_{2}H_{1}-\frac{\vec{%
\kappa }_{1}}{\sqrt{m_1^2+{\vec{\kappa}_{1}}^{2}}}\cdot \int d^{3}\sigma (%
\vec{E}_{\perp S} \times \vec{B}_S)(\tau ,\vec \sigma )=  \nonumber \\
&=&H-Q_{1}Q_{2}h_{1}=H_{0}+Q_{1}Q_{2}H_{1}-Q_{1}Q_{2}h_{1}.  \label{d9}
\end{eqnarray}
Thus letting
\begin{equation}
H_{1}={\frac{1}{4\pi \eta }}+\tilde{H}_{1},  \label{d10}
\end{equation}
and using [see Eqs.(\ref{VI4}) and (\ref{VI5}) for the expressions of $h_1$
and ${\vec h}_1$]

\begin{eqnarray}
h_{1}(\dot{\vec{\eta}_{1}},\dot{\vec{\eta}_{2}},\vec{\eta}) &=&h_{1}({\frac{%
\vec{\kappa}_{1}}{\sqrt{m_1^2+{\vec{\kappa}_{1}}^{2}}}},{\frac{\vec{ \kappa}%
_{2}}{\sqrt{m_2^2+{\vec{\kappa}_{2}}^{2}}}},\vec{\eta} )+O(Q_{1}Q_{2}),
\nonumber \\
\vec{h}_{1}(\dot{\vec{\eta}_{1}},\dot{\vec{\eta}_{2}},\vec{\eta}) &=&\vec{h}
_{1}({\frac{\vec{\kappa}_{1}}{\sqrt{m_1^2+{\vec{\kappa}_{1}}^{2}}}},{\ \frac{%
\vec{\kappa}_{2}}{\sqrt{m_2^2+{\vec{\kappa}_{2}}^{2}}}},\vec{\eta}
)+O(Q_{1}Q_{2}),  \label{d11}
\end{eqnarray}
where

\begin{equation}
\int d^{3}\sigma (\vec{E}_{\perp S}\times \vec{B}_{S})(\tau ,\vec{\sigma}%
):=Q_{1}Q_{2}\vec{h}_{1}(\dot{\vec{\eta}_{1}},\dot{\vec{\eta}_{2}},\vec{\eta}%
),  \label{d12}
\end{equation}
then we obtain the following differential equation for $\tilde{H}_{1}$
\begin{eqnarray}
&&\tilde{H}_{1}-{\frac{\partial \tilde{H}_{1}}{\partial \vec{\kappa}_{1}}}%
\cdot \vec{\kappa}_{1}{\frac{(m_{1}^{2}+\vec{\kappa}_{1}^{2})}{m_{1}^{2}}}-{%
\ \frac{\partial \tilde{H}_{1}}{\partial \vec{\kappa}_{2}}}\cdot \vec{\kappa}%
_{2}{\frac{(m_{2}^{2}+\vec{\kappa}_{2}^{2})}{m_{2}^{2}}}=  \nonumber \\
&=&h_{1}({\frac{\vec{\kappa}_{1}}{\sqrt{m_{1}^{2}+{\vec{\kappa}_{1}}^{2}}}},{%
\ \frac{\vec{\kappa}_{2}}{\sqrt{m_{2}^{2}+{\vec{\kappa}_{2}}^{2}}}},\vec{\eta%
})-  \nonumber \\
&-&{\frac{(m_{1}^{2}+\vec{\kappa}_{1}^{2})}{m_{1}^{2}}}\vec{\kappa}_{1}\cdot
\frac{\partial }{\partial \vec{\kappa}_{1}}(\frac{\vec{\kappa}_{2}}{\sqrt{%
m_{2}^{2}+{\vec{\kappa}_{2}}^{2}}}\cdot \vec{h}_{1}({\frac{\vec{\kappa}_{1}}{%
\sqrt{m_{1}^{2}+{\vec{\kappa}_{1}}^{2}}}},{\frac{\vec{\kappa}_{2}}{\sqrt{%
m_{2}^{2}+{\vec{\kappa}_{2}}^{2}}}},\vec{\eta}))-  \nonumber \\
&-&{\frac{(m_{2}^{2}+\vec{\kappa}_{2}^{2})}{m_{2}^{2}}}\vec{\kappa}_{2}\cdot
\frac{\partial }{\partial \vec{\kappa}_{2}}(\frac{\vec{\kappa}_{1}}{\sqrt{%
m_{1}^{2}+{\vec{\kappa}_{1}}^{2}}}\cdot \vec{h}_{1}({\frac{\vec{\kappa}_{1}}{%
\sqrt{m_{1}^{2}+{\vec{\kappa}_{1}}^{2}}}},{\frac{\vec{\kappa}_{2}}{\sqrt{%
m_{2}^{2}+{\vec{\kappa}_{2}}^{2}}}},\vec{\eta})+O(Q_{1}Q_{2}).  \label{d13}
\end{eqnarray}
The $O(Q_{1}Q_{2})$ term gives a vanishing contribution as both sides are
multiplied by $Q_{1}Q_{2}$. Using this and ${\vec{\nabla}}_{\eta }^{2}|\vec{%
\eta}|^{l}=l(l-1)|\vec{\eta}|^{l-2}$ we obtain, in addition to the
expression given for $h_{1}$ in Eq.(\ref{VI12}),
\begin{eqnarray}
&&\int d^{3}\sigma ({\vec{E}}_{\perp S}\times {\vec{B}}_{S})(\tau ,\vec{%
\sigma})=Q_{1}Q_{2}\vec{h}_{1}({\frac{\vec{\kappa}_{1}}{\sqrt{m_{1}^{2}+{%
\vec{\kappa}_{1}}^{2}}}},{\frac{\vec{\kappa}_{2}}{\sqrt{m_{2}^{2}+{\vec{%
\kappa}_{2}}^{2}}}},\vec{\eta})=  \nonumber \\
&=&{\frac{Q_{1}Q_{2}}{4\pi }}\sum_{m=0}^{\infty }\sum_{n=0}^{\infty }\Big({%
\vec{\nabla}}_{\eta }\Big[{\frac{\vec{\kappa}_{1}}{\sqrt{m_{1}^{2}+{\vec{%
\kappa}_{1}}^{2}}}}\cdot {\frac{\vec{\kappa}_{2}}{\sqrt{m_{2}^{2}+{\vec{%
\kappa}_{2}}^{2}}}}\times   \nonumber \\
&&{\frac{\Big(({\frac{\vec{\kappa}_{1}}{\sqrt{m_{1}^{2}+{\vec{\kappa}_{1}}%
^{2}}}}\cdot {\vec{\nabla}}_{\eta })^{2m+1}({\frac{\vec{\kappa}_{2}}{\sqrt{%
m_{2}^{2}+{\vec{\kappa}_{2}}^{2}}}}\cdot {\vec{\nabla}}_{\eta })^{2n}+({%
\frac{\vec{\kappa}_{1}}{\sqrt{m_{1}^{2}+{\vec{\kappa}_{1}}^{2}}}}\cdot {\vec{%
\nabla}}_{\eta })^{2m}({\frac{\vec{\kappa}_{2}}{\sqrt{m_{2}^{2}+{\vec{\kappa}%
_{2}}^{2}}}}\cdot {\vec{\nabla}}_{\eta })^{2n+1}\Big)\,\eta ^{2n+2m+1}}{%
(2n+2m+2)!}}-  \nonumber \\
&-&{\frac{\Big(({\frac{{\vec{\kappa}_{1}}\sqrt{m_{1}^{2}+{\vec{\kappa}_{1}}%
^{2}}}{\cdot }{\vec{\nabla}}_{\eta })^{2m+2}({\frac{\vec{\kappa}_{2}}{\sqrt{%
m_{2}^{2}+{\vec{\kappa}_{2}}^{2}}}}\cdot {\vec{\nabla}}_{\eta })^{2n+1}+({%
\frac{\vec{\kappa}_{1}}{\sqrt{m_{1}^{2}+{\vec{\kappa}_{1}}^{2}}}}\cdot {\vec{%
\nabla}}_{\eta })^{2m+1}}({\frac{\vec{\kappa}_{2}}{\sqrt{m_{2}^{2}+{\vec{%
\kappa}_{2}}^{2}}}}\cdot {\vec{\nabla}}_{\eta })^{2n+2}\Big)\,\eta ^{2n+2m+3}%
}{(2n+2m+4)!}}]\Big]\Big).  \nonumber \\
&&{}  \label{d14}
\end{eqnarray}
The differential equation for $\tilde{H}_{1}$ is of the form
\begin{equation}
\tilde{H}_{1}-{\cal O}\tilde{H}_{1}=h_{1}({\frac{\vec{\kappa}_{1}}{\sqrt{%
m_{1}^{2}+{\vec{\kappa}_{1}}^{2}}}},{\frac{\vec{\kappa}_{2}}{\sqrt{m_{2}^{2}+%
{\vec{\kappa}_{2}}^{2}}}},\vec{\eta})-{\cal \vec{O}}\cdot \vec{h}_{1}({\
\frac{\vec{\kappa}_{1}}{\sqrt{m_{1}^{2}+{\vec{\kappa}_{1}}^{2}}}},{\frac{%
\vec{\kappa}_{2}}{\sqrt{m_{2}^{2}+{\vec{\kappa}_{2}}^{2}}}},\vec{\eta}),
\label{d15}
\end{equation}
where we define the linear operator ${\cal O}$ by
\begin{equation}
{\cal O}={\frac{(m_{1}^{2}+\vec{\kappa}_{1}^{2})}{m_{1}^{2}}}\vec{\kappa}%
_{1}\cdot {\frac{\partial }{\partial \vec{\kappa}_{1}}}+{\frac{(m_{2}^{2}+%
\vec{\kappa}_{2}^{2})}{m_{2}^{2}}}\vec{\kappa}_{2}\cdot {\frac{\partial }{%
\partial \vec{\kappa}_{2}}},  \label{d16}
\end{equation}
and the\ linear operator ${\cal \vec{O}}$ by
\begin{equation}
{\cal \vec{O}}={\frac{(m_{1}^{2}+\vec{\kappa}_{1}^{2})}{m_{1}^{2}}}\vec{%
\kappa}_{1}\cdot \frac{\partial }{\partial \vec{\kappa}_{1}}\frac{\vec{\kappa%
}_{2}}{\sqrt{m_{2}^{2}+{\vec{\kappa}_{2}}^{2}}}+{\frac{(m_{2}^{2}+\vec{\kappa%
}_{2}^{2})}{m_{2}^{2}}}\vec{\kappa}_{2}\cdot \frac{\partial }{\partial \vec{%
\kappa}_{2}}\frac{\vec{\kappa}_{1}}{\sqrt{m_{1}^{2}+{\vec{\kappa}_{1}}^{2}}}.
\label{d17}
\end{equation}
In order to solve for $\tilde{H}_{1}$ we need first to work out the right
hand side of its differential equation. \ First note
\begin{equation}
{\cal \vec{O}}{\frac{\vec{\kappa}_{1}}{\sqrt{m_{1}^{2}+{\vec{\kappa}_{1}}^{2}%
}}}\cdot {\frac{\vec{\kappa}_{2}}{\sqrt{m_{2}^{2}+{\vec{\kappa}_{2}}^{2}}}}=(%
{\frac{\vec{\kappa}_{1}}{\sqrt{m_{1}^{2}+{\vec{\kappa}_{1}}^{2}}}}+{\frac{%
\vec{\kappa}_{2}}{\sqrt{m_{2}^{2}+{\vec{\kappa}_{2}}^{2}}})\frac{\vec{\kappa}%
_{1}}{\sqrt{m_{1}^{2}+{\vec{\kappa}_{1}}^{2}}}}\cdot {\frac{\vec{\kappa}_{2}%
}{\sqrt{m_{2}^{2}+{\vec{\kappa}_{2}}^{2}}}},  \label{d18}
\end{equation}
while
\begin{eqnarray}
&&{\cal \vec{O}}({\frac{\vec{\kappa}_{1}}{\sqrt{m_{1}^{2}+{\vec{\kappa}_{1}}%
^{2}}}}\cdot {\vec{\nabla}}_{\eta })^{2m+1}({\frac{\vec{\kappa}_{2}}{\sqrt{%
m_{2}^{2}+{\vec{\kappa}_{2}}^{2}}}}\cdot {\vec{\nabla}}_{\eta })^{2n+1}=
\nonumber \\
&=&[(2n+1){\frac{\vec{\kappa}_{1}}{\sqrt{m_{1}^{2}+{\vec{\kappa}_{1}}^{2}}}}%
+(2m+1){\frac{\vec{\kappa}_{2}}{\sqrt{m_{2}^{2}+{\vec{\kappa}_{2}}^{2}}}]}%
\times   \nonumber \\
&&({\ \frac{\vec{\kappa}_{1}}{\sqrt{m_{1}^{2}+{\vec{\kappa}_{1}}^{2}}}}\cdot
{\vec{\nabla}}_{\eta })^{2m+1}({\frac{\vec{\kappa}_{2}}{\sqrt{m_{2}^{2}+{%
\vec{\kappa}_{2}}^{2}}}}\cdot {\vec{\nabla}}_{\eta })^{2n+1}.  \label{d19}
\end{eqnarray}
So Eq.(\ref{VI5}) imply
\begin{eqnarray}
&&{\cal \vec{O}}\cdot Q_{1}Q_{2}\vec{h}_{1}({\frac{\vec{\kappa}_{1}}{\sqrt{%
m_{1}^{2}+{\vec{\kappa}_{1}}^{2}}}},{\frac{\vec{\kappa}_{2}}{\sqrt{m_{2}^{2}+%
{\vec{\kappa}_{2}}^{2}}}},\vec{\eta})={\frac{Q_{1}Q_{2}}{4\pi }}%
\sum_{m=0}^{\infty }\sum_{n=0}^{\infty }\Big(\Big[{(2n+1)}({\frac{\vec{\kappa%
}_{1}}{\sqrt{m_{1}^{2}+{\vec{\kappa}_{1}}^{2}}}}\cdot {\vec{\nabla}}_{\eta
})^{2}+  \nonumber \\
&&+(2n+2m+4)({\frac{\vec{\kappa}_{1}}{\sqrt{m_{1}^{2}+{\vec{\kappa}_{1}}^{2}}%
}}\cdot {\vec{\nabla}}_{\eta })({\frac{\vec{\kappa}_{2}}{\sqrt{m_{2}^{2}+{%
\vec{\kappa}_{2}}^{2}}}}\cdot {\vec{\nabla}}_{\eta })+(2m+1)({\frac{\vec{%
\kappa}_{2}}{\sqrt{m_{2}^{2}+{\vec{\kappa}_{2}}^{2}}}}\cdot {\vec{\nabla}}%
_{\eta })^{2}\Big]\times   \nonumber \\
&&\times \Big[{\frac{\vec{\kappa}_{1}}{\sqrt{m_{1}^{2}+{\vec{\kappa}_{1}}^{2}%
}}}\cdot {\frac{\vec{\kappa}_{2}}{\sqrt{m_{2}^{2}+{\vec{\kappa}_{2}}^{2}}}}%
\frac{({\frac{\vec{\kappa}_{1}}{\sqrt{m_{1}^{2}+{\vec{\kappa}_{1}}^{2}}}}%
\cdot {\vec{\nabla}}_{\eta })^{2m}({\frac{\vec{\kappa}_{2}}{\sqrt{m_{2}^{2}+{%
\vec{\kappa}_{2}}^{2}}}}\cdot {\vec{\nabla}}_{\eta })^{2n}}{(2n+2m+2)!}\eta
^{2n+2m+1}-  \nonumber \\
&&-\frac{({\frac{\vec{\kappa}_{1}}{\sqrt{m_{1}^{2}+{\vec{\kappa}_{1}}^{2}}}}%
\cdot {\vec{\nabla}}_{\eta })^{2m+1}({\frac{\vec{\kappa}_{2}}{\sqrt{%
m_{2}^{2}+{\vec{\kappa}_{2}}^{2}}}}\cdot {\vec{\nabla}}_{\eta })^{2n+1}}{%
(2n+2m+4)!}\eta ^{2n+2m+3}\Big]\Big).  \label{d20}
\end{eqnarray}
In analogy to our decomposition of the field energy integral we decompose
this into single and double sum pieces giving
\begin{eqnarray}
&&{\cal \vec{O}\cdot }Q_{1}Q_{2}\vec{h}_{1}={\frac{Q_{1}Q_{2}}{4\pi }%
\sum_{m=1}^{\infty }{\frac{1}{(2m)!}}[{\frac{\vec{\kappa}_{2}}{\sqrt{%
m_{2}^{2}+{\vec{\kappa}_{2}}^{2}}}}\cdot \frac{\vec{\kappa}_{1}}{\sqrt{%
m_{1}^{2}+{\vec{\kappa}_{1}}^{2}}}}\times   \nonumber \\
&&\left( ({\frac{\vec{\kappa}_{1}}{\sqrt{m_{1}^{2}+{\vec{\kappa}_{1}}^{2}}}}%
\cdot {\vec{\nabla}}_{\eta })^{2m}+({\frac{\vec{\kappa}_{2}}{\sqrt{m_{2}^{2}+%
{\vec{\kappa}_{2}}^{2}}}}\cdot {\vec{\nabla}}_{\eta })^{2m}\right) )\eta
^{2m-1}-  \nonumber \\
&&-{\frac{1}{(2m+2)!}}\Big({\frac{\vec{\kappa}_{2}}{\sqrt{m_{2}^{2}+{\vec{%
\kappa}_{2}}^{2}}}\cdot }{\vec{\nabla}}_{\eta }({\frac{\vec{\kappa}_{1}}{%
\sqrt{m_{1}^{2}+{\vec{\kappa}_{1}}^{2}}}}\cdot {\vec{\nabla}}_{\eta
})^{2m+1}+  \nonumber \\
&&+{\frac{\vec{\kappa}_{1}}{\sqrt{m_{1}^{2}+{\ \vec{\kappa}_{1}}^{2}}}\cdot }%
{\vec{\nabla}}_{\eta }({\frac{\vec{\kappa}_{2}}{\sqrt{m_{2}^{2}+{\vec{\kappa}%
_{2}}^{2}}}}\cdot {\vec{\nabla}}_{\eta })^{2m+1}\Big)\eta ^{2m+1}]+
\nonumber \\
&&+\sum_{m=0}^{\infty }\sum_{n=0}^{\infty }\big((2n+2m+4)[{\frac{\vec{\kappa}%
_{1}}{\sqrt{m_{1}^{2}+{\vec{\kappa}_{1}}^{2}}}}\cdot {\frac{\vec{\kappa}_{2}%
}{\sqrt{m_{2}^{2}+{\vec{\kappa}_{2}}^{2}}}}  \nonumber \\
&&\frac{({\frac{\vec{\kappa}_{1}}{\sqrt{m_{1}^{2}+{\vec{\kappa}_{1}}^{2}}}}%
\cdot {\vec{\nabla}}_{\eta })^{2m+1}({\frac{\vec{\kappa}_{2}}{\sqrt{%
m_{2}^{2}+{\vec{\kappa}_{2}}^{2}}}}\cdot {\vec{\nabla}}_{\eta })^{2n+1}}{%
(2n+2m+2)!}\eta ^{2n+2m+1}-  \nonumber \\
&&-\frac{({\frac{\vec{\kappa}_{1}}{\sqrt{m_{1}^{2}+{\vec{\kappa}_{1}}^{2}}}}%
\cdot {\vec{\nabla}}_{\eta })^{2m+2}({\frac{\vec{\kappa}_{2}}{\sqrt{%
m_{2}^{2}+{\vec{\kappa}_{2}}^{2}}}}\cdot {\vec{\nabla}}_{\eta })^{2n+2}}{%
(2n+2m+4)!}\eta ^{2n+2m+3}]+  \nonumber \\
&&+(2n+2m+6)[{\frac{\vec{\kappa}_{1}}{\sqrt{m_{1}^{2}+{\vec{\kappa}_{1}}^{2}}%
}}\cdot {\frac{\vec{\kappa}_{2}}{\sqrt{m_{2}^{2}+{\vec{\kappa}_{2}}^{2}}}}%
\frac{({\frac{\vec{\kappa}_{1}}{\sqrt{m_{1}^{2}+{\vec{\kappa}_{1}}^{2}}}}%
\cdot {\vec{\nabla}}_{\eta })^{2m+2}({\frac{\vec{\kappa}_{2}}{\sqrt{%
m_{2}^{2}+{\vec{\kappa}_{2}}^{2}}}}\cdot {\vec{\nabla}}_{\eta })^{2n+2}}{%
(2n+2m+4)!}\eta ^{2n+2m+3}-  \nonumber \\
&&-\frac{({\frac{\vec{\kappa}_{1}}{\sqrt{m_{1}^{2}+{\vec{\kappa}_{1}}^{2}}}}%
\cdot {\vec{\nabla}}_{\eta })^{2m+3}({\frac{\vec{\kappa}_{2}}{\sqrt{%
m_{2}^{2}+{\vec{\kappa}_{2}}^{2}}}}\cdot {\vec{\nabla}}_{\eta })^{2n+3}}{%
(2n+2m+6)!}\eta ^{2n+2m+5}]\big)\}.  \label{d21}
\end{eqnarray}
Combining like terms we obtain
\begin{eqnarray}
&&h_{1}-{\cal \vec{O}}\cdot \vec{h}_{1}={\frac{Q_{1}Q_{2}}{4\pi }\{}\left( {%
\frac{\vec{\kappa}_{1}}{\sqrt{m_{1}^{2}+{\vec{\kappa}_{1}}^{2}}}}\cdot {\
\frac{\vec{\kappa}_{2}}{\sqrt{m_{2}^{2}+{\vec{\kappa}_{2}}^{2}}}\frac{1}{%
\eta }-}({\frac{\vec{\kappa}_{1}}{\sqrt{m_{1}^{2}+{\vec{\kappa}_{1}}^{2}}}}%
\cdot {\vec{\nabla}}_{\eta })({\frac{\vec{\kappa}_{2}}{\sqrt{m_{2}^{2}+{\vec{%
\kappa}_{2}}^{2}}}}\cdot \nabla )\frac{\eta }{2}\right) +  \nonumber \\
&&+\sum_{m=0}^{\infty }\sum_{n=0}^{\infty }\big(-(2n+2m+3)[{\frac{\vec{\kappa%
}_{1}}{\sqrt{m_{1}^{2}+{\vec{\kappa}_{1}}^{2}}}}\cdot {\frac{\vec{\kappa}_{2}%
}{\sqrt{m_{2}^{2}+{\vec{\kappa}_{2}}^{2}}}}\times   \nonumber \\
&&\frac{({\frac{\vec{\kappa}_{1}}{\sqrt{m_{1}^{2}+{\vec{\kappa}_{1}}^{2}}}}%
\cdot {\vec{\nabla}}_{\eta })^{2m+1}({\frac{\vec{\kappa}_{2}}{\sqrt{%
m_{2}^{2}+{\vec{\kappa}_{2}}^{2}}}}\cdot {\vec{\nabla}}_{\eta })^{2n+1}}{%
(2n+2m+2)!}\eta ^{2n+2m+1}-  \nonumber \\
&&-\frac{({\frac{\vec{\kappa}_{1}}{\sqrt{m_{1}^{2}+{\vec{\kappa}_{1}}^{2}}}}%
\cdot {\vec{\nabla}}_{\eta })^{2m+2}({\frac{\vec{\kappa}_{2}}{\sqrt{%
m_{2}^{2}+{\vec{\kappa}_{2}}^{2}}}}\cdot {\vec{\nabla}}_{\eta })^{2n+2}}{%
(2n+2m+4)!}\eta ^{2n+2m+3}]-  \nonumber \\
&&-(2n+2m+5)[{\frac{\vec{\kappa}_{1}}{\sqrt{m_{1}^{2}+{\vec{\kappa}_{1}}^{2}}%
}}\cdot {\frac{\vec{\kappa}_{2}}{\sqrt{m_{2}^{2}+{\vec{\kappa}_{2}}^{2}}}}%
\frac{({\frac{\vec{\kappa}_{1}}{\sqrt{m_{1}^{2}+{\vec{\kappa}_{1}}^{2}}}}%
\cdot {\vec{\nabla}}_{\eta })^{2m+2}({\frac{\vec{\kappa}_{2}}{\sqrt{%
m_{2}^{2}+{\vec{\kappa}_{2}}^{2}}}}\cdot {\vec{\nabla}}_{\eta })^{2n+2}}{%
(2n+2m+4)!}\eta ^{2n+2m+3}-  \nonumber \\
&&-\frac{({\frac{\vec{\kappa}_{1}}{\sqrt{m_{1}^{2}+{\vec{\kappa}_{1}}^{2}}}}%
\cdot {\vec{\nabla}}_{\eta })^{2m+3}({\frac{\vec{\kappa}_{2}}{\sqrt{%
m_{2}^{2}+{\vec{\kappa}_{2}}^{2}}}}\cdot {\vec{\nabla}}_{\eta })^{2n+3}}{%
(2n+2m+6)!}\eta ^{2n+2m+5}]\big)\}.  \label{d22}
\end{eqnarray}
Based on the above expression we assume a particular solution of the form
\begin{eqnarray}
&&Q_{1}Q_{2}H_{1}(\vec{\kappa}_{1},\vec{\kappa}_{2},\vec{\eta})=k{\frac{%
Q_{1}Q_{2}}{4\pi }}\times   \nonumber \\
&&\left( {\frac{\vec{\kappa}_{1}}{\sqrt{m_{1}^{2}+{\vec{\kappa}_{1}}^{2}}}}%
\cdot {\frac{\vec{\kappa}_{2}}{\sqrt{m_{2}^{2}+{\vec{\kappa}_{2}}^{2}}}\frac{%
1}{\eta }-}({\frac{\vec{\kappa}_{1}}{\sqrt{m_{1}^{2}+{\vec{\kappa}_{1}}^{2}}}%
}\cdot {\vec{\nabla}}_{\eta })({\frac{\vec{\kappa}_{2}}{\sqrt{m_{2}^{2}+{%
\vec{\kappa}_{2}}^{2}}}}\cdot {\vec{\nabla}}_{\eta })\frac{\eta }{2}\right) +
\nonumber \\
&&+{\frac{Q_{1}Q_{2}}{4\pi }}\sum_{m=0}^{\infty }\sum_{n=0}^{\infty }[a_{mn}{%
\ \frac{\vec{\kappa}_{1}}{\sqrt{m_{1}^{2}+{\vec{\kappa}_{1}}^{2}}}}\cdot {\
\frac{\vec{\kappa}_{2}}{\sqrt{m_{2}^{2}+{\vec{\kappa}_{2}}^{2}}}}\times
\nonumber \\
&&{\frac{({\ \frac{\vec{\kappa}_{1}}{\sqrt{m_{1}^{2}+{\vec{\kappa}_{1}}^{2}}}%
}\cdot {\vec{\nabla}}_{\eta })^{2m+1}({\frac{\vec{\kappa}_{2}}{\sqrt{%
m_{2}^{2}+{\vec{\kappa}_{2}}^{2}}}}\cdot {\vec{\nabla}}_{\eta })^{2n+1}\eta
^{2n+2m+1}}{(2n+2m+2)!}}-  \nonumber \\
&&-b_{mn}{\frac{({\frac{\vec{\kappa}_{1}}{\sqrt{m_{1}^{2}+{\vec{\kappa}_{1}}%
^{2}}}}\cdot {\vec{\nabla}}_{\eta })^{2m+2}({\frac{\vec{\kappa}_{2}}{\sqrt{%
m_{2}^{2}+{\vec{\kappa}_{2}}^{2}}}}\cdot {\vec{\nabla}}_{\eta })^{2n+2}\eta
^{2n+2m+3}}{(2n+2m+4)!}}+  \nonumber \\
&&+c_{mn}{\frac{\vec{\kappa}_{1}}{\sqrt{m_{1}^{2}+{\vec{\kappa}_{1}}^{2}}}}%
\cdot {\frac{\vec{\kappa}_{2}}{\sqrt{m_{2}^{2}+{\vec{\kappa}_{2}}^{2}}}}{\
\frac{({\frac{\vec{\kappa}_{1}}{\sqrt{m_{1}^{2}+{\vec{\kappa}_{1}}^{2}}}}%
\cdot {\vec{\nabla}}_{\eta })^{2m+2}({\frac{\vec{\kappa}_{2}}{\sqrt{%
m_{2}^{2}+{\vec{\kappa}_{2}}^{2}}}}\cdot {\vec{\nabla}}_{\eta })^{2n+2}\eta
^{2n+2m+3}}{(2n+2m+4)!}}-  \nonumber \\
&&-d_{mn}{\frac{({\frac{\vec{\kappa}_{1}}{\sqrt{m_{1}^{2}+{\vec{\kappa}_{1}}%
^{2}}}}\cdot {\vec{\nabla}}_{\eta })^{2m+3}({\frac{\vec{\kappa}_{2}}{\sqrt{%
m_{2}^{2}+{\vec{\kappa}_{2}}^{2}}}}\cdot {\vec{\nabla}}_{\eta })^{2n+3}\eta
^{2n+2m+5}}{(2n+2m+6)!}}].  \label{d23}
\end{eqnarray}
Using
\begin{equation}
{\cal O}{\frac{\vec{\kappa}_{1}}{\sqrt{m_{1}^{2}+{\vec{\kappa}_{1}}^{2}}}}%
\cdot {\frac{\vec{\kappa}_{2}}{\sqrt{m_{2}^{2}+{\vec{\kappa}_{2}}^{2}}}}=2{\
\frac{\vec{\kappa}_{1}}{\sqrt{m_{1}^{2}+{\vec{\kappa}_{1}}^{2}}}}\cdot {\
\frac{\vec{\kappa}_{2}}{\sqrt{m_{2}^{2}+{\vec{\kappa}_{2}}^{2}}}},
\label{d24}
\end{equation}
and
\begin{eqnarray}
&&{\cal O}({\frac{\vec{\kappa}_{1}}{\sqrt{m_{1}^{2}+{\vec{\kappa}_{1}}^{2}}}}%
\cdot {\vec{\nabla}}_{\eta })^{2m+1}({\frac{\vec{\kappa}_{2}}{\sqrt{%
m_{2}^{2}+{\vec{\kappa}_{2}}^{2}}}}\cdot \nabla )^{2n+1}=  \nonumber \\
&=&(2n+2m+2)({\frac{\vec{\kappa}_{1}}{\sqrt{m_{1}^{2}+{\vec{\kappa}_{1}}^{2}}%
}}\cdot {\vec{\nabla}}_{\eta })^{2m+1}({\frac{\vec{\kappa}_{2}}{\sqrt{%
m_{2}^{2}+{\vec{\kappa}_{2}}^{2}}}}\cdot {\vec{\nabla}}_{\eta })^{2n+1},
\label{d25}
\end{eqnarray}
the left \ hand side of the differential equation for $\tilde{H}_{1}(\vec{%
\kappa}_{1},\vec{\kappa}_{2},\vec{\eta})$ becomes
\begin{eqnarray}
&&(1-{\cal O})Q_{1}Q_{2}\tilde{H}_{1}(\vec{\kappa}_{1},\vec{\kappa}_{2},\vec{%
\eta})=-k{\frac{Q_{1}Q_{2}}{8\pi }}\times   \nonumber \\
&&\left( {\frac{\vec{\kappa}_{1}}{\sqrt{m_{1}^{2}+{\ \vec{\kappa}_{1}}^{2}}}}%
\cdot {\frac{\vec{\kappa}_{2}}{\sqrt{m_{2}^{2}+{\vec{\kappa}_{2}}^{2}}}\frac{%
1}{\eta }-}({\frac{\vec{\kappa}_{1}}{\sqrt{m_{1}^{2}+{\vec{\kappa}_{1}}^{2}}}%
}\cdot {\vec{\nabla}}_{\eta })({\frac{\vec{\kappa}_{2}}{\sqrt{m_{2}^{2}+{%
\vec{\kappa}_{2}}^{2}}}}\cdot {\vec{\nabla}}_{\eta })\eta \right) =
\nonumber \\
&=&{\frac{Q_{1}Q_{2}}{4\pi }}\sum_{m=0}^{\infty }\sum_{n=0}^{\infty
}[-a_{mn}(2n+2m+3){\frac{\vec{\kappa}_{1}}{\sqrt{m_{1}^{2}+{\vec{\kappa}_{1}}%
^{2}}}}\cdot {\frac{\vec{\kappa}_{2}}{\sqrt{m_{2}^{2}+{\vec{\kappa}_{2}}^{2}}%
}}\times   \nonumber \\
&&{\frac{({\frac{\vec{\kappa}_{1}}{\sqrt{m_{1}^{2}+{\vec{\kappa}_{1}}^{2}}}}%
\cdot {\vec{\nabla}}_{\eta })^{2m+1}({\frac{\vec{\kappa}_{2}}{\sqrt{%
m_{2}^{2}+{\vec{\kappa}_{2}}^{2}}}}\cdot {\vec{\nabla}}_{\eta })^{2n+1}\eta
^{2n+2m+1}}{(2n+2m+2)!}}+  \nonumber \\
&&+b_{mn}(2n+2m+3){\frac{({\frac{\vec{\kappa}_{1}}{\sqrt{m_{1}^{2}+{\vec{%
\kappa}_{1}}^{2}}}}\cdot {\vec{\nabla}}_{\eta })^{2m+2}({\frac{\vec{\kappa}%
_{2}}{\sqrt{m_{2}^{2}+{\vec{\kappa}_{2}}^{2}}}}\cdot {\vec{\nabla}}_{\eta
})^{2n+2}\eta ^{2n+2m+3}}{(2n+2m+4)!}}-  \nonumber \\
&&-c_{mn}(2n+2m+5){\frac{\vec{\kappa}_{1}}{\sqrt{m_{1}^{2}+{\vec{\kappa}_{1}}%
^{2}}}}\cdot {\frac{\vec{\kappa}_{2}}{\sqrt{m_{2}^{2}+{\vec{\kappa}_{2}}^{2}}%
}}\times   \nonumber \\
&&{\frac{({\frac{\vec{\kappa}_{1}}{\sqrt{m_{1}^{2}+{\vec{\kappa}_{1}}^{2}}}}%
\cdot {\vec{\nabla}}_{\eta })^{2m+2}({\frac{\vec{\kappa}_{2}}{\sqrt{%
m_{2}^{2}+{\vec{\kappa}_{2}}^{2}}}}\cdot {\vec{\nabla}}_{\eta })^{2n+2}\eta
^{2n+2m+3}}{(2n+2m+4)!}}+  \nonumber \\
&&+d_{mn}(2n+2m+5)({\frac{({\frac{\vec{\kappa}_{1}}{\sqrt{m_{1}^{2}+{\vec{%
\kappa}_{1}}^{2}}}}\cdot {\vec{\nabla}}_{\eta })^{2m+3}({\frac{\vec{\kappa}%
_{2}}{\sqrt{m_{2}^{2}+{\vec{\kappa}_{2}}^{2}}}}\cdot {\vec{\nabla}}_{\eta
})^{2n+3}\eta ^{2n+2m+5}}{(2n+2m+6)!}}].  \label{d26}
\end{eqnarray}
Comparing the two sides of the equation leads to
\begin{equation}
k=-1;\,\,\,\,a_{mn}=b_{mn}=c_{mn}=d_{mn}=1,  \label{d27}
\end{equation}
and thus
\begin{eqnarray}
&&Q_{1}Q_{2}\tilde{H}_{1}(\vec{\kappa}_{1},\vec{\kappa}_{2},\vec{\eta})=-{\
\frac{Q_{1}Q_{2}}{4\pi }}\times   \nonumber \\
&&\left( {\frac{\vec{\kappa}_{1}}{\sqrt{m_{1}^{2}+{\vec{\kappa}_{1}}^{2}}}}%
\cdot {\frac{\vec{\kappa}_{2}}{\sqrt{m_{2}^{2}+{\vec{\kappa}_{2}}^{2}}}\frac{%
1}{\eta }-}({\frac{\vec{\kappa}_{1}}{\sqrt{m_{1}^{2}+{\vec{\kappa}_{1}}^{2}}}%
}\cdot {\vec{\nabla}}_{\eta })({\frac{\vec{\kappa}_{2}}{\sqrt{m_{2}^{2}+{%
\vec{\kappa}_{2}}^{2}}}}\cdot {\vec{\nabla}}_{\eta })\frac{\eta }{2}\right) +
\nonumber \\
&&+{\frac{Q_{1}Q_{2}}{4\pi }}\sum_{m=0}^{\infty }\sum_{n=0}^{\infty }[{\frac{%
\vec{\kappa}_{1}}{\sqrt{m_{1}^{2}+{\vec{\kappa}_{1}}^{2}}}}\cdot {\frac{\vec{%
\kappa}_{2}}{\sqrt{m_{2}^{2}+{\vec{\kappa}_{2}}^{2}}}}{\frac{({\frac{\vec{%
\kappa}_{1}}{\sqrt{m_{1}^{2}+{\vec{\kappa}_{1}}^{2}}}}\cdot {\vec{\nabla}}%
_{\eta })^{2m+1}({\ \frac{\vec{\kappa}_{2}}{\sqrt{m_{2}^{2}+{\vec{\kappa}_{2}%
}^{2}}}}\cdot {\vec{\nabla}}_{\eta })^{2n+1}\eta ^{2n+2m+1}}{(2n+2m+2)!}}-
\nonumber \\
&&-{\frac{({\frac{\vec{\kappa}_{1}}{\sqrt{m_{1}^{2}+{\vec{\kappa}_{1}}^{2}}}}%
\cdot {\vec{\nabla}}_{\eta })^{2m+2}({\frac{\vec{\kappa}_{2}}{\sqrt{%
m_{2}^{2}+{\vec{\kappa}_{2}}^{2}}}}\cdot {\vec{\nabla}}_{\eta })^{2n+2}\eta
^{2n+2m+3}}{(2n+2m+4)!}}+  \nonumber \\
&&+{\frac{\vec{\kappa}_{1}}{\sqrt{m_{1}^{2}+{\vec{\kappa}_{1}}^{2}}}}\cdot {%
\frac{\vec{\kappa}_{2}}{\sqrt{m_{2}^{2}+{\vec{\kappa}_{2}}^{2}}}}{\frac{({\
\frac{\vec{\kappa}_{1}}{\sqrt{m_{1}^{2}+{\vec{\kappa}_{1}}^{2}}}}\cdot {\vec{%
\nabla}}_{\eta })^{2m+2}({\frac{\vec{\kappa}_{2}}{\sqrt{m_{2}^{2}+{\vec{%
\kappa}_{2}}^{2}}}}\cdot {\vec{\nabla}}_{\eta })^{2n+2}\eta ^{2n+2m+3}}{%
(2n+2m+4)!}}-  \nonumber \\
&&-{\frac{({\frac{\vec{\kappa}_{1}}{\sqrt{m_{1}^{2}+{\vec{\kappa}_{1}}^{2}}}}%
\cdot {\vec{\nabla}}_{\eta })^{2m+3}({\frac{\vec{\kappa}_{2}}{\sqrt{%
m_{2}^{2}+{\vec{\kappa}_{2}}^{2}}}}\cdot {\vec{\nabla}}_{\eta })^{2n+3}\eta
^{2n+2m+5}}{(2n+2m+6)!}}],  \label{d28}
\end{eqnarray}
which agrees exactly with the Darwin portion of $M$ obtained earlier Eq.(\ref
{VI14}).

\vfill\eject

\section{Schild-Like Solution for the Two-Body Problem in the Case of Equal
Masses}

\bigskip

Here we present the semiclassical Hamilton equations for the equal mass case
restricted to circular orbits (we suppress the tilde notation). \ First note
that the Hamiltonian Eq.(\ref{VI37})
\begin{eqnarray}
&&H=2\sqrt{m^{2}+\vec{\kappa}^{2}}+\frac{Q_{1}Q_{2}}{4\pi \eta }+{\frac{%
Q_{1}Q_{2}}{8\pi \eta }}\times   \nonumber \\
&&\frac{[m^{2}(3\vec{\kappa}^{2}+(\vec{\kappa}\cdot \hat{\eta})^{2}]-2\vec{%
\kappa}^{2}[\vec{\kappa}^{2}-3(\vec{\kappa}\cdot \hat{\eta})^{2}]\sqrt{\frac{%
m^{2}+\vec{\kappa}^{2}}{m^{2}+(\vec{\kappa}\cdot \hat{\eta})^{2}}}-2[\vec{%
\kappa}^{2}+(\vec{\kappa}\cdot \hat{\eta})^{2}][m^{2}+(\vec{\kappa}\cdot
\hat{\eta})^{2}]}{(m^{2}+\vec{\kappa}^{2})[m^{2}+(\vec{\kappa}\cdot \hat{\eta%
})^{2}]}.  \label{e1}
\end{eqnarray}
is of the form
\begin{equation}
H=2\sqrt{m^{2}+\vec{\kappa}^{2}}+\frac{Q_{1}Q_{2}}{4\pi \eta }+{\frac{%
Q_{1}Q_{2}}{8\pi \eta }}f({\vec{\kappa}}^{2},(\vec{\kappa}\cdot \hat{\eta}%
)^{2}).  \label{e2}
\end{equation}
Thus, Hamilton's equations are [$f_{,1}=\partial f/\partial {\vec{\kappa}}%
^{2}$, $f_{,2}=\partial f/\partial (\vec{\kappa}\cdot \hat{\eta})^{2}$]
\begin{eqnarray}
-{\dot{\vec{\kappa}}}\,\stackrel{\circ }{=}\frac{\partial H}{\partial \vec{%
\eta}} &=&-\frac{Q_{1}Q_{2}}{4\pi \eta ^{3}}\vec{\eta}-\frac{Q_{1}Q_{2}}{%
8\pi \eta ^{3}}\vec{\eta}f({\vec{\kappa}}^{2},(\vec{\kappa}\cdot \hat{\eta}%
)^{2})+\frac{Q_{1}Q_{2}}{4\pi \eta }f,_{2}({\vec{\kappa}}^{2},(\vec{\kappa}%
\cdot \hat{\eta})^{2})\,\vec{\kappa}(\vec{\kappa}\cdot \hat{\eta}),
\nonumber \\
\dot{\vec{\eta}}\, &&\stackrel{\circ }{=}\frac{\partial H}{\partial \vec{%
\kappa}}=\frac{2\vec{\kappa}}{\sqrt{\vec{\kappa}^{2}+m^{2}}}+\frac{Q_{1}Q_{2}%
}{4\pi \eta }[f,_{1}({\vec{\kappa}}^{2},(\vec{\kappa}\cdot \hat{\eta})^{2})\,%
\vec{\kappa}+f,_{2}({\vec{\kappa}}^{2},(\vec{\kappa}\cdot \hat{\eta})^{2})\,%
\vec{\eta}(\vec{\kappa}\cdot \hat{\eta})].  \label{e3}
\end{eqnarray}
From this we can see that circular orbits defined by $\dot{\vec{\eta}}\cdot
\vec{\eta}=0$ are implied by $\vec{\kappa}\cdot \vec{\eta}=0$. \ This
furthermore implies that $-\dot{\vec{\kappa}}\cdot \vec{\kappa}=0.$ \ Thus
not only is $\vec{\eta}^{2}=const.$ but also $\vec{\kappa}^{2}=const.$ \
Imposing these conditions on the above Hamilton equations we can simplify
our equations above to \
\begin{eqnarray}
-\dot{\vec{\kappa}}\, &\stackrel{\circ }{=}\,&\frac{\partial H}{\partial
\vec{\eta}}=-\frac{Q_{1}Q_{2}}{4\pi \eta ^{3}}\vec{\eta}-\frac{Q_{1}Q_{2}}{%
8\pi \eta ^{3}}\,\vec{\eta}f({\vec{\kappa}}^{2},0)=B\vec{\eta},  \nonumber \\
\dot{\vec{\eta}}\, &\stackrel{\circ }{=}\,&\frac{\partial H}{\partial \vec{%
\kappa}}=\frac{2\vec{\kappa}}{\sqrt{\vec{\kappa}^{2}+m^{2}}}+\frac{Q_{1}Q_{2}%
}{4\pi \eta }f,_{1}({\vec{\kappa}}^{2},0)\,\vec{\kappa}=A\vec{\kappa}.
\label{e4}
\end{eqnarray}
with $B$ and $A$ constants. \ Combine the two equations and we find
\begin{equation}
\dot{\vec{\kappa}}=\frac{\ddot{\vec{\eta}}}{A}=-B\vec{\eta},  \label{e5}
\end{equation}
so that
\begin{equation}
\ddot{\vec{\eta}}=-AB\vec{\eta}:=-\Omega ^{2}\vec{\eta},  \label{e6}
\end{equation}
with
\begin{eqnarray}
\Omega ^{2} &=&-\frac{Q_{1}Q_{2}}{4\pi \eta ^{3}\sqrt{\vec{\kappa}^{2}+m^{2}}%
}[1+f(\vec{\kappa}^{2},0)]=  \nonumber \\
&=&-\frac{Q_{1}Q_{2}}{4\pi \eta ^{3}\sqrt{\vec{\kappa}^{2}+m^{2}}}[1-\vec{%
\kappa}^{2}\frac{2\vec{\kappa}^{2}\sqrt{\vec{\kappa}^{2}+m^{2}}-m^{3}}{%
2m^{3}(m^{2}+\vec{\kappa}^{2})}].  \label{e7}
\end{eqnarray}
The frequency is defined by the initial data $\eta $ and $|\vec{\kappa}|$
and is real \ for $Q_{1}Q_{2}<0$ for $0\leq \vec{\kappa}^{2}\leq \vec{\kappa}%
_{\max }^{2}$ in which $\vec{\kappa}_{\max }^{2}$ is the value at which $%
\Omega ^{2}=0$. Let us remark that at the semiclassical level $Q_{i}$ are
Grassmann variables: therefore $\Omega \sim \sqrt{Q_{1}Q_{2}}$ is to be
interpreted as an even algebraic object satisfying $\Omega ^{4}=0$, $%
Q_{1}\Omega ^{2}=Q_{2}\Omega ^{2}=0$.

We also find
\begin{equation}
H=2\sqrt{\vec{\kappa}^{2}+m^{2}}[1-\frac{1}{4}\Omega ^{2}\vec{\eta}^{2}].
\label{e8}
\end{equation}
Our solution to Eq.(\ref{e6}) is
\begin{eqnarray}
\vec{\eta}(\tau ) &=&\vec{\alpha}\cos \Omega \tau +\vec{\beta}\sin \Omega
\tau =  \nonumber \\
&=&\vec{\alpha}(1-\frac{1}{2!}\Omega ^{2}\tau ^{2})+\vec{\beta}(\Omega \tau -%
\frac{1}{3!}\Omega ^{3}\tau ^{3}),  \label{e9}
\end{eqnarray}
and thus
\begin{equation}
\dot{\vec{\eta}}(\tau )=-\vec{\alpha}\Omega ^{2}\tau +\vec{\beta}(\Omega -%
\frac{1}{2}\Omega ^{3}\tau ^{2}).  \label{e10}
\end{equation}
But $\dot{\vec{\eta}}\cdot \vec{\eta}=0.$ \ This implies that
\[
(\vec{\beta}^{2}-\vec{\alpha}^{2})\Omega ^{2}\tau +\vec{\alpha}\cdot \vec{%
\beta}\Omega (1-2\Omega ^{2}\tau ^{2})=0,
\]
so \ that
\begin{equation}
\vec{\beta}^{2}=\vec{\alpha}^{2},\,\,\,\,\vec{\alpha}\cdot \vec{\beta}=0,
\label{e11}
\end{equation}
and therefore
\begin{equation}
\vec{\alpha}^{2}=\vec{\eta}^{2}.  \label{e12}
\end{equation}
Otherwise the vectors $\vec{\alpha}$ and $\vec{\beta}$ are arbitrary. \

\vfill\eject

\end{document}